\documentclass[twoside,fleqn]{report}

\title{Emerging Jordan forms, with applications to critical statistical models and conformal field theory}
\date{14 March 2024}
\author{Lawrence Liu}

\usepackage[margin=1in]{geometry}
\usepackage{amsmath}
\usepackage{amssymb}
\usepackage{amsthm}
\usepackage{newpxtext}
\usepackage[vvarbb]{newpxmath}
\usepackage{hyperref}
\usepackage[titletoc]{appendix}
\usepackage[style=numeric,sorting=none,maxbibnames=99]{biblatex}
\usepackage[asterism]{sectionbreak}
\usepackage{xcolor}
\usepackage{graphicx}

\DeclareCiteCommand{\citeauthor}
 {\boolfalse{citetracker}%
 \boolfalse{pagetracker}%
 \usebibmacro{prenote}}
 {\ifciteindex
 {\indexnames{labelname}}
 {}%
 \printtext[bibhyperref]{\printnames{labelname}}}
 {\multicitedelim}
 {\usebibmacro{postnote}}

\let\inodot\i
\newcommand{\overbar}[1]{\mkern 1.5mu\overline{\mkern-1.5mu#1\mkern-1.5mu}\mkern 1.5mu}
\newcommand{\munderbar}[1]{\mkern 1.5mu\underline{\mkern-1.5mu#1\mkern-1.5mu}\mkern 1.5mu}
\def\e{\mathrm e}
\def\i{\mathrm i}
\def\d{\mathrm d}
\DeclareMathOperator{\sgn}{sgn}
\DeclareMathOperator{\diag}{diag}
\DeclareMathOperator{\Span}{span}
\DeclareMathOperator{\tr}{tr}
\DeclareMathOperator{\rank}{rank}
\newtheorem{conjecture}{Conjecture}
\newtheorem{proposition}{Proposition}
\newtheorem{theorem}{Theorem}
\newtheorem{definition}{Definition}

\usepackage{tikz-cd}
\usetikzlibrary{calc}
\usepackage{booktabs}
\usepackage{multirow}
\usepackage{makecell}

\newcommand{\arcsize}{0.2}
\newcommand{\loopU}{\vcenter{\hbox{
\begin{tikzpicture}
\draw[thick, dotted] (-\arcsize -0.05,-\arcsize - 0.05) -- (-\arcsize-0.05, 0.05);
\draw[thick, dotted] (3*\arcsize + 0.05,-\arcsize - 0.05) -- (3*\arcsize+ 0.05, 0.05);
\draw[thick, dotted] (-\arcsize - 0.05, 0.05) -- (3*\arcsize+ 0.05, 0.05);
\draw[thick, dotted] (-\arcsize - 0.05, -\arcsize - 0.05) -- (3*\arcsize+ 0.05, -\arcsize - 0.05);
\draw[thick] (0,0) arc (-180:0:\arcsize);
\end{tikzpicture}}}}
\newcommand{\loopJL}{\vcenter{\hbox{
\begin{tikzpicture}
\draw[thick, dotted] (-\arcsize -0.05,-\arcsize - 0.05) -- (-\arcsize-0.05, 0.05);
\draw[thick, dotted] (3*\arcsize + 0.05,-\arcsize - 0.05) -- (3*\arcsize+ 0.05, 0.05);
\draw[thick, dotted] (-\arcsize - 0.05, 0.05) -- (3*\arcsize+ 0.05, 0.05);
\draw[thick, dotted] (-\arcsize - 0.05, -\arcsize - 0.05) -- (3*\arcsize+ 0.05, -\arcsize - 0.05);
\draw[thick] (0,0) arc (0:-90:\arcsize);
\draw[thick] (\arcsize*2,0) arc (-180:-90:\arcsize);
\end{tikzpicture}}}}
\newcommand{\loopII}{\vcenter{\hbox{
\begin{tikzpicture}
\draw[thick, dotted] (-\arcsize -0.05,-\arcsize - 0.05) -- (-\arcsize-0.05, 0.05);
\draw[thick, dotted] (3*\arcsize + 0.05,-\arcsize - 0.05) -- (3*\arcsize+ 0.05, 0.05);
\draw[thick, dotted] (-\arcsize - 0.05, 0.05) -- (3*\arcsize+ 0.05, 0.05);
\draw[thick, dotted] (-\arcsize - 0.05, -\arcsize - 0.05) -- (3*\arcsize+ 0.05, -\arcsize - 0.05);
\draw[thick] (0,0) -- (0,-\arcsize);
\draw[thick] (\arcsize*2,0) -- (\arcsize*2,-\arcsize);
\end{tikzpicture}}}}

\usepackage{mathtools}
\DeclareMathOperator{\im}{Im}
\DeclareMathOperator{\re}{Re}
\usepackage{subfig}
\usepackage{xparse}

\DeclareDocumentCommand\ket{ s m }
{ 
	\IfBooleanTF{#1}
	{\lvert{#2}\rangle} 
	{\left\lvert{#2}\right\rangle} 
}

\DeclareDocumentCommand\innerproduct{ s m g }
{ 
	\IfBooleanTF{#1}
	{ 
		\IfNoValueTF{#3}
		{\langle{#2}\vert{#2}\rangle}
		{\langle{#2}\vert{#3}\rangle}
	}
	{ 
		\IfNoValueTF{#3}
		{\left\langle{#2}\middle\vert{#2}\right\rangle}
		{\left\langle{#2}\middle\vert{#3}\right\rangle}
	}
}
\DeclareDocumentCommand\braket{}{\innerproduct} 
\DeclareDocumentCommand\ip{}{\innerproduct} 
\DeclareDocumentCommand\expectationvalue{ s s m g }
{ 
	\IfNoValueTF{#4}
	{
		\IfBooleanTF{#1}
		{\langle{#3}\rangle} 
		{\left\langle{#3}\right\rangle} 
	}
	{
		\IfBooleanTF{#1}
		{
			\IfBooleanTF{#2}
			{\left\langle{#4}\middle\vert{#3}\middle\vert{#4}\right\rangle} 
			{\langle{#4}\vert{#3}\vert{#4}\rangle} 
		}
		{\left\langle{#4}\middle\vert\smash{#3}\middle\vert{#4}\right\rangle} 
	}
}

\DeclareDocumentCommand\matrixelement{ s s m m m }
{ 
	\IfBooleanTF{#1}
	{
		\IfBooleanTF{#2}
		{\left\langle{#3}\middle\vert{#4}\middle\vert{#5}\right\rangle} 
		{\langle{#3}\vert{#4}\vert{#5}\rangle} 
	}
	{\left\langle{#3}\middle\vert\smash{#4}\middle\vert{#5}\right\rangle} 
}

\DeclareDocumentCommand\mel{}{\matrixelement} 

\DeclareDocumentCommand\ev{}{\expectationvalue}

\DeclareDocumentCommand\qqtext{ s m }{\IfBooleanTF{#1}{}{\quad}\text{#2}\quad}
\DeclareDocumentCommand\qq{}{\qqtext}

\DeclareDocumentCommand\partialderivative{ s o m g g d() }
{ 
	\IfBooleanTF{#1}
	{\let\fractype\flatfrac}
	{\let\fractype\frac}
	\IfNoValueTF{#4}
	{
		\IfNoValueTF{#6}
		{\fractype{\partial \IfNoValueTF{#2}{}{^{#2}}}{\partial #3\IfNoValueTF{#2}{}{^{#2}}}}
		{\fractype{\partial \IfNoValueTF{#2}{}{^{#2}}}{\partial #3\IfNoValueTF{#2}{}{^{#2}}} \argopen(#6\argclose)}
	}
	{
		\IfNoValueTF{#5}
		{\fractype{\partial \IfNoValueTF{#2}{}{^{#2}} #3}{\partial #4\IfNoValueTF{#2}{}{^{#2}}}}
		{\fractype{\partial^2 #3}{\partial #4 \partial #5}}
	}
}

\DeclareDocumentCommand\pdv{}{\partialderivative} 

\newenvironment{dedication}
 {\vspace{30ex}\begin{quotation}\begin{center}\begin{em}}
 {\par\end{em}\end{center}\end{quotation}}

\usepackage{float}

\usepackage[ruled,linesnumbered]{algorithm2e}
\usepackage{fancyvrb}
\fvset{tabsize=4}

\begin{document}
\maketitle

\begin{abstract}
Two novel frameworks for handling mathematical and physical problems are introduced. The first, the emerging Jordan form, generalizes the concept of the Jordan canonical form, a well-established tool of linear algebra. The second, dual Jordan quantum physics, generalizes the framework of quantum physics to one in which the hermiticity postulate is considerably relaxed. These frameworks are then used to resolve some long-outstanding problems in theoretical physics, coming from critical statistical models and conformal field theory. I describe these problems and the difficulties involved in finding satisfactory solutions, then show how the concepts of emerging Jordan forms and dual Jordan quantum physics are naturally suited to overcoming these difficulties. Although their applications in this work are limited in scope to rather specific problems, the frameworks themselves are completely general, and I describe ways in which they may be used in other areas of mathematics and physics. Several appendices close the work, which include improvements to a widely used computational algorithm and corrections to some published data.
\end{abstract}

\begin{dedication}
Dedicatio

Universis preceptoribus meis, praeteritum, presentibus et futuris, scolarem ad uitam remanere
\end{dedication}

\tableofcontents

\chapter*{Introduction} \label{introduction}
\addcontentsline{toc}{chapter}{Introduction}

\emph{``We who cut mere stones must always be envisioning cathedrals.''---medieval quarry workers's creed }\cite{HuntThomas1999,Newport2016}

\sectionbreak

Two novel frameworks for handling mathematical and physical problems are introduced. The first, the emerging Jordan form, generalizes the concept of the Jordan canonical form, a well-established tool of linear algebra. The second, dual Jordan quantum physics, generalizes the framework of quantum physics to one in which the hermiticity postulate is considerably relaxed. These frameworks are then used to resolve some long-outstanding problems in theoretical physics, coming from critical statistical models and conformal field theory. I describe these problems and the difficulties involved in finding satisfactory solutions, then show how the concepts of emerging Jordan forms and dual Jordan quantum physics are naturally suited to overcoming these difficulties. Although their applications in this work are limited in scope to rather specific problems, the frameworks themselves are completely general, and I describe ways in which they may be used in other areas of mathematics and physics. Several appendices close the work, which include improvements to a widely used computational algorithm and corrections to some published data.

\sectionbreak

The results discussed in this work began from a project whose purpose was to gain understanding of the action of conformal symmetry on lattice models, and to use this understanding to determine multi-point correlation functions in a variety of systems of physical interest. As will be evident, a thorough study of these problems requires a mixture of new techniques that draw inspiration from ideas in mathematics, physics, and numerical analysis, many of which came about while carrying out this undertaking.

The topic of conformal field theory has grown manifold since its beginnings in the early 1990s, and is relevant now to condensed matter physics as well as high-energy physics and mathematics, in two dimensions as well as higher numbers of dimensions. Despite this impressive progress, some of the very questions that prompted the topic in the first place remain unsolved. There is a fundamental reason for this failure: most of the successes of conformal invariance so far have been based on ``top-down'' approaches, where results were largely guessed based on symmetry considerations, and then checked to be correct. Such top-down approaches work well in the context of ordinary quantum field theories, but fail for many problems of physical interest in condensed matter physics, like polymers, percolation, or disordered systems. In technical terms, this is because the field theories describing long-distance properties of statistical problems at criticality are not necessarily unitary when considered as one-dimensional quantum field theories. This technical problem translates into the fact that the action of the conformal symmetry---that is, of the Virasoro algebra---is not necessarily simple: modules can appear which are not fully reducible, or nontrivial zero-norm-square states can occur that are in fact physical. The associated mathematical problems are daunting when considered abstractly: the representation theory of the Virasoro when the constraint of unitarity is removed is technically wild, and hope for progress using a top-down approach is very limited.

The goal then is to solve the difficulty via a ``bottom-up'' approach, by moving from the usual abstract symmetry approach to a much more concrete one by studying directly the action of the conformal symmetry on lattice models. Using a mix of analytical and numerical techniques, one can determine what kinds of modules of the Virasoro algebra are relevant to, for instance, the percolation problem, and what kinds of degeneracy equations physical observables (such as the probability to belong to a certain cluster) may satisfy. This knowledge will be combined with the direct determination of the spectrum of percolation in the $s$ and $t$ channels of four-point correlation functions, using representation theory of the associated lattice algebra, the Temperley--Lieb algebra. Finally, the information thus gained will be used as starting points for a bootstrap approach, where, by implementing crossing symmetry in particular, four-point (and higher) correlation functions will be built as sums over conformal blocks.

While the project is ambitious, it will be based on significant achievements of the last few years. On the one hand, the idea of studying the conformal symmetry on the lattice by building discretized versions of the Virasoro generators, which was pioneered by \textcite{KooSaleur1994}, has attracted much attention recently, both in the mathematics community \cite{ZiniWang2018}, and in the quantum information community \cite{MilstedVidal2017}. Exciting new technical possibilities (based in part on the matrix product states technique, as well as the form-factors technique) have appeared that should be exploited very quickly. On the other hand, significant progress has also occurred both in the abstract bootstrap approach \cite{PiccoRibaultSantachiara2016} and in the determination of spectra of geometrical models \cite{JacobsenSaleur2019}. The topic seems ripe for a breakthrough, which, it is hoped, this work will describe.

Apart from fundamental progress in our understanding of geometrical problems, the completion of this program would also constitute a significant step forward in the solution of nonunitary conformal field theories such as those occurring in the description of critical points in different universality classes of topological insulators (like the plateaux transition in the integer quantum Hall effect). In a different direction, the issue of discretizing the Virasoro algebra is of importance in quantum computing, where many researchers are trying to imagine quantum computers and algorithms able to simulate quantum field theories \cite{Freedman1998}. Finally, any study of quantum field theory based on lattice discretizations is of high interest to mathematicians, since quantum fields and path integrals prove so difficult to define rigorously.

Over the course of carrying out this program, on the lattice side, my work and ideas, and those of many of my collaborators, seemed to lead inexorably to the two frameworks presented here, particularly the emerging Jordan block. At first, the emerging Jordan block simply served a purpose to allow us to compute $b$ numbers where we otherwise could not. Unexpectedly, the results of some of these measurements were so successful (in Figure \ref{W11bs}, for instance, the measured $b^{(1)}_{11}$ and the theoretical $b_{11}$ nearly overlap exactly for $c < 1/2$), that we wondered whether there was something more to our method. Considerable effort was expended in figuring out some of the more important properties of and structures within these emerging Jordan blocks, and along the way we found counterexamples to some of our earlier conjectures, leading to their refinement and further insight. We have assembled our observations into the beginnings of a coherent mathematical structure that we genuinely believe to be of general practical use, subject to the validity of Conjectures \ref{conjecture_emerging} and \ref{conjecture_infinite}. And when we look back at our earlier attempts at our problems, we essentially see that the emerging Jordan block came looking for us. To me, this is what the interplay between physics and mathematics is about.

\section*{Outline of this work}
\addcontentsline{toc}{section}{Outline of this work}

This work is divided into three primary parts, corresponding to background information, novel theoretical frameworks, and finally applications that put together the first two parts.

Part \ref{background} begins with two chapters on mathematics. The first of these consists of a collection of concepts from mathematics, all well-established in the mathematical canon and part of the core education for mathematicians. They are written, however, using the conventions most often encountered in physics, which include departures in terminology and notations. A second chapter on the representation theory of Temperley--Lieb algebras consists of more recent and specialized results, but which have a bearing on the physical models studied in Chapter \ref{physics_lattice} and Part \ref{applications}. The two chapters on physics follow the same scheme, with the second one pertaining to the study of conformal field theory using discrete lattice models. Finally, Chapter \ref{computational} discusses some aspects of matrix computations. Ordinarily used as a black-box tool, sophisticated numerical algorithms and software packages typically produce accurate and reliable results without needing fine-tuning from the user. However, overcoming some of the shortcomings of these computational tools produced insights that helped to develop approaches for handling some of the physics problems. It is important to distinguish when an unexpected result is a false positive due to the numerical tool itself, or whether it is a genuine aspect of the mathematics. I therefore describe algorithms for the most frequently used matrix computations in this work, as they are an integral part of a thorough study of the models contained herein, and to provide context for an improvement described in Appendix \ref{DMA}.

Many treatises with a substantial segment devoted to background material claim to include these details so that the work is essentially self-contained, assuming sufficient mathematical and logical sophistication on the part of the reader. I have yet to see this executed in a convincing way, except possibly the first volume of Bourbaki's \emph{Elements of Mathematics} series \cite{Bourbaki_2006,Bourbaki_2004} and Whitehead and Russell's \emph{Principia Mathematica} \cite{WhiteheadRussell1935,WhiteheadRussell1927}. Let me then disclaim the self-contained nature of this work, and state instead that the purposes of the background chapters of Part \ref{background} are to provide context for the new techniques developed in Part \ref{theory} and their applications in Part \ref{applications}, and to establish notations and terminology used throughout.

New methods to handling the mathematical and physical questions studied are described in Part \ref{theory}. While the development of these methods took place in the context of studying the problems of Chapters \ref{TL_algebra} and \ref{physics_lattice} and Part \ref{applications}, they are easily abstracted and apply very generally. The purpose of this part is to present them as self-contained and coherent frameworks for studying problems within their domains of application. Chapter \ref{emerging} considers the limit of a sequence of diagonalizable linear operators (or the limit of a [continuous] function whose codomain is a space of linear operators) and the question of whether the limit is diagonalizable. While the answer turns out to be very simple, my collaborators and I were surprised to find that the approach described therein had not been tried before, as far as we could find. Chapter \ref{biorthogonal} develops a systematic method to compute eigenspace projection operators for operators that are not normal. In the standard formulation of quantum physics the Hamiltonian operator is postulated to be hermitian (with respect to a positive-definite inner product), thus normal, and its eigenstates, assumed to form a complete set, furnish a convenient basis in terms of which to express a general state. The components of a general state in this basis are easily computed---formally, one uses the eigenspace projection operators constructed from the eigenstates. This chapter therefore represents the analogue of this basic procedure for a framework of quantum physics in which the assumption of hermiticity is lifted.

Assembling all of these admittedly elaborate pieces, I finally turn to new results and implications in physics in Part \ref{applications}. I demonstrate that the sophistication of these new methods pays off when insight is gained in long-outstanding problems whose progress has stalled in recent years. I also hope that the reader sees that these methods were necessary, and essentially unavoidable in any path towards a complete solution---they are tailor-made to overcome the difficulties encountered in previous attempts at solution. Chapter \ref{Virasoro} describes how indecomposable structures, and hence Jordan blocks, arise in the continuum limit of the loop formulation of the Potts model, and verifies some signatures of that indecomposability. Chapter \ref{Jordan_loop} then describes attempts to observe this Jordan block structure directly---first through correlation functions and OPEs, then via emerging Jordan blocks. Chapter \ref{sl21_chapter} describes the application of the new projection operators to the $s\ell(2|1)$ spin chain. While many problems remain unresolved, the results presented there demonstrate the internal consistency of the construction of the dual Jordan projection operators. In Chapter \ref{mixing}, I describe how some lattice quantities do not directly correspond to a clear continuum analogue, and my attempts at disentangling them.

Six appendices close the work. Appendix \ref{DMA} describes what I call ``dynamic multiplicity adjustment,'' a simple enhancement to the widely used implicitly restarted Arnoldi method that drastically improves convergence for highly degenerate problems, like those encountered in this work. Appendix \ref{mathematica} contains efficient code in Mathematica that the reader may use as a starting point to replicate the numerical results described here. Appendix \ref{subquotient} describes an algorithm for verifying subquotient structures in indecomposable modules, which has not yet been used in the study of the problems described in this work. Appendix \ref{corrections} lists corrections to the data in the first two tables of \textcite{KooSaleur1994}, so that they may serve as a useful check for those attempting to replicate the results described there. Finally, Appendices \ref{more_calculations} and \ref{more_tables} contain additional calculations, tables, and figures that supplement the main text, but are not otherwise essential.

Parts of the material in Chapters \ref{TL_algebra}, \ref{physics_lattice}, \ref{Virasoro}, and \ref{Jordan_loop} are adapted from \textcite{Grans-Samuelsson2020,Grans-Samuelsson2021} (in which ``et al.''\ includes myself, and where the latter version contains another appendix added post-publication). However, one important change that I apply uniformly compared to the published article is that for consistency with the remainder of the present work, I retain the notations $(\cdot,\cdot)$ and $(\cdot|\cdot)$ for the standard complex scalar product, and $\langle\cdot,\cdot\rangle$ and $\ip*{\cdot}$ for the conformal (loop) scalar product. I also modify other notations based on uniformity and personal preference. Sorry for the possible confusion, but this is my territory. Some of the description of the representation theory of the affine Temperley--Lieb algebra and the two Jones--Temperley--Lieb algebras comes from \textcite{Gainutdinov2015}. I have likewise rewritten much of the exposition, made editorial changes and corrections, and changed notations. Computations resulting in the data presented in the tables of Section \ref{singlet_states} were carried out by Jesper Jacobsen. All errors are mine.

\section*{Acknowledgments}
\addcontentsline{toc}{section}{Acknowledgments}

Much support during this research project came from family and friends. A debt of gratitude has been paid, in part, by including them in the dedication page, since they are certainly among that group.

Special thanks go to Hubert Saleur, for entrusting me with challenging problems, to L'Institut de Physique Théorique at CEA Saclay and its community for their hospitality, and to Jesper Jacobsen for comments on earlier drafts of this work.

Some of the research leading to the findings in Chapter \ref{Virasoro} and its subsequent publication \cite{Grans-Samuelsson2020,Grans-Samuelsson2021} was supported by a Chateaubriand Fellowship offered by the Office for Science and Technology of the Embassy of France in the United States. The Center for Advanced Research Computing (CARC) at the University of Southern California (\url{https://carc.usc.edu}) provided computing resources that have contributed to the research results reported within this work.

I am also grateful for the musicians whose works filled my offices and other working spaces. Non-exhaustively, they include: Isaac Albéniz, Charles-Valentin Alkan, George Antheil, Johann Sebastian Bach, Mily Balakirev, Ludwig van Beethoven, Aleksandr Borodin, Sergei Bortkiewicz, Lili Boulanger, Johannes Brahms, Ferruccio Busoni, Fryderyk Chopin, Mikhail Glinka, Leopold Godowsky, Enric Granados, Joseph Haydn, Adolf von Henselt, Nikolai Kapustin, György Ligeti, Ferenc Liszt, Gustav Mahler, Nikolay Medtner, Felix Mendelssohn, Olivier Messiaen, Moritz Moszkowski, Wolfgang Amadeus Mozart, Sergey Prokofiev, Sergei Rachmaninoff, Einojuhani Rautavaara, Jean-Henri Ravina, Christopher Rouse, Franz Schubert, Aleksandr Scriabin, Karol Szymanowski, Pyotr Tchaikovsky, Sigismond Thalberg, Iannis Xenakis. The cognoscenti will recognize the obvious bias that appears in this list.

\part{Background} \label{background}
\chapter{Mathematics: basic concepts} \label{mathematics_concepts}
\section{Inner products} \label{inner_products}

I assume the reader is already familiar with vector spaces and inner products. This purpose of this section is to establish the conventions used for the remainder of the work, as there are a number of competing ones used throughout mathematics and physics.
\begin{definition}
Let $V$ be a vector space over $k$, where $k = \mathbb R$ or $\mathbb C$. An \emph{inner product} is a map $f: V\times V \to k$ such that for all $u,v,w\in V$ and $\alpha\in k$,
\begin{subequations}
\begin{gather}
f(u,v+w) = f(u,v) + f(u,w), \\
f(u,\alpha v) = \alpha f(u,v), \qq{and} \\
f(u,v) = f(v,u)^*. \label{eq:conjugate}
\end{gather}
If, additionally, for every nonzero $v\in V$,
\begin{equation} \label{eq:positive_ip}
f(v,v) > 0,
\end{equation}
\end{subequations}
$f$ is called a \emph{positive-definite inner product}. Finally, an inner product is said to be \emph{non-degenerate} if $f(u,v) = 0$ for all $v\in V$ implies that $u = 0$.
\end{definition}

Note that I follow the physics convention of linearity in the second argument, and that an inner product is not explicitly required to satisfy \eqref{eq:positive_ip}. Nevertheless, I will occasionally use the term ``indefinite'' inner product to call attention to the fact that \eqref{eq:positive_ip} is violated. The squared norm of $v\in V$ is $f(v,v)$. By property \eqref{eq:conjugate}, this number is real, though not necessarily positive. Occasionally, in the physics literature, ``norm'' is used to mean the squared norm $f(v,v)$, and in particular, ``negative norm'' means $f(v,v) < 0$ (used, for instance, in the discussion of ghosts in string theory \cite{GSW1987,West2012}). But to avoid such imprecision, $f(v,v)$ and equivalent notations in this work will always be called ``squared norm,'' ``norm squared,'' or some unambiguous variant.

Alternative notations for $f(u,v)$, which I will use exclusively hereafter, are $(u,v)$, $(u|v)$, $\langle u, v\rangle$, and $\ip*{u}{v}$.

Mathematicians may be willing to forgive the use of ``inner product'' for the general indefinite inner product, which they would call an ``hermitian form.'' But ``squared norm'' for a possibly negative quantity is undoubtedly a more difficult proposition to accept. I offer some appeasement by having used $k$ to denote a field, although I cannot discuss field extensions here (except to note that $\mathbb C$ is a field extension of $\mathbb R$).

\section{The spectral theorem for normal operators} \label{spectral_theorem}

Throughout this work, ``operator'' always means ``linear operator.''

Given a non-degenerate inner product, to every linear operator $A$ there corresponds its adjoint (or hermitian adjoint, or hermitian conjugate) $A^\dagger$, such that for any $u,v\in V$,
\begin{equation}
(A^\dagger u,v) = (u,Av).
\end{equation}
It is necessary that the inner product be non-degenerate for $A^\dagger$ to be well-defined and unique \cite[p.~533]{Lang2002}. The reader is familiar with the standard inner product for a finite-dimensional vector space (in some orthonormal basis), $(u,v) = \sum_i u_i^* v_i$, for which the hermitian adjoint is the conjugate-transpose, $A^\dagger = A^*$. A linear operator $A$ such that $A = A^\dagger$ is called self-adjoint, or hermitian. The concept of adjoint may be extended to vectors. For a vector $u$, its adjoint is an element of the dual space, the linear functional $u^\dagger$ such that $u^\dagger v = (u,v)$ for any $v\in V$.

A standard theorem of linear algebra states that every hermitian operator $H$ is unitarily diagonalizable, with real eigenvalues (the additional assumption that $H$ be compact is necessary in infinite dimensions). That is, there exists a basis $\{u_i\}$ such that
\begin{subequations}
\begin{gather}
H u_i = \lambda_i u_i, \\
(u_i, u_j) = \delta_{ij},
\end{gather}
\end{subequations}
for a set of eigenvalues $\{\lambda_i\} \subset \mathbb R$. The machinery of the standard formulation of quantum mechanics rests upon this result (see Section \ref{quantum_physics}). Nevertheless, its proof relies on property \eqref{eq:positive_ip}, and fails in the more general context.

A number of physical problems involve indefinite inner products which arise naturally (i.e., the correct physics obtains when calculations are carried out using these inner products), whose Hamiltonian operators are hermitian with respect to these inner products, but cannot be unitarily diagonalized. I will thus reserve the term ``hermitian'' to mean hermitian with respect to some positive-definite inner product henceforth. A non-hermitian operator is therefore not hermitian with respect to any positive-definite inner product.

I give two statements of the spectral theorem as used in this work. The theorem is stated for normal operators $A$, which satisfy $[A,A^\dagger] = 0$, and therefore subsume hermitian operators. Its analogue for non-normal finite-dimensional operators with real spectra is treated in Chapter \ref{biorthogonal}.

\begin{theorem}[Spectral theorem in terms of eigenvectors] \label{spectral_theorem_eigenvectors}
If $A$ is a normal operator on a finite-dimensional complex inner product space $V$, where the inner product is positive-definite, then $V$ has an orthonormal basis consisting of eigenvectors of $A$. Therefore, a normal operator on such a space is diagonalizable.
\end{theorem}

\begin{theorem}[Spectral theorem in terms of projection operators]
Let $A$ be a normal operator on a finite-dimensional complex inner product space $V$, where the inner product is positive-definite, and $\{\lambda_1,\ldots,\lambda_p\}$ its distinct eigenvalues. Then
\begin{equation}
V = \bigoplus_{i=1}^p V_i,
\end{equation}
where $V_i$ is the eigenspace associated to eigenvalue $\lambda_i$. If $\Pi_i$ is the hermitian projection operator onto $V_i$, then
\begin{subequations}
\begin{gather}
\sum_{i=1}^p \Pi_i = I, \\
\Pi_i \Pi_j = \delta_{ij}\Pi_i, \qq{and} \\
A = \sum_{i=1}^p \lambda_i \Pi_i.
\end{gather}
\end{subequations}
\end{theorem}

While I focus exclusively on discrete spectra, an operator, necessarily infinite-dimensional, with a discrete and continuous spectrum can be treated compactly using measure-theoretic ideas \cite{AkhiezerGlazman1963}. For example, to each self-adjoint operator $\Omega$, there exists a unique spectral family $\{E_\omega\}$ such that
\begin{equation}
\Omega = \int_{-\infty}^\infty\!\omega\,\d E_\omega.
\end{equation}
The spectral family $\{E_\omega\}$ is a one-parameter family of orthogonal projection operators such that
\begin{enumerate}
\item the family is monotonically increasing: $E_{\omega'}E_\omega = E_\omega E_{\omega'} = E_\omega$ when $\omega' \ge \omega$;
\item the family is right-continuous: $\lim_{\epsilon\to 0^+}E_{\omega+\epsilon} = E_\omega$;
\item $\lim_{\omega\to-\infty}E_\omega = 0$ and $\lim_{\omega\to\infty}E_\omega = I$.
\end{enumerate}
The integral can be restricted to the real line because the self-adjoint property implies a real spectrum.

\section{The Jordan canonical form} \label{Jordan}

In general, not all operators are diagonalizable, even over the algebraically closed field of complex numbers. The closest one can get for such operators is the following theorem, whose statement is adapted from \textcite[p.~222]{HoffmanKunze1971}.

\begin{theorem} \label{ND}
Let $A$ be a linear operator on the finite-dimensional complex vector space $V$. Then there is a diagonalizable operator $D$ on $V$ and a nilpotent operator $N$ on $V$ such that $A = D + N$ and $DN = ND$. The diagonalizable operator $D$ and the nilpotent operator $N$ are uniquely determined and each of them is a polynomial in $A$.
\end{theorem}

The simplest explicit matrix representation of the operator decomposition in Theorem \ref{ND} is the Jordan canonical form [\emph{Id.}\ at Section 7.3] (also ``Jordan normal form'').

\begin{theorem}[Jordan canonical form]
Let $A$ be a linear operator on the finite-dimensional complex vector space $V$. There exists a basis $B$ of $V$ such that the matrix of $A$ in this basis is block diagonal:
\begin{subequations} \label{eq:Jordan_canonical_form}
\begin{equation}
A_B = \begin{pmatrix}
A_1 \\
 & A_2 & \\
 & & \ddots \\
 & & & A_p
\end{pmatrix},
\end{equation}
where $A_1,\ldots,A_p$ are matrices. Each $A_i$ is of the form
\begin{equation}
A_i = \begin{pmatrix}
J_{i1} \\
 & J_{i2} \\
 & & \ddots \\
 & & & J_{in_i}
\end{pmatrix},
\end{equation}
where each $J_{ij}$ is an elementary Jordan block with eigenvalue $\lambda_i$:
\begin{equation} \label{eq:elementary_Jordan_block}
J_{ij} = \begin{pmatrix}
\lambda_i & 1 \\
 & \lambda_i & 1 \\
 & & \ddots & \ddots \\
 & & & \lambda_i & 1 \\
 & & & & \lambda_i
\end{pmatrix}.
\end{equation}
\end{subequations}
\end{theorem}
By the primary decomposition theorem [\emph{Id.}\ at p.~220], the vector space $V = \bigoplus_{i=1}^p V_i$ is a direct sum of invariant subspaces, where $V_i$ is the null space of $(A - \lambda_i I)^{r_i}$, $r_i$ being the multiplicity of $\lambda_i$ as a root of the minimal polynomial for $A$. Furthermore, $r_i$ is the rank of the largest elementary Jordan block in $A_i$.

The Jordan canonical form subsumes the diagonalization of diagonalizable operators. When an operator is diagonalizable, all of the elementary Jordan blocks are $1\times 1$ matrices, with no space for off-diagonal elements, and the Jordan form is a diagonal matrix. If all of the elementary Jordan blocks have rank 1, then the Jordan canonical form will be called \emph{trivial}. The process of finding the Jordan form of an operator---its elementary Jordan blocks and a basis in which the operator has such a form---will still be called \emph{diagonalization}, for brevity.

The general expression given for an elementary Jordan block has ones above the main diagonal. This means that to each elementary Jordan block $J_{ij}$, there is a proper eigenvector $u_{ij1}$ such that $Au_{ij1} = \lambda_i u_{ij1}$ and a set of generalized eigenvectors, $\{u_{ijk} | 2 \le k \le m_{ij}\}$, where $m_{ij}$ is the dimension or rank of $J_{ij}$, such that $A u_{ijk} = \lambda_i u_{ijk} + u_{ij,k-1}$. By defining $u_{ij0} = 0$, the preceding equation can be used with $1 \le k \le m_{ij}$.

By a trivial rearrangement of the basis vectors, the ones in the elementary Jordan block can be moved below the main diagonal, which leads to an equivalent statement of the Jordan canonical form. We will have use for both of these versions, so when the distinction is important, I will call a matrix in Jordan canonical form with elementary Jordan blocks of the form in Eq.~\eqref{eq:elementary_Jordan_block} \emph{upper Jordan} and \emph{lower Jordan} when the ones are below the main diagonal---the transpose of Eq.~\eqref{eq:elementary_Jordan_block}.

The fact that the off-diagonal elements are equal to $1$ in particular is not terribly important. Each off-diagonal element can be changed to an arbitrary nonzero scalar by rescaling the elements of the basis. With this in mind, any matrix of the form described by Eq.~\eqref{eq:Jordan_canonical_form} or its transpose, where the off-diagonal elements are not necessarily equal to $1$, will be considered to be in \emph{Jordan form}. I will generically refer to the nonzero off-diagonal elements as \emph{couplings} or \emph{Jordan couplings}, and I will reserve the term ``Jordan \emph{canonical} form'' for the special case in which all of the Jordan couplings are equal to $1$.

One more piece of terminology: the collection of eigenvectors (proper and generalized) that form a basis for an elementary Jordan block (of rank $n$) will be called a \emph{tower} (of height $n$), especially when the rank of the Jordan block is $3$ or greater. Although this terminology is not standard in this context, I have found that it gives the right intuition when thinking about the action of an operator with a nontrivial Jordan form.

While I focus almost exclusively on operators acting on finite-dimensional vector spaces, some interesting subtleties arise in the infinite-dimensional case. For instance, the eigenvalues of a finite-dimensional operator are found on the main diagonal of its Jordan form. In the infinite-dimensional case, this may not be true. The most familiar examples of infinite Jordan towers to physicists, though not usually thought of as such, come from the creation and annihilation operators $a^\dagger$ and $a$ for the harmonic oscillator. In the basis of energy eigenstates $\ket*{n} = (a^\dagger)^n/\sqrt{n!}\,\ket*{0}$, $a$ is upper Jordan and $a^\dagger$ is lower Jordan, both with only zeros appearing on the main diagonal. The annihilation operator $a$ has infinitely many eigenstates parametrized by $z\in\mathbb C$, the coherent states $\ket*{z} = \e^{za^\dagger}\ket*{0}$, while the creation operator has no eigenstates (here, I am allowing infinite linear combinations). This may be contrasted with, say, the raising and lowering operators $L_+$ and $L_- = L_+^\dagger$ in finite-dimensional highest-weight representations of $\mathfrak{su}(2)$. Another example is the differentiation map acting on the space of polynomials $k[x]$ (with $k$ a field of characteristic zero), which, in the basis $\{x^n/n!\}$, is represented by an infinite elementary Jordan block with eigenvalue $0$.

\section{Representation theory of algebras}

Representation theory is a huge enterprise of its own \cite{FultonHarris2004,ASS2006}, and so I will not even attempt to give an introduction to this subject. I only use the present section to state some definitions as I use them.

\begin{definition} \label{def_representation}
Let $k$ be a field. A \emph{$k$-algebra} is a ring with identity $A$ such that $A$ has a $k$-vector space structure compatible with the multiplication operation of the ring. A \emph{left $A$-module} is a pair $(M,\cdot)$ consisting of a $k$-vector space $M$ and a multiplication $\cdot:A\times M \to M$ such that $\cdot$ is compatible with the vector space operations in $M$ (left and right distributivity, associativity, unitality). A \emph{representation} of $A$ is a left $A$-module.
\end{definition}

\begin{definition} \label{def_indecomposable}
Let $k$ be an algebraically closed field, let $A$ be a $k$-algebra, and let $M$ be an $A$-module. $M$ is \emph{simple} or \emph{irreducible} if $M$ is nonzero and the only submodules of $M$ are the zero module and $M$. $M$ is \emph{semisimple} if $M$ is a direct sum of simple $A$-modules. $M$ is \emph{indecomposable} if $M$ is nonzero, $M$ is not simple, and $M$ has no module direct sum decomposition $M \cong M_1 \oplus M_2$ in terms of nonzero $A$-modules. The $k$-algebra $A$ is \emph{semisimple} if every $A$-module is semisimple.
\end{definition}

Note that I have adopted the convention that an indecomposable module is not irreducible (not simple). In standard terminology, an irreducible or simple module is indeed indecomposable. However, because so much discussion centers on modules that are indecomposable but not irreducible, I have specifically excluded it from my definition to avoid overusing the phrase ``indecomposable but not irreducible.'' (Indeed, I count over 25 uses of the term in this work.) ``$k$-algebra'' will be shortened to ``algebra'' (in the remainder of this work, $k = \mathbb R$ or $\mathbb C$), ``$A$-module'' will be shortened to ``module'' when the algebra is clear from context, and ``module'' will always mean ``left module.'' Otherwise, there should be nothing controversial about the definitions, adapted from \textcite[Chapter 1]{ASS2006}.

Unfortunately, ``representation'' is such a ubiquitous word that many instances of it in this work do not refer to the object of Definition \ref{def_representation}, but also the correspondence between mathematical symbols and abstract or concrete objects.
 
\chapter{Mathematics: representation theory of various Temperley--Lieb algebras} \label{TL_algebra}

\section{Generators and relations for Temperley--Lieb algebras} \label{generators}

In terms of generators and relations, various algebras to which the name ``Temperley--Lieb'' (TL) is associated begin with generators $\{e_j\}$ subject to relations
\begin{subequations} \label{eq:TL_relations}
\begin{gather}
e^2_j = me_j, \label{eq:TL_m}\\
e_j e_{j\pm 1}e_j = e_j, \qq{and} \\
e_j e_k = e_k e_j, \qquad (|j - k| > 1)
\end{gather}
\end{subequations}
where the indices take the values $1,\ldots,N-1$, with $N$ a fixed integer. Each value of $N$ defines a different algebra. These relations are typically illustrated using diagrams for the generators. On two horizontal rows of $N$ points each, connect the neighboring pairs at $j$ and $j+1$, and vertically match all other pairs:
\begin{equation} \label{eq:e_j_diagram}
e_j =
\raisebox{-5mm}{
\begin{tikzpicture}[scale=0.6]
 \draw[thick] (0,-1)--(0,1) node[above]{\scriptsize $1$};
\end{tikzpicture}}
\quad\cdots\quad
\raisebox{-5mm}{\begin{tikzpicture}[scale=0.6]
 \draw[thick] (-1,-1)--(-1,1);
 \draw[thick] (2,-1)--(2,1);
 \draw[thick] (0,-1)--(0,-0.5) arc(180:0:5mm and 4mm)--(1,-1);
 \draw[thick] (0,1) node[above]{\scriptsize $j$}--(0,0.5) arc(180:360:5mm and 4mm)--(1,1) node[above]{\scriptsize $j+1$};
\end{tikzpicture}}
\quad\cdots\quad
\raisebox{-5mm}{\begin{tikzpicture}[scale=0.6]
 \draw[thick] (0,-1)--(0,1) node[above]{\scriptsize $N$};
\end{tikzpicture}}
\end{equation}
When two generators are multiplied, the corresponding diagram is obtained following certain rules: stack the diagrams from top to bottom (corresponding to left to right multiplication), replace closed loops with a numerical factor $m$, and straighten out the curves that connect vertically. The ambiguity in the last rule can be avoided by defining elements of the algebra as equivalence classes of such diagrams under isotopy, but we will not use this description. In any diagram of the algebra, the lines that connect from top to bottom (rather than between two sites in the same horizontal row) are called \emph{through-lines}, and the sites to which they are connected are called \emph{free}. The identity element $1$ consists of a diagram with all $N$ vertical pairs of sites connected by straight through-lines. It can be shown formally that this algebra of diagrams is isomorphic to the abstract Temperately--Lieb algebra defined by generators and relations \cite{RidoutSaintAubin2014}.

A simple generalization is to adjoin a generator $e_N$ and site $N+1$, and identify site $N+1$ with site $1$, thereby obtaining a periodic Temperley--Lieb algebra. The $N$ sites in the top and bottom rows can be thought of as being placed on the inner and outer boundaries of an annulus, evenly spaced, and multiplication consists of nesting two annuli after rescaling one so that its inner boundary matches the outer boundary of the other. In both cases the accessible diagrams consist of planar pairings of points using curves that do not intersect. As before, the curves that join points on the inner boundary to points on the outer boundary of the annulus are called through-lines.

With periodic boundary conditions, it is natural to introduce a translation $\tau$. In diagrammatic terms, using the labels of the right hand side of Eq.~\eqref{eq:e_j_diagram}, $\tau$ consists of the diagram where site $i$ of the bottom row is connected to site $i+1 \pmod{N}$ of the top row for $i = 1,\ldots, N$. The following additional defining relations are then obeyed diagrammatically:
\begin{subequations} \label{eq:TL_relations_affine}
\begin{gather}
\tau e_j \tau^{-1} = e_{j+1}, \qq{and} \label{eq:TL_translation_1} \\
\tau^2 e_{N-1} = e_1\cdots e_{N-1}.
\end{gather}
\end{subequations}
Moreover, $\tau^{\pm N}$ are central elements. The algebra generated by the generators $e_i$ and $\tau^{\pm 1}$ together with these relations is usually called the affine Temperley--Lieb algebra $T^{\text a}_N(m)$.

One may restrict to even powers of $\tau$. This requires $N = 2L$ to be an even number, which defines $L$, and in the remainder of this work we consider only this situation (i.e., even $N$). Formally, the elements $\tau^{\pm 2}$ are adjoined to the algebra instead of $\tau^{\pm 1}$, and Eq.~\eqref{eq:TL_translation_1} is replaced by
\begin{equation} \label{eq:TL_translation_2}
\tau^2 e_j \tau^{-2} = e_{j+2}.
\end{equation}
For both the affine Temperley--Lieb algebra and the present context, it is possible for noncontractible loops to be generated that circle the annulus, and that cannot be shrunk to a point via a homotopy without colliding with the inner boundary of the annulus. In the present setting, we assign these noncontractible loops the same weight---they are replaced by a factor of $m$. Finally, we take a quotient by the ideal generated by $\tau^N - 1$. In terms of annular diagrams, one annular diagram becomes equivalent to another obtained from the first by rotating the inner boundary through a full rotation, with the sites and attached curves following along with it. The resulting object with the modifications described in this paragraph is called the \emph{augmented Jones--Temperley--Lieb algebra} $JTL^{\text{au}}_N(m)$, and it is finite-dimensional \cite{Gainutdinov2015}. $JTL^{\text{au}}_N(m)$ is slightly larger than the \emph{Jones--Temperley--Lieb algebra} $JTL_N(m)$, introduced by \textcite{ReadSaleur2007} and further studied by \textcite{GRS2013b}. The difference is entirely in the ideal with zero through-lines---i.e., the annular diagrams in which points of the outer boundary are paired together and points of the inner boundary are paired together (again, using curves that do not intersect), and no point of the inner boundary is paired with a point of the outer boundary. In $JTL^{\text{au}}_N(m)$, all such diagrams are allowed. In $JTL_N(m)$, only diagrams that can be drawn without crossing the periodic ``boundary'' between both pairs of sites $N$ and $1$ are allowed---in other words, the diagrams that can just as well be drawn on the (nonperiodic) rectangle. This ideal (the subalgebra of $JTL_N(m)$ with zero through-lines) is also known as the oriented Jones annular subalgebra in the Brauer algebra \cite{Jones1994}. Formally, there is a covering homomorphism (surjection) of algebras $\psi: JTL^{\text{au}}_N(m) \to JTL_N(m)$, which acts nontrivially only in the zero through-lines subalgebra (of $JTL^{\text{au}}_N(m)$). Its action is best understood through an example:
\begin{equation}\label{eq:psi_ex}
\psi: \quad
 \raisebox{-6mm}{\begin{tikzpicture}
 	\draw[thick, dotted] (-0.05,0.5) arc (0:10:0 and -7.5);
 	\draw[thick, dotted] (-0.05,0.55) -- (2.65,0.55);
 	\draw[thick, dotted] (2.65,0.5) arc (0:10:0 and -7.5);
	\draw[thick, dotted] (-0.05,-0.85) -- (2.65,-0.85);
	\draw[thick] (0,0.1) arc (-90:0:0.5 and 0.4);
	\draw[thick] (0,-0.1) arc (-90:0:0.9 and 0.6);
	\draw[thick] (2.6,-0.1) arc (-90:0:-0.9 and 0.6);
	\draw[thick] (2.6,0.1) arc (-90:0:-0.5 and 0.4);
	\draw[thick] (0.5,-0.8) arc (0:90:0.5 and 0.5);
	\draw[thick] (1.8,-0.8) arc (0:180:0.5 and 0.5);
	\draw[thick] (2.1,-0.8) arc (0:90:-0.5 and 0.5);
	\end{tikzpicture}}
	\quad\mapsto\quad
 \raisebox{-6mm}{\begin{tikzpicture}
 	\draw[thick, dotted] (-0.05,0.5) arc (0:10:0 and -7.5);
 	\draw[thick, dotted] (-0.05,0.55) -- (2.65,0.55);
 	\draw[thick, dotted] (2.65,0.5) arc (0:10:0 and -7.5);
	\draw[thick, dotted] (-0.05,-0.85) -- (2.65,-0.85);
	\draw[thick] (0.5,0.5) arc (-180:0:0.8 and 0.56);
	\draw[thick] (0.8,0.5) arc (-180:0:0.5 and 0.4);
	\draw[thick] (2.1,-0.8) arc (0:180:0.8 and 0.56);
	\draw[thick] (1.8,-0.8) arc (0:180:0.5 and 0.4);
	\end{tikzpicture}}
\end{equation}
In Eq.~\eqref{eq:psi_ex} and what follows, dotted ``framing'' rectangles are understood to have their left and right boundaries identified, as if they were cut from an annulus. The homomorphism $\psi$ essentially redraws (uniquely) the curves that keep the same pairs matched, but without crossing the boundary of the framing rectangle.

The algebra $JTL^{\text{au}}_N(m)$ has dimension
\begin{equation}
\dim(JTL^{\text{au}}_N(m)) = \binom{N}{N/2}^2 + \sum_{j=1}^{N/2}\binom{N}{N/2-j}^2,
\end{equation}
while $JTL_N(m)$ has dimension
\begin{equation}
\dim(JTL_N(m)) = \left[\binom{N}{N/2} - \binom{N}{N/2-1}\right]^2 + \sum_{j=1}^{N/2}\binom{N}{N/2-j}^2.
\end{equation}

\section{Cellular algebras} \label{cellular}

A unified framework for discussing the representation theory of Temperley--Lieb and other algebras is the concept of a cellular algebra. Cellular algebras were first introduced by \textcite{GrahamLehrer1996} in the course of finding a ``systemic understanding of the non-semisimple [specializations] of Hecke algebras and of a variety of other algebras with geometric connections''---to include some of the various Temperley--Lieb algebras of Section \ref{generators}.

\begin{definition}[{\textcite[Eq.~(1.1)]{GrahamLehrer1996}}] \label{cellular_algebra}
Let $R$ be a commutative ring with identity. A \emph{cellular algebra} over $R$ is an associative (unital) algebra $A$, together with cell datum $(\Lambda, W, C, *)$ where
\begin{enumerate}
\item $\Lambda$ is a partially ordered set and for each $\lambda\in\Lambda$, $W(\lambda)$ is a finite set (the set of ``tableaux of type $\lambda$'') such that $C: \coprod_{\lambda\in\Lambda} W(\lambda)\times W(\lambda) \to A$ is an injective map with image an $R$-basis of $A$.

\item If $\lambda\in\Lambda$ and $i,j\in W(\lambda)$, write $C(i,j) = C^\lambda_{ij}\in A$. Then $*$ is an $R$-linear anti-involution of $A$ such that $C^\lambda_{ij} = C^\lambda_{ji}$.

\item If $\lambda\in\Lambda$ and $i,j\in W(\lambda)$ then for any element $a\in A$ we have
\begin{equation}
aC^\lambda_{ij} \equiv \sum_{k\in W(\lambda)} r_a(i,k)C^\lambda_{kj} \pmod{A(<\lambda)}
\end{equation}
where $r_a(i,k)\in R$ is independent of $j$ and where $A(<\lambda)$ is the $R$-submodule of $A$ generated by $\{C^\mu_{ij}|\mu < \lambda; i,j\in W(\mu)\}$.
\end{enumerate}
\end{definition}
This definition of cellular algebras appears as it was introduced by \citeauthor{GrahamLehrer1996} [\emph{Id.}], with some notations exchanged to correspond to that of \textcite{Gainutdinov2015} and the rest of the present work. The algebras $JTL^{\text{au}}_N(m)$ and $JTL_N(m)$ are cellular algebras, but $T^{\text a}_N(m)$ is not.

A particular set of representations of a given cellular algebra exists as a natural consequence of the axioms.
\begin{definition}[{\textcite[Eq.~(2.1)]{GrahamLehrer1996}}] \label{cell_module}
Let $A$ be a cellular algebra with cell datum $(\Lambda, W, C, *)$. For each $\lambda\in\Lambda$ define the (left) $A$-module $\mathscr W(\lambda)$ as follows: $\mathscr W(\lambda)$ is a free $R$-module with basis $\{C_i|i\in W(\lambda)\}$ and $A$-action defined by
\begin{equation}
aC_i = \sum_{j\in M(\lambda)}r_a(i,j)C_j \qquad (a\in A, i \in M(\lambda))
\end{equation}
where $r_a(i,j)$ is the element of $R$ defined in Definition \ref{cellular_algebra}. $\mathscr W(\lambda)$ is called the \emph{cell representation}, or \emph{cell module} of $A$ corresponding to $\lambda\in\Lambda$.
\end{definition}
Similarly, co-cell representations or co-cell modules $\widetilde{\mathscr W}(\lambda)$ are duals to the cell modules---the spaces of linear functionals on $\mathscr W(\lambda)$ with the naturally induced action of the algebra $A$.

\section{Standard and co-standard modules} \label{standard}

We specialize now to the various Temperley--Lieb algebras, beginning first with the affine Temperley--Lieb algebra $T^{\text a}_N(m)$. Although it is not cellular, we are interested in some of its finite-dimensional representations, which turn out to be the cell modules of $JTL^{\text{au}}_N(m)$ and $JTL_N(m)$. Define $\mathfrak q$, a quantum group parameter, via $m = \mathfrak q + \mathfrak q^{-1}$. For the generic case where $\mathfrak q$ is not a root of unity, there are irreducible representations $\mathscr W_{j,\e^{\i\phi}}$ parametrized by two numbers. (The number of sites $N$ is assumed fixed and does not appear in the notation for the representations $\mathscr W_{j,\e^{\i\phi}}$, although, strictly speaking, there is a different representation for each value of $N$.)

In terms of diagrams, the first is the number of through-lines $2j$, with $j = 1,\ldots,L$; recall that $N = 2L$ is assumed even (Section \ref{generators}, discussion surrounding Eq.~\eqref{eq:TL_translation_2}). Consider a particular diagram with $2j$ through-lines within this representation. Anticipating that the parameter $j$ will be involved in the set of weights $\Lambda$, whenever the natural action of the algebra---the stacking or nesting of diagrams in Section \ref{generators}---decreases the number of through-lines, we stipulate that the result is zero; i.e., whenever the action contracts two or more free sites. Compare with Definition \ref{cell_module}, where all of the basis elements of a cell module have the same weight. 

Furthermore, for a given value of $j$, it is possible, using the action of the algebra, to cyclically permute the free sites: this gives rise to the introduction of a pseudomomentum, which is parametrized by $\phi$. Whenever a diagram has $2j$ through-lines winding counterclockwise around the annulus $l$ times, this diagram may be replaced by one with the same connections but with the through-lines unwound, multiplied by a numerical factor $\e^{\i j l \phi}$. Stated more simply, there is a phase $\e^{\i\phi/2}$ for each complete counterclockwise winding of a through-line; however, it is impossible to unwind a single through-line on its own without colliding with other through-lines. Similarly, for $2j$ lines winding clockwise $l$ times, the numerical factor is $\e^{-\i j l \phi}$ when they are unwound, or $\e^{-\i\phi/2}$ per through-line per complete winding.

We will often use an equivalent representation, where the phase is evenly graduated over the course of unwinding through-lines, instead of abruptly applying it to each complete winding. Instead of a phase $\e^{\pm\i\phi/2}$ when a through-line makes a complete winding, we instead apply a phase $\e^{\pm\i\phi/2N}$ for each site a through-line moves right or left. This formulation preserves invariance under the translation operator.

A concrete visualization of this representation, $\mathscr W_{j,\e^{\i\phi}}$, can be obtained via the following consideration. Since the free sites are not allowed to be contracted, the pairwise connections between non-free sites on the inner boundary cannot be changed by the action of the algebra. This part of the diagrammatic information is thus extraneous and must be omitted to avoid overcounting the basis, and to ensure the representation is irreducible. It suffices to focus on the (isotopically distinct) upper halves of the affine diagrams, obtained by cutting each affine diagram across its $2j$ through-lines, and discarding the bottom half. Each upper half is then called a \emph{link state}, and for simplicity the ``half'' through-lines attached to the free sites on the outer boundary (or top boundary of the framing rectangle) are still called through-lines. The phase $\e^{\pm\i\phi/2}$ is now attributed each time one of these through-lines moves through the periodic boundary condition of the framing rectangle in the rightward or leftward direction, or, equivalently, $\e^{\pm\i\phi/2N}$ for each step right or left, as before. With these conventions, we will use this formulation to study the representations $\mathscr W_{j,\e^{\i\phi}}$, with the Temperley--Lieb algebra action obtained by stacking the affine diagrams on top of the link states.

The dimensions of these modules $\mathscr W_{j,\e^{\i\phi}}$ are then easily found by counting the link states. They are given by
\begin{equation} \label{eq:W_j_dimension}
\hat d_j = \binom{N}{N/2 + j}.
\end{equation}
Note that these dimensions do not depend on $\phi$, but representations with different $\e^{\i\phi}$ are not isomorphic. These cell modules $\mathscr W_{j,\e^{\i\phi}}$ are also known as \emph{standard modules}. (Again, we have abused terminology by calling them cell modules in the sense of Definition \ref{cell_module} even though $T^{\text a}_N(m)$ is not a cellular algebra---they are, however, cell modules of $JTL^{\text{au}}_N(m)$ and $JTL_N(m)$. We will thus use ``standard module'' henceforth.)

The standard modules $\mathscr W_{j,\e^{\i\phi}}$ are irreducible for generic values of $\mathfrak q$ and $\phi$. However, degeneracies appear whenever the following \emph{resonance criterion} is satisfied \cite{MartinSaleur1993,GrahamLehrer1998}:
\begin{equation} \label{eq:resonance}
\exists k \in \mathbb N^* : \e^{\i\phi} = \mathfrak q^{2j+2k}.
\end{equation}
The representation $\mathscr W_{j,\mathfrak q^{2j+2k}}$ then becomes reducible, and contains a submodule isomorphic to $\mathscr W_{j+k,\mathfrak q^{2j}}$. The quotient $\mathscr W_{j,\mathfrak q^{2j+2k}}/\mathscr W_{j+k,\mathfrak q^{2j}}$ is generically irreducible, with dimension
\begin{equation} \label{eq:W_j_irreducible_dimension}
\overbar d_j \equiv \hat d_j - \hat d_{j+k}.
\end{equation}

There are also standard representations for $j = 0$---i.e., no through-lines. There is no pseudomomentum, but representations are still characterized by a second parameter, which now specifies the weight given to noncontractible loops, which are not possible for $j > 0$. Parametrizing this weight as $z + z^{-1}$, the corresponding standard module is denoted $\mathscr W_{0,z^2}$. This module is isomorphic to $\mathscr W_{0,z^{-2}}$, so we denote both henceforth by $\mathscr W_{0,z^{\pm 2}}$. If we identify $z = \e^{\i\phi/2}$, the resonance criterion of Eq.~\eqref{eq:resonance} still applies.

It is natural to require that $z + z^{-1} = m$, so that contractible and noncontractible loops get the same weight. Imposing this requirement leads to the module $\mathscr W_{0,\mathfrak q^{\pm 2}}$, which is reducible even for generic $\mathfrak q$. Indeed, Eq.~\eqref{eq:resonance} is satisfied with $j = 0$, $k = 1$, and hence $\mathscr W_{0,\mathfrak q^{\pm 2}}$ contains a submodule isomorphic to $\mathscr W_{11}$. Taking the quotient $\mathscr W_{0,\mathfrak q^{\pm 2}}/\mathscr W_{11}$ leads to a simple module for generic $\mathfrak q$, denoted by $\overbar{\mathscr W}_{\!\!0,\mathfrak q^{\pm 2}}$. It has dimension
\begin{equation}
\overbar d_0 = \hat d_0 - \hat d_1 = \binom{N}{N/2} - \binom{N}{N/2 + 1},
\end{equation}
in agreement with Eqs.~\eqref{eq:W_j_dimension} and \eqref{eq:W_j_irreducible_dimension}, so that we may extend the definitions of both to $j = 0$.

The difference between $\mathscr W_{0,\mathfrak q^{\pm 2}}$ and $\overbar{\mathscr W}_{\!\!0,\mathfrak q^{\pm 2}}$ is the analogous to the difference between $JTL^{\text{au}}_N(m)$ and $JTL_N(m)$. In $\mathscr W_{0,\mathfrak q^{\pm 2}}$, curves that pair sites in the link states can cross the periodic ``boundary,'' while in $\overbar{\mathscr W}_{\!\!0,\mathfrak q^{\pm 2}}$ they cannot. (A curve pairing sites in $\mathscr W_{0,\mathfrak q^{\pm 2}}$ cannot cross the boundary more than once [on net]; otherwise the curve would intersect itself. The same is true for the subalgebra of $JTL^{\text{au}}_N(m)$ with zero through-lines.) There exists a projection mapping defining the quotient whose action on link states in $\mathscr W_{0,\mathfrak q^{\pm 2}}$ is analogous to that of Eq.~\eqref{eq:psi_ex} on diagrams in $JTL^{\text{au}}_N(m)$.

Some concrete examples of link states are, for $N = 4$:
\begin{equation}
\vcenter{\hbox{\begin{tikzpicture}
\newcommand{\dist}{0.2}
\draw[thick] (0,0) arc (-180:0:\dist);
\draw[thick] (4*\dist,0) -- (4*\dist,-2*\dist);
\draw[thick] (6*\dist,0) -- (6*\dist,-2*\dist);
\draw[thick, dotted] ($(current bounding box.north east) + (0.05+\dist,0.05)$) rectangle ($(current bounding box.south west)+ (-0.05-\dist,-0.05)$);
\end{tikzpicture}}}
\qquad
\vcenter{\hbox{\begin{tikzpicture}
\newcommand{\dist}{0.2}
\draw[thick] (4*\dist,0) arc (-180:0:\dist);
\draw[thick] (2*\dist,0) -- (2*\dist,-2*\dist);
\draw[thick] (0,0) -- (0,-2*\dist);
\draw[thick, dotted] ($(current bounding box.north east) + (0.05+\dist,0.05)$) rectangle ($(current bounding box.south west)+ (-0.05-\dist,-0.05)$);
\end{tikzpicture}}}
\qquad
\vcenter{\hbox{\begin{tikzpicture}
\newcommand{\dist}{0.2}
\draw[thick] (\dist,-\dist) arc (-90:0:\dist);
\draw[thick] (4*\dist,0) arc (-180:0:\dist);
\draw[thick] (9*\dist,-\dist) arc (-90:-180:\dist);
\draw[thick, dotted] ($(current bounding box.north east) + (0.05,0.05)$) rectangle ($(current bounding box.south west)+ (-0.05,-0.05-1*\dist)$);
\end{tikzpicture}}}
\qquad
\vcenter{\hbox{\begin{tikzpicture}
\newcommand{\dist}{0.2}
\draw[thick] (2*\dist,0) arc (-180:0:\dist);
\draw[thick] (0,0) arc (-180:0:3*\dist);
\draw[thick, dotted] ($(current bounding box.north east) + (0.05+\dist,0.05)$) rectangle ($(current bounding box.south west)+ (-0.05-\dist,-0.05)$);
\end{tikzpicture}}} 
\end{equation}
In $\overbar{\mathscr W}_{\!\!0,\mathfrak q^{\pm 2}}$, the latter two diagrams are identified. Two examples of link states for $N = 6$ are:
\begin{equation}
\vcenter{\hbox{\begin{tikzpicture}
\newcommand{\dist}{0.2}
\draw[thick] (2*\dist,0) arc (-180:0:\dist);
\draw[thick] (0,0) arc (-180:0:3*\dist);
\draw[thick] (8*\dist,0) -- (8*\dist,-3*\dist);
\draw[thick] (10*\dist,0) -- (10*\dist,-3*\dist);
\draw[thick, dotted] ($(current bounding box.north east) + (0.05+\dist,0.05)$) rectangle ($(current bounding box.south west)+ (-0.05-\dist,-0.05)$);
\end{tikzpicture}}} 
\qquad
\vcenter{\hbox{\begin{tikzpicture}
\newcommand{\dist}{0.2}
\draw[thick] (2*\dist,0) arc (-180:0:\dist);
\draw[thick] (6*\dist,0) arc (-180:0:\dist);
\draw[thick] (0,0) -- (0,-2*\dist);
\draw[thick] (10*\dist,0) -- (10*\dist,-2*\dist);
\draw[thick, dotted] ($(current bounding box.north east) + (0.05+\dist,0.05)$) rectangle ($(current bounding box.south west)+ (-0.05-\dist,-0.05)$);
\end{tikzpicture}}}
\end{equation}

At this point I introduce the following convenient notation for link states, useful both as a compact notation and for computation (Appendix \ref{mathematica}). A link state is represented by a collection of ordered pairings and singletons. The explicit notation of singletons is optional, and a singleton $(i)$ represents a through-line at site $i$. A pairing $(ij)$ or $(i,j)$ denotes a curve that starts from site $i$ and goes forward to $j$ in the rightward direction. When $i > j$ such a curve starts from site $i$, goes past the last site $N$ and crosses the boundary, returns periodically from behind site $1$, then terminates at site $j$. For the six link states exhibited graphically, their representations in my notation are $(12)(3)(4)$, $(1)(2)(34)$, $(23)(41)$, $(14)(23)$, $(14)(23)(5)(6)$, and $(1)(23)(45)(6)$. Since singletons are optional, they may well be denoted $(12)$, $(34)$, $(23)(41)$, $(14)(23)$, $(14)(23)$, and $(23)(45)$. Of course, $N$ must be understood from context to avoid ambiguities, such as the two instances of $(14)(23)$, but I will point out that exactly the same ambiguity arises in the representations of elements of the symmetric group $S_N$ by disjoint cycles when fixed points are omitted. And, as with $S_N$, the order of the pairings and singletons is immaterial, though the order of sites within a pairing is important. In this notation, the action of $e_j$ on a state with singletons $(j)(j+1)$ is $e_j(j)(j+1)(\cdots) = (j,j+1)(\cdots)$, where $(\cdots)$ stands for all other pairs, and $e_j(j,j+1)(\cdots) = m(j,j+1)(\cdots)$. The action of $e_j$ on a general state can be reduced to several cases (Appendix \ref{mathematica}). When using this notation, one must also be clear about the phase incurred when through-lines wind under the action of the algebra---the pairing notation does not expressly incorporate the parameter $\e^{\i\phi}$.

\section{$\mathscr W_{0,\mathfrak q^{\pm 2}}$ and indecomposability} \label{indecomposability}
The standard module $\mathscr W_{0,\mathfrak q^{\pm 2}}$ most simply illustrates the indecomposability we study in detail in Chapters \ref{Virasoro} and \ref{Jordan_loop}. Consider the case $N = 2$---i.e., the loop formulation of the Potts model for a two-site system (see Section \ref{Potts_description}), in the sector with no through-lines and with noncontractible loops given the same weight $m = \mathfrak q + \mathfrak q^{-1}$ as contractible ones.

Let us first write the two elements of the Temperley--Lieb algebra in the basis of the two link states $v_1 = (12)$ and $v_2 = (21)$ (pictorially, $\loopU$\, and $\loopJL$\,):
\begin{subequations}
\begin{gather}
e_1 = m \begin{pmatrix}
1 & 1 \\
0 & 0
\end{pmatrix}, \qq{and} \\
e_2 = m \begin{pmatrix}
0 & 0 \\
1 & 1 \end{pmatrix}.
\end{gather}
\end{subequations}
Clearly, $e_1(v_1 - v_2) = e_2(v_1 - v_2) = 0$. Meanwhile, at $N = 2$ the action of $e_1$ and $e_2$ on the single state $(1)(2)$ (pictorially, $\loopII$\,) in $\mathscr W_{11}$ is zero by definition, since the number of through-lines would decrease. Thus we see that $\mathscr W_{0,\mathfrak q^{\pm 2}}$ admits a submodule, generated by $v_1 - v_2$, that is isomorphic to $\mathscr W_{11}$. Diagrammatically, using what is technically called a \emph{Loewy diagram}, we have
\begin{equation} \label{eq:standard}
\mathscr W_{0,\mathfrak q^{\pm 2}}: \begin{tikzcd}
\overbar{\mathscr W}_{\!\!0,\mathfrak q^{\pm 2}} \arrow[d] \\
\mathscr W_{11} 
\end{tikzcd}.
\end{equation}
In such a diagram, the bottom module is a submodule, while the top module is a quotient module. The arrow indicates that within the standard module $\mathscr W_{0,\mathfrak q^{\pm 2}}$ a state in $\mathscr W_{11}$ can be reached from a state in $\overbar{\mathscr W}_{\!\!0,\mathfrak q^{\pm 2}}$ through the action of the Temperley--Lieb algebra, but the opposite is impossible.

\section{Cell modules for root-of-unity values of $\mathfrak q$} \label{root_of_unity}

When $\mathfrak q$ is a root of unity, there are infinitely many solutions $k$ to Eq.~\eqref{eq:resonance}, leading to a complex pattern of degeneracies. Now specializing to the Jones--Temperley--Lieb (JTL) algebras $JTL^{\text{au}}_N(m)$ and $JTL_N(m)$, the rule that winding through-lines can simply be unwound (from the quotient by $\tau^N - 1$) means that the pseudomomentum must satisfy \cite{Jones1994} 
\begin{equation} \label{eq:momentum_quantization}
j\phi \equiv 0 \pmod{2\pi}.
\end{equation}
All possible values of the parameter $z^{\pm 2} = \e^{\pm\i\phi}$ are thus $j$th roots of unity, satisfying $z^{2j} = 1$. The kernel of the homomorphism $\psi$ described by Eq.~\eqref{eq:psi_ex} acts trivially on the standard modules if $j > 0$. The standard modules $\mathscr W_{j,\e^{\i\phi}}$ and $\overbar{\mathscr W}_{\!\!0,\mathfrak q^{\pm 2}}$ are thus the cell modules of $JTL^{\text{au}}_N(m)$ and $JTL_N(m)$ for generic $\mathfrak q$.

In the terminology of cellular algebras, the basis $C^\lambda_{ij}$ consists of the annular diagrams, and the anti-involution $*$ is an inversion of the annulus (or a reflection of the framing rectangle about its horizontal axis). The set of weights $\Lambda$ consists of pairs $(j,\e^{\i\phi})$ satisfying Eq.~\eqref{eq:momentum_quantization} and $(0,\mathfrak q^2)$:
\begin{equation}
\Lambda = \{(0,\mathfrak q^2)\} \cup \{(j,\e^{2\pi\i l/j}) | 1 \le j \le L, 0 \le l \le j-1\}.
\end{equation}
The finite sets $W(\lambda)$ and the basis of the cell representations can be identified with the link states with $2j$ through-lines, where $\lambda = (j,\e^{\i\phi})$ (to include $j = 0$). There is a partial order $\preceq$ on $\Lambda$ due to \textcite{GrahamLehrer1998}: first, $(j_1, k_1) \preceq (j_2, k_2)$ if $j_1 \le j_2$ as integers and
\begin{equation}
k_1 = \mathfrak q^{2\epsilon j_2} \qq{and} k_2 = \mathfrak q^{2\epsilon j_1}, \qquad (\epsilon = \pm 1)
\end{equation}
then $\preceq$ is extended to a partial order by transitivity.

Since not all pairs of weights in $\Lambda$ are comparable, the partial order $\preceq$ induces equivalence classes in $\Lambda$---two weights are in the same equivalence class if and only if there exists a weight to which both are comparable. A result of \citeauthor{GrahamLehrer1998} [\emph{Id.}] is that there exist nontrivial homomorphisms only between cell modules with weights from the same equivalence class. There are nontrivial classes containing two or more weights only when $\mathfrak q$ is a root of unity. Then the cell modules $\mathscr W_{j,\e^{\i\phi}}$ whose weights belong to a nontrivial class are indecomposable, except for weights that are maximal with respect to $\preceq$. Denote the top simple subquotient---the quotient by its maximal submodule---by $[j,\e^{\i\phi}]$. Using the partial order, the simple-module content of these cell modules can be deduced [\emph{Id.}]:
\begin{equation}
\mathscr W_{j,k} \simeq \bigoplus_{\substack{(j',k')\in\Lambda \\ (j,k)\preceq(j',k')}} [j',k'],
\end{equation}
where $\simeq$ denotes equality as vector spaces, but not as $JTL^{\text{au}}_N(m)$-modules, as the action of the algebra connects simple modules with differing weights. $JTL^{\text{au}}_N(m)$ is thus non-semisimple.

Hereafter in this work, we will only be concerned with the two Jones--Temperley--Lieb algebras $JTL^{\text{au}}_N(1)$ and $JTL_N(1)$, with $m = 1$. (For the affine Temperley--Lieb algebra, we consider the range $m \in (-2,2]$.)
With $m = 1$, $\mathfrak q = \e^{\i\pi/3}$, and we may now draw the corresponding Loewy diagram for $\mathscr W_{0,\mathfrak q^2}$:
\begin{equation} \label{eq:W_0_structure}
\mathscr W_{0,\mathfrak q^2}: 
\begin{tikzcd}
 & {[0,\mathfrak q^2]}\arrow[dl]\arrow[dr] & \\
{[1,1]}\arrow[to=3-3]\arrow[d] & & {[2,1]}\arrow[to=3-1]\arrow[d] \\
{[3,\mathfrak q^2]} \arrow[to=4-3]\arrow[d] & & {[3,\mathfrak q^{-2}]}\arrow[to=4-1]\arrow[d] \\
{[4,1]}\arrow[to=5-3]\arrow[d] & & {[5,1]}\arrow[to=5-1]\arrow[d] \\
{[6,\mathfrak q^2]} \arrow[to=6-3]\arrow[d] & & {[6,\mathfrak q^{-2}]}\arrow[to=6-1]\arrow[d] \\
\mathord{\makebox[\widthof{\([6,\mathfrak q^2]\)}]{\vdots}} & & \mathord{\makebox[\widthof{\([6,\mathfrak q^2]\)}]{\vdots}}
\end{tikzcd}
\end{equation}
Of course, for finite $L$, the tower terminates. The structure at the bottom depends on the value of $L\ \mathrm{mod}\ 3$, and these cases are exhibited separately by \textcite{Gainutdinov2015} (in which may also be found detailed diagrams of the partial order $\preceq$ that leads to the above Loewy diagram). Other modules $\mathscr W_{j,\e^{\i\phi}}$ whose weights appear as simple modules $[j,\e^{\i\phi}]$ in Eq.~\eqref{eq:W_0_structure} are given by diagrams that ``emanate'' from the corresponding simple modules, and which are identical in structure (except for the length of the tower) but with different labels at each node.

\chapter{Physics: basic concepts}

This chapter summarizes the basic concepts of physics relevant to the rest of the work, and exhibits two physics problems to motivate the theoretical study of the models therein. Most of the general material is assumed to be familiar to the reader, and serves to prime the reader for the remainder of the work. Although I am rather concise in this section, nothing presented here should be controversial. The reader will have their own preferred sources for this material, and I draw from \textcite{Shankar1994,NielsenChuang2016,Kardar2007,DFMS1997,Cardy1996}.

\section{Quantum and statistical physics} \label{quantum_physics}

To oversimplify matters, quantum physics is a framework for physical theories that attempt to explain the behavior of physical systems at a small scale where classical physics fails. There are several equivalent formulations of the postulates of quantum physics.

\sectionbreak

\noindent\textbf{Postulates of quantum physics (Schr\"odinger formulation)}
\begin{enumerate}
\item The state of a system is described by an element $\Psi$ in a Hilbert space with a positive-definite inner product.

\item Observables are represented by hermitian operators. For quantum observables $A$ and $B$ that have classical analogues $\alpha$ and $\beta$, $[A,B] = \i\hbar\{\alpha,\beta\}$, where $[A,B]$ is the commutator of $A$ and $B$, and $\{\alpha,\beta\}$ is the classical Poisson bracket of $\alpha$ and $\beta$.

\item The measurement of an observable $\Omega$ yields one of its the eigenvalues, $\omega$. If the system is in the state $\Psi$, then $P(\omega) = \mel*{\Psi}{\Pi_\omega}{\Psi}/\ip{\Psi}$, where $\Pi_\omega$ is the projection operator onto the eigenspace associated to $\omega$, is the probability of obtaining the value $\omega$ if $\omega$ is part of the discrete spectrum of $\Omega$, or the probability density at $\omega$ if $\omega$ is part of the continuous spectrum.

\item $\Psi$ evolves in time according to
\begin{equation} \label{eq:schrodinger}
\i\hbar\pdv{\Psi}{t} = H\Psi,
\end{equation}
where $H$ is the Hamiltonian operator for the system.

\end{enumerate}

\noindent\textbf{Postulates of quantum physics (Heisenberg formulation)}
\begin{enumerate}
\item The state of a system is described by an element $\Psi$ in a Hilbert space with a positive-definite inner product.

\item Observables are represented by hermitian operators. For quantum observables $A$ and $B$ that have classical analogues $\alpha$ and $\beta$, $[A,B] = \i\hbar\{\alpha,\beta\}$, where $[A,B]$ is the commutator of $A$ and $B$, and $\{\alpha,\beta\}$ is the classical Poisson bracket of $\alpha$ and $\beta$.

\item The measurement of an observable $\Omega$ yields one of its the eigenvalues, $\omega$. If the system is in the state $\Psi$, then $P(\omega) = \mel*{\Psi}{\Pi_\omega}{\Psi}/\ip{\Psi}$, where $\Pi_\omega$ is the projection operator onto the eigenspace associated to $\omega$, is the probability of obtaining the value $\omega$ if $\omega$ is part of the discrete spectrum of $\Omega$, or the probability density at $\omega$ if $\omega$ is part of the continuous spectrum.

\item The operators $\Omega$ evolve in time according to
\begin{equation}
\i\hbar\pdv{\Omega}{t} = [\Omega,H],
\end{equation}
where $H$ is the Hamiltonian operator for the system.

\end{enumerate}

\noindent\textbf{Postulates of quantum physics (von Neumann formulation)}
\begin{enumerate}
\item The state of a system is described by a positive hermitian \emph{density operator} $\rho$ acting on a Hilbert space with a positive-definite inner product. It satisfies $\tr\rho = 1$.

\item Observables are represented by hermitian operators. For quantum observables $A$ and $B$ that have classical analogues $\alpha$ and $\beta$, $[A,B] = \i\hbar\{\alpha,\beta\}$, where $[A,B]$ is the commutator of $A$ and $B$, and $\{\alpha,\beta\}$ is the classical Poisson bracket of $\alpha$ and $\beta$.

\item If an observable $\Omega$ is measured many times in the state $\rho$, the average (expected) value of these measurements is $\mathrm E[\Omega] = \tr(\Omega\rho)$.

\item The density operator $\rho$ evolves in time according to
\begin{equation}
\i\hbar\pdv{\rho}{t} = [H,\rho],
\end{equation}
where $H$ is the Hamiltonian operator for the system.

\end{enumerate}

\sectionbreak

Which formulation one ultimately chooses as an analytical and computational framework is largely a matter of convenience. For instance, the Heisenberg formulation is largely used in quantum field theory as the only state one really treats explicitly is the ground state, and the von Neumann formulation is tailor-made for mixed ensembles. Note, however, that in each of these formulations the postulate of observables being represented by hermitian operators remains the same. A rich seam of physics is uncovered when this assumption is lifted, and much of the rest of this work is devoted to the beginnings of a systematic treatment of quantum physics with this postulate appropriately replaced.

A unifying element for all of the three formulations presented is the propagator $U$. Using the propagator to advance arbitrary states forward in time,
\begin{equation}
\Psi(t) = U(t)\Psi(0),
\end{equation}
$U$ itself satisfies Eq.~\eqref{eq:schrodinger} with initial condition $U(0) = I$. For a time-independent Hamiltonian, which encompasses many cases of interest, and an assumption retained throughout this work, one has
\begin{equation}
U(t) = \e^{-\i H t/\hbar}.
\end{equation}
By considerations involving the independence of $\mel*{\Psi(t)}{\Omega(t)}{\Psi(t)}$ and $\tr[\Omega(t)\rho(t)]$ across different formulations of quantum physics, we have similarly
\begin{gather}
\Omega(t) = U^\dagger(t)\Omega(0)U(t) \qq{and} \\
\rho(t) = U(t)\rho(0)U^\dagger(t)
\end{gather}
in their respective formulations. Each of these expressions is simplified when $U$ is expressed in terms of the projection operators associated with the spectral decomposition $H = \sum_i E_i \Pi_i$:
\begin{equation}
U(t) = \sum_j \e^{-\i E_j t/\hbar}\Pi_j.
\end{equation}
These facts show that the primary task in any quantum mechanical problem is essentially to find the spectral decomposition of the Hamiltonian.

Finally, an alternative expression for the propagator is given in terms of Feynman's path integral (or functional integral). If $q$ generically represents the degrees of freedom for the system, then the matrix elements of the propagator are
\begin{equation}
\braket*{q_f(t)}{q_i(0)} = \mel*{q_f}{U(t)}{q_i} = \int_{(q_i,0)}^{(q_f,t)} \! [\d q]\, \e^{\i S[q]/\hbar}.
\end{equation}
The integration is over all paths in phase space starting at $q_i$, ending at $q_f$, and taking place over the time interval $[0,t]$; $S[q]$ is the (classical) action associated with such a parametrized path. The integration measure may be written
\begin{equation} \label{eq:dq_measure}
[\d q] = \lim_{N\to\infty}\sqrt{\frac{Nm}{2\pi\i \hbar t}}\left(\prod_{j=1}^{N-1} \sqrt{\frac{Nm}{2\pi\i\hbar t}}\,\d q_j\right),
\end{equation}
where $q_j = q(jt/N)$.

Quantum field theory is a quantum mechanical description of systems whose basic degrees of freedom are fields $\phi$, rather than point particles. Once a Lagrangian is specified, the path integral provides the quickest generalization for the transition amplitudes:
\begin{equation}
\braket*{\phi_f(q, t_f)}{\phi_i(q, t_i)} = \int\![\d\phi(q,t)] \e^{\i S[\phi]}.
\end{equation}
In quantum field theory the terms ``field'' and ``operator'' are used nearly synonymously. In the path integral formalism, the fields appearing in a path integral take on all allowed classical field configurations, possibly subject to constraints in the integration measure. In the canonical quantization of fields, which more closely parallels the postulates discussed at the beginning, the fields become operator-valued distributions. Both formalisms are useful and provide complementary theoretical perspectives, and so I will also use these terms interchangeably in the contexts where they are commonly used.

A primary quantity of interest is the correlation function, as it is directly related to the experimentally measurable scattering amplitudes. For a point particle, the $n$-point correlation function is
\begin{equation}
\ev*{q(t_1)q(t_2)\cdots q(t_n)} = \ev*{\mathrm T[q(t_1)q(t_2)\cdots q(t_n)]}{0}.
\end{equation}
In this equation, $\ket*{0}$ is the ground state, and $\mathrm T$ the time-ordering operator. The time-ordering operator may be avoided by using the path integral formalism:
\begin{equation}
\ev*{q(t_1)q(t_2)\cdots q(t_n)} = \lim_{\epsilon\to 0} \frac{\displaystyle \int \! [\d q]\, q(t_1)q(t_2)\cdots q(t_n)\e^{\i S_\epsilon[q]/\hbar}}{\displaystyle \int \! [\d q]\, \e^{\i S_\epsilon[q]/\hbar}},
\end{equation}
where $S_\epsilon$ is the action obtained by replacing $t$ by $t(1-\i\epsilon)$. This ``$\epsilon$ prescription'' is essential in proving the equality of the two expressions given for the $n$-point function. It is useful to push this prescription further, and define correlation functions entirely in imaginary time $t = -\i\tau$ ($\tau\in\mathbb R$). In doing so, the action in real time becomes the Euclidean action in imaginary time: $\i S[q(t\to-\i\tau)] = -S_E[q(\tau)]$. The change to imaginary time also causes the kinetic energy term in the Lagrangian to change sign, becoming instead the (real-time) Hamiltonian. This Euclidean formalism is used hereafter, switching $\tau$ back to $t$ and dropping the subscript $E$.

Consider a correlation function $\ev*{\phi_i(z_i)\phi_j(z_j)\Phi}$, where $\Phi$ denotes an arbitrary product of operators far away from $z_i$ and $z_j$. Invoking ideas of locality and completeness of the operators, the \emph{operator product expansion} (OPE) postulates that this correlation function may be replaced by a sum of the form
\begin{equation}
\ev*{\phi_i(z_i)\phi_j(z_j)\Phi} = \sum_k C_{ijk}(z_i - z_j)\ev*{\phi_k(z_j)\Phi}.
\end{equation}
Importantly, the coefficients $C_{ijk}$ do not depend on the arbitrary product $\Phi$. An OPE such as this one is frequently written without the expectation value and the arbitrary fields as
\begin{subequations} \label{eq:OPE_basic}
\begin{gather}
\phi_i(z_i)\phi_j(z_j) = \sum_k C_{ijk}(z_i - z_j)\phi_k(z_j) \qq{or} \\
\phi_i(z_i)\phi_j(z_j) \sim \sum_k C_{ijk}(z_i - z_j)\phi_k(z_j),
\end{gather}
\end{subequations}
it being understood that both sides of an OPE appear within a correlation function, possibly multiplied by other fields. The equalities do not make sense on their own, since, for instance, in the path integral formalism, the fields must be allowed to take all possible values independently. While the existence of OPEs is axiomatic for generic quantum field theories, they may be formalized in special cases via vertex algebras and vertex operator algebras, particularly for two-dimensional conformal field theory.

\sectionbreak

Statistical physics is a framework of physics that attempts to infer macroscopic properties of complex physical systems with only knowledge of the Hamiltonian in terms of microscopic degrees of freedom. From Boltzmann's ergodic hypothesis and Laplace's suggestion that one should apply a uniform distribution to unknown events according to the ``principle of insufficient reason,'' one obtains the Boltzmann distribution for a system in thermal contact with its surroundings at temperature $T$---the probability that the microscopic degrees of freedom are in the state $i$ is proportional to $\e^{-\beta E_i}$:
\begin{equation}
p_i = \frac{\e^{-\beta E_i}}{Z},
\end{equation}
where $E_i$ is the energy of the system in configuration $i$ and $\beta = 1/kT$. As the total probability that the system be found in any state is $1$, the normalizing factor $Z$, the partition function, must be
\begin{equation} \label{eq:Z_sum}
Z = \sum_i \e^{-\beta E_i}.
\end{equation}
The sum may be discrete or continuous according to the nature of the states describing the system. The preceding expression holds whether the system is described using classical or quantum physics. For a quantum system, a convenient expression that compactly incorporates the sum is
\begin{equation}
Z = \tr \e^{-\beta H}.
\end{equation}
The primary goal of statistical physics is the calculation of the partition function, as all macroscopic quantities may be derived from it.

Suppose our system is quantum-mechanical, with degrees of freedom $q$. The trace in the partition function may be written as
\begin{equation}
Z = \int\!\d q\,\mel*{q}{\e^{-\beta H}}{q} = \int_{(q,0)}^{(q,\beta\hbar)}\![\d q]\,\e^{-S[q(t)]}.
\end{equation}
Here, we have used the fact that the propagator $U = \e^{-\i H t/\hbar}$ becomes $\e^{-Ht/\hbar}$ in the Euclidean formalism, allowing us to identify $\beta$ with $t/\hbar$ and use the path integral.

This analogy may be generalized to a system with a continuum of degrees of freedom. In all instances, the partition function of a $d$-dimensional quantum system may be written as a path integral where time is imaginary and takes on the character of a spatial variable, thus giving rise to a $(d+1)$-dimensional classical system. This equivalence underlies the study of two-dimensional classical problems, such as percolation and the two-dimensional Potts model, using one-dimensional quantum systems.

\section{Conformal field theory}

A conformal field theory (CFT) is a quantum field theory with conformal symmetry. Systems with conformal symmetry are invariant, or covariant, under conformal transformations. Conformal transformations are essentially scale transformations combined with rotations, with scale factors and rotations that may vary smoothly from point to point. They thus preserve the angles between two arbitrary curves crossing at some point. In two dimensions, conformal invariance takes a new meaning. Any analytic mapping of the complex plane onto itself is conformal, and thus furnishes a local conformal transformation. It is this local conformal invariance that allows for exact solutions to two-dimensional CFTs, which will be the only CFTs we consider hereafter. Because of the close connection between analytic and conformal mappings, we use complex coordinates $z$ and $\overbar z$ on the complex plane.

To each field, we may associate a \emph{holomorphic conformal dimension} (or holomorphic conformal weight) $h$ and an \emph{antiholomorphic conformal dimension} (antiholomorphic conformal weight) $\overbar h$. Alternative terminology for ``holomorphic'' includes ``left'' and ``chiral,'' and alternative terminology for ``antiholomorphic'' includes ``right'' and ``antichiral.'' If, under an \emph{invertible} conformal transformation $z \to w(z)$ and $\overbar z \to \overbar w(\overbar z)$, a field $\phi$ transforms as
\begin{equation}
\phi(z,\overbar z) \to \left(\frac{\d w}{\d z}\right)^{-h}\left(\frac{\d\overbar w}{\d\overbar z}\right)^{-\overbar h}\phi(z,\overbar z),
\end{equation}
$\phi$ is called \emph{quasi-primary}. If $\phi$ transforms the same way under \emph{arbitrary} conformal maps, it is called \emph{primary}. This behavior fully determines the form of two- and three-point functions:
\begin{subequations}
\begin{gather}
\ev*{\phi_1(z_1,\overbar z_1)\phi_2(z_2,\overbar z_2)} = \begin{cases} \frac{\displaystyle c_{12}}{\displaystyle z_{12}^{2h}\overbar z_{12}^{2\overbar h}} & h_1 = h_2 = h\qq{and} \overbar h_1 = \overbar h_2 = \overbar h \\[3ex]
0 & \text{otherwise} \end{cases}, \\
\ev*{\phi_1(z_1,\overbar z_1)\phi_2(z_2,\overbar z_2)\phi_3(z_3, \overbar z_3)} = \frac{c_{123}}{z_{12}^{h_1 + h_2 - h_3} z_{23}^{h_2 + h_3 - h_1} z_{13}^{h_3 + h_1 - h_2}\overbar z_{12}^{\overbar h_1 + \overbar h_2 - \overbar h_3} \overbar z_{23}^{\overbar h_2 + \overbar h_3 - \overbar h_1} \overbar z_{13}^{\overbar h_3 + \overbar h_1 - \overbar h_2}},
\end{gather}
\end{subequations}
where $(h_i, \overbar h_i)$ are the left and right conformal weights of $\phi_i$, $z_{ij} = z_i - z_j$, and $\overbar z_{ij} = \overbar z_i - \overbar z_j$. The coefficient $c_{123}$ is closely associated to the function $C_{123}$ (in these notations) from the OPE in Eq.~\eqref{eq:OPE_basic}. Four-point functions are not so uniquely constrained, since there are \emph{anharmonic ratios} of four points that are conformally invariant. In general, the four-point function has the form
\begin{equation}
\ev*{\phi_1(z_1,\overbar z_1)\phi_2(z_2,\overbar z_2)\phi_3(z_3, \overbar z_3)\phi_4(z_4, \overbar z_4)} = f\Big(\frac{z_{12}z_{34}}{z_{13}z_{24}},\frac{\overbar z_{12}\overbar z_{34}}{\overbar z_{13}\overbar z_{24}}\Big)\prod_{\substack{i,j = 1 \\ i<j}}^4 z_{ij}^{h/3 - h_i - h_j}\overbar z_{ij}^{\overbar h/3 - \overbar h_i - \overbar h_j},
\end{equation}
where $h = \sum_{i=1}^4 h_i$ and $\overbar h = \sum_{i=1}^4 \overbar h_i$, and $f$ is a function not determined by conformal invariance. Four-point functions may be reduced to sums of three-point functions using the OPE; one frequently encounters four-point functions given by ``sums over conformal blocks,'' which encapsulates this process. If one assumes that the conformal blocks are constrained by ``crossing symmetry,'' a natural assumption, then they can be found, in principle. The method of computing conformal blocks by simply assuming crossing symmetry is called the \emph{bootstrap approach}. The bootstrap approach must be validated by checking self-consistency conditions, since it may potentially over-constrain the theory. Numerical studies may also be used to corroborate the conclusions of the bootstrap approach (Chapters \ref{physics_lattice}, \ref{Virasoro}, and \ref{Jordan_loop}).

\emph{Ward identities} express the consequence of symmetries on the correlation functions of field theories. The most important Ward identity involves the \emph{stress--energy tensor} (or energy--momentum tensor) $T$ and its antiholomorphic counterpart $\overbar T$ (more precisely, $T$ and $\overbar T$ are fields, but normalized components of a genuine rank-2 tensor), here written as OPEs:
\begin{subequations} \label{eq:Ward_identity}
\begin{gather}
T(z)X = \sum_{i=1}^n \left[\frac{h_iX}{(z-w_i)^2} + \frac{\partial_{w_i}X}{z-w_i}\right] + \text{reg.} \qq{and} \\
\overbar T(\overbar z)X = \sum_{i=1}^n \left[\frac{\overbar h_iX}{(\overbar z-\overbar w_i)^2} + \frac{\partial_{\overbar w_i}X}{\overbar z-\overbar w_i}\right] + \text{reg.},
\end{gather}
\end{subequations}
where $X = \phi_1(w_1,\overbar w_1)\cdots \phi_n(w_n,\overbar w_n)$ and ``reg.'' stands for a holomorphic function of $z$ (antiholomorphic function of $\overbar z$). An infinitesimal conformal map $z \to z + \epsilon(z)$, $\overbar z \to \overbar z + \overbar\epsilon(\overbar z)$ gives the following \emph{conformal Ward identity}:
\begin{equation}
\delta_{\epsilon,\overbar\epsilon}X = -\frac{1}{2\pi\i}\oint_C\!\d z\,\epsilon(z)T(z)X + \frac{1}{2\pi\i}\oint_C\!\d\overbar z\,\overbar\epsilon(\overbar z)\overbar T(\overbar z)X,
\end{equation}
where the contour $C$ encloses all of the points $(w_i,\overbar w_i)$. This expression does not require the fields in $X$ to be primary, merely local.

The conformal Ward identity establishes $T$ and $\overbar T$ as the generators of conformal transformations of quantum fields. They may be expanded in modes:
\begin{subequations}
\begin{gather}
T(z) = \sum_{n=-\infty}^\infty z^{-n-2} L_n \\
\overbar T(\overbar z) = \sum_{n=-\infty}^\infty \overbar z^{-n-2} \overbar L_n.
\end{gather}
\end{subequations}
The mode generators $L_n$ obey the \emph{Virasoro algebra}:
\begin{subequations} \label{eq:virasoro_algebra}
\begin{gather}
[L_m, L_n] = (m-n)L_{m+n} + \frac{c}{12}m(m^2 - 1)\delta_{m,-n}, \\
[L_m, \overbar L_n] = 0, \\
[\overbar L_m, \overbar L_n] = (m-n)\overbar L_{m+n} + \frac{c}{12}m(m^2 - 1)\delta_{m,-n}.
\end{gather}
\end{subequations}
It is a direct sum of two identical algebras, the summands of which are also called the Virasoro algebra. To be explicit, we will call the algebra with generators $L_n$ and $\overbar L_n$ obeying Eq.~\eqref{eq:virasoro_algebra} $\mathrm{Vir}\oplus\overbar{\mathrm{Vir}}$. The bar on the right summand does not imply a modification, but stands as a reminder that it signifies the antiholomorphic counterpart to the left one. The generators $L_{-1}$ and $\overbar L_{-1}$ are often denoted by $\partial = \partial_z$ and $\overbar\partial = \partial_{\overbar z}$, as they have the same effect within OPEs and correlation functions. There is a subalgebra of $\mathrm{Vir}$ spanned by $L_{-1}$, $L_0$, and $L_1$; they generate the invertible conformal transformations.

The Virasoro algebra is the unique central extension of the Witt algebra. The coefficient $c$ in the central term is called the \emph{central charge}, and it is given by the following OPEs between $T$ and $\overbar T$:
\begin{subequations}
\begin{gather}
T(z)T(w) = \frac{c/2}{(z-w)^4} + \frac{2T(w)}{(z-w)^2} + \frac{\partial T(w)}{z-w}, \\
T(z)\overbar T(\overbar w) = 0, \\
\overbar T(\overbar z)\overbar T(\overbar w) = \frac{c/2}{(\overbar z-\overbar w)^4} + \frac{2\overbar T(\overbar w)}{(\overbar z-\overbar w)^2} + \frac{\overbar \partial \overbar T(\overbar w)}{\overbar z-\overbar w}.
\end{gather}
\end{subequations}
Thus $T$ and $\overbar T$ are not primary. However, they are quasi-primary, with conformal dimensions $(2,0)$ and $(0,2)$ (by comparison with Eq.~\eqref{eq:Ward_identity}).

Here, I collect some parameters and notations used ubiquitously in generic CFTs. The central charge $c$ that describes a given CFT comes from the central term in the Virasoro algebra commutation relations, or from the most singular term in the OPE of $T$ with itself. A parameter, $x$, is defined via
\begin{equation}
c = 1 - \frac{6}{x(x+1)}.
\end{equation}
The inverse relation, valid for $-\infty < c < 1$, is given by
\begin{equation}
x = -\frac{1-c\pm\sqrt{(1-c)(25-c)}}{2(1-c)}.
\end{equation}
The lower sign is taken so that $x > 0$. These relations will be used frequently to convert between $c$ and $x$ as independent variables without explicit mention. The \emph{Kac formula},
\begin{equation}
h_{rs} = \frac{[r(x+1)-sx]^2-1}{4x(x+1)},
\end{equation}
uses two labels to describe conformal weights. Particularly for integers $r$ and $s$, the Kac formula gives a compact representation of many conformal weights important to a given theory (specified by $c$).

Highest-weight representations of $\mathrm{Vir}$ are constructed by fixing a highest-weight state $\ket*{h}$ satisfying $L_0\ket*{h} = h\ket*{h}$ and $L_n\ket*{h} = 0$ for $n > 0$, then declaring the \emph{descendant states}
\begin{equation}
L_{-k_1}L_{-k_2}\cdots L_{-k_n}\ket*{h} \qquad (1 \le k_1 \le \cdots \le k_n)
\end{equation}
to be linearly independent. Each descendant state above corresponds to an eigenvalue $h' = h + k_1 + k_2 + \cdots + k_n$ of $L_0$ (or conformal weight $h'$), and a linear combination of descendant states all with conformal weights $h + N$ is called a \emph{descendant at level $N$}. These highest-weight representations are often called \emph{Verma modules}. An inner product may be defined on the Verma module by $\braket*{h}{h} = 1$ and $L_n^\dagger = L_{-n}$.

A representation of $\mathrm{Vir}$ is \emph{unitary} if it has no negative norm squared states. From
\begin{equation}
\mel*{h}{L_nL_{-n}}{h} = 2hn + \frac{c}{12}n(n^2-1)
\end{equation}
it follows that representations with $c < 0$ or highest weight $h < 0$ are nonunitary (these conditions are sufficient, but not necessary).
 
The structure of Verma modules over the Virasoro algebras has been determined by \textcite{FeiginFuchs1984}. For $c \le 1$, a condition that will hold for the remainder of the work, a reducible Verma module either admits a filtration by simple modules, or has a braid-type submodule structure. For $x \ge 2$ a positive integer, the Verma module shows the braid-type structure
\begin{equation}
\begin{tikzcd}
 & (r,s)\arrow[dl]\arrow[dr] & \\
(r,2x-s)\arrow[to=3-3]\arrow[d] & & (r+x-1,x-s)\arrow[to=3-1]\arrow[d] \\
(r,2x+s) \arrow[to=4-3]\arrow[d] & & (r+2(x-1),s)\arrow[to=4-1]\arrow[d] \\
\mathord{\makebox[\widthof{\((r,2x-s)\)}]{\vdots}}\arrow[to=5-3]\arrow[d] & & \mathord{\makebox[\widthof{\((r,2x-s)\)}]{\vdots}}\arrow[to=5-1]\arrow[d] \\
(r,kx+(-1)^ks+[1-(-1)^k]x/2) \arrow[to=6-3]\arrow[d] & &(r+k(x-1),(-1)^ks+[1-(-1)^k]x/2)\arrow[to=6-1]\arrow[d] \\
\mathord{\makebox[\widthof{\((r,2x-s)\)}]{\vdots}} & & \mathord{\makebox[\widthof{\((r,2x-s)\)}]{\vdots}}
\end{tikzcd}
\end{equation}
where each vertex $(p,q)$ represents a Verma module built on the highest-weight $h_{pq}$, and an arrow $A \to B$ means $B$ is a submodule of $A$, with transitivity implied. To obtain the corresponding Loewy diagram, replace the module $(p,q)$ by its simple subquotient, obtained by taking a quotient by the sum of its two maximal submodules indicated by the arrows. This structure may be compared with that of Eq.~\eqref{eq:W_0_structure}, showing that the JTL algebra is useful to study many structures of the continuum CFT that have analogues at finite size---the subject of Chapter \ref{physics_lattice}.

The modules $(p,q)$ at each vertex are generated by \emph{singular vectors}, which are particular zero-norm descendants of the field of highest weight that may be constructed systematically. The existence of singular vectors implies a (hypergeometric-type) differential equation for the undetermined function in the four-point function, whose solution may also be used as a check for the bootstrap approach.

\section{Two concrete physics problems}
\subsection{The quantum Hall effect}

Here I give the briefest possible introduction to the quantum Hall effect, which is now a huge enterprise in physics \cite{PrangeGirvin1987}. I follow \textcite{Tong2016} with the presentation condensed and inessential details swept under the rug.

\begin{figure}
\centering
\includegraphics[width=0.6\textwidth]{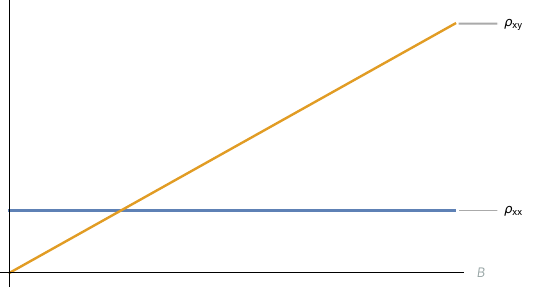}
\caption{The Hall effect.}
\label{classicalhall}
\end{figure}

From classical electromagnetism, we are familiar with Ohm's law (upgraded from the humble $V = IR$),
\begin{equation}
\rho\mathbf J = \mathbf E,
\end{equation}
where $\rho$ is the resistivity tensor, $\mathbf J$ is the current density, and $\mathbf E$ is the electric field. Now focus on two dimensions, the $xy$-plane, and put a magnetic field $\mathbf B = B\hat{\mathbf z}$ through the system. A calculation in the framework of classical electromagnetism predicts that
\begin{subequations}
\begin{gather}
\rho_{xx} = \frac{m}{ne^2\tau} \qq{and} \\
\rho_{xy} = \frac{B}{ne},
\end{gather}
\end{subequations}
where $m$ is the electron mass, $n$ is the density of free electrons, $\tau$ is a parameter that measures the average scattering time for electrons, and $e$ is the elementary electric charge. This is the essence of the (classical) Hall effect: an electric field in the $x$-direction gives rise to a current with a component in the $y$-direction in the presence of a magnetic field perpendicular to the system. The transverse resistivity $\rho_{xy}$ increases linearly with the strength of the magnetic field $B$ and $\rho_{xx}$ remains constant. Graphically, we expect something like Figure \ref{classicalhall}.

\begin{figure}
\centering
\includegraphics[width=0.6\textwidth]{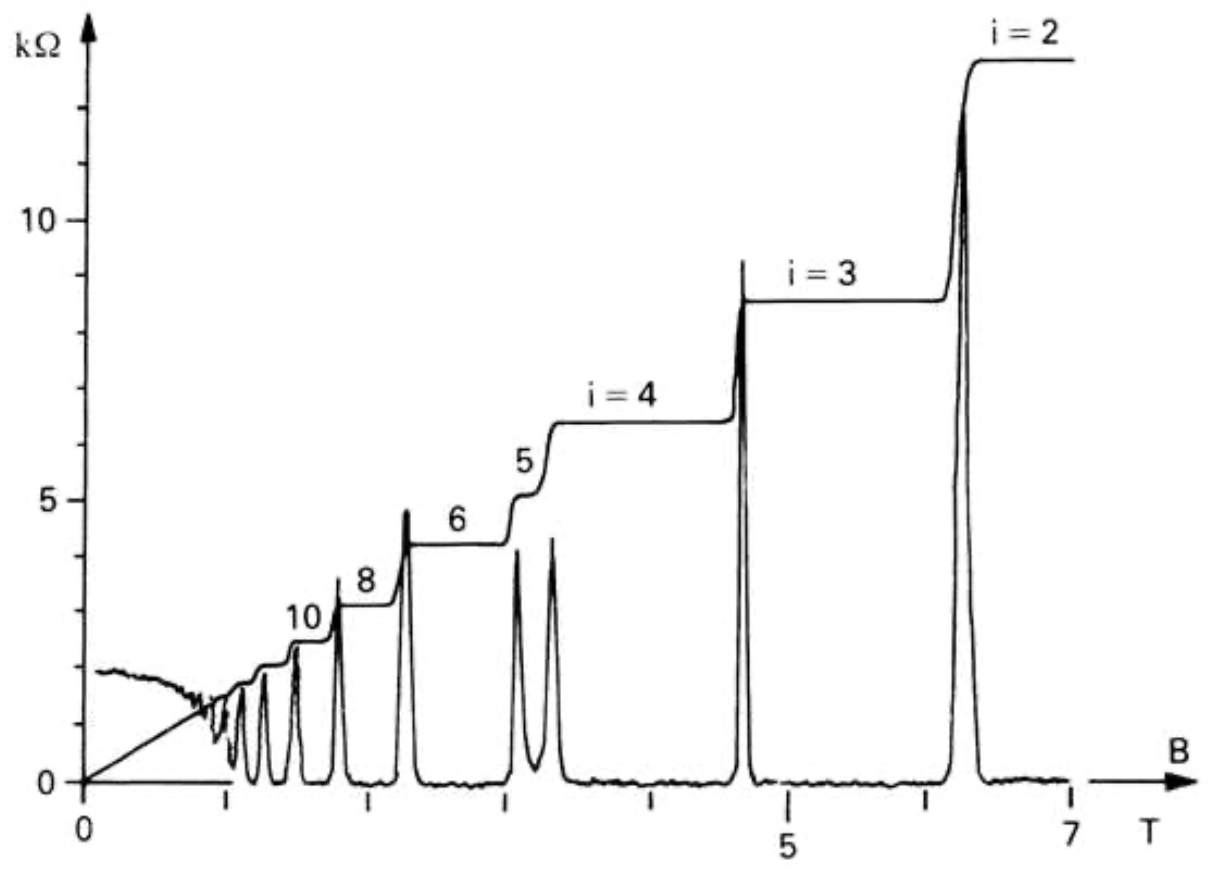}
\caption{The quantum Hall effect. From \textcite{Tong2016}.}
\label{quantumhall}
\end{figure}

At low temperatures and strong magnetic fields, quantum effects become important. Experimental measurements yield a graph that is shown in Figure \ref{quantumhall}. The staircase represents $\rho_{xy}$ and the noisy peaks are $\rho_{xx}$. This is the (integer) quantum Hall effect.

The most obvious feature is the existence of plateaux, intervals of magnetic field strengths which give the same resistivity within each. The resistivity plateaux have the values
\begin{equation}
\rho_{xy} = \frac{h}{\nu e^2},
\end{equation}
where $h$ is Planck's constant and $\nu$ has been measured to be an integer to an accuracy of 1 part in about $10^9$. At the center of each of these plateaux, the magnetic field is
\begin{equation}
B = \frac{nh}{\nu e}.
\end{equation}
Perhaps counterintuitively, the underlying mechanism for the existence of the plateaux is disorder. In the limiting case of zero disorder, corresponding to a perfectly clean sample, the plateaux disappear and we return to Figure \ref{classicalhall}. (There is an intermediate case where plateaux appear at some rational values $\nu$, appropriately termed the fractional quantum Hall effect.) In fact, at increasing disorder (up to a point) the plateaux stabilize and grow wider.

The physics challenge here is to explain how the presence of strong disorder gives rise to something as exact and pure as an integer, and the necessity of this disorder.

\subsection{Percolation}

\begin{figure}
\centering
\includegraphics[width=0.5\textwidth]{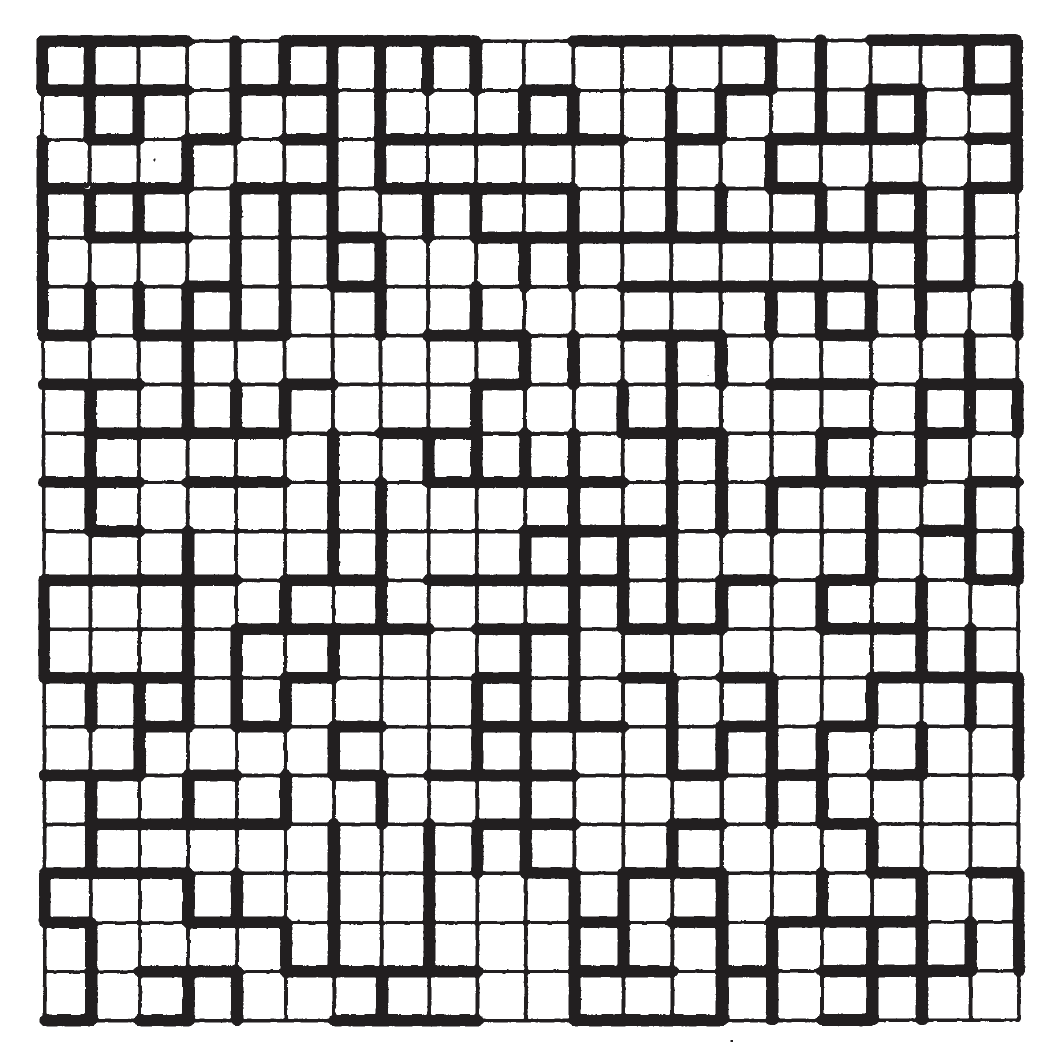}
\caption{A configuration illustrating bond percolation. From \textcite{DFMS1997}.}
\label{percolation}
\end{figure}

There are a number of formulations of percolation. As the simplest case, I describe bond percolation on the two-dimensional square lattice. A square lattice (say, a connected subset of $\mathbb Z^2$) has bonds, or links, that can independently be occupied with probability $p$ or empty with probability $1-p$ (Figure \ref{percolation}). The most basic question regarding this setup is whether there is a path along occupied bonds from one side to the other. For the limiting case of the infinite square lattice $\mathbb Z^2$, it is known that there is a critical value $p = p_{\text c} = 1/2$, above which the probability of such a path existing is 1, and below which the probability is 0 \cite{BollobasRiordan2006}.

\begin{figure}
\centering
\includegraphics[width=0.6\textwidth]{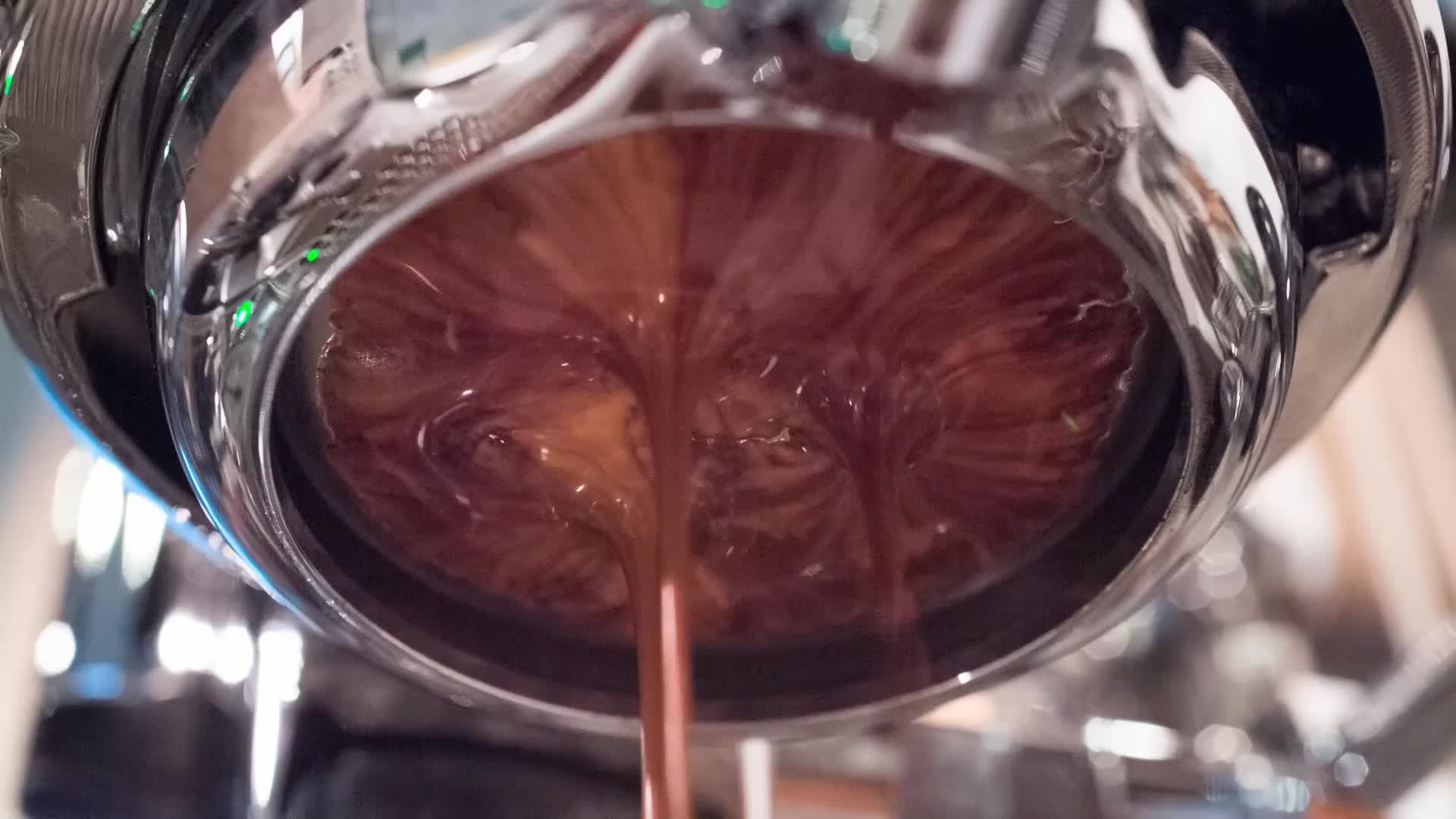}
\caption{Percolation of water through a network of firmly packed, finely ground coffee. Photo by Scott Schiller, used under CC BY 2.0.}
\label{coffee}
\end{figure}

A physical realization of percolation in three dimensions is the extraction of espresso (Figure \ref{coffee}). At the top, highly pressurized water (15 bar) makes its way through densely packed, finely ground coffee, emerging at the bottom. The goal in this scenario is to evenly saturate the grounds, allowing for a full, even extraction.

Aside from the basic presence or absence of a given bond, there are a number of more sophisticated observables one can consider \cite{Gainutdinov2015}. Cluster connectivities can be defined as the probability that a given set of points belongs to the same cluster. For two points, it is known that this probability decays with their separation $r$ as $r^{-5/24}$, for $p = p_{\text c}$. It is also possible to consider refined connectivities, such as the probability that a set of points belongs to the same cluster and also are connected via two non-intersecting paths on this cluster. Of interest later will be percolation hulls, the probability that two points belong to the boundary of the same cluster, which decays as $r^{-1/2}$. This can be generalized to observables where $k$ cluster boundaries come together, with the probability decaying as $r^{-(4k^2 - 1)/6}$ \cite{SaleurDuplantier1987}.

\sectionbreak

These two problems are, in fact, closely related. We understand that the existence of plateaux is due to disorder, and the integers are in fact Chern numbers, arising from topological considerations \cite{Tong2016}. On the other hand, our analytical understanding of the plateaux transition is still lacking \cite{Zirnbauer1997}. It is expected that this transition is described by some nonunitary conformal field theory, a candidate of which has been proposed \cite{Zirnbauer1999}. At a jump, a percolating path develops and the electrons undergo a localization--delocalization transition. The essential features of this quantum percolation transition are summarized in the Chalker--Coddington random network model, which has resisted analytical treatment. However, the Chalker--Coddington model may be systematically truncated and studied, leading to a series of two-dimensional loop models over a Temperley--Lieb algebra \cite{IFC2011}. The loop model is discussed as a formulation of the $Q$-state Potts model in Section \ref{Potts_model}, and its algebraic aspects are studied there and in Part \ref{applications}.

\chapter{Physics: conformal field theory on the lattice} \label{physics_lattice}

\section{Discretization of continuum theories}

\subsection{Necessity of discretization}

Numerical study of field theories on a computer necessarily involves reducing the continuum of degrees of freedom to a finite number of them. At the same time, the degrees of freedom that are retained must be sufficient to probe the properties of the field theory accurately. For local field theories, at least, a natural choice is the collection of field amplitudes at a discrete set of positions. The problem of percolation is nonlocal, but it may be traded for a local formulation (at the cost of nonunitarity) \cite{Nienhuis1984}.

Even outside of the particular context of lattice models, note that signatures of discretization are present in the field theories themselves. For instance, the path integral measure, Eq.~\eqref{eq:dq_measure}, is defined via a limiting process in which the time interval $[0,t]$ is treated as a sequence of discrete values $jt/N$, with $0 \le j \le N$ and $N\to\infty$.

\subsection{Identification of fields on the lattice} \label{field_identification}

In order to study a continuum theory as a limit of a lattice-discretized theory, we must identify those objects in the continuum theory that have a corresponding realization in the discrete model.

The first question is how one can realize the cornerstone of two-dimensional conformal field theory, the Virasoro algebra, using the algebra generators available in the lattice model. For generators $e_i$ obeying the Temperley--Lieb relations, it has been conjectured \cite{KooSaleur1994} that the Koo--Saleur generators
\begin{subequations}
\begin{gather}
L_n = -\frac{N}{4\pi v_{\text F}} \sum_{j=1}^N \e^{2\pi\i j n/N}\left(e_j - e_\infty + \frac{\i}{v_{\text F}}[e_j, e_{j+1}]\right) + \frac{c}{24}\delta_{n0}, \\
\overbar L_n = -\frac{N}{4\pi v_{\text F}} \sum_{j=1}^N \e^{2\pi\i j n/N}\left(e_j - e_\infty - \frac{\i}{v_{\text F}}[e_j, e_{j+1}]\right) + \frac{c}{24}\delta_{n0},
\end{gather}
\end{subequations}
furnish representations of $\mathrm{Vir}$ and $\overbar{\mathrm{Vir}}$ in the limit $N\to\infty$, when restricted to scaling states. While the precise nature of the convergence is subtle and under active study (Section \ref{evidence_lattice}, see also \textcite{MilstedVidal2017}), numerical and analytical checks have demonstrated many of the expected properties of the Virasoro algebra \cite{KooSaleur1994} (see also Appendix \ref{corrections}).

Thus, an eigenstate of $L_0$ ($\overbar L_0$) with eigenvalue $h(N)$ ($\overbar h(N)$) can be viewed as a lattice representation on $N$ sites of a field with left (right) conformal dimension $h \equiv \lim_{N\to\infty} h(N)$ ($\overbar h \equiv \lim_{N\to\infty}\overbar h(N)$). The scaling dimensions $\Delta(N) = h(N) + \overbar h(N)$ are obtained by diagonalizing $H_0 = L_0 + \overbar L_0$. The precise identification of conformal fields is a subtle process because the convergence $h(N) \to h$ and $\overbar h(N) \to \overbar h$ can be slow, because these values are discretely parametrized by $N$ rather than a continuous parameter, making them more difficult to track, and because the number of states proliferates rapidly with increasing $N$. To label the scaling states correctly, one must carefully follow sequences of eigenvalues $h(N)$ with increasing $N$, paying attention to other characteristics such as momentum and parity. The general methodology is explained by \textcite[Appendix A.5]{JacobsenSaleur2019}. For our present purposes, we will assume that this assignment of scaling fields can be done, and we will check that our identifications are self-consistent by showing that the lattice versions have properties expected of their corresponding continuum fields. Examples of these scaling field assignments appear in Section \ref{singlet_states}.

\section{Potts model} \label{Potts_model}

\subsection{General description} \label{Potts_description}

The $Q$-state Potts model is central to much of this work. It may be defined on an arbitrary graph $G$ as a collection of spins $\sigma_i$ living on the vertices of the graph, labeled by $i$. The spins take (at first) integer values from $1$ to $Q$. For a graph with $N$ vertices the number of possible configurations is $Q^N$, and the energy of a particular configuration is
\begin{equation}
E[\sigma] = -\sum_{(ij)\in E_G}\alpha_{ij}\delta_{\sigma_i \sigma_j},
\end{equation}
where $E_G$ is the set of edges in the graph $G$ and $\alpha_{ij}$ are edge-dependent coupling constants. With all couplings equal, the partition function (Eq.~\eqref{eq:Z_sum}) reads
\begin{equation}
Z = \sum_{\{\sigma\}}\prod_{(ij)\in E_G} \e^{K\delta_{\sigma_i\sigma_j}},
\end{equation}
with $K = -\beta\alpha \equiv -\beta\alpha_{ij}$. Now write $\e^{K\delta_{\sigma_i\sigma_j}} = 1 + (\e^K - 1)\delta_{\sigma_i\sigma_j}$, expand the product, and perform the sum over all spins to obtain
\begin{equation}
Z = \sum_{A\subset E_G} (\e^K - 1)^{|A|}Q^{k(A)},
\end{equation}
where the sum is over all $2^{|E_G|}$ subsets of $E_G$, and $|X|$ is the cardinality of $X$. $k(A)$ denotes the number of connected components (also called Fortuin--Kasteleyn clusters or FK clusters) in the subgraph $G_A$ with the same vertices as $G$ but only the edges $A$. $Z$ is now written as a polynomial in $Q$, and extending it to values of $Q$ that are not an integer defines the partition function for the Potts model at arbitrary $Q$.

The simplest case is to take the vertices of the graph to be a regular lattice, with edges going between only nearest neighbors, and to take all couplings to be equal. Henceforth we take $G$ to be the square lattice in two dimensions. Under these further assumptions, there is a critical value of $K$, $K_{\text c}$, given by $\e^{K_{\text c}} - 1 = \sqrt{Q}$, so that the model is conformally invariant in the continuum limit.

An equivalent formulation of $Z$ when $G$ is the two-dimensional square lattice is given by the loop model on the medial lattice $M_G = (V_M, E_M)$. The vertices $V_M$ of $M_G$ are situated at the midpoints of the original edges $E_G$, and two vertices in $V_M$ are connected by an edge in $E_M$ whenever the former stand on edges in $E_G$ that are incident on a common vertex from $G$. In particular, when $G$ is a square lattice, $M_G$ is just another square lattice, tilted through an angle $\pi/4$ and scaled down by a factor of $\sqrt 2$. There is a bijection between edge subsets $A\subset E_G$ and completely-packed loops on $M_G$. The loops are defined so that they turn around the FK clusters and their internal cycles. One has then, using the Euler relation,
\begin{equation} \label{eq:loop_Z}
Z = Q^{|V|/2}\sum_{A\subset E_G} \left(\frac{\e^K-1}{\sqrt Q}\right)^{|A|} Q^{l(A)/2},
\end{equation}
where $l(A)$ is the number of loops as defined by the bijection. The loop fugacity is $\sqrt Q \equiv \mathfrak q + \mathfrak q^{-1}$, which defines the quantum group parameter $\mathfrak q$.

The loop--cluster formulation gives rise to a representation---in the technical sense of a representation of an associative algebra---of $T^{\text a}_N$, as we now explain. In practice, states in the transfer matrix must be defined so as to allow the bookkeeping of the nonlocal quantities $k(A)$ or $l(A)$. In the cluster picture, a state is a set partition of the $L$ sites in a row, with two vertices belonging to the same block in the partition if and only if they are connected via the part of the FK clusters seen below that row. Equivalently, in the loop picture, a state is a pairwise matching of $N = 2L$ medial sites in a row, with each site seeing either a vertex of $G$ on its left and a dual vertex of $M_G$ on its right, or vice versa. The bijection between cluster and loop configurations provides as well a bijection between the corresponding cluster and loop states. The transfer matrix evolves the loop states by the relations of the affine Temperley--Lieb algebra $T^{\text a}_N$ given in Eqs.~\eqref{eq:TL_relations} and \eqref{eq:TL_relations_affine}, and to match the loop weights between Eqs.~\eqref{eq:TL_m} and \eqref{eq:loop_Z} we must identify
\begin{equation}
m = \sqrt Q.
\end{equation}

To account also for the computation of correlation functions, a few modifications must be made. The case of four-point functions has been expounded by \textcite{JacobsenSaleur2019}, but in the present discussion it suffices to consider the simpler case of two-point functions. These can be computed in the cylinder geometry by placing one point at each extremity of the cylinder. The issue is then ensuring the propagation of $j$ distinct clusters between the two extremities in a setup compatible with the transfer matrix formalism. This can be done, on the one hand, in the cluster picture by letting the states be $L$-site set partitions including $j$ marked blocks, and, on the other hand, in the loop picture by letting the states be $N$-site pairwise matchings including $2j$ defect lines---precisely the through-lines already encountered in the discussion of $T^{\text a}_N$. The sum over states must then be restricted so as to ensure that the marked clusters or defect lines propagate all along the cylinder. Moreover, it turns out to be necessary to keep track of the windings of either type of marked object around the periodic direction of the cylinder. Fortunately, in the loop picture, these considerations lead directly to the definition of a type of representation---the affine Temperley--Lieb standard module---which is well-known in the algebra literature, and which has been covered in Chapter \ref{TL_algebra}.

A special point must be made in connection with the present discussion: there is sometimes a confusion related to the type of object one may wish to consider as part of ``the'' Potts model CFT. By such a CFT we mean here the field theory describing long-distance properties of observables which are built locally in terms of Potts spins for integral $Q$, then continued to real $Q$ using the FK expansion. Examples include the spins themselves but also the energy and many more observables as discussed, for instance, by \textcite{VJS2012,VasseurJacobsen2014,CJV2017}. Other objects have been defined and studied in the literature, in particular those describing the properties of domain walls, boundaries of domains where the Potts spins take identical values \cite{DJS2010a,DJS2010b}. These are not local with respect to the Potts spin variables. Whether there is a ``bigger'' CFT containing \emph{all} of these observables at once remains an open question---see \textcite{VasseurJacobsen2012} for an attempt in this direction.

To have a better idea of the observables pertaining to the Potts model CFT for generic $Q$, one can start with the torus partition function, which has been determined in the continuum limit \cite{diFrancesco1987,ReadSaleur2001,RichardJacobsen2007}. Parametrizing
\begin{equation}
\sqrt Q = 2\cos\left(\frac{\pi}{x+1}\right), \qquad (x \in (0,\infty])
\end{equation}
the central charge is
\begin{equation} \label{eq:central_charge}
c = 1 - \frac{6}{x(x+1)},
\end{equation}
while the Kac formula reads
\begin{equation}
h_{rs} = \frac{[r(x+1)-sx]^2-1}{4x(x+1)}.
\end{equation}
The continuum-limit partition function is then given by
\begin{equation} \label{eq:torus_Z}
Z_Q = F_{0,\mathfrak q^{\pm 2}} + \frac{Q-1}{2}F_{0,-1} + \sum_{j>0} D_{j0}F_{j1} + \sum_{\substack{j>0,k>1\\k|j}} \sum_{\substack{0<p<k \\ p\wedge k = 1}} D_{j,\pi p/k} F_{j,\e^{2\pi\i p/k}},
\end{equation}
where $p\wedge k$ is the greatest common divisor of $p$ and $k$. The coefficients $D_{jK}$ can be thought of as ``multiplicities,'' although for generic $Q$, they are not integers. Their interpretation in terms of symmetries is beyond the scope of this discussion \cite{Gainutdinov2015,JRS2022}. They are given by
\begin{equation}
D_{jK} = \frac{1}{j}\sum_{r=0}^{j-1} \e^{2\i Kr} w(j,j\wedge r),
\end{equation}
($j\wedge 0 = j$ by definition) and
\begin{equation}
w(j,d) = \mathfrak q^{2d} + \mathfrak q^{-2d} + \frac{Q-1}{2}(\i^{2d} + \i^{-2d}) = \mathfrak q^{2d} + \mathfrak q^{-2d} + (-1)^d(Q-1).
\end{equation}
The $F_{j,\e^{\i\phi}}$ are given by sums,
\begin{equation} \label{eq:F_characters}
F_{j,\e^{\i\phi}} = \frac{q^{-c/24}\overbar q^{-c/24}}{P(q)P(\overbar q)} \sum_{e\in\mathbb Z} q^{h_{e-e_\phi,-j}}\overbar q^{h_{e-e_\phi,j}},
\end{equation}
in which
\begin{equation}
P(q) = \prod_{n=1}^\infty (1-q^n)
\end{equation}
and $e_\phi = \phi/2\pi$. As usual, $q$ and $\overbar q$ are the modular parameters of the torus. $P(q)$, the inverse generating function for the partition numbers, can be written in terms of the Dedekind eta function as $P(q) = q^{-1/24}\eta(q)$.

Equations \eqref{eq:torus_Z} and \eqref{eq:F_characters} encode the operator content of the $Q$-state Potts model CFT. The conformal weights arising from the last term in Eq.~\eqref{eq:torus_Z} are of the form
\begin{equation}
(h_{e-p/k,j},h_{e-p/k,-j}). \qquad (e\in\mathbb Z)
\end{equation}
The first two terms must be handled slightly differently. Using the identity
\begin{equation}
F_{0,\mathfrak q^{\pm 2}} - F_{11} = \sum_{n=1}^\infty K_{n1}\overbar K_{n1} \equiv \overbar F_{0,\mathfrak q^{\pm 2}}
\end{equation}
with the Kac character
\begin{equation}
K_{rs} = q^{h_{rs}-c/24}\frac{1-q^{rs}}{P(q)},
\end{equation}
we see that we get the set of diagonal fields
\begin{equation}
(h_{n1},h_{n1}). \qquad (n\in\mathbb N^*)
\end{equation}
The partition function can then be rewritten as
\begin{equation} \label{eq:torus_Z1}
Z_Q = \overbar F_{0,\mathfrak q^{\pm 2}} + \frac{Q-1}{2}F_{0,-1} + F_{11} + \sum_{j>0} D_{j0}F_{j1} + \sum_{\substack{j>0,k>1\\k|j}} \sum_{\substack{0<p<k \\ p\wedge k = 1}} D_{j,\pi p/k} F_{j,\e^{2\pi\i p/k}}.
\end{equation}

We notice now that $D_{10} = \mathfrak q^2 + \mathfrak q^{-2} - (Q-1) = Q - 2 - (Q - 1) = -1$. Hence $F_{11}$ disappears, in fact, from the partition function. $F_{11}$ corresponds geometrically to the so-called hull operator \cite{SaleurDuplantier1987}---related to the indicator function that a point is at the boundary of an FK cluster---with corresponding conformal weights $(h_{01},h_{01})$. Therefore, this operator is absent from the partition function. We will, nevertheless, continue to consider $\mathscr W_{11}$, since this module does appear in related models, such as the ``ordinary'' loop model or the ``$U(m)$'' model \cite{ReadSaleur2001}. We note meanwhile that the higher hull operators---related to the indicator function that $j > 1$ distinct hulls come close together at the scale of the lattice spacing---with conformal weights $(h_{0j},h_{0j})$ in $F_{j1}$ do appear in the partition function, including the Potts case.

The decomposition of the Potts-model partition function in Eq.~\eqref{eq:torus_Z1} for generic $Q$ is, in fact, in one-to-one correspondence with an algebraic decomposition of the Hilbert space $\mathscr H_Q$ in terms of modules of the affine Temperley--Lieb algebra that is exact in finite size \cite{ReadSaleur2007}. This decomposition formally reads
\begin{equation} \label{eq:torus_H}
\mathscr H_Q = \overbar{\mathscr W}_{\!\!0,\mathfrak q^{\pm 2}} + \frac{Q-1}{2}{\mathscr W}_{0,-1} + {\mathscr W}_{11} + \sum_{j>0} D_{j0}{\mathscr W}_{j1} + \sum_{\substack{j>0,k>1\\k|j}} \sum_{\substack{0<p<k \\ p\wedge k = 1}} D_{j,\pi p/k} {\mathscr W}_{j,\e^{2\pi\i p/k}}.
\end{equation}
Eq.~\eqref{eq:torus_H} is only formal in the sense that, for generic $Q$, the multiplicities are not integers, and $\mathscr H_Q$ cannot be interpreted as a proper vector space. By contrast, the modules $\mathscr W_{j,\e^{\i\phi}}$ are well-defined spaces with integer dimension \emph{independent of $Q$}, as discussed in Section \ref{standard}. Note that one must take into account the fact that the sums must be properly truncated for a finite lattice system.

The torus partition function of Eq.~\eqref{eq:torus_Z} is obtained by a trace over $\mathscr H_Q$,
\begin{equation}
Z_Q = \tr_{\mathscr H_Q} \e^{-\beta_R H} \e^{-\i\beta_I P},
\end{equation}
where the real parameters $\beta_R > 0$ and $\beta_I$ determine the size of the torus, while $H$ and $P$ denote respectively the lattice Hamiltonian and momentum operators. Introducing the (modular) parameters
\begin{subequations}
\begin{gather}
q = \exp\left[-\frac{2\pi}{N}(\beta_R + \i\beta_I)\right], \\
\overbar q = \exp\left[-\frac{2\pi}{N}(\beta_R - \i\beta_I)\right],
\end{gather}
\end{subequations}
we have, in the limit where the size of the system $N\to\infty$, with $\beta_R,\beta_I\to\infty$ such that $q$ and $\overbar q$ remain finite,
\begin{equation}
\tr_{\mathscr W_{j,\e^{\i\phi}}} \e^{-\beta_R H} \e^{-\i\beta_I P} \overset{N\to\infty}{\rightsquigarrow} F_{j,\e^{\i\phi}}.
\end{equation}
In the $j = 0$ case the relevant continuum limit is
\begin{equation}
\tr_{\overbar{\mathscr W}_{\!\!0,\mathfrak q^{\pm 2}}}\e^{-\beta_R H} \e^{-\i\beta_I P} \overset{N\to\infty}{\rightsquigarrow} \overbar F_{0,\mathfrak q^{\pm 2}}
\end{equation}
The nature of the convergence of the limits in these final expressions is subtle and will be discussed in great detail in Chapter \ref{Virasoro}.

\subsection{Discrete Virasoro algebra in the Potts model}

While the Potts model is often defined as an isotropic lattice model on the square lattice, a point of view we share in Section \ref{Potts_description}, it is well known that the corresponding universality class extends to a critical manifold with properly related horizontal and vertical couplings. The case of an infinitely large vertical coupling (the vertical direction is taken as imaginary time---see Section \ref{quantum_physics}) leads to the Hamiltonian limit where the model dynamics is described by a Hamiltonian instead of a transfer matrix. We restrict to this limit in what follows, in order to match as closely as possible the lattice model to the formalism of radial quantization of the continuum CFT.

The Hamiltonian describing the $Q$-state Potts model can be expressed using Temperley--Lieb generators \cite{KooSaleur1994} as
\begin{equation} \label{eq:Potts_Hamiltonian}
H = -\frac{1}{v_{\text F}}\sum_{j=1}^N(e_j - e_\infty)
\end{equation}
for even $N$. Here, the prefactor $v_{\text F}$, the \emph{Fermi velocity} (sometimes \emph{sound velocity}), is chosen to ensure relativistic invariance at low energy. Its value is $v_{\text F} = \pi\sin\gamma/\gamma$, where $\gamma \in [0,\pi)$ is defined via $\mathfrak q = \e^{\i\gamma}$, so that $m = \sqrt Q \in (-2,2]$. $e_\infty$ is a constant energy density added to cancel out extensive contributions to the ground-state energy. Its value is given by
\begin{equation} \label{eq:e_inf}
e_\infty = \sin\gamma\int_{-\infty}^\infty\!\frac{\sinh[(\pi-\gamma)t]}{\sinh(\pi t)\cosh(\gamma t)}\,\d t.
\end{equation}
In Eq.~\eqref{eq:Potts_Hamiltonian}, the generators $e_j$ can be taken to act on different representations of $T^{\text a}_N(m)$. The original representation, used for integer $Q$, uses matrices of dimension $Q^L\times Q^L$, corresponding to a chain of $L = N/2$ Potts spins. The FK formulation of the Potts model for real $Q$ can be obtained by using the loop formulation instead.

Note that when taking one of the standard modules $\mathscr W_{j,\e^{\i\phi}}$ as the representation of choice, the energy density $e_\infty$ is independent of $\phi$.

It is also known that the XXZ or vertex model representation of $T^{\text a}_N$ could be used instead of the loop representation with ``very similar results.'' This point has to be considered with caution, however: while the algebra $T^{\text a}_N$ is always the same, the representations (i.e., using loops/clusters or spins/arrows in the transfer matrix) are not necessarily isomorphic. Section \ref{XXZ} discusses this point in more detail.

Following Eq.~\eqref{eq:Potts_Hamiltonian}, we define the Hamiltonian density as $h_j = -e_j/v_{\text F}$. From the Hamiltonian density we then construct a lattice momentum density $p_j = \i[h_j, h_{j-1}] = -\i [e_{j-1},e_j]/v_{\text F}^2$ using energy conservation \cite{MilstedVidal2017}. We can then introduce a momentum operator $P$ as
\begin{equation}
P = -\frac{\i}{v_{\text F}^2} \sum_{j=1}^N [e_j, e_{j+1}].
\end{equation}
(This notation will not be used in the sequel, where we reserve $P$ for parity.)

From $h_j$ and $p_j$ we build components of a discretized stress tensor as
\begin{subequations}
\begin{gather}
T_j = \frac12(h_j + p_j), \\
\overbar T_j = \frac12(h_j - p_j),
\end{gather}
\end{subequations}
from which we construct discretized versions of the Virasoro generators as the Fourier modes [\emph{Id.}]. This construction gives rise to the Koo--Saleur generators
\begin{subequations} \label{eq:KS_generators}
\begin{gather}
L_n = -\frac{N}{4\pi v_{\text F}} \sum_{j=1}^N \e^{2\pi\i j n/N}\left(e_j - e_\infty + \frac{\i}{v_{\text F}}[e_j, e_{j+1}]\right) + \frac{c}{24}\delta_{n0}, \\
\overbar L_n = -\frac{N}{4\pi v_{\text F}} \sum_{j=1}^N \e^{2\pi\i j n/N}\left(e_j - e_\infty - \frac{\i}{v_{\text F}}[e_j, e_{j+1}]\right) + \frac{c}{24}\delta_{n0},
\end{gather}
\end{subequations}
which were first derived via other means by \textcite{KooSaleur1994} (see also Appendix \ref{corrections}). The crucial additional ingredient is the central charge, given by Eq.~\eqref{eq:central_charge}. Note that the identification of the central charge is actually a subtle question, and may be affected by boundary conditions, as discussed further by \textcite{GSJS2021}.

\subsection{A note on the XXZ representation} \label{XXZ}

In the XXZ representation, the generators $e_j$ act on $(\mathbb C^2)^{\otimes N}$ as
\begin{equation}
e_j = -\sigma_j^- \sigma_{j+1}^+ - \sigma_j^+ \sigma_{j+1}^- - \frac{\cos\gamma}{2} \sigma_j^z\sigma_{j+1}^z - \frac{\i\sin\gamma}{2}(\sigma_j^z - \sigma_{j+1}^z) + \frac{\cos\gamma}{2},
\end{equation}
where the operators $\sigma_j$ are the usual Pauli matrices, so the Hamiltonian is that of the familiar XXZ spin chain
\begin{equation} \label{eq:XXZ_Hamiltonian}
H = \frac{1}{2v_{\text F}}\sum_{j=1}^N[\sigma_j^x \sigma_{j+1}^x + \sigma_j^y \sigma_{j+1}^y + \cos\gamma(\sigma_j^z \sigma_{j+1}^z - 1) + 2\e_\infty],
\end{equation}
with anisotropy parameter $\Delta = \cos\gamma$. In the usual computational basis where $(1,0)$ (read as a column vector) corresponds to spin up in the $z$ direction at a given site, the Temperley--Lieb generator $e_j$ acts on spins $j$ and $j+1$ (with periodic boundary conditions) as
\begin{equation}
e_j = I \otimes \cdots\otimes I \otimes \begin{pmatrix}
0 & 0 & 0 & 0 \\
0 & \mathfrak q^{-1} & -1 & 0 \\
0 & -1 & \mathfrak q & 0 \\
0 & 0 & 0 & 0
\end{pmatrix} \otimes I \otimes \cdots \otimes I.
\end{equation}

It is also possible to introduce twisted boundary conditions in the spin chain without changing the Hamiltonian of Eq.~\eqref{eq:Potts_Hamiltonian}, by modifying the expression of the Temperley--Lieb generator acting between the first and last spins with a twist parametrized by $\phi$. In terms of the Pauli operators, this twist imposes the boundary conditions $\sigma_{N+1}^z = \sigma_1^z$ and $\sigma_{N+1}^\pm = \e^{\mp\i\phi}\sigma_1^\pm$. In the generic case, the XXZ model with magnetization $S_z = \sum_{i=1}^N S_i^z  = \sum_{i=1}^N \sigma_i^z/2 = j$ and twist $\e^{\i\phi}$ provides a representation of the module $\mathscr W_{j,\e^{\i\phi}}$. This is not true in the non-generic case.

The XXZ and loop representations share many common features. Most importantly, the value of the ground-state energy is the same for both, and so is the value of the Fermi velocity determining the correct multiplicative normalization of the Hamiltonian (cf.\ Eqs.~\eqref{eq:Potts_Hamiltonian} and \eqref{eq:XXZ_Hamiltonian}). This occurs because the ground state is found in the same module $\mathscr W_{j,\e^{\i\phi}}$ for both models, or in closely related modules for which the extensive part of the ground-state energy (and thus, the constant $e_\infty$) is the same. In general, the XXZ and loop representations involve mostly \emph{different modules}. For the XXZ chain, the modules appearing in the spin chain depend on the twist angle $\phi$. For the loop model, the modules depend on the rules one adopts in treating non-contractible loops, or lines winding around the system. If everything were always both generic and non-degenerate, a study of the physics in each irreducible module $\mathscr W_{j,\e^{\i\phi}}$ would suffice to answer all questions about all $T^{\text a}_N(m)$ models (as well as the corresponding Virasoro modules obtained in the scaling limit---see Section \ref{Potts_Virasoro}). However, it turns out that degenerate cases are always relevant to the physical problems at hand, and the modules can ``break up'' or ``get glued'' differently.

To illustrate the latter point, we consider the XXZ representation with $S_z = 0$ and twisted boundary conditions, with twist parameter $\e^{\i\phi} = \mathfrak q^{-2}$. A basis of this sector is $u = \ket*{\uparrow\downarrow}$ and $v = \ket*{\downarrow\uparrow}$. We have then
\begin{equation}
e_1 = \begin{pmatrix}
\mathfrak q^{-1} & -1 \\
-1 & \mathfrak q
\end{pmatrix}, \quad e_2 = \begin{pmatrix}
\mathfrak q & -\mathfrak q^2 \\
-\mathfrak q^{-2} & \mathfrak q^{-1}
\end{pmatrix}.
\end{equation}
We find that $e_1(u + \mathfrak q^{-1}v) = e_2(u + \mathfrak q^{-1}v) = 0$, while $e_1(u - \mathfrak q v) = (\mathfrak q + \mathfrak q^{-1})(u - \mathfrak q v)$ and $e_2(u - \mathfrak q v) = (\mathfrak q + \mathfrak q^{-1})(u - \mathfrak q v) + (\mathfrak q^3 + \mathfrak q^{-1})(u + \mathfrak q^{-1} v)$. Now consider the module $\mathscr W_{11}$, which is the spin $S_z = 1$ sector with no twist, where $e_1 = e_2 = 0$. By comparison, we see that $u + \mathfrak q^{-1} v$ generates a module isomorphic to $\mathscr W_{11}$. Meanwhile, $u - \mathfrak q v$ does not generate a submodule, since $e_2$ acting on this vector yields a component along $u + \mathfrak q^{-1} v$. However, if we quotient by the subspace generated by $u + \mathfrak q^{-1} v$, we get a one-dimensional module where $e_1$ and $e_2$ act as $\mathfrak q + \mathfrak q^{-1}$, which is precisely the module $\overbar{\mathscr W}_{\!\!0,\mathfrak q^{\pm 2}}$. We thus get the same result as in the loop model---i.e., the structure \eqref{eq:standard} of the standard module.

Considering instead $\e^{\i\phi} = \mathfrak q^2$, we have
\begin{equation}
e_1 = \begin{pmatrix}
\mathfrak q^{-1} & -1 \\
-1 & \mathfrak q
\end{pmatrix}, \quad e_2 = \begin{pmatrix}
\mathfrak q & -\mathfrak q^{-2} \\
-\mathfrak q^2 & \mathfrak q^{-1}
\end{pmatrix}.
\end{equation}
We see that $e_1(u - \mathfrak q v) = e_2(u - \mathfrak q v) = (\mathfrak q + \mathfrak q^{-1})(u - \mathfrak q v)$, while $e_1(u + \mathfrak q^{-1}v) = 0$ and $e_2(u + \mathfrak q^{-1}v) = (\mathfrak q - \mathfrak q^{-3})(u - \mathfrak q v)$. Hence this time we get a proper submodule isomorphic to $\overbar{\mathscr W}_{\!\!0,\mathfrak q^{\pm 2}}$, while we only get $\mathscr W_{11}$ as a quotient module. The corresponding structure can be represented as
\begin{equation} \label{eq:costandard}
\widetilde{\mathscr W}_{0,\mathfrak q^{\pm 2}}: \begin{tikzcd}
\overbar{\mathscr W}_{\!\!0,\mathfrak q^{\pm 2}} \\
\mathscr W_{11} \arrow[u]
\end{tikzcd}.
\end{equation}
Observe that the shapes in Eqs.~\eqref{eq:standard} and \eqref{eq:costandard} are related by inverting the (unique in this case) arrows; the module in Eq.~\eqref{eq:costandard} is called \emph{co-standard}, and we indicate this duality by placing a tilde on top of the usual notation for the standard module.

In summary, from this short exercise we see that while in the generic case the loop and spin representations are isomorphic, this equivalence breaks down in the non-generic case, where $\phi$ is such that the resonance criterion (Eq.~\eqref{eq:resonance}) is met. Only standard modules are encountered in the loop model while in the XXZ spin chain both standard and co-standard modules are encountered. This feature extends to larger $N$ \cite{GSJS2021}. We note that in the case where $\mathfrak q$ is also a root of unity, the distinction between the two representations becomes even more pronounced: the modules in the XXZ chain are no longer isomorphic to standard \emph{or} co-standard modules. This idea will be further explored in a subsequent work \cite{GSJLS}.

\subsection{The choices of metric, and duality}

The XXZ chain can be considered in a precise way as a lattice analogue of the twisted free boson theory \cite{GSJS2021}. It is well known in the latter case that two natural scalar products can be defined. The first one---which is positive definite---corresponds to the continuum limit of the ``native'' positive-definite scalar product for the spin chain, and, in terms of the free boson current modes, corresponds to choosing the hermitian conjugate $a_n^* = a_{-n}$. A crucial observation is that for this scalar product $L_n^* \ne L_{-n}$. This means that squared norms of descendants cannot be obtained using Virasoro algebra commutation relations.

The second scalar product corresponds to one where the hermitian conjugate of the Virasoro generator $L_n$ is simply given by $L^\dagger_n = L_{-n}$. This ``conformal scalar product'' is known \cite{VJS2011,DJS2010c,VGJS2012} to correspond, on the lattice, to a modified scalar product in the XXZ spin chain where $\mathfrak q$ is treated as a formal, self-conjugate parameter \cite{CJS2017}.

The loop model can be naturally equipped with two scalar products as well. Choosing basic link states to be mutually orthogonal and of unit squared norm defines a ``native'' positive-definite scalar product for which the Temperley--Lieb generators, the transfer matrix, and the Hamiltonian are not self-adjoint, while for the lattice Virasoro generators $L_n^* \ne L_{-n}$. We will denote this scalar product by $(\cdot| \cdot)$.

Meanwhile, we can also introduce the ``loop scalar product'' $\ip*{\cdot}$ within a standard module, obtained by gluing the mirror image of one link state on top of the other and evaluating the result according to certain rules that we now describe. First, unless all through-lines connect from top to bottom the result is zero. We also take into account the weight of straightening the connected through-lines, using the idea of the graduated phase: a through-line that has moved to the right (left) is assigned the weight $\e^{\i\phi/2N}$ ($\e^{-\i\phi/2N}$) for each step. Each contractible loop carries the weight $m = \mathfrak q + \mathfrak q^{-1}$, while each non-contractible loop carries the weight $\e^{\i\phi/2} + \e^{-\i\phi/2}$. To illustrate this scalar product we take the following examples, where the solid lines around the rightmost diagrams signify that we assign them a value according to these rules (the notation on the left hand side is defined in Section \ref{standard}):
\begin{subequations} \label{eq:ip_examples}
\begin{gather}
\braket*{(12)(3)(4)}{(1)(2)(34)} = \left\langle\;\vcenter{\hbox{\begin{tikzpicture}
\newcommand{\dist}{0.2}
\draw[thick] (0,0) arc (-180:0:\dist);
\draw[thick] (4*\dist,0) -- (4*\dist,-2*\dist);
\draw[thick] (6*\dist,0) -- (6*\dist,-2*\dist);
\draw[thick, dotted] ($(current bounding box.north east) + (0.05+\dist,0.05)$) rectangle ($(current bounding box.south west)+ (-0.05-\dist,-0.05)$);
\end{tikzpicture}}}
\;\middle|\;
\vcenter{\hbox{\begin{tikzpicture}
\newcommand{\dist}{0.2}
\draw[thick] (4*\dist,0) arc (-180:0:\dist);
\draw[thick] (2*\dist,0) -- (2*\dist,-2*\dist);
\draw[thick] (0,0) -- (0,-2*\dist);
\draw[thick, dotted] ($(current bounding box.north east) + (0.05+\dist,0.05)$) rectangle ($(current bounding box.south west)+ (-0.05-\dist,-0.05)$);
\end{tikzpicture}}}
\;\right\rangle
=\;
\vcenter{\hbox{\begin{tikzpicture}
\newcommand{\dist}{0.2}
\begin{scope}[yscale=-1,xscale=1]
\draw[thick] (0,0) arc (-180:0:\dist);
\draw[thick] (4*\dist,0) -- (4*\dist,-2*\dist);
\draw[thick] (6*\dist,0) -- (6*\dist,-2*\dist);
\end{scope}
\draw[thick] (4*\dist,0) arc (-180:0:\dist);
\draw[thick] (2*\dist,0) -- (2*\dist,-2*\dist);
\draw[thick] (0,0) -- (0,-2*\dist);
\draw ($(current bounding box.north east) + (0.05+\dist,0.05)$) rectangle ($(current bounding box.south west)+ (-0.05-\dist,-0.05)$);
\end{tikzpicture}}}
\; = 0 ,\\
\begin{aligned}
\braket*{(14)(23)(5)(6)}{(1)(23)(45)(6)} &= \left\langle\;\vcenter{\hbox{\begin{tikzpicture}
\newcommand{\dist}{0.2}
\draw[thick] (2*\dist,0) arc (-180:0:\dist);
\draw[thick] (0,0) arc (-180:0:3*\dist);
\draw[thick] (8*\dist,0) -- (8*\dist,-3*\dist);
\draw[thick] (10*\dist,0) -- (10*\dist,-3*\dist);
\draw[thick, dotted] ($(current bounding box.north east) + (0.05+\dist,0.05)$) rectangle ($(current bounding box.south west)+ (-0.05-\dist,-0.05)$);
\end{tikzpicture}}}
\;\middle|\;
\vcenter{\hbox{\begin{tikzpicture}
\newcommand{\dist}{0.2}
\draw[thick] (2*\dist,0) arc (-180:0:\dist);
\draw[thick] (6*\dist,0) arc (-180:0:\dist);
\draw[thick] (0,0) -- (0,-2*\dist);
\draw[thick] (10*\dist,0) -- (10*\dist,-2*\dist);
\draw[thick, dotted] ($(current bounding box.north east) + (0.05+\dist,0.05)$) rectangle ($(current bounding box.south west)+ (-0.05-\dist,-0.05)$);
\end{tikzpicture}}}
\;\right\rangle \\
&= \;\vcenter{\hbox{\begin{tikzpicture}
\newcommand{\dist}{0.2}
\begin{scope}[yscale=-1,xscale=1]
\draw[thick] (2*\dist,0) arc (-180:0:\dist);
\draw[thick] (0,0) arc (-180:0:3*\dist);
\draw[thick] (8*\dist,0) -- (8*\dist,-3*\dist);
\draw[thick] (10*\dist,0) -- (10*\dist,-3*\dist);
\end{scope}
\draw[thick] (2*\dist,0) arc (-180:0:\dist);
\draw[thick] (6*\dist,0) arc (-180:0:\dist);
\draw[thick] (0,0) -- (0,-2*\dist);
\draw[thick] (10*\dist,0) -- (10*\dist,-2*\dist);
\draw ($(current bounding box.north east) + (0.05+\dist,0.05)$) rectangle ($(current bounding box.south west)+ (-0.05-\dist,-0.05)$);
\end{tikzpicture}}}
\;= \e^{4\i\phi/2N} m,
\end{aligned} \\
\braket*{(23)(14)}{(14)(23)} = \left\langle\;\vcenter{\hbox{\begin{tikzpicture}
\newcommand{\dist}{0.2}
\draw[thick] (\dist,-\dist) arc (-90:0:\dist);
\draw[thick] (4*\dist,0) arc (-180:0:\dist);
\draw[thick] (9*\dist,-\dist) arc (-90:-180:\dist);
\draw[thick, dotted] ($(current bounding box.north east) + (0.05,0.05)$) rectangle ($(current bounding box.south west)+ (-0.05,-0.05-1*\dist)$);
\end{tikzpicture}}}
\;\middle|\;
\vcenter{\hbox{\begin{tikzpicture}
\newcommand{\dist}{0.2}
\draw[thick] (2*\dist,0) arc (-180:0:\dist);
\draw[thick] (0,0) arc (-180:0:3*\dist);
\draw[thick, dotted] ($(current bounding box.north east) + (0.05+\dist,0.05)$) rectangle ($(current bounding box.south west)+ (-0.05-\dist,-0.05)$);
\end{tikzpicture}}}
\;\right\rangle
= \;
\vcenter{\hbox{\begin{tikzpicture}
\newcommand{\dist}{0.2}
\begin{scope}[xshift=2*\dist cm]
\draw[thick] (2*\dist,0) arc (-180:0:\dist);
\draw[thick] (0,0) arc (-180:0:3*\dist);
\end{scope}
\begin{scope}[yscale=-1,xscale=1]
\draw[thick] (\dist,-\dist) arc (-90:0:\dist);
\draw[thick] (4*\dist,0) arc (-180:0:\dist);
\draw[thick] (9*\dist,-\dist) arc (-90:-180:\dist);
\end{scope}
\draw ($(current bounding box.north east) + (0.05,0.05)$) rectangle ($(current bounding box.south west)+ (-0.05,-0.05)$);
\end{tikzpicture}}}
= (\e^{\i\phi/2}+\e^{-\i\phi/2}) m .
\end{gather}
\end{subequations}
This loop scalar product is then extended by sesquilinearity to the whole space of loop states. The adjoint $U^\dagger$ of a word $U$ in the Temperley--Lieb algebra can be defined similarly by reflecting the diagram representing it about a horizontal line:
\begin{equation}
\vcenter{\hbox{\begin{tikzpicture}
\newcommand{\dist}{0.2}
\draw[thick] (2*\dist,0) arc (-180:0:\dist);
\draw[thick] (0,0) arc (-180:0:3*\dist);
\draw[thick] (8*\dist,0) -- (8*\dist,-3*\dist);
\draw[thick] (10*\dist,0) -- (10*\dist,-3*\dist);
\begin{scope}[xshift=0cm,yshift=-5*\dist cm]
\begin{scope}[yscale=-1,xscale=1]
\draw[thick] (4*\dist,0) arc (-180:0:\dist);
\draw[thick] (0,0) arc (-180:0:\dist);
\draw[thick] (8*\dist,0) -- (8*\dist,-3*\dist);
\draw[thick] (10*\dist,0) -- (10*\dist,-3*\dist);
\end{scope}
\end{scope}
\draw[thick,dotted] ($(current bounding box.north east) + (0.05+\dist,0.05)$) rectangle ($(current bounding box.south west)+ (-0.05-\dist,-0.05)$);
\end{tikzpicture}}}^{\,\dagger} = \;\vcenter{\hbox{\begin{tikzpicture}
\newcommand{\dist}{0.2}
\begin{scope}[yscale=-1,xscale=1]
\draw[thick] (2*\dist,0) arc (-180:0:\dist);
\draw[thick] (0,0) arc (-180:0:3*\dist);
\draw[thick] (8*\dist,0) -- (8*\dist,-3*\dist);
\draw[thick] (10*\dist,0) -- (10*\dist,-3*\dist);
\begin{scope}[xshift=0cm,yshift=-5*\dist cm]
\begin{scope}[yscale=-1,xscale=1]
\draw[thick] (4*\dist,0) arc (-180:0:\dist);
\draw[thick] (0,0) arc (-180:0:\dist);
\draw[thick] (8*\dist,0) -- (8*\dist,-3*\dist);
\draw[thick] (10*\dist,0) -- (10*\dist,-3*\dist);
\end{scope}
\end{scope}
\end{scope}
\draw[thick,dotted] ($(current bounding box.north east) + (0.05+\dist,0.05)$) rectangle ($(current bounding box.south west)+ (-0.05-\dist,-0.05)$);
\end{tikzpicture}}} \;.
\end{equation}From this definition it is clear that the generators $e_i$ are themselves self-adjoint with respect to this inner product, and consequently $L_n^\dagger = L_{-n}$. It is well known that the loop scalar product is invariant with respect to the Temperley--Lieb action: $\ip*{x}{Uy} = \ip*{U^\dagger x}{y}$. The loop scalar product is, of course, not positive definite. It is not degenerate, however, provided $m \ne 0$. Moreover, it is known to go over to the conformal scalar product in the continuum limit \cite{DJS2010c}.

For a given module $\mathscr W$, we can define the dual (conjugate) module $\widetilde{\mathscr W}$ by the map $u\mapsto \ip*{u}{\cdot}$; i.e., by taking mirror images. In general, we have an isomorphism $\widetilde{\mathscr W}_{j,\e^{\i\phi}} \cong \mathscr W_{j,\e^{-\i\phi}}$. When $\mathscr W_{j,\e^{\i\phi}}$ is reducible but indecomposable, the corresponding Loewy diagram has its arrows reversed, as illustrated by a comparison of Eqs.~\ref{eq:standard} and \ref{eq:costandard}. The modules $\mathscr W_{j1}$ are self-dual.

An important point is that, if a Temperley--Lieb module is self-dual, then since the Hamiltonian itself is, as well as the definition of scaling states, the action of the continuum limit of the Koo--Saleur generators should define an action on the scaling limit of the module that is also invariant under duality in the CFT. If both the Temperley--Lieb module and the $\mathrm{Vir}\oplus\overbar{\mathrm{Vir}}$-module are irreducible, this has no useful consequences. We will see (Chapters \ref{Virasoro} and \ref{Jordan_loop}), however, that the modules $\mathscr W_{j1}$, while irreducible, have a continuum limit which is not so. Their self-duality implies invariance of the Loewy diagrams for the continuum limit with respect to reversal of the $\mathrm{Vir}\oplus\overbar{\mathrm{Vir}}$ arrows, with very interesting consequences.

\section{$s\ell(2|1)$ superspin chain} \label{sl21_section}

The periodic $s\ell(2|1)$ superspin chain with alternating fundamental and conjugate fundamental representations is closely related to the properties of the hulls of percolation clusters. It can be argued to have a conformally invariant continuum limit, which must have central charge $c = 0$ and be logarithmic \cite{Gainutdinov2015}. It is most conveniently defined in terms of creation and annihilation operators on a periodic lattice of length $N = 2L$. Each site carries a bosonic space of dimension 2 along with a fermionic space of dimension 1, with the restriction that each site has only one particle. That is, the Hilbert space on site $i$ is $\Span\{b^\dagger_{i1}\ket*{0}, b^\dagger_{i2}\ket*{0}, f^\dagger_i\ket*{0}\} \equiv \Span\{\ket*{0},\ket*{1},\ket*{2}\}$. In this section and any other involving this particular model, $\ket*{0}$ may variously denote the vacuum state, the first basis vector $b^\dagger_{i1}\ket*{0}$, or the ground state, and it will be clear in context what is meant. By analogy with qubits or two-level systems, this basis will be called the \emph{computational basis}.

On each site the representations alternate. In terms of the commutation relations of the creation and annihilation operators, they read
\begin{subequations}
\begin{gather}
[b_{ia},b^\dagger_{jb}] = \delta_{ij}\delta_{ab}, \\
\{f_i, f^\dagger_j\} = (-1)^i\delta_{ij}. \label{eq:sl21_fermion}
\end{gather}
\end{subequations}
The literature sometimes uses $b^\dagger_{i1}$, $b^\dagger_{i2}$, $f^\dagger_i$, $b_{i1}$, $b_{i2}$, and $f_i$ to denote creation and annihilation operators for the fundamental representation on even sites and $\overbar b^\dagger_{i1}$, $\overbar b^\dagger_{i2}$, $\overbar f^\dagger_i$, $\overbar b_{i1}$, $\overbar b_{i2}$, and $\overbar f_i$ for the conjugate representation on odd sites. I will just use the unbarred versions, keeping in mind the factor $(-1)^i$ in Eq.~\eqref{eq:sl21_fermion} to remember that odd sites carry the conjugate representation.

Because of the sign in the fermionic anticommutation relations, the bilinear form $\ip*{\cdot}$ is indefinite. One can calculate the squared norm of a basis vector by writing out the product with its hermitian conjugate and using the commutation relations to annihilate the vacuum. The shortcut is to locate the fermions, and then the squared norm is $\prod_{i_f} (-1)^{i_f}$, where $i_f$ is a site where a fermion is located. For example, on $N = 4$ sites, $\ip*{2200} = (-1)^1 \times (-1)^2 = -1$, and $\ip*{2222} = (-1)^1\times(-1)^2\times(-1)^3\times(-1)^4 = 1$. As in the loop model, one may also naturally consider a positive-definite inner product, the standard inner product of $\mathbb C^n$ where different computational basis states are declared to be orthogonal and of unit norm squared: $(i|j) = \delta_{ij}$.

If we define
\begin{equation} \label{eq:sl21_e_i}
e_i = [b^\dagger_{i1} b^\dagger_{i+1,1} + b^\dagger_{i2} b^\dagger_{i+1,2} + (-1)^i f^\dagger_i f^\dagger_{i+1}][b_{i1} b_{i+1,1} + b_{i2} b_{i+1,2} - (-1)^i f_i f_{i+1}], \qquad (1\le i \le N)
\end{equation}
we obtain a representation of the Temperley--Lieb algebra with loop weight $m = 1$. (Periodicity implies that site $N + 1$ should be identified with site 1.) In the basis $\{\ket*{00},\ket*{01},\ldots,\ket*{21},\ket*{22}\}$ of the Hilbert space of sites $i$ and $i+1$, this generator has the matrix representation
\begin{equation}
e_i = \begin{pmatrix}
1 & 0 & 0 & 0 & 1 & 0 & 0 & 0 & (-1)^{i+1} \\
0 & 0 & 0 & 0 & 0 & 0 & 0 & 0 & 0 \\
0 & 0 & 0 & 0 & 0 & 0 & 0 & 0 & 0 \\
0 & 0 & 0 & 0 & 0 & 0 & 0 & 0 & 0 \\
1 & 0 & 0 & 0 & 1 & 0 & 0 & 0 & (-1)^{i+1} \\
0 & 0 & 0 & 0 & 0 & 0 & 0 & 0 & 0 \\
0 & 0 & 0 & 0 & 0 & 0 & 0 & 0 & 0 \\
0 & 0 & 0 & 0 & 0 & 0 & 0 & 0 & 0 \\
(-1)^i & 0 & 0 & 0 & (-1)^i & 0 & 0 & 0 & -1
\end{pmatrix}. \qquad (1 \le i \le N - 1)
\end{equation}
In the same basis of the Hilbert space of sites $N$ and $1$, in that order, the very last generator has the matrix form
\begin{equation}
e_N = \begin{pmatrix}
1 & 0 & 0 & 0 & 1 & 0 & 0 & 0 & (-1)^f \\
0 & 0 & 0 & 0 & 0 & 0 & 0 & 0 & 0 \\
0 & 0 & 0 & 0 & 0 & 0 & 0 & 0 & 0 \\
0 & 0 & 0 & 0 & 0 & 0 & 0 & 0 & 0 \\
1 & 0 & 0 & 0 & 1 & 0 & 0 & 0 & (-1)^f \\
0 & 0 & 0 & 0 & 0 & 0 & 0 & 0 & 0 \\
0 & 0 & 0 & 0 & 0 & 0 & 0 & 0 & 0 \\
0 & 0 & 0 & 0 & 0 & 0 & 0 & 0 & 0 \\
-(-1)^f & 0 & 0 & 0 & -(-1)^f & 0 & 0 & 0 & -1
\end{pmatrix},
\end{equation}
where $f$ is the number of fermions in the rest of the sites. For example, $\ket*{0021}$, $\ket*{0022}$, and $\ket*{2022}$ all have $f = 1$ since the fermions in the first and last sites aren't counted, and $\ket*{0220}$ has $f = 2$. A few more examples should clarify the action of $e_N$ on $N = 4$ sites:
\begin{subequations}
\begin{gather}
e_N\ket*{2022} = -\ket*{0020} - \ket*{1021} - \ket*{2022}, \\
e_N\ket*{2012} = \ket*{0010} + \ket*{1010} - \ket*{2012}, \\
e_N\ket*{0000} = \ket*{0000} + \ket*{1001} - \ket*{2002} = e_N\ket*{1001} = e_N\ket*{2002}, \\
e_N\ket*{0020} = \ket*{0020} + \ket*{1021} + \ket*{2022} = e_N\ket*{1021}.
\end{gather}
\end{subequations}
Adjoining the translation operator by 2 sites, $\tau^2$, satisfying $\tau^N = 1$, we obtain a representation of $JTL^{\text{au}}_N(1)$.

As with the loop model, we obtain the Hamiltonian
\begin{equation} \label{eq:sl21_Hamiltonian}
H = -\frac{1}{v_{\text F}}\sum_{j=1}^N(e_j - e_\infty).
\end{equation}
In this case, since this model only furnishes a representation of the TL algebra for $m = 1$, we have $c = 0$, $x = 2$, and evaluating the integral of Eq.~\eqref{eq:e_inf}, $e_\infty = 1$.

$H$ commutes with $B$ and $Q^z$, where
\begin{subequations}
\begin{gather}
B = \sum_{i=1}^N \left(f^\dagger_i f_i + \frac{(-1)^i}{2} (b_{i1}^\dagger b_{i1} + b_{i2}^\dagger b_{i2})\right), \\
Q^z = \sum_{i=1}^N \frac{(-1)^i}{2} (b_{i1}^\dagger b_{i1} - b_{i2}^\dagger b_{i2}).
\end{gather}
\end{subequations}
The allowed quantum numbers are $-N/4 \le B \le N/4$ in steps of $1/2$, and $-N/2 + |B| \le Q^z \le N/2 - |B|$ in unit steps. The subspace where $B = Q^z = 0$ is of primary importance, as it contains the ground state. On four sites, this sector has dimension 15, and is spanned by $\ket*{0000}$, $\ket*{0011}$, $\ket*{0022}$, $\ket*{0110}$, $\ket*{0220}$, $\ket*{1001}$, $\ket*{1100}$, $\ket*{1111}$, $\ket*{1122}$, $\ket*{1221}$, $\ket*{2002}$, $\ket*{2112}$, $\ket*{2200}$, $\ket*{2211}$, and $\ket*{2222}$. For $N = 6$ it has dimension 93, for $N = 8$ it has dimension 639, and for $N = 10$ it has dimension 4653. In general, the dimension is given by \cite{OEIS1}
\begin{equation}
\dim(N) = \sum_{k=0}^{N/2}\binom{N/2}{k}^2\binom{2k}{k}.
\end{equation}
This dimension grows asymptotically as $\dim(N) \sim O(3^N/N)$. In fact, I computed $\dim(N) \approx 0.8269933\times 3^N/N$ by computing these quantities for $N = 10^7$. It turns out also that $v_{\text F}/\pi = 3\sqrt 3/2\pi = 0.8269933\ldots$, where $v_{\text F}$ is as defined after Eq.~\eqref{eq:Potts_Hamiltonian}, which may be a fortuitous coincidence or may signal something important. It is often hard to tell at first glance.

By representing every integer from $0$ to $3^N - 1$ as length $N$ base 3 strings, it is simple via Mathematica (or any reasonable computing language) to pick out the strings corresponding to a basis of any sector corresponding to some combination of $B$ and $Q^z$ (see Appendix \ref{mathematica}). Since the action of $e_i$ replaces pairs of adjacent matching digits with a linear combination of the same (see Eq.~\eqref{eq:sl21_e_i}), it is also simple to construct the action of the $e_i$ as matrices, whether just in a particular sector (conserved by all of the $e_i$ operators) or on the whole Hilbert space. $e_N$ requires some care because of the fermion number condition, but keeping this in mind, it is also doable. Then one simply takes their sum to form the Hamiltonian, and the Koo--Saleur generators.

The idea that the action of $e_i$ is to replace adjacent pairs with a linear combination of pairs gives a way to generate the basis vectors of the smallest TL submodule containing the ground state. For symmetry reasons, we expect the ground state to contain a component along the vector $\ket*{0\cdots 0}$, consisting of a type-1 boson on each site. Start with this state, and generate new basis vectors by acting with all of the $e_i$ operators until no new states are generated. For an example on four sites, beginning with $\ket*{0000}$, the action of $e_1$ to $e_4$ generates $\ket*{1100}$, $\ket*{2200}$, $\ket*{0110}$, $\ket*{0220}$, $\ket*{0011}$, $\ket*{0022}$, $\ket*{1001}$, and $\ket*{2002}$. Repeating the procedure on all of these new states gives $\ket*{1111}$, $\ket*{1122}$, $\ket*{2211}$, $\ket*{2222}$, $\ket*{2112}$, $\ket*{1221}$, and no new states are generated after that. In practice it takes fewer iterations when starting with $\ket*{0\cdots0}$, $\ket*{1\cdots1}$, and $\ket*{2\cdots2}$ as the seed. Note that this does not generate the full $B = Q^z = 0$ sector, but rather an invariant submodule of it, which will be called the \emph{identity module} or \emph{vacuum sector}, as it contains the ground state, corresponding to the identity field. The dimension of this submodule on $N$ sites is the number of walks of length $N$ on the 3-regular tree beginning and ending at some fixed vertex \cite{OEIS2}. Numerically, I find that the dimension grows like $9.5746(\sqrt 8)^N/N^{3/2} \approx 12\sqrt{2/\pi}(\sqrt 8)^N/N^{3/2}$. This asymptotic behavior should follow from the generating function $f(x) = 4/(1+3\sqrt{1-8x})$.
 
Because the Hamiltonian in Eq.~\eqref{eq:sl21_Hamiltonian} exhibits Jordan blocks at finite size \cite{Gainutdinov2015}, it is a useful model to study the indecomposability parameters of logarithmic conformal field theory (Chapter \ref{sl21_chapter}).

\chapter{Computational methods} \label{computational}
\section{Matrix diagonalization}

Matrix diagonalization is a process to find a basis in which a given matrix is diagonal (where possible---see Theorem \ref{ND}). If $A$ is a matrix, one seeks a matrix $U$ and a diagonal matrix $D$ such that $U^{-1}AU = D$. This is often written as $AU = UD$, to easily account for partial diagonalizations, which are not invertible. For a diagonalizable matrix, the problem may be solved by computing its eigenvalues and eigenvectors---if $U$ is the matrix whose columns are the eigenvectors of $A$, then $D$ is diagonal with its elements being the eigenvalues of $A$.

Complete diagonalization requires solving the characteristic equation of $A$ completely, then determining the eigenvectors by setting up a system of linear equations for each eigenvalue. Although straightforward in principle, this procedure is computationally expensive for large matrices, and subject to compounded numerical errors and instabilities.

Fortunately, in practice, one does not require the complete spectrum of $A$, but only some information about a few of its extremal eigenvalues, or a small subset of the spectrum determined by some other criteria. This relaxed requirement opens up possibilities for other techniques of calculating these subsets of eigenvalues and eigenvectors, which I refer to generically as \emph{partial diagonalization}.

The simplest example for calculating the single largest eigenvalue (by magnitude) of $A$ is the \emph{power method}. Suppose $\lambda_1$ and $\lambda_2$ are the two largest eigenvalues of $A$, and $|\lambda_1| > |\lambda_2|$. By taking an arbitrary vector $v_0$ and recursively defining $v_{n+1} = Av_n/\|Av_n\|$, then $v_N^*Av_N$ is approximately equal to $\lambda_1$ with a relative error of order $|\lambda_2/\lambda_1|^N$ (unless $v_0$ has no component along the corresponding eigenvector, a possibility with vanishing probability). Here, the norm is the 2-norm and $*$ the conjugate-transpose. If, furthermore, $A$ is hermitian, one may extend this method to calculate a few more eigenvectors, taking into account properties such as orthogonality and parity (and other techniques used in the variational method). The smallest eigenvalues and their corresponding eigenvectors may be accomplished (at least in principle, if not always efficiently) by applying the power method to $A^{-1}$. If $A$ is not invertible because of a zero eigenvalue, then one may ``shift and invert,'' and consider $(A - \mu I)^{-1}$ with $\mu$ appropriately chosen. This technique also works to find the eigenvalues closest to $\mu$, as they correspond to the largest eigenvalues of the shifted and inverted matrix.

Aside from the difficulties in dealing with more than one eigenvector, a downside to the power method is that the information obtained is refined only through a computation involving multiplication by $A$, which can be expensive if $A$ is large. Furthermore, the sequence $v_0,\ldots,v_{N-1}$ may contain useful information, but that information is effectively discarded as earlier iterations are not expressly taken into account.

\section{Arnoldi methods} \label{Arnoldi}

For large sparse matrices, the implicitly restarted Arnoldi method is the dominant algorithm of choice to determine the desired part of the spectrum and their associated eigenvectors. More precisely, it determines a basis for the subspace spanned by these eigenvectors and the matrix form of $A$ restricted to this subspace and in this basis. This latter matrix, much smaller than the initial matrix, can then be diagonalized using more traditional or black-box tools.

The implicitly restarted Arnoldi method was developed by Sorensen, and a careful mathematical exposition showing how it produces a reliable output can be found in \textcite{Sorensen1992,Sorensen1997}. Here, I have gathered the appropriate definitions and algorithms that suffice to allow the reader to implement them. Furthermore, I have modified the algorithms found therein to work for complex matrices; as given in the published articles, they only work for real matrices. Throughout this section, $*$ denotes the conjugate-transpose.

If $A$ is an $n\times n$ complex matrix, then a \emph{$k$-step Arnoldi factorization} of $A$ is a relation of the form
\begin{equation} \label{eq:arnoldi_k}
AV_k = V_k H_k + f_k e_k^T,
\end{equation}
where $V_k$ is an $n\times k$ matrix with orthonormal columns, $V_k^* f_k = 0$, and $H_k$ is a $k\times k$ upper Hessenberg matrix. (An upper Hessenberg matrix is upper triangular, except that it also allows nonzero elements on the first subdiagonal.) The idea is that if the residual vector $f_k$ is (nearly) zero, then the columns of $V_k$ span an invariant (or nearly invariant) subspace and $H_k$ is the projection of $A$ to that subspace. The $k$-step Arnoldi factorization is accomplished with Algorithm \ref{arnoldi_k}. The input vector $v$ is arbitrary and can be generated randomly. In line 1, $V_1 = (v_1)$ means to initiate an $n\times 1$ matrix $V_1$ with column entries given by $v_1$, and $H_1 = (\alpha_1)$ means to initiate a $1\times 1$ matrix $H_1$ with entry $\alpha_1$. The notation $(M, v)$ of lines 4 and 8 mean to adjoin a column to $M$, with entries given by $v$, and, similarly, $\hat H_j$ is given by adjoining a row to $H_j$.

\begin{algorithm} 
\KwData{$n\times n$ matrix $A$, $n$-dimensional column vector $v$, $k \le n$}
\KwResult{$V_k$, $H_k$, $f_k$ as in Eq.~\eqref{eq:arnoldi_k}}
$v_1 = v/\|v\|$; $w = Av_1$; $\alpha_1 = v_1^* w$; $f_1 = w - v_1\alpha_1$; $V_1 = (v_1)$; $H_1 = (\alpha_1)$\;
\For{$j = 1$ \KwTo $k-1$}{
$\beta_j = \|f_j\|$; $v_{j+1} = f_j/\beta_j$\;
$V_{j+1} = (V_j, v_{j+1})$; $\hat H_j = \begin{pmatrix} H_j \\ \beta_j e_j^T \end{pmatrix}$\;
$z = Av_{j+1}$\;
$h = V_j^* z$; $f_{j+1} = z - V_{j+1}h$\;
optional: $s = V_{j+1}^* f_{j+1}$; $f_{j+1} \leftarrow f_{j+1} - V_{j+1}s$; $h \leftarrow h+s$\;
$H_{j+1} = (\hat H_j, h)$\;
}
\caption{$k$-step Arnoldi factorization}
\label{arnoldi_k}
\end{algorithm}

The $k$-step Arnoldi factorization is the first step to a partial diagonalization. However, the resulting matrix $H_k$ does not yet reflect the criteria set for the eigenvalues sought. In a nutshell, the idea of the implicitly restarted Arnoldi algorithm is to extend the Arnoldi factorization, and use the additional data, along with some input from the user, to filter out ``unwanted'' eigenvalues and keep the ``wanted'' eigenvalues. The implicitly restarted Arnoldi method is given in Algorithm \ref{IRAM}. The descriptor ``implicit'' refers to the fact that upon each iteration, an Arnoldi factorization is extended from $k$ steps to $m$ steps, then truncated back to $k$ steps, in contrast with ``explicitly'' restarted Arnoldi methods that generate new starting vectors from which to compute Arnoldi factorizations from scratch.

\begin{algorithm}
\KwData{$n\times n$ matrix $A$, $n$-dimensional column vector $v$, $k,p$ such that $k+p\le n$}
\KwResult{$V_k$, $H_k$, $f_k$ as in Eq.~\eqref{eq:arnoldi_k}}
$m = k + p$\;
compute a $k$-step Arnoldi factorization using Algorithm \ref{arnoldi_k}\;
\Repeat{convergence}{
beginning with the $k$-step Arnoldi factorization $(V_k, H_k, f_k)$, apply $p$ additional steps of the Arnoldi process to obtain an $m$-step Arnoldi factorization $(V_m, H_m, f_m)$ [i.e., run Algorithm \ref{arnoldi_k}, lines 2--9, with line 2 replaced by ``for $j = k$ to $k + p - 1$ do'']\;
compute $\sigma(H_m)$ and select a set of $p$ shifts $\mu_1$, $\mu_2$, $\ldots$, $\mu_p$ based on $\sigma(H_m)$ or other information\;
$Q = I_m$\;
\For{$j = 1$ \KwTo $p$}{
$(Q_j, R_j) = \text{QR}(H_m - \mu_j I_m)$\;
$H_m \leftarrow Q_j^* H_m Q_j$; $Q \leftarrow QQ_j$\;
}
$V_m \leftarrow V_mQ$; $v = V_m e_{k+1}$\;
$f_k \leftarrow v(e_{k+1}^TH_m e_k) + f_m (e_{k+p}^TQe_k)$; $V_k \leftarrow V_m(1:n,1:k)$; $H_k \leftarrow H_m(1:k,1:k)$\;
}
\caption{Implicitly restarted Arnoldi method}
\label{IRAM}
\end{algorithm}

For Algorithm \ref{IRAM}, the QR factorization of line 8 is a standard algorithm that is built into many computing languages, and can be found in many libraries and packages. It refers to the fact that any complex square matrix can be written as a product of a unitary matrix $Q$ and an upper triangular matrix $R$. The convergence of line 13 is determined by the residual $\|f_k\|$ being smaller than some set tolerance. In line 5, the spectrum of $H_m$, $\sigma(H_m)$, furnishes an approximation to a subset of the spectrum of $A$. Sorting $\sigma(H_m) = \{\lambda_1,\lambda_2,\ldots,\lambda_k,\lambda_{k+1},\ldots,\lambda_{k+p}\}$ by ``importance,'' a natural choice for the shifts in line 5 is $\{\mu_1,\ldots,\mu_p\} = \{\lambda_{k+1},\ldots,\lambda_{k+p}\}$, which are ``filtered'' out of $H_k$ and $V_k$. For example, if one wants the $k$ largest eigenvalues in magnitude of $A$, then $\{\mu_1,\ldots,\mu_p\}$ are the smallest elements of $\sigma(H_m)$ in magnitude. Finally, the notation $M(1:m,1:n)$ refers to the submatrix of $M$ consisting of its first $m$ rows and $n$ columns.

\section{Numerical computation of the Jordan canonical form} \label{jordan_algorithm}

Because of the instability of the Jordan canonical form with respect to arbitrarily small perturbations, a computational algorithm to find it must necessarily adapt the theory of the Jordan form to a finite-precision situation. I have found the algorithm of \textcite{KagstromRuhe1980} to be fairly reliable for the problems studied in Part \ref{applications}. In this section I summarize their algorithm and give some brief remarks about my particular implementation. However, the reader looking to implement this algorithm will need to refer to the original article.

\begin{enumerate}
\item Suppose $A$ is a matrix to be diagonalized (which itself often comes from the output of Algorithm \ref{IRAM}). The algorithm of \textcite{KagstromRuhe1980} first calls for $A$ to be upper triangularized: $A = STS^{-1}$, where $T$ is upper triangular. The authors use a sequence of library routines; I use a pre-built Schur decomposition algorithm to accomplish this step.

\item Sort the diagonal elements of $T$ so that multiple eigenvalues are adjacent. The sorting is accomplished using a sequence of unitary matrices that swap two adjacent diagonal elements. The result is $T = UT_1U^*$, where $T_1$ is still upper triangular, but its diagonal elements have been sorted.

\item Decide which groups of eigenvalues along the main diagonal correspond to numerical multiple eigenvalues. If there are $p$ distinct numerical eigenvalues, partition the matrix $T_1$ as
\begin{equation}
T_1 = \begin{pmatrix}
T_{11} & T_{12} & \cdots & T_{1p} \\
 & T_{22} & \cdots & T_{2p} \\
 & & \ddots & \vdots \\
 & & & T_{pp}
\end{pmatrix},
\end{equation}
with each diagonal block $T_{kk}$ corresponding to a different numerical eigenvalue. Suppose group $k$ has $t_k$ elements, so that $T_{kk}$ is a $t_k\times t_k$ matrix.

\item Use elimination matrices to eliminate the elements above the main block diagonal---the blocks $T_{ij}$ with $i < j$, working upwards in rows and then from left to right in columns. The result is
\begin{equation}
T_2 = M^{-1}T_1 M = \diag\{T_{11},T_{22},\ldots,T_{pp}\}.
\end{equation}
The result thus far may be summarized as
\begin{equation}
A = YT_2Y^{-1},
\end{equation}
with $Y = SUM$. Partition the matrix $Y$ into groups of columns, corresponding to the diagonal blocks $T_{kk}$:
\begin{equation}
AY_k = Y_k T_{kk}.\qquad (1 \le k \le p)
\end{equation}

\item For each $k$, write $Y_k = U_k R_{kk}$, where $U_k$ has orthonormal columns and $R_{kk}$ is square and upper triangular---i.e., the QR decomposition for non-square matrices.

\item $T'_{kk} = R_{kk}T_{kk}R^{-1}_{kk}$ is the restriction of $A$ to the invariant subspace spanned by the columns of $U_k$ (of dimension $t_k$). It is supposed to correspond to one numerical multiple eigenvalue, which is taken to be the average of the diagonal elements:
\begin{equation}
\lambda_k = \frac{1}{t_k}\tr T'_{kk}.
\end{equation}
Iterated singular value decomposition of $T'_{kk}$ and its submatrices determines the number of towers and their heights, and from this a nilpotent matrix $B_k$ is constructed that is unitarily similar to $T'_{kk} - \lambda_k I_{t_k}$ up to some numerical error. $B_k$ is strictly upper triangular.

\item Gaussian elimination on $B_k$ finally gives the Jordan form, up to a permutation of the basis vectors. The basis vectors---the eigenvectors and generalized eigenvectors---are the columns of the total similarity transformation obtained by accumulating all of the steps. In general, the resulting Jordan form does not have unit couplings, but it is simple enough to rescale the basis vectors to obtain the Jordan canonical form.

\end{enumerate}

The most important considerations involve deciding when two numerical values on the diagonal of $T$ in step 2 represent the same multiple eigenvalue. The simplest choice is to declare two diagonal elements $\alpha$ and $\beta$ to represent the same multiple eigenvalue if $|\alpha - \beta| < \delta$ for some eigenvalue tolerance $\delta$. The choice of $\delta$ depends on the numerical precision of the calculation. A perturbation to a matrix element of order $\epsilon$ in a Jordan block of rank $2$ causes an order $\epsilon^{1/2}$ shift in the exact eigenvalues (see beginning of Section \ref{emerging_applications}). More generally, a perturbation of order $\epsilon$ in a Jordan block of rank $n$ shifts the eigenvalues by an amount of order $\epsilon^{1/n}$. Thus, the eigenvalue tolerance will in part depend on the ranks of the Jordan blocks one is expecting. For a computation with working precision $\epsilon \sim 10^{-16}$ and Jordan blocks of rank at most 2, I have found $10^{-7} \lesssim \delta \lesssim 10^{-6}$ to work well---knowing also that the correct eigenvalues have a separation greater than this. Similarly, when Jordan blocks of rank 3 are expected to appear, I take $10^{-5} \lesssim \delta \lesssim 10^{-4}$.

\part{Theory} \label{theory}

\chapter{Emerging Jordan forms} \label{emerging}

This chapter describes how the formation of Jordan blocks can be observed in certain limits, when they are absent except possibly in the limit. More precisely, suppose an operator $A(x)$ parametrized by $x$ is generically diagonalizable, except possibly at $x = x_0$. Can we infer properties of $A(x_0)$ without studying it directly, but instead by looking at the limit $\lim_{x\to x_0}A(x)$?

The mathematical framework to be described in this chapter came about while investigating the formation of Jordan blocks in the Hamiltonians and transfer matrices of lattice regularizations of logarithmic conformal field theories as the number of sites on the lattice increased. For some representations, Jordan blocks can be found in the Hamiltonian or transfer matrix of a finite-size lattice, and computations could be carried out by working directly with the eigenvectors and generalized eigenvectors. In this situation many questions of interest can be answered on the lattice and then extrapolated to the continuum theory, as described in Chapter \ref{physics_lattice}.

There are also indecomposable structures known or conjectured to appear in the scaling limit that are not manifest on the lattice. Clearly, the scaling limit is not directly accessible computationally. We must then hope that the Jordan block in the continuum theory ``builds up'' in some sense from the lattice model, rather than appearing suddenly and discontinuously, and if so, whether there are signatures of this build-up that can be detected.

As we will see, such hopes are auspicious, and this kind of approach is possible. One strategy is to investigate the effect of an indecomposable structure on four-point functions and measure the amplitudes that would be affected by the presence or absence of such a structure. I describe how this strategy was applied successfully to the loop model in Section \ref{amplitudes}.

Nevertheless, one may reasonably ask whether a direct signature of an emerging Jordan block exists---one that would hold in a much more general setting, and not just limited to four-point functions in logarithmic conformal field theories. As it turns out, the answer is affirmative, and almost embarrassingly simple, while being completely general. While there are some involved algebraic expressions, nothing in this chapter is conceptually complicated. Yet neither I nor my collaborators have been able to find this framework in the mathematical literature. The lesson here is that it pays to ask exactly the right questions.

Because of the generality of the framework, I describe the theory of emerging Jordan blocks in a fully mathematical setting, devoid of any physical context, except for some references to its application.

\section{The $J$ measure}

Consider a one-parameter continuous family of linear operators or matrices, $A(x)$, acting on a finite-dimensional vector space $V$ over $\mathbb R$ or $\mathbb C$. The eigenvalues of a generic operator are all distinct, and thus such an operator is diagonalizable. Whenever an operator is diagonalizable, it has a basis of eigenvectors. As the operator changes continuously, it is possible to choose a basis that changes continuously with it, so long as the operator remains diagonalizable. Even if one encounters a situation of non-diagonalizability, the limit of the continuously parametrized set of basis vectors is well-defined, although it may not be a basis, and it will not be a basis of the limiting matrix.

Let us concretely consider two eigenvalues of $A(x)$, $\lambda(x)$ and $\mu(x)$, with associated eigenvectors $u(x)$ and $v(x)$, and suppose that at $x = x_0$, $\lambda(x_0) = \mu(x_0)$. Does $A(x_0)$ remain diagonalizable? If so, then the geometric eigenspace associated to the eigenvalue $\lambda(x_0) = \mu(x_0)$ has dimension 2, and is spanned by $u(x_0)$ and $v(x_0)$. In particular, $u(x_0)$ and $v(x_0)$ remain linearly independent. If, however, $A(x_0)$ is not diagonalizable, with $\lambda(x_0)$ being a defective eigenvalue, then the geometric eigenspace associated to $\lambda(x_0)$ has dimension 1, yet it is still spanned by $u(x_0)$ and $v(x_0)$. This means that $u(x_0)$ and $v(x_0)$ lie along the same line in $V$, or that they are parallel (or antiparallel).

For a such a pair of eigenvectors $\{u, v\}$ (hereafter we do not explicitly notate the dependence on $x$), one can consider the quotient
\begin{equation}
J(u,v) = \frac{(u|v)}{\sqrt{(u|u)(v|v)}},
\end{equation}
where $(\cdot|\cdot)$ is any positive-definite inner product. For simplicity, we will take it to be the standard inner product. Note that by the Cauchy--Schwarz inequality, $|(u|v)|^2 \le (u|u)(v|v)$ implies $|J(u,v)| \le 1$. The proof of the Cauchy--Schwarz inequality also shows that it is an equality if and only if the two vectors are parallel: $u = \alpha v$. If $J(u,v)$ approaches a number of unit modulus, $\e^{\i\theta}$, we are approaching a non-diagonalizable matrix, since a basis is by definition a linearly independent set, which cannot contain parallel vectors. Thus I will consider $|J(u,v)| \to 1$ as $x \to x_0$ to signify that a Jordan block (of rank 2) forms in that limit, but is not otherwise present.

To be sure, $J$ must be applied to a pair of eigenvectors of $A$ for which the eigenvalues are still distinct, and studied in the limit as the eigenvalues become equal. If it is applied to two independent eigenvectors with the same eigenvalue, then the test is not meaningful because the two eigenvectors can be redefined by choosing a different spanning set for the eigenspace. If it is applied to an eigenvector/generalized eigenvector pair for a defective eigenvalue, then the two are not parallel. Furthermore, the generalized eigenvector can always be redefined by adding a scalar multiple of the eigenvector, thus changing the value of $J$.

Finally, it seems that the diagnostic $|J| \to 1$ does not tell us anything about the generalized eigenvector; both $u$ and $v$ converge to the same proper eigenvector, up to a scale. We will see how this information comes about.

Let me concretely illustrate the foregoing discussion on a test example. The matrix
\begin{equation} \label{eq:test_2}
M(x) = \begin{pmatrix}
0 & 1 \\
0 & x
\end{pmatrix}
\end{equation}
has (unnormalized) eigenvectors $(1,0)$ and $(1,x)$. As $x \to 0$, they become equal, and a Jordan block obviously forms in the limiting matrix $M(0)$. If $u$ and $v$ are the two eigenvectors, then
\begin{equation}
J(u,v) = \frac{(u|v)}{\sqrt{(u|u)(v|v)}} = \frac{1}{\sqrt{1+x^2}},
\end{equation}
which is clearly less than $1$ for $x \ne 0$ and $1$ for $x = 0$.

Next, I turn to the construction of the generalized eigenvector at $x = 0$, which is $(0,1)$. Actually, it can be anything of the form $(y,1)$, and I have made the choice that is orthogonal. One way to construct the generalized eigenvector from the proper eigenvectors as $x \to 0$ is to perform Gram--Schmidt orthogonalization. For example, if we start with the pair $(u,v)$, subtract off the component of $v$ along $u$, and renormalize, we obtain $u$ and $\tilde v = (0,1)$, which is exactly what we want. Of course, for generic $x$, $\tilde v$ is no longer an eigenvector. Because both $u$ and $v$ converge to the same vector at $x = 0$, we should be able to treat them on an equal footing. If we apply the Gram--Schmidt procedure to the ordered pair $(v,u)$, the result is
\begin{equation}
(v,u) \overset{\text{GS}}{\to} (v,\tilde u) \equiv \left((1,x),\left(\frac{|x|}{\sqrt{1+x^2}}, -\frac{\sgn x}{\sqrt{1+x^2}}\right)\right),
\end{equation}
and we have
\begin{equation}
\lim_{x\to 0}(v,\tilde u) = ((1,0),(0,\mp 1)),
\end{equation}
depending on whether the limit is approached from above or below. Either way, we recover the eigenvector and the generalized eigenvector up to a sign, which is easily corrected.

Giving a label to the preceding situation, it seems natural to say that $M(x)$ is an \emph{emerging Jordan block} of rank-2 in the limit $x \to 0$.

In a sense, the emerging Jordan block with the Gram--Schmidt-orthogonalized vector is the only possible outcome if the two eigenvectors become parallel. Consider two eigenvectors with distinct eigenvalues that become parallel in some limit. They span a two-dimensional subspace $W \subset V$. If we track this subspace, as the limit is taken, it remains two-dimensional, but we seem to have only one basis vector for it as the eigenvectors have converged to become parallel. Equally well, we can take as a basis of $W$ one of the eigenvectors and an orthonormalized one constructed from the other. The second vector is no longer an eigenvector, but it remains well-defined as the limit is taken. It does not become a proper eigenvector, either, because if we had two linearly independent eigenvectors in the limit then they would not have become parallel. The limiting subspace can only be the eigenspace associated to a single eigenvalue, because of the primary decomposition theorem (Section \ref{Jordan}). The Gram--Schmidt vector must then be a generalized eigenvector, and the operator restricted to this subspace has a nontrivial Jordan form.

\sectionbreak

Let us now consider the rank-3 case. Consider the matrix
\begin{equation}
M(x,y) = \begin{pmatrix}
0 & 1 & 0 \\
0 & x & 1 \\
0 & 0 & y
\end{pmatrix},
\end{equation}
for which we consider the limit $x,y\to 0$. The eigenvectors are
\begin{subequations}
\begin{gather}
u = (1,0,0), \\
v = (1,x), \\
w = (1,y,y(y-x)),
\end{gather}
\end{subequations}
corresponding to eigenvalues $0$, $x$, and $y$. The $J$ values are
\begin{subequations}
\begin{gather}
J(u,v) = \frac{1}{\sqrt{1+x^2}}, \\
J(u,w) = \frac{1}{\sqrt{1+y^2+y^2(y-x)^2}}, \\
J(v,w) = \frac{1+xy}{\sqrt{(1+x^2)(1+y^2+y^2(y-x)^2)}}.
\end{gather}
\end{subequations}
As $x,y\to 0$, $u,v,w \to (1,0,0)$ and become parallel, all three $J$ measures become $1$, and we obtain a rank-3 Jordan block for the eigenvalue zero. We can also consider separately the three cases where:
\begin{enumerate}
\item $x \to 0$, with $y$ fixed.
\item $y \to 0$, with $x$ fixed.
\item $y \to x$, with $x$ fixed to a nonzero value.
\end{enumerate}
Case 1 (resp. 2, 3) leads to $J(u,v) \to 1$ (resp. $J(u,w) \to 1$, $J(v,w) \to 1$), with the other two tending to a limit that is less than 1. The limiting matrix then displays a rank-2 Jordan block for the eigenvalue $0$ (resp. $0$, $x$), and a single eigenvalue $y$ (resp. $x$, $0$) that has a geometric eigenvector.

Applying the Gram--Schmidt procedure to the ordered triple $(u,v,w)$, we obtain the standard basis 
\begin{equation}
(u,v,w) \overset{\text{GS}}{\to} ((1,0,0), (0,1,0), (0,0,1)),
\end{equation} which is exactly the geometric eigenvector along with the two generalized eigenvectors of the Jordan block. If we consider any other ordering for the three eigenvectors and take the limit $x,y\to 0$, the same basis results, up to some signs in the vectors, which are easily corrected. That the resulting basis gives the Jordan \emph{canonical} form of $M(0,0)$ is fortuitous; we will see how values of the Jordan couplings emerge in Section \ref{coupling}. In particular, normalization to unity in the Gram--Schmidt procedure is not necessary, so long as norms remain nonzero in the limit.

As before, I will say that $M(x,y)$ is an emerging Jordan block of rank 3 in the limit $x,y\to 0$. But we can also see that $M(x,y)$ contains an emerging Jordan block of rank 2 in any of the three limits $x \to 0$, $y \to 0$, or $y\to x$.

These observations naturally suggest the following conjecture.
\begin{conjecture}[Emerging Jordan block of rank $n$] \label{conjecture_emerging}
Let $A:X \to L(V)$ be a map whose codomain is the space of linear operators on the $N$-dimensional vector space $V$ (over $\mathbb R$ or $\mathbb C$), with $X$ a metric space. Suppose $A(\xi)$ is diagonalizable for $\xi\in X$, with distinct eigenvalues $\{\lambda_1(\xi), \ldots, \lambda_N(\xi)\}$ and associated eigenvectors $\{u_1(\xi), \ldots, u_N(\xi)\}$. Let $\xi_0\in\partial X \subset \overbar X$, where $\partial X$ is the boundary of $X$ and $\overbar X$ the closure of $X$. The domain of $A$ may be extended to $\overbar X$, since $L(V)$ is complete. Suppose, as $\xi \to \xi_0$,
\begin{enumerate}
\item A subset of the eigenvalues, $\{\lambda_1(\xi),\ldots,\lambda_n(\xi)\}$, all tend to the same limit $\lambda$;
\item for all pairs $\{i,j\} \subset \{1,\ldots,n\}$, $\lim_{\xi\to\xi_0}|J(u_i(\xi),u_j(\xi))| = 1$;
\item for all pairs $\{i,j\} \subset \{1,\ldots,N\}$ such that $i \le n < j \le N$, $\lim_{\xi\to\xi_0}|J(u_i(\xi),u_j(\xi))| < 1$.
\end{enumerate}
Then $A(\xi_0)$ contains a Jordan block of rank $n$ for the eigenvalue $\lambda$. A basis of the generalized eigenspace is given by performing the Gram--Schmidt orthonormalization procedure on the set $\{u_1(\xi),\ldots,u_n(\xi)\}$, in any order, and taking the limit $\xi\to\xi_0$. The matrix of $A(\xi_0)$ restricted to this eigenspace, in the basis of the limiting Gram--Schmidt vectors, is upper triangular. We say that $A(\xi)$ contains an emerging Jordan block of rank-$n$ in the limit $\xi \to \xi_0$, or if $n = N$, then $A(\xi)$ is an emerging Jordan block in the same limit.
\end{conjecture}
As formal as this conjecture sounds, the assumptions are not overly restrictive, and merely abstract the two concrete examples I exhibited in the most general terms I could imagine. In this formal setting, the rank-2 example corresponds to $X = \mathbb R\backslash \{0\}$, the punctured real line, and the rank-3 example corresponds to $X = \mathbb R^2\backslash \{(0,0)\}$, the punctured plane. I will prove this for the cases $n = 2$ and $3$ with $\overbar X = \mathbb R^n$, and give expressions that suggest how to approach the problem and prove it for all positive integers $n$ in Section \ref{coupling}.

I also note that the Gram--Schmidt procedure must produce orthonormal vectors (in particular, no zero vectors) since the starting vectors are linearly independent, being eigenvectors with distinct eigenvalues.

All of the examples thus far have dealt with finite-dimensional vector spaces, and these most directly inform Conjecture \ref{conjecture_emerging}. Certainly many subtleties exist in infinite-dimensional vector spaces that make many conclusions harder to establish or false. But it seems natural here to boldly proclaim an obvious generalization.

\begin{conjecture}[Emerging Jordan blocks in infinite dimensions] \label{conjecture_infinite}
Conjecture \ref{conjecture_emerging} holds when suitably generalized to infinite-dimensional vector spaces and arbitrary-rank emerging Jordan blocks. The dimension may be countably infinite or uncountable.
\end{conjecture}

To some extent some of Conjecture \ref{conjecture_infinite} must be true if we are to study separable Hilbert spaces, the domains of our conformal field theories, using a sequence of finite-dimensional vector spaces to extrapolate their properties accurately.

The framework described in this section may also be applied to a sequence of operators $(A(n))_{n=1}^\infty$ in the limit $n \to \infty$, thinking of the eigenvalues and $J$ measures at discrete parameter values $n$ being smoothly interpolated between the discrete values, even if the operator itself does not exist for noninteger $n$.

\sectionbreak

In general, an operator may have multiple emerging Jordan blocks as a limit is taken. The limiting operator will then have a Jordan form that contains multiple blocks, all emerging simultaneously. Thus an appropriate name for the formalism as a whole is that of the \emph{emerging Jordan form}.

\section{Rates of convergence and the Jordan coupling} \label{coupling}

The examples of the previous section may have seemed artificial in that I began with a matrix whose limiting form was obviously a Jordan block, and then proceeded to verify that fact. In fact, those observations hold in a more general setting where the conclusion is not evident in the setup, as I demonstrate for ranks 2 and 3. Unfortunately, the generalization to arbitrary $n$ is not as straightforward as allowing all indices to run from $1$ to $n$; we will see what complications arise in the rank-3 case compared to the rank-2 case, and what it might mean for the general rank-$n$ case.

The primary recurring theme of this section is that the way in which $J$ depends on the differences among the eigenvalues has implications for the rank and form of the emerging Jordan blocks.

I believe the propositions and conjectures in this chapter are of quite general validity. However, to deal with the simplest cases, in this section \emph{all vector spaces are real}. Certainly the reader should be able to imagine that a generalization exists for complex vector spaces, although the expressions will pick up complex conjugates in some places, and signs become phases. But the real case, which I have studied the most thoroughly (especially in the applications of Part \ref{applications}), should suffice to validate the ideas proposed here.

\subsection{Rank 2}

Consider more generally the matrix
\begin{equation} \label{eq:matrix_2}
M(x,y) = \begin{pmatrix}
x & a \\
0 & y
\end{pmatrix}.
\end{equation}
(The conclusions do not change if $a = a(x,y)$ depends continuously on the parameters.) By changing the sign of one of the basis vectors, we may assume without loss of generality that $a \ge 0$. The eigenvectors corresponding to eigenvalues $x$ and $y$ are now $u = (1,0)$ and $v = (a,y-x)$. Their $J$ measure is
\begin{equation} \label{eq:rank_2_Juv}
J \equiv J(u,v) = \frac{a}{\sqrt{a^2 + (y-x)^2}}.
\end{equation}
The $J$ measure correctly identifies that as long as $a \ne 0$, $M(x,y)$ is an emerging Jordan block of rank 2 as $y \to x$. Taking series expansions in $y-x$, we find that
\begin{subequations} \label{eq:ratio_series_a}
\begin{gather}
J = 1 - \frac{(y-x)^2}{2a^2} + O((y-x)^4), \\
\sqrt{2(1-J)} = \frac{|y-x|}{a} + O((y-x)^3).
\end{gather}
\end{subequations}
These expressions are invariant under rescaling of the matrix (up to a sign). If $M \to cM$ then the new eigenvalues are $cx$ and $cy$, while the eigenvectors remain the same, and, for instance, $\sqrt{2(1-J)} = |cy - cx|/ca$, where $ca$ is the new off-diagonal matrix element. Turning this line of thought around, the relative rates of convergence of $J \to 1$ and the difference of the eigenvalues to $0$ may be able to tell us about the Jordan coupling.

This hunch turns out to be true, as the following calculation shows. Suppose we have an operator $M$ and two vectors such that
\begin{subequations} \label{eq:rank_2_eigenvectors}
\begin{gather}
Mu = xu, \\
Mv = yv,
\end{gather}
\end{subequations}
and we are given that
\begin{equation} \label{eq:rank_2_J}
\frac{(u|v)}{\sqrt{(u|u)(v|v)}} = J,
\end{equation}
where everything is potentially dependent on $x$ and $y$, but not denoted explicitly. Note also that we have not assumed $M$ to act on a 2-dimensional vector space. This situation obtains, for instance, when one partially diagonalizes a larger matrix but does not know a priori the form of $M$ when restricted to these two vectors. By changing the sign of $v$, if necessary, we may assume that $J \ge 0$. Create the new vector (by orthogonalizing $v$ from $u$, then normalizing the result),
\begin{equation}
\tilde v = \frac{\displaystyle v - \frac{(u|v)}{(u|u)}u}{\displaystyle\sqrt{\left(v - \frac{(u|v)}{(u|u)}u\middle|v - \frac{(u|v)}{(u|u)}u\right)}} = \frac{\hat v - J\hat u}{\sqrt{1-J^2}}.
\end{equation}
By routine calculation we find that
\begin{equation}
M\tilde v = y\tilde v + \frac{(y-x)J}{\sqrt{1-J^2}}\hat u.
\end{equation}
Therefore, in the basis $B = (\hat u,\tilde v)$, the matrix of the operator $M$ is
\begin{equation}
M_B = \begin{pmatrix}
x & \displaystyle \frac{(y-x)J}{\sqrt{1-J^2}} \\
0 & y
\end{pmatrix},
\end{equation}
If $\lim_{y\to x}J<1$, then
\begin{equation}
\lim_{y\to x}\frac{(y-x)J}{\sqrt{1-J^2}} = 0.
\end{equation}
This means that if the limiting eigenvectors $u$ and $v$ are not parallel, the limiting matrix representing $M$ is diagonalizable, with $M_B = xI_2$. If, on the other hand, $\lim_{y\to x} J = 1$, then, assuming analyticity, we may write
\begin{equation}
J = 1 - \frac{(y-x)^2}{2a^2} + \cdots,
\end{equation}
with $a > 0$. The linear term disappears since $J$ must have a maximum at $y = x$. The limiting off-diagonal matrix element is then
\begin{equation} \label{eq:rank_2_coupling}
\lim_{y\to x}\frac{(y-x)J}{\sqrt{1-J^2}} = \lim_{y\to x}\frac{y-x}{\sqrt{1-J^2}/J} = a,
\end{equation}
and we obtain a rank-2 Jordan block for the eigenvalue $x$, with Jordan coupling $a$. The central expression can be interpreted as the ratio of the difference in eigenvalues to a function of $J$. To order $x^2$, it may be checked that $\sqrt{2(1-J)} = \sqrt{1-J^2}/J$, so there is no conflict with Eq.~\eqref{eq:ratio_series_a}. Thus, the Jordan coupling is given by the relative rates of convergence. If $a$ is large then $J \to 1$ faster, meaning that the vectors become parallel faster. So the term ``coupling'' is particularly appropriate.

Turning things around yet again, given $M$, $x$, $y$, $u$, $v$, and $J$ as in Eqs.~\eqref{eq:rank_2_eigenvectors} and \eqref{eq:rank_2_J}, we may \emph{define} the emerging Jordan coupling as $a = (y-x)J/\sqrt{1-J^2}$. In these terms, $M$ is an emerging Jordan block of rank 2 if and only if the emerging Jordan coupling tends to a nonzero value. This last characterization is basis-independent, and depends only on knowledge of the eigenvalues and eigenvectors in any basis from which it is possible to calculate $J$.

We have found, in an arbitrary operator $M$, an emerging rank-2 Jordan block based solely on the diagnostic that $J \to 1$. With some simple calculations, we are also able to extract the emerging Jordan coupling. I summarize the preceding discussion in the following result.

\begin{proposition}[Characterizations of emerging rank-2 Jordan blocks]
Let $M$ be a linear operator and $u$ and $v$ two of its eigenvectors, with $x$ and $y$ being their associated eigenvalues. In the limit $y \to x$, the following are equivalent:
\begin{enumerate}
\item $J(u,v) \to 1$.
\item $a \equiv (y-x)J(u,v)/\sqrt{1-J(u,v)^2}$ tends to a nonzero value.
\item $M$ contains an emerging Jordan block of rank 2 for the eigenvalue $x$ with emerging Jordan coupling $a$. 
\end{enumerate}
\end{proposition}

All of the concepts of this section are fully illustrated on a concrete, less artificial example coming from physics in Section \ref{finite_Jordan}, using the matrix in Eq.~\eqref{eq:loop_H_4}. The reader may turn to that example now, which ends with the paragraph containing Eq.~\eqref{eq:Jordan_coupling_56}, as it is also written in a purely mathematical context.

\subsection{Rank 3}

Let us try the same exercise with three vectors and try to obtain the corresponding result for an emerging rank-3 Jordan block. Since the resulting expressions are not those one would immediately guess by analogy with the rank-2 case, it is worth seeing this case written out in detail. Throughout this section I suppress explicit notation of functional dependencies on the real parameters $a$, $b$, $c$, $\lambda_1$, $\lambda_2$, and $\lambda_3$.

Begin with the matrix
\begin{equation} \label{eq:matrix_3}
M = \begin{pmatrix}
\lambda_1 & a & b \\
0 & \lambda_2 & c \\
0 & 0 & \lambda_3
\end{pmatrix}.
\end{equation}
By adjusting the signs of the three basis vectors, we can make the three off-diagonal elements nonnegative. The eigenvectors are
\begin{subequations}
\begin{gather}
u_1 = (1,0,0), \\
u_2 = (a,\lambda_2-\lambda_1,0), \\
u_3 = (ac+b(\lambda_3-\lambda_2), c(\lambda_3-\lambda_1), (\lambda_3-\lambda_1)(\lambda_3-\lambda_2)),
\end{gather}
\end{subequations}
corresponding to eigenvalues $\lambda_1$, $\lambda_2$, and $\lambda_3$. The values $J_{ij} \equiv J(u_i,u_j)$ are
\begin{subequations} \label{eq:rank_3_Jijs}
\begin{gather}
J_{12} = \frac{a}{\sqrt{a^2 + (\lambda_2-\lambda_1)^2}}, \\
J_{13} = \frac{ac+b(\lambda_3-\lambda_2)}{\sqrt{a^2 c^2 + c^2(\lambda_3-\lambda_1)^2 + b^2(\lambda_3-\lambda_2)^2 + (\lambda_3-\lambda_1)^2(\lambda_3-\lambda_2)^2 + 2abc(\lambda_3-\lambda_2)}}, \\
J_{23} = \frac{a^2c + ab(\lambda_3-\lambda_2) + c(\lambda_3-\lambda_1)(\lambda_3-\lambda_2)}{\sqrt{(a^2 + (\lambda_2-\lambda_1)^2)(a^2c^2+c^2(\lambda_3-\lambda_1)^2 + b^2(\lambda_3-\lambda_2)^2 + (\lambda_3-\lambda_1)^2(\lambda_3-\lambda_2)^2 + 2abc(\lambda_3-\lambda_2))}}.
\end{gather}
\end{subequations}
It may be checked that as $\lambda_2,\lambda_3 \to \lambda_1$, all three $J$ values approach $1$ if and only if the Jordan canonical form of the limiting matrix has a rank-3 Jordan block. Whether this situation obtains depends on the values of $a$, $b$, and $c$. It may also be checked that in the limit of one pair of eigenvalues becoming equal, with the third fixed to a different value, the $J$ value of the respective pair of eigenvectors tends to 1 if and only if the Jordan canonical form of the limiting matrix has a rank-2 Jordan block. The verification of these statements requires a tedious examination of cases and is best done by computer algebra.

Now suppose the differences in eigenvalues are all small ($\lambda_2-\lambda_1 = O(\epsilon)$, $\lambda_3-\lambda_2 = O(\epsilon)$, $\lambda_3-\lambda_1 = O(\epsilon)$), and for convenience assume $\lambda_3 > \lambda_2 > \lambda_1$. Taking series expansions gives
\begin{subequations}
\begin{gather}
J_{12} = 1 - \frac{(\lambda_2-\lambda_1)^2}{2a^2} + O(\epsilon^4), \\
J_{13} = 1 - \frac{(\lambda_3-\lambda_1)^2}{2a^2} + \frac{b(\lambda_3-\lambda_1)^2(\lambda_3-\lambda_2)}{a^3c} + O(\epsilon^4), \\
J_{23} = 1 - \frac{(\lambda_3-\lambda_2)^2}{2a^2} + \frac{b(\lambda_3-\lambda_2)^2(\lambda_3-\lambda_1)}{a^3c} + O(\epsilon^4).
\end{gather}
\end{subequations}
Perhaps surprisingly, all of the leading nontrivial coefficients only give information about $a$. To get the ratio $b/c$, we must use the third-order term, and to get the values of $b$ and $c$ separately we need the fourth-order term (not shown here). Based on the expressions, it does not seem productive to fiddle around with the three $J$ measures and guess at the functions of them that yield the three parameters $a$, $b$, and $c$, especially when they arise more readily in an upcoming analysis.

However, the fact that the leading terms have the same dependence on $a$ yields a useful test. Suppose, among a set of eigenvectors, we are looking for three of them that form an emerging rank-3 Jordan block. Instead of studying every subset of three vectors in detail, one may simply calculate the $J$ measures between all pairs, and extract the (approximate) value of $a$ for that pair (for instance, by computing $(\lambda_j - \lambda_i)J_{ij}/\sqrt{1-J_{ij}^2}$, as in Eq.~\eqref{eq:rank_2_coupling}). All three pairs among the candidate vectors forming an emerging rank-3 Jordan block must display the same coefficient. 

Conversely, suppose we are given an operator $M$ and three vectors such that
\begin{subequations}
\begin{gather}
Mu_1 = \lambda_1 u_1, \\
Mu_2 = \lambda_2 u_2, \\
Mu_3 = \lambda_3 u_3,
\end{gather}
\end{subequations}
and suppose we know the values $J_{ij} \equiv J(u_i,u_j)$. We assume the $\lambda_i$ are distinct and study the situation in the limit where they all tend to the same value. Since the vector spaces are assumed real, $J_{ji} = J_{ij}$. Let us orthonormalize the eigenvectors following the Gram--Schmidt procedure, and obtain
\begin{subequations} \label{eq:rank_3_GS}
\begin{gather}
e_1 = \hat u_1, \\
e_2 = \frac{\hat u_2 - J_{12}\hat u_1}{\sqrt{D_2}}, \\
e_3 = \frac{(1-J_{12}^2)\hat u_3 - (J_{13} - J_{12}J_{23})\hat u_1 - (J_{23} - J_{12}J_{13})\hat u_2}{\sqrt{D_2D_3}},
\end{gather}
\end{subequations}
where $D_2 = 1 - J_{12}^2$ and $D_3 = 1 - J_{12}^2 - J_{13}^2 - J_{23}^2 + 2J_{12}J_{13}J_{23}$. In this basis, $B = (e_1, e_2, e_3)$, the operator $M$ has the matrix representation
\begin{equation} \label{eq:emerging_rank_3}
M_B = \begin{pmatrix}
\lambda_1 & \displaystyle\frac{(\lambda_2 - \lambda_1)J_{12}}{\sqrt{D_2}} & \displaystyle\frac{(\lambda_3 - \lambda_1)(J_{13} - J_{12}J_{23}) + (\lambda_3 - \lambda_2)(J_{23} - J_{12}J_{23})J_{12}}{\sqrt{D_2D_3}} \\[2ex]
0 & \lambda_2 & \displaystyle\frac{(\lambda_3 - \lambda_2)(J_{23} - J_{12}J_{13})}{\sqrt{D_3}} \\[2ex]
0 & 0 & \lambda_3
\end{pmatrix}.
\end{equation}
The differences in the eigenvalues again play an important role.

It may be checked that substitution of Eq.~\eqref{eq:rank_3_Jijs} into Eq.~\eqref{eq:emerging_rank_3} yields Eq.~\eqref{eq:matrix_3}, when $\lambda_3 > \lambda_2 > \lambda_1$. Lifting this assumption, we still obtain the same matrix, up to possible signs appearing in front of $a$, $b$, and $c$. The verification of this assertion is yet again another tedious exercise in algebra that is best turned over to a computer.

\subsection{Rank $n$}

As of this writing, the general formalism for emerging rank-$n$ blocks has yet to be completed in a systematic manner. Instead, I collect some facts and considerations that seem unavoidable in the process of completing this theory.

In the rank-2 and rank-3 cases I adjusted the signs of the eigenvectors so that the off-diagonal matrix elements would come out to be nonnegative. The reason is that through various computational explorations, it seemed that only the absolute values of the off-diagonal elements were relevant, appearing in expressions as $a^2$ or $|a|$, for instance. This was always possible because the number of off-diagonal elements, $n(n-1)/2$, was at most the dimension of the space $n$. For $n \ge 4$, it does not seem immediately obvious that the same assumption constitutes no loss of generality, and that nonnegative off-diagonal elements are always possible through adjusting signs. At the same time, it is not obvious that there is an easy counterexample, either. There exist constraints among the $J$ values, so that not all of them are independent. For example, it is intuitively obvious that
\begin{equation}
|J_{ik}| = 1 \implies |J_{ij}| = |J_{jk}|,
\end{equation}
which follows easily from the definitions and the Cauchy--Schwarz inequality.

One such constraint follows from the unremarkable observation that $J_{ij} \equiv \cos\theta_{ij}$ is just the cosine of the angle between $u_i$ and $u_j$ in the geometry of the space determined by the inner product, where $0 \le \theta \le \pi$. From the geometric fact $\theta_{ij} + \theta_{jk} \ge \theta_{ik} \ge |\theta_{ij} - \theta_{jk}|$ it is not hard to show
\begin{equation}
\frac12\left(\sqrt{(1+J_{ij})(1+J_{jk})} + \sqrt{(1-J_{ij})(1-J_{jk})}\right) \ge \sqrt{\frac{1+J_{ik}}{2}} \ge \frac12\left(\sqrt{(1+J_{ij})(1+J_{jk})} - \sqrt{(1-J_{ij})(1-J_{jk})}\right).
\end{equation}

Suppose we begin with an operator $M$ and $n$ eigenvectors $u_i$ such that $Mu_i = \lambda_i u_i$. Following the steps of the rank-2 and rank-3 analyses, the first step is to create an orthonormal basis. The result of the Gram--Schmidt process can be written in closed form, using formal determinants. Starting with the ordered set of vectors $(u_1,\ldots,u_n)$, the result is
\begin{subequations}
\begin{equation}
e_j = \frac{1}{\sqrt{\Delta_{j-1}\Delta_j}} \det \begin{pmatrix}
(u_1|u_1) & (u_2|u_1) & \cdots & (u_j|u_1) \\
(u_1|u_2) & (u_2|u_2) & \cdots & (u_j|u_2) \\
\vdots & \vdots & \ddots & \vdots \\
(u_1|u_{j-1}) & (u_2|u_{j-1}) & \cdots & (u_j|u_{j-1}) \\
u_1 & u_2 & \cdots & u_j
\end{pmatrix},
\end{equation}
with
\begin{equation}
\Delta_j = \det\begin{pmatrix}
(u_1|u_1) & (u_2|u_1) & \cdots & (u_j|u_1) \\
(u_1|u_2) & (u_2|u_2) & \cdots & (u_j|u_2) \\
\vdots & \vdots & \ddots & \vdots \\
(u_1|u_j) & (u_2|u_j) & \cdots & (u_j|u_j)
\end{pmatrix}.
\end{equation}
\end{subequations}
Formal determinants containing vectors are defined by cofactor expansion. The result of the Gram--Schmidt procedure is unaffected by rescaling of all vectors by positive constants. We may therefore replace $u_i$ with $\hat u_i$ and obtain
\begin{subequations}
\begin{equation}
e_j = \frac{1}{\sqrt{D_{j-1}D_j}} \det \begin{pmatrix}
J_{11} & J_{21} & \cdots & J_{j1} \\
J_{12} & J_{22} & \cdots & J_{j2} \\
\vdots & \vdots & \ddots & \vdots \\
J_{1,j-1} & J_{2,j-1} & \cdots & J_{j,j-1} \\
\hat u_1 & \hat u_2 & \cdots & \hat u_j
\end{pmatrix},
\end{equation}
with
\begin{equation}
D_j = \det(J_{rs})_{r,s=1}^j,
\end{equation}
\end{subequations}
using the fact that $(\hat u_i|\hat u_j) = J_{ij}$. As a check, the first three values give the unit vectors of Eq.~\eqref{eq:rank_3_GS}, using symmetry and $J_{ii} = 1$.

Because $e_j$ is a sum of terms, each of which contains exactly one $\hat u_i$, the action of $M$ on $e_j$ is given by the same expression for $e_j$ with $\hat u_i$ replaced by $\lambda_i\hat u_i$:
\begin{equation}
Me_j = \frac{1}{\sqrt{D_{j-1}D_j}} \det \begin{pmatrix}
J_{11} & J_{21} & \cdots & J_{j1} \\
J_{12} & J_{22} & \cdots & J_{j2} \\
\vdots & \vdots & \ddots & \vdots \\
J_{1,j-1} & J_{2,j-1} & \cdots & J_{j,j-1} \\
\lambda_1\hat u_1 & \lambda_2\hat u_2 & \cdots & \lambda_j\hat u_j
\end{pmatrix}
\end{equation}
To see this, imagine performing the cofactor expansion along the bottom row, applying $M$, then putting the result back into formal determinant form.

\section{Next steps}
\subsection{Further study of emerging Jordan blocks} \label{further_study}

The typical view of the Jordan canonical form today is that it is primarily a conceptual tool, and not useful for real computations \cite{Boyd2008}. Either there is a 1 in the superdiagonal of the Jordan canonical form or there is a 0, and this 1 can be destroyed by an infinitesimal perturbation of a matrix element. However, in this chapter I have built an approximation to the Jordan form that remains stable to small perturbations. This fact allows the tools of analysis---limits and continuity---to be brought into a problem that had previously been thought to be discontinuously all-or-nothing.

For the rank-2 case, there is a curious observation. Let $M(x)$ be continuously parametrized by $x$, and $u(x)$ and $v(x)$ the corresponding eigenvectors that lead to an emerging Jordan block in some limit. In each case we have seen so far, one of the eigenvectors is independent of the parameter $x$ (more generally, a meaningful combination of parameters that tends to zero, as shown in the following examples). For the eigenvectors of the test matrix in Eq.~\eqref{eq:test_2}, the eigenvector $(1,0)$ is obviously independent of $x$ as it forms an emerging Jordan block with $(1,x)$. More generally, for $M(x,y)$ in Eq.~\eqref{eq:matrix_2}, $u$ is independent of $y - x$. In the example of Section \ref{finite_Jordan} beginning with Eq.~\eqref{eq:loop_H_4}, eigenvector $v_5$ of Eq.~\eqref{eq:eigenvectors_generic_m} does not depend on $m$. In the test matrix of Eq.~\eqref{eq:matrix_3} we can see three examples of this pattern: if $\lambda_j \to \lambda_i$ there is an emerging Jordan block between $u_i$ and $u_j$, with $u_i$ being independent of $\lambda_j - \lambda_i$, where $1 \le i < j \le 3$. So I wonder if this is a general pattern that can be proved.

I have worked out the rank-2 and rank-3 emerging Jordan blocks in full detail. While the general rank-$n$ case will likely be fairly involved and certainly less commonly used than the lowest rank cases, it would be desirable to complete this derivation and give a sense of coherence to the framework as well as links to other fields of mathematics where it might be profitably used. Pushing this to its limit, an emerging countably infinite Jordan tower may also find practical application and provide new perspectives, as I alluded to in Section \ref{Jordan} with the harmonic oscillator.

Finally, I have assumed that all vector spaces were real, which simplified many calculations as $J_{ij} = J_{ji}$. What might we find for complex inner product spaces, or inner product spaces over other complete topological fields? More generally, suppose $R$ is a commutative ring with identity, and $J$ a proper ideal. Then the concept of Cauchy sequence can be defined \cite[p.~162]{Lang2002}, and the $J$-adic completion of $R$ is (isomorphic to) the inverse limit $\varprojlim R/J^n$. Modules over the $J$-adic completion could potentially furnish another setting for emerging Jordan blocks, since there is now the concept of the limit of a sequence tending to the unit element.

\subsection{Possible applications of emerging Jordan blocks} \label{emerging_applications}

The matrix
\begin{equation}
\begin{pmatrix}
1 & 1 \\
\epsilon^2 & 1
\end{pmatrix}
\end{equation}
has eigenvalues $1+\epsilon$ and $1-\epsilon$, with respective eigenvectors $(1,\epsilon)$, $(1,-\epsilon)$. You and I agree that this would be a fairly asinine basis to use for practical computations when $\epsilon$ is small, and especially when $\epsilon$ is not certain, but a random variable following a distribution with zero mean and some small variance. This latter situation is common, such as that resulting from compounded numerical errors. In this case, the emerging Jordan block would be a more useful form of the matrix, and the coefficients of other vectors expanded in the orthogonal basis obtained from the Gram--Schmidt procedure would be stable to small perturbations in $\epsilon$.

A well-known result from the theory of second-order linear differential equations (with generalizations to higher-order equations) involves a so-called circuit matrix \cite[Theorem 15.2.1]{Hassani2013}.
\begin{theorem} \label{SOLDE}
If $p(z)$ and $q(z)$ are analytic for $r_1 < |z - z_0| < r_2$, then the second-order linear differential equation $w'' + p(z)w' + q(z)w = 0$ admits a basis of solutions $\{w_1,w_2\}$ in the neighborhood of the singular point $z_0$, where either
\begin{subequations}
\begin{gather}
w_1(z) = (z-z_0)^\alpha f(z) \qq{and} \\
w_2(z) = (z-z_0)^\beta g(z),
\end{gather}
\end{subequations}
or, in exceptional cases (when the circuit matrix is not diagonalizable),
\begin{subequations}
\begin{gather}
w_1(z) = (z-z_0)^\alpha f(z) \qq{and} \\
w_2(z) = w_1(z)[g_1(z) + \log(z-z_0)].
\end{gather}
\end{subequations}
The functions $f(z)$, $g(z)$, and $g_1(z)$ are analytic and single-valued in the annular region.
\end{theorem}
In Theorem \ref{SOLDE}, the exponents $\alpha$ and $\beta$ are the eigenvalues of the rank-2 circuit matrix. If, however, $\alpha$ and $\beta$ are very near each other, the solutions $w_1$ and $w_2$ in the theorem could look very similar to each other, and may not furnish as useful a basis. Theorem \ref{SOLDE} could be reformulated in terms of an emerging Jordan block for the circuit matrix, and one would see how the logarithm emerges as well.

The fact that logarithms appear in the correlation functions of logarithmic conformal field theory is well-connected to the fact that the dilation operator $L_0$ is not diagonalizable. These logarithmic terms were initially found by subtracting divergences and taking limits, a procedure that probably also has a parallel to subtracting components along earlier vectors in the Gram--Schmidt process.

\chapter{Biorthogonal and dual Jordan quantum physics} \label{biorthogonal}

One of the postulates of essentially all formulations of quantum physics involves the stipulation that observables are represented by hermitian operators. This assumption has the important consequences that their eigenvalues, representing the possible values of measurements, are real, and that the time evolution $\e^{-\i Ht}$ associated with an hermitian Hamiltonian is unitary. Furthermore, the eigenstates are orthogonal (Theorem \ref{spectral_theorem_eigenvectors}).

Non-hermitian Hamiltonians have appeared in the physics literature, and are usually dealt with on an ad hoc basis. For instance, Hamiltonians with eigenvalues that have a negative imaginary part have been used to describe decay processes, and a complex eigenvalue automatically renders the Hamiltonian to be non-hermitian. To illustrate how such a Hamiltonian implies decay, suppose $H\psi = (E-\i\Gamma)\psi$, with $\Gamma > 0$. Then the squared norm of the initial state decays as
\begin{equation}
\psi^\dagger(t)\psi(t) = (\e^{-\i Ht}\psi)^\dagger \e^{-\i Ht}\psi = \e^{-2\Gamma t}\psi^\dagger\psi,
\end{equation}
with $\Gamma$ proportional to the decay rate, or inversely proportional to the lifetime. Nevertheless, the use of a non-hermitian Hamiltonian can be regarded as an approximation to a more fundamental theory, since decay processes should really be treated in the formalism of quantum field theory, which allows for the creation and annihilation of particles.

In another context, \textcite{Bender2007} and collaborators \cite{BenderBoettcher1998} have developed what is now known as $PT$-symmetric quantum mechanics. These involve Hamiltonians with $PT$ symmetry rather than hermiticity. Under certain restrictions, $PT$-symmetric Hamiltonians can be shown to have real eigenvalues, thus preserving the measurement interpretation. However, calculations in this formalism are done by deforming the geometry of the problem (wave functions in one dimension, for instance, are no longer defined on the real line but in ``asymptotic wedges'' in the complex plane) and using the $CPT$ inner product $\ip*{\psi}{\chi}^{CPT}$ (Eq.~(78) in \textcite{Bender2007}), which renders the Hamiltonian to be hermitian again (recall that hermiticity is defined with respect to some inner product [Section \ref{inner_products}], so changing the inner product changes the notion of hermiticity).

Other uses for non-hermitian Hamiltonians have been described within some of the references \cite{Bender2007,BenderBoettcher1998,Brody2013}.

Single Jordan blocks of small rank can be handled as they arise exceptionally in some physical problems. However, there is a growing class of models where Jordan blocks in the Hamiltonian proliferate, and physically relevant operators couple different blocks. Thus they cannot be handled in isolation. The rich indecomposable structures of logarithmic conformal field theory, for instance, can be traced to the nondiagonalizability of the dilation operator, which plays the role of a Hamiltonian. Likewise, nonunitarity is natural in geometric problems where observables are not local with respect to the basic degrees of freedom, and in systems where one has to average over disorder.

In the absence of an analytical framework, or when setting one up becomes intractable, numerical studies can provide a valuable pathway for progress, and inspire the correct analytical approach. Though not conceptually difficult, carrying out accurate and efficient numerical studies requires the synergy of a number of computational techniques. Some of these techniques particularly suited to the problem at hand are described in Chapter \ref{computational}. Taken as a whole, this work should allow the reader to carry out the procedures described in this chapter beginning with an arbitrary operator.

The spectral theorem for normal operators (Section \ref{spectral_theorem}) and the measurement postulate of quantum mechanics (Section \ref{quantum_physics}) may be framed in terms of orthogonal projection operators. The purpose of this chapter is to find these orthogonal projection operators in a situation where the assumption of hermiticity has been lifted.

In this chapter we assume that vector spaces are finite-dimensional.

\section{Biorthogonal projection operators}
The first step is to consider Hamiltonians that are non-hermitian but nevertheless diagonalizable with real eigenvalues. The exposition of this section follows \textcite{Brody2013}, where many additional elementary results are spelled out for the non-degenerate case, and I add in some of my own observations. Because of the primacy of the Hamiltonian in quantum physics, throughout this section, I use ``Hamiltonian'' to mean any linear operator which may be non-hermitian, but is diagonalizable with real eigenvalues. The eigenvalues may have any pattern of degeneracies.

To fix notation, let $H$ be a Hamiltonian. Then $\{u_i, v_i\}$ is a biorthogonal basis if
\begin{subequations}
\begin{gather}
Hu_i = \lambda_i u_i, \\
v_i H = \lambda_i v_i,
\end{gather}
\end{subequations}
and $\{u_i\}$ and $\{v_i\}$ are linearly independent sets. Note that in the matrix picture, $u_i$ is a column vector, and $v_i$ is a row vector. More abstractly, $u_i$ belongs to the underlying vector space and $v_i$ is a linear functional on the vector space. I will compactly denote this situation with the notation ${}_VH = H_U = \Lambda = \diag\{\lambda_i\}$. That is, $VHV^{-1} = U^{-1}HU = \Lambda$, where $U$ is the basis matrix with columns $u_i$ and $V$ is the basis matrix with rows $v_i$. When $H$ is hermitian the bases are identified as $v_i = u_i^\dagger$.

I am intentionally avoiding the use of Dirac's bra-ket notation and being careful with the meaning of the symbol $\dagger$. In practice, whether numerically or algebraically, $\{v_i\}$ is determined via $H^*v_i^* = \lambda_i v_i^*$, with $*$ denoting the conjugate-transpose. Nevertheless, the relation
\begin{equation}
v_i H = \lambda_i v_i \iff H^\dagger v_i^\dagger = \lambda_i v_i^\dagger
\end{equation}
should be true for any anti-involution $\dagger$, and, in particular, it should cause no issue to use the conjugate-transpose of $H$ to compute $\{v_i^*\}$, and consequently $\{v_i\}$, regardless of the proper hermitian adjoint determined by the inner product.

Put another way, for a generic Hamiltonian, $H$ and $H^*$ are not simply related. The conjugate-transpose, expressed in the symbol $*$, loses its significance as a physically meaningful operation, and becomes a computational tool to determine $\{v_i\}$, which contains the information relevant to the physics. Once determined, all physically meaningful quantities should be expressed in terms of $H$, $\{\lambda_i\}$, $\{u_i\}$, and $\{v_i\}$, which are all non-starred.

Thus particularly careful attention must be paid to the ordering of factors since $v_i u_j$ is a row-column matrix product that evaluates to a scalar and $u_i v_j$ is an outer product that results in a linear operator or a matrix.

When $H$ is non-degenerate, $v_i H u_j = \lambda_i v_i u_j$ and $v_i H u_j = \lambda_j v_i u_j$ depending on whether $H$ acts to the left or to the right. Hence $(\lambda_i - \lambda_j)v_i u_j = 0$ implies $v_i u_j = \delta_{ij} v_i u_i$, or that $VU$ is diagonal. If $H$ is degenerate, $M_{ij} = v_i u_j$ is block diagonal with block sizes corresponding to the pattern of degeneracies. Diagonalize $M = SDS^{-1}$. Then $\{u_i', v_i'\}$ is a biorthogonal basis satisfying $v_i' u_j' = \delta_{ij} v_i' u_i'$ with $U' = US$ and $V' = S^{-1}V$. Sending $VU \to (VK^*)(KU)$ where $K$ is block unitary (with the block sizes in the same pattern as the degeneracies, and unitary with respect to the standard inner product) preserves the eigenspaces and the fact that $(VK^*)(KU)$ is diagonal. Additionally, the eigenvectors can be rescaled as $u_i \to c_i u_i$, $v_i \to d_i v_i$ ($c_i,d_i\ne 0$) without affecting this diagonal structure.

The completeness relation is now
\begin{equation}
\sum_i \frac{u_iv_i}{v_iu_i} = 1.
\end{equation}
The projectors
\begin{equation}
\Pi_i = \frac{u_iv_i}{v_iu_i}
\end{equation}
satisfy $\Pi_i\Pi_j = \delta_{ij}\Pi_i$. Since $VU$ is diagonal and the product of invertible matrices, $v_iu_i \ne 0$, so the projectors are well-defined. Thus, for any row vector $u$,
\begin{equation}
u = \sum_i \Pi_i u = \sum_i\left(\frac{v_i u}{v_i u_i}\right)u_i \equiv \sum_i a_i u_i
\end{equation}
is its expansion in the basis $\{u_i\}$. Similarly, for a column vector $v$,
\begin{equation}
v = \sum_i v\Pi_i = \sum_i\left(\frac{vu_i}{v_iu_i}\right)v_i \equiv \sum_i b_i v_i
\end{equation}
is its expansion in the basis $\{v_i\}$.

One may ask about the status of the operators na\"\inodot vely constructed by applying the expression from hermitian quantum mechanics,
\begin{equation}
\Phi_i = \frac{u_i u_i^*}{u_i^* u_i}.
\end{equation}
They are no longer orthogonal projectors as $\Phi_i\Phi_j \ne \delta_{ij}\Phi_i$, but merely idempotent: $\Phi_i^2 = \Phi_i$. However,
\begin{subequations}
\begin{gather}
\Pi_i\Phi_i = \Phi_i, \\
\Phi_i\Pi_i = \Pi_i,
\end{gather}
\end{subequations}
which might be termed ''pairwise idempotent'' (I think these expressions are correct and Eq.~(11) in \textcite{Brody2013} is erroneous). The first of these relations can be strengthened to
\begin{equation}
\Pi_i\Phi_j = \delta_{ij}\Phi_i.
\end{equation}

\section{Another diagnostic for emerging Jordan blocks} \label{diagnostic}
One of the facts demonstrated in the preceding section was that if $\{u_i, v_i\}$ is a biorthogonal basis, then $VU$ is diagonal, or could be diagonalized, and is also invertible. If we now continuously parametrize the Hamiltonian and the biorthogonal basis, we may consider the diagonal elements of $v_i u_i$ as continuous functions of the parameter. The limit $v_i u_i \to 0$ for some $i$ should then signal a nontrivial Jordan block at the eigenvalue $\lambda_i$. This approach has yet to be explored, since the approach in Chapter \ref{emerging} was much simpler, dealing only with right eigenvectors. But completion of this program may yield additional insights and computational tools.

\section{Dual Jordan projection operators} \label{dual_Jordan}

Now suppose furthermore that $H$ is non-diagonalizable, with a nontrivial Jordan structure. There is a column basis $\{u_{ij} | 1\le i\le n, 1\le j\le m_i\}$ and a row basis $\{v_{ij}|1\le i\le n, 1\le j\le m_i\}$ such that
\begin{subequations} \label{eq:dual_Jordan_basis}
\begin{gather}
Hu_{i1} = \lambda_i u_{i1}, \\
Hu_{ij} = \lambda_i u_{ij} + u_{i,j-1}, \quad(j>1) \\
v_{i1}H = \lambda_i v_{i1}, \\
v_{ij}H = \lambda_i v_{ij} + v_{i,j-1}. \quad(j>1)
\end{gather}
\end{subequations}
The index $i$ labels the Jordan block (of rank $m_i$) and $j$ the position of the vector within that tower. By defining $u_{i0} = 0$ we can also use just the second equation in each pair without any restriction on $j$. The $\lambda_i$ do not have to be distinct---there can be an eigenvalue with multiple Jordan blocks. However, the eigenvalues are assumed real. As before, I use the notation $H_U = J_+$, ${}_VH = J_-$, where $J_+$ is upper Jordan and $J_-$ is lower Jordan. As before, the row basis is usually determined via $H^*v_{ij}^* = \lambda_i v_{ij}^* + v_{i,j-1}^*$. Call such a pair $\{u_{ij}, v_{ij}\}$ a dual Jordan basis.

I will build up to the general result in three stages.

\subsection{Single Jordan block}
For the simplest case, let $H_U = J_+$, where $J_+$ consists of a single rank-$n$ Jordan block with eigenvalue $\lambda$. Thus $Hu_1 = \lambda u_1$, and $Hu_j = \lambda u_j+u_{j-1}$ when $2\le j\le n$. As above, we define $u_0 = 0$, and the preceding equation then holds without restriction on $j$.

What degrees of freedom remain in altering the $u_i$ without affecting the Jordan block $J_+$? First, it is clear that rescaling all of the $u_i$ by the same scalar constant will preserve $J_+$. However, they cannot be rescaled independently as in the biorthogonal case, as the relative normalizations are fixed by the superdiagonal couplings in $J_+$. Second, choose a set of scalars $\{c_i|1\le i\le n-1\}$. Then in the basis $\{u_i'\}$, where
\begin{subequations}
\begin{equation}
u_i' = u_i + \sum_{j=1}^{i-1} c_{i-j}u_j,
\end{equation}
$H$ still has the form $J_+$, which is easily verified by direct calculation. These two properties may be condensed by extending the list of coefficients to $\{c_i|0 \le i \le n-1\}$ and setting
\begin{equation} \label{eq:u_shift}
u_i' = u_i + \sum_{j=1}^i c_{i-j}u_j,
\end{equation}
\end{subequations}
which represents a rescaling of all the vectors by a factor $c_0 + 1$ on top of taking linear combinations with other vectors. The $n$ numbers $\{c_i\}$ represent all of the degrees of freedom in linearly shifting the basis while preserving $J_+$. The proof of this assertion follows from the $n = 1$ case in the more general result to be discussed in Section \ref{multiple_blocks}.

We could use these $n$ degrees of freedom to set $u_1^* u_1 = 1$, and to make the highest vector, $u_n$, orthogonal to all the rest: $u_n^* u_j = \delta_{nj}u_n^*u_n$. Generically, the matrix of inner products is still fairly dense:
\begin{equation}
U^*U = (u^*_i u_j) = \begin{pmatrix}
1 & * & \cdots & * & 0 \\
* & * & \cdots & * & 0 \\
\vdots & \vdots & \ddots & \vdots & 0 \\
* & * & \cdots & * & 0 \\
0 & 0 & \cdots & 0 & *
\end{pmatrix}.
\end{equation}
This is the nicest we can make $U^*U$ look. The same remarks apply to the row basis $\{v_i\}$, where $v_i H = \lambda v_i + v_{i-1}$ and $v_0 = 0$. But as we have seen, $u_i^* u_j$ and $v_i v_j^*$ are not particularly meaningful quantities. We will abandon this convention henceforth and adopt another one shortly.

In contrast with the biorthogonal case, the matrix $VU$ is not diagonal, but lower Hankel; the entries are constant across each anti-diagonal and the upper left half of the matrix is empty:
\begin{equation} \label{eq:VU_matrix}
VU = (v_i u_j) = \begin{pmatrix}
0 & 0 & \cdots & 0 & a_1 \\
0 & \reflectbox{$\ddots$} & \reflectbox{$\ddots$} & a_1 & a_2 \\
\vdots & \reflectbox{$\ddots$} & \reflectbox{$\ddots$} & \reflectbox{$\ddots$} & \vdots \\
0 & a_1 & \reflectbox{$\ddots$} & \reflectbox{$\ddots$} & \vdots \\
a_1 & a_2 & \cdots & \cdots & a_n
\end{pmatrix}
\end{equation}
Put differently, $v_i u_j = z_{i+j}$ depends only on $i+j$, where $z_{n+i} \equiv a_i$ as in the matrix above and $z_j = 0$ for $j \le n$. To prove this, simply note that $v_i u_j = v_n(H - \lambda)^{2n-i-j}u_n$, which manifestly depends only on the total $i + j$. If $i+j\le n$ then $v_i u_j = v_{i+j}u_0 = 0$ since $u_0 = 0$, or note also that $(H-\lambda)^{2n-i-j} = 0$ for $i+j\le n$. This computation also implies that $VHU$ is lower Hankel: $v_i H u_j = \lambda v_i u_j + v_i u_{j-1}$, and both terms on the right are represented by lower Hankel matrices. If we reorder the $U$ basis backwards (and form the matrix $\tilde U$), then $H_{\tilde U} = J_-$ and $V\tilde U$ is lower Toeplitz. Similarly, ${}_{\tilde V}H = J_+$ and $\tilde VU$ is upper Toeplitz.

Now we are going to renormalize and shift the $u_i$ to obtain the completeness relation. Define $u_i'$ as in Eq.~\eqref{eq:u_shift} and $a_i$ as in Eq.~\eqref{eq:VU_matrix}. Then we demand
\begin{equation}
v_n u_i' = a_i + \sum_{j=1}^i c_{i-j}a_j = \delta_{i1}. \quad (1\le i\le n)
\end{equation}
Invert the above equations to find a set of coefficients $c_i$ satisfying them, and let $u_i' \to u_i$, dropping the primes. This is possible as long as $a_1 \ne 0$: the $i = 1$ equation reads $a_1(1+c_0) = 1 \implies c_0 = 1/a_1 - 1$, and inductively having solved the $i$th equation, substituting the values up to $c_{i-1}$ into the $(i+1)$th equation yields a linear relation for $c_i$ of the form $a_1 c_i + x = 0$. Furthermore, $a_1 = 0$ cannot happen: $V$ and $U$ are both invertible, since their rows and columns, respectively, form bases, and $\det VU = \pm a_1^n \ne 0$, as can be seen by rotating $VU$ into Toeplitz form.

Having done this, $v_i u_j = \delta_{i+j,n+1}$, which has the nice matrix form (remember that $VU$ is still lower Hankel after redefining $U$ in this way)
\begin{equation} \label{eq:matrix_completeness}
VU = (v_i u_j) = \begin{pmatrix}
0 & \cdots & 0 & 1 \\
\vdots & \reflectbox{$\ddots$} & \reflectbox{$\ddots$} & 0 \\
0 & \reflectbox{$\ddots$} & \reflectbox{$\ddots$} & \vdots \\
1 & 0 & \cdots & 0
\end{pmatrix}.
\end{equation}
There is still a remaining complex degree of freedom, which may be used to scale jointly $u_i \to cu_i$, $v_i \to c^{-1}v_i$. The completeness relation is
\begin{equation}
\sum_{i=1}^n u_i v_{n+1-i} = 1.
\end{equation}
The preceding results look nicer if we flip one of the bases, remembering that doing this also transposes the Jordan form in the new basis:
\begin{subequations}
\begin{gather}
\tilde v_i u_j = v_i \tilde u_j = \delta_{ij} \iff \tilde VU = V\tilde U = 1, \\
\sum_{i=1}^n \tilde u_i v_i = \sum_{i=1}^n u_i \tilde v_i = 1.
\end{gather}
\end{subequations}
The projection operators
\begin{equation}
\Pi_i \equiv u_i v_{n+1-i} = u_i\tilde v_i = \tilde u_{n+i-1}v_{n+i-1}
\end{equation}
satisfy the usual $\Pi_i\Pi_j = \delta_{ij}\Pi_i$ and are therefore orthogonal.

\subsection{Multiple Jordan blocks for a single eigenvalue} \label{multiple_blocks}
The next level of complexity, building on the previous section, is to consider a matrix with a unique eigenvalue $\lambda$. Let $\{u_{ij},v_{ij}|1\le i\le n, 1\le j \le m_1\}$ be a dual Jordan basis such that $H_U = J_+$ and ${}_VH = J_-$. The total dimension of $H$ is the algebraic multiplicity of $\lambda$, $d \equiv \sum_{i=1}^n m_i$, and Eq.~\eqref{eq:dual_Jordan_basis} holds with all $\lambda_i = \lambda$. As before, let $u_{i0} = 0$ for convenience.

Let us follow the same procedure as in the previous section and ask again: what degrees of freedom remain in altering the $u_{ij}$ without affecting the Jordan block $J_+$? Suppose $Hu_{ij}' = \lambda u_{ij}' + u_{i,j-1}'$, with
\begin{equation}
u_{ij}' = u_{ij} + \sum_{k=1}^n \sum_{l=1}^{m_k} c^{ik}_{jl}u_{kl}.
\end{equation}
Explicit calculation yields
\begin{equation}
Hu_{ij}' = \lambda\Big(u_{ij} + \sum_{k=1}^n \sum_{l=1}^{m_k} c^{ik}_{jl}u_{kl}\Big) + u_{i,j-1} + \sum_{k=1}^n \sum_{l=1}^{m_k} c^{ik}_{jl}u_{k,l-1} = \lambda u_{ij}' + u_{i,j-1} + \sum_{k=1}^n \sum_{l=1}^{m_k} c^{ik}_{j,l+1}u_{kl}.
\end{equation}
We must identify the last two terms as $u_{i,j-1}'$:
\begin{equation}
u_{i,j-1}' = u_{i,j-1} + \sum_{k=1}^n \sum_{l=1}^{m_k} c^{ik}_{j,l+1}u_{kl}.
\end{equation}
Thus
\begin{equation}
\sum_{k=1}^n \sum_{l=1}^{m_k} c^{ik}_{j-1,l}u_{kl} = \sum_{k=1}^n \sum_{l=1}^{m_k-1} c^{ik}_{j,l+1}u_{kl}
\end{equation}
implies $c^{ik}_{j-1,l} = c^{ik}_{j,l+1}$ as the $u_{kl}$ form a basis, or that
\begin{equation}
c^{ik}_{jl} \equiv c^{ik}_{j-l}
\end{equation}
depends only on the difference $j-l$. Next, considering
\begin{equation}
Hu_{i1}' = \lambda\Big(u_{i1} + \sum_{k=1}^n \sum_{l=1}^{m_k} c^{ik}_{1-l}u_{kl}\Big) + \sum_{k=1}^n \sum_{l=1}^{m_k} c^{ik}_{1-l}u_{k,l-1} = \lambda u_{i1}' + \sum_{k=1}^n \sum_{l=1}^{m_k} c^{ik}_{1-l}u_{k,l-1}
\end{equation}
implies $c^{ik}_{1-l} = 0$ for $l > 1$, or
\begin{equation}
c^{ik}_p = 0. \quad (p < 0)
\end{equation}
Thus the ranges of the indices are $\{c^{ij}_k|1 \le i \le n, 1 \le j \le n, 0 \le k \le m_i - 1\}$. The number of free parameters is thus $nd$.

As before, $z^{ik}_{j+l} \equiv v_{ij}u_{kl} = v_{im_i}(H-\lambda)^{m_i+m_k-j-l}u_{km_k}$ depends only on the sum $j+l$. Furthermore, $z^{ik}_{j+l} = 0$ if $j+l \le M_{ik} \equiv \max\{m_i,m_k\}$. Once again, in order to obtain the completeness relation, we shift $u_{kl} \to u'_{kl} = u_{kl} + \sum_{pq}c^{kp}_{l-q} u_{pq}$ and demand
\begin{equation} \label{eq:multiple_Jordan_shift}
v_{ij}u_{kl}' = v_{ij}u_{kl} + v_{ij}\sum_{p=1}^n \sum_{q=1}^{m_p} c^{kp}_{l-q} u_{pq} = z^{ik}_{j+l} + \sum_{p=1}^n \sum_{q=1}^{m_p} c^{kp}_{l-q} z^{ip}_{j+q} = \delta_{ik}\delta_{j+l,m_i+1}.
\end{equation}
For fixed values of $i$ and $k$, the range of $j+l$ is $M_{ik} + 1 \le j+l \le m_i + m_k$. The number of equations is then $m_i + m_k - M_{ik} = m_{ik} \equiv \min\{m_i,m_k\}$. Ranging over $1\le i,k\le n$, the total number of constraints is bounded by $nd$:
\begin{equation}
\sum_{i=1}^n\sum_{k=1}^n m_{ik} \le \sum_{i=1}^n \sum_{k=1}^n m_k = \sum_{i=1}^n d = nd.
\end{equation}
It is thus possible to solve the linear equations Eq.~\eqref{eq:multiple_Jordan_shift} for the $nd$ parameters $\{c^{kp}_l\}$, though not necessarily uniquely, so long as $z^{ik}_{j+l}$ is invertible when regarded as a matrix with rows labeled by the composite index $\alpha = (ij)$ and columns labeled by the composite index $\beta = (kl)$. That this is true follows from $z_{\alpha\beta} = v_\alpha u_\beta$ being a matrix product of invertible matrices. When all Jordan blocks have the same size, $m_i = m$ for all $i$, then the coefficients are uniquely determined as follows.

Define $z_j$ and $c_j$ to be matrices with entries $(z_j)_{ik} = z^{ik}_j$ and $(c_j)_{ik} = c^{ik}_j$. Eq.~\eqref{eq:multiple_Jordan_shift} for $j = m$ in matrix form is
\begin{equation}
z_{m+l}^T + \sum_{q=1}^l c_{l-q}z^T_{m+q} = \delta_{l1}I_n.
\end{equation}
The $l = 1$ equation reads
\begin{equation}
z_{m+1}^T + c_0 z_{m+1}^T = I \implies c_0 = (z_{m+1}^T)^{-1} - I_n.
\end{equation}
Iteratively, having solved the $l$th equation to obtain $c_0,\ldots,c_{l-1}$, the $(l+1)$th equation has the form
\begin{equation}
c_l z_{m+1}^T + z_{m+l+1}^T + \sum_{q=2}^{l+1}c_{l+1-q}z_{m+q}^T \equiv c_l z_{m+1}^T + x = 0 \implies c_l = -x(z_{m+1}^T)^{-1},
\end{equation}
where $x$ is composed of known quantities. Thus, the coefficients $\{c_l^{ik}\}$ are determined.

Redefine $u_{ij} \to u'_{ij}$ using the coefficients $\{c^{ik}_l\}$ and drop the primes (no longer assuming all $m_i = m$, so a set of coefficients must be obtained by other means if this doesn't hold). Then as before, $VU$ is block diagonal with blocks of size $m_i$ all of the form given in Eq.~\eqref{eq:matrix_completeness}. The completeness relation reads
\begin{equation}
\sum_{i=1}^n \sum_{j=1}^{m_i} u_{ij}v_{i,m_i + 1 - j} = 1.
\end{equation}
The projection operators
\begin{equation}
\Pi_{ij} = u_{ij}v_{i,m_i + 1 - j}
\end{equation}
satisfy $\Pi_{ij}\Pi_{kl} = \delta_{ik}\delta_{jl}\Pi_{ij}$ and are thus mutually orthogonal.

\subsection{The general Jordan canonical form}
Having handled Jordan forms with a single repeated eigenvalue, results for the general case fall neatly into place. Let $\{u_{ij}, v_{ij}|1\le i\le n, 1\le j \le m_i\}$ be a dual Jordan basis.

I will show first that $v_{ij}u_{kl} = 0$ whenever $\lambda_i \ne \lambda_k$. Thus a semblance of orthogonality among different eigenspaces is retained. The proof follows by double induction on $j$ and $l$. First, I will show that $v_{i1}u_{k1} = 0$. Next, I will show that $v_{i,j-1}u_{k1} = 0$ implies $v_{ij}u_{k1} = 0$ and $v_{i1}u_{k,l-1} = 0$ implies $v_{i1}u_{kl} = 0$. Finally I will show that if $v_{ij}u_{k,l-1} = v_{i,j-1}u_{kl} = 0$, then it follows that $v_{ij}u_{kl} = 0$. The principle of double induction then implies that $v_{ij}u_{kl} = 0$ for all $j,l$ where $1 \le j\le m_i$, $1 \le l \le m_k$.

For the base case, note that $v_{i1}Hu_{k1} = \lambda_i v_{i1}u_{k1} = \lambda_k v_{i1}u_{k1}$ depending on whether $H$ acts to the left or to the right. Hence $(\lambda_i - \lambda_k)v_{i1}u_{k1} = 0$ implies $v_{i1}u_{k1} = 0$ as $\lambda_i \ne \lambda_k$ by assumption.

For the first inductive step, assume $v_{i,j-1}u_{k1} = 0$. Then
\begin{subequations}
\begin{equation}
v_{ij}Hu_{k1} = \lambda_iv_{ij}u_{k1} + v_{i,j-1}u_{k1} = \lambda_iv_{ij}u_{k1}
\end{equation}
when $H$ acts to the left and
\begin{equation}
v_{ij}Hu_{k1} = \lambda_k v_{ij}u_{k1}
\end{equation}
\end{subequations}
when $H$ acts to the right. Thus, once again, $(\lambda_i - \lambda_k)v_{ij}u_{k1} = 0$ and it follows that $v_{ij}u_{k1} = 0$. That $v_{i1}u_{k,l-1} = 0$ implies $v_{i1}u_{kl} = 0$ is similar.

For the second inductive step, assume $v_{ij}u_{k,l-1} = v_{i,j-1}u_{kl} = 0$ for some $j$ and $l$. Then
\begin{subequations}
\begin{equation}
v_{ij}Hu_{kl} = \lambda_i v_{ij}u_{kl} + v_{i,j-1}u_{kl} = \lambda_i v_{ij}u_{kl}
\end{equation}
when $H$ acts to the left and
\begin{equation}
v_{ij}Hu_{kl} = \lambda_k v_{ij}u_{kl} + v_{ij}u_{k,l-1} = \lambda_k v_{ij}u_{kl}
\end{equation}
\end{subequations}
when $H$ acts to the right. Yet again, $(\lambda_i - \lambda_k)v_{ij}u_{kl} = 0$ implies $v_{ij}u_{kl} = 0$, and the proof is complete.

Since eigenspaces with distinct eigenvalues are orthogonal, the procedure of the preceding section can be applied to each eigenvalue to obtain the orthogonality and completeness relations,
\begin{subequations}
\begin{gather}
v_{ij}u_{kl} = \delta_{ik}\delta_{j+l,m_i+1}, \\
\sum_{i=1}^n \sum_{j=1}^{m_i} u_{ij} v_{i,m_1+1-j} = 1.
\end{gather}
\end{subequations}

For purposes of practical computation, it is useful to introduce another index to label the distinct eigenvalues. Following this idea, let $\{u_{ijk}, v_{ijk}|1\le i\le p, 1\le j\le n_i, 1\le k\le m_{ij}\}$ be a dual Jordan basis, where $p$ is the number of distinct eigenvalues $\{\lambda_1,\ldots,\lambda_p\}$, $n_i$ is the number of Jordan blocks for the eigenvalue $\lambda_i$, and $m_{ij}$ is the rank of the $j$th Jordan block for the eigenvalue $\lambda_i$. In terms of this set of indices, the previous results read as follows. The orthogonality of different eigenspaces is simply expressed as $v_{ijk}u_{lmn} = \delta_{il}v_{ijk}u_{imn}$. This last quantity has a dependence only on $k+n$, and not their values separately: $v_{ijk}u_{lmn} = \delta_{il}z^{i;jm}_{k+n}$. To obtain orthogonality and completeness, let $u'_{lmn} = u_{lmn} + \sum_{rs}c^{l;mr}_{n-s} u_{lrs}$ and solve for coefficients $\{c^{l;mr}_q\}$ such that
\begin{equation}
v_{ijk}u'_{lmn} = \delta_{il}\delta_{jm}\delta_{k+n,m_{ij}+1}.
\end{equation}
The completeness relation is then
\begin{equation}
\sum_{i=1}^p\sum_{j=1}^{n_i}\sum_{k=1}^{m_{ij}}u_{ijk}v_{i,j,m_{ij}+1-k} = 1,
\end{equation}
having dropped the primes on $u'_{ijk}$. The projection operators, the summands in the completeness relation, are mutually orthogonal.

\section{Next steps}

In this chapter I have described the basic task of finding the coefficients in the expansion of a general state as a sum of generalized eigenstates of a Hamiltonian, which suffices for the purposes of the rest of the work. There is much more to be done, of course. Conspicuously absent is a discussion of time evolution, which, even if we maintain the same postulate (number 4 in the three formulations of Section \ref{quantum_physics}), takes on a profoundly different character---it is not unitary. Another natural question to ask concerns the nature of measurement. My intuition tells me that the right place to look for this is in terms of the positive operator-valued measurement \cite{NielsenChuang2016}.

The procedure of Section \ref{dual_Jordan} assumed a nontrivial Jordan form for the Hamiltonian under study. Given that I wrote Chapter \ref{emerging} on emerging Jordan forms, since Jordan forms may only appear in some limit, can the procedure be adapted to that situation? It would have to differ from the biorthogonal formalism since the end result of the emerging Jordan form is a matrix form for the Hamiltonian that is not diagonal.

Another possible extension of the theory here is a density matrix formulation. Since the left and right Jordan bases are treated asymmetrically (one is shifted while keeping the other fixed in order to find the projection operators) it may be the case that the projection operators rather than left or right states are the more fundamental object. This extension would also bring in the possibility of applying the density matrix renormalization group (DMRG) technique to systems with nonunitary dynamics---I have been told that DMRG for the nonunitary statistical models discussed in Part \ref{applications} have failed because of the nature of the states \cite{HJcomm}.

\part{Applications} \label{applications}

\chapter{The action of the Virasoro algebra in the two-dimensional Potts and loop models} \label{Virasoro}

The spectrum of conformal weights for the CFT describing the two-dimensional critical $Q$-state Potts model (or its close cousin, the dense loop model) has been known for more than 30 years \cite{diFrancesco1987}. However, the exact nature of the corresponding $\mathrm{Vir}\oplus\overbar{\mathrm{Vir}}$ representations has remained unknown up to now. Here, we solve the problem for generic values of $\mathfrak q$, whose relation to $Q$ is described in Section \ref{Potts_description}. This is achieved by a mixture of different techniques: a careful study of the Koo--Saleur generators (Section \ref{evidence_lattice}), combined with measurements of four-point amplitudes, on the numerical side, and OPEs and the four-point amplitudes recently determined using the ``interchiral conformal bootstrap'' \cite{He2020} on the analytical side. We find that null-descendants of diagonal fields having weights $(h_{r1}, h_{r1})$ (with $r \in \mathbb N^*$) are truly zero, so these fields come with simple $\mathrm{Vir}\oplus\overbar{\mathrm{Vir}}$ (Kac) modules. Meanwhile, fields with weights $(h_{rs}, h_{r,-s})$ and $(h_{r,-s},h_{rs})$ (with $r,s\in \mathbb N^*$) come in indecomposable but not completely reducible representations mixing four simple $\mathrm{Vir}\oplus\overbar{\mathrm{Vir}}$-modules with a familiar diamond shape. The top and bottom fields in these diamonds have weights $(h_{r,-s}, h_{r,-s})$, and form a two-dimensional Jordan block for $L_0$ and $\overbar L_0$. This establishes, among other things, that the Potts-model CFT is logarithmic for generic $\mathfrak q$. Unlike the case of non-generic (root of unity) values of $\mathfrak q$, these indecomposable structures are not present in finite size, but we can nevertheless show from the numerical study of the lattice model how the rank-two Jordan blocks build up in the infinite-size limit.

\section{Overview}

The full solution of the CFT describing the critical $Q$-state Potts model for generic $\mathfrak q$ (or its cousins, the critical and dense $O(n)$ models) in two dimensions still eludes us, more than 30 years after the pioneering work of \textcite{DotsenkoFateev1984}. While most critical exponents of interest were quickly determined (for some, even before the advent of CFT, using Coulomb-gas techniques) \cite{denNijs1983,Nienhuis1984,Saleur1987}, the non-rationality of the theory (for generic $\mathfrak q$) as well as its nonunitarity (inherited from the geometrical nature of the lattice model) made further progress using top-down approaches (such as the one used for minimal unitary models \cite{FQS1985}) considerably more difficult (see also \nameref{introduction}). Several breakthroughs took place, however, in the last decade. First, many three-point functions were determined using connections with Liouville theory at $c < 1$ \cite{DelfinoViti2011,PSVD2013,IJS2016}. Second, a series of attempts using conformal bootstrap ideas \cite{JacobsenSaleur2019,He2020,GoriViti2018,PRS2016,PRS2019,HGSJS2020} led to the determination of some of the most fundamental four-point functions in the problem (namely, those defined geometrically, and hence for generic $\mathfrak q$), also shedding light on the OPE algebra and the relevance of the partition functions determined by \textcite{diFrancesco1987}. In particular, the set of operators---the so-called spectrum---required to describe the partition function \cite{diFrancesco1987} and correlation functions \cite{JacobsenSaleur2019} was settled. While the picture remains incomplete, a complete solution of the problem now appears within reach.

An intriguing aspect of the spectrum proposed by \textcite{diFrancesco1987,JacobsenSaleur2019} is the appearance of fields with conformal weights given by the Kac formula $h_{rs}$, with $r,s\in\mathbb N^*$. We call these weights \emph{degenerate}. It is known that for some of these fields---such as the energy operator with weights $(h_{21}, h_{21})$---the null-state descendants are truly zero, and the corresponding four-point functions obey the Belavin--Polyakov--Zamolodchikov (BPZ) differential equations \cite{BPZ1984}. It is also expected that this does not hold for \emph{all} fields with degenerate weights. In fact, it was suggested by \textcite{He2020,JacobsenSaleur2019} that, in the Potts-model case, \emph{only} fields with weights $(h_{r1},h_{r1})$ give rise to null descendants. Since the spectrum of the model is expected to contain non-diagonal fields with weights $(h_{rs}, h_{r,-s})$ and $(h_{r,-s}, h_{rs})$ for $r,s\in\mathbb N^*$, this means that the theory should contain fields with degenerate (left or right) weights whose null descendants are nonzero, even though their two-point function vanishes. It has been well understood since the work of \textcite{Gurarie1993} that in this case, ``logarithmic partners'' must be invoked to compensate for the corresponding divergences occurring in the OPEs. Such partners give rise to Jordan blocks for $L_0$ or $\overbar L_0$, and make the theory a logarithmic CFT---i.e., a theory where the action of the product of left and right Virasoro algebras $\mathrm{Vir}\oplus\overbar{\mathrm{Vir}}$ is not fully reducible. This, in turn, is made possible by the theory not being unitary in the first place \cite{GRR2013}.

A great deal of our understanding of the fields with degenerate weights in the Potts model comes from indirect arguments, such as the solution of the bootstrap equations for correlation functions and the presence of an underlying ``interchiral'' algebra, responsible for relations between some of the conformal-block amplitudes \cite{He2020}. This chapter explores the issue much more directly using the lattice regularization of $\mathrm{Vir}\oplus\overbar{\mathrm{Vir}}$ first introduced by \textcite{KooSaleur1994}. My collaborators perform a parallel analysis of the XXZ spin chain in a companion paper \cite{GSJS2021}, also for the non-rational case. Other applications have demonstrated the utility of the lattice approach \cite{MilstedVidal2017,ZMV2018,ZMV2020}.

This chapter relates to the remainder of the work as follows. Basic facts about the two-dimensional Potts model and its CFT appeared in Section \ref{Potts_model}. The algebra of local energy and momentum densities turns out to be the affine Temperley--Lieb algebra. Various incarnations of this algebra and its representation theory were discussed in Chapter \ref{TL_algebra}. While somewhat technical, the results for the affine Temperley--Lieb algebra are crucial, since they will be used as a starting point to understand the corresponding representations of $\mathrm{Vir}\oplus\overbar{\mathrm{Vir}}$ in the continuum limit. Returning to the CFT, general strategies to study the action of $\mathrm{Vir}\oplus\overbar{\mathrm{Vir}}$ starting from lattice models appeared in Chapter \ref{physics_lattice}. Putting all these pieces together, the first results appear in Section \ref{vir_degenerate} where we argue for the existence of indecomposable modules of $\mathrm{Vir}\oplus\overbar{\mathrm{Vir}}$ in the continuum limit of the Potts model for generic $\mathfrak q$. Our main results are given as Propositions \ref{loop_0} and \ref{loop_j}. These results are to be contrasted with the non-degenerate case, a reminder of which is presented first in \ref{vir_nondegenerate}. Evidence from the lattice that supports our main results is discussed in Section \ref{evidence_lattice}.

The primary theme of this chapter is the presence of indecomposable structures in the continuum theory. In the following chapter, we associate quantities to these indecomposable structures and describe methods of measuring them on the lattice, where, strictly speaking, they are absent.

\section{$\mathrm{Vir}\oplus\overbar{\mathrm{Vir}}$-modules in the Potts model CFT} \label{Potts_Virasoro}

\subsection{The non-degenerate case} \label{vir_nondegenerate}

We recall once more that in this chapter $\mathfrak q$ is assumed to take generic values (not a root of unity). Whenever $\phi$ is such that the resonance criterion Eq.~\eqref{eq:resonance} is not met we say that $\phi$ is \emph{generic}, and when Eq.~\eqref{eq:resonance} is satisfied $\phi$ is referred to as \emph{non-generic}.

Since $\mathfrak q$ is generic throughout, both $c$ and its parametrization $x$ in Eq.~\eqref{eq:central_charge} take generic, irrational values. The conformal weights may or may not be degenerate, depending on the lattice parameters. In the non-degenerate case, which corresponds to generic lattice parameters (the opposite does not always hold), it is natural to expect that the Temperley--Lieb module decomposes accordingly into a direct sum of Verma modules:
\begin{equation} \label{eq:nondegenerate_limit}
\mathscr W_{j,\e^{\i\phi}} \rightsquigarrow \bigoplus_{e\in\mathbb Z} V_{e-e_\phi,-j} \otimes \overbar V_{e-e_\phi,j}.
\end{equation}
The symbol $\rightsquigarrow$ means that the action of the lattice Virasoro generators \emph{restricted to scaling states} on $\mathscr W_{j,\e^{\i\phi}}$ corresponds to the decomposition on the right hand side when $N \to \infty$. This statement is discussed in considerable detail by \textcite{GSJS2021}.

Recall that a Verma module is a highest-weight representation of the Virasoro algebra,
\begin{equation} \label{eq:Virasoro_algebra}
[L_m, L_n] = (m-n)L_{m+n} + \frac{c}{12} m(m^2 - 1)\delta_{m+n,0},
\end{equation}
generated by a highest-weight vector $\ket*{h}$ satisfying $L_n\ket*{h} = 0$ for $n > 0$, and for which all the descendants $L_{-n_1}\cdots L_{-n_k}\ket*{h}$, with $0 < n_1 \le n_2 \le \cdots \le n_k$ and $k > 0$, are considered as independent, subject only to the commutation relations of Eq.~\eqref{eq:Virasoro_algebra}. In the non-degenerate case where the Verma module is irreducible, it is the only kind of module that can occur, motivating the identification in Eq.~\eqref{eq:nondegenerate_limit}. We note that this identification is independent of whether we consider the loop model or the XXZ spin chain.

\subsection{The degenerate case} \label{vir_degenerate}

In the degenerate case the conformal weights may take degenerate values $h = h_{rs}$ with $r,s\in\mathbb N^*$, in which case a singular vector appears in the Verma module. By definition, a singular vector is a vector that is both a descendant and a highest-weight state. For instance, starting with $\ket*{h_{11}} = \ket*{0}$ we see, by using the Virasoro commutation relations of Eq.~\eqref{eq:Virasoro_algebra}, that
\begin{equation}
L_1(L_{-1}\ket*{0}) = 2L_0\ket*{0} = 0,
\end{equation}
while $L_n(L_{-1}\ket*{0}) = 0$ for $n > 1$. Hence $L_{-1}\ket*{0}$ is a singular vector. Under the action of the Virasoro algebra, this vector generates a submodule. For generic $\mathfrak q$, this submodule is irreducible, and we have the decomposition
\begin{equation}
V^{(\text d)}_{11}: \begin{tikzcd}
X_{11} \arrow[d] \\
V_{1,-1}
\end{tikzcd}
\end{equation}
where we have introduced the notation $V^{(\text d)}$ to denote the degenerate Verma module, and we also denote by $X_{rs}$ the irreducible Virasoro module (in this case, technically a ``Kac module''), with generating function of levels
\begin{equation}
K_{rs} = q^{h_{rs} - c/24} \frac{1 - q^{rs}}{P(q)}.
\end{equation}
The subtraction of the singular vector at level $rs$ gives rise to a quotient module.

In cases of degenerate conformal weights, there is more than one possible module that could appear, and the identification in Eq.~\eqref{eq:nondegenerate_limit} may no longer hold. Furthermore, the identification now depends on the representation of $T^{\text a}_N(m)$ one considers. We restrict here to the loop--cluster representation, while corresponding results about the XXZ representation are discussed by \textcite{GSJS2021}.

For the modules $\mathscr W_{0,\mathfrak q^{\pm 2}}$, without through-lines, the Verma structure is seen even at finite size---see Eq.~\eqref{eq:standard}. Using the numerical methods described in Section \ref{evidence_lattice}, we find that the corresponding loop states are never annihilated by the $A_{n1}$ or $\overbar A_{n1}$ combinations of Virasoro generators.

We recall now from Section \ref{standard} that the module $\mathscr W_{0,\mathfrak q^{\pm 2}}$ appears in the loop model by keeping track of how points are connected across the periodic boundary. However, the Potts model where non-contractible loops have the same weight $m$ as contractible ones naturally involves the quotient $\overbar{\mathscr W}_{\!\!0,\mathfrak q^{\pm 2}}$ for which there are no degenerate states on the lattice. The spectrum generating function for this module in the continuum limit is then
\begin{equation}
\overbar F_{0,\mathfrak q^{\pm 2}} = F_{0,\mathfrak q^{\pm 2}} - F_{11} = \sum_{n=1}^\infty K_{n1}\overbar K_{n1},
\end{equation}
which involves only Kac modules. It is thus natural to formulate the following conclusion for the scaling limit.
\begin{conjecture}[Quotient loop-model module without through-lines] \label{loop_0}
We have the scaling limit
\begin{equation}
\overbar{\mathscr W}_{\!\!0,\mathfrak q^{\pm 2}} \rightsquigarrow \bigoplus_{n=1}^\infty X_{n1}\otimes \overbar X_{n1}.
\end{equation}
\end{conjecture}
Note that this structure implies that the corresponding highest-weight states $\ket*{h,\overbar h}$ are now annihilated:
\begin{equation}
A_{n1}\ket*{h_{n1},h_{n1}} = \overbar A_{n1}\ket*{h_{n1},h_{n1}} = 0.
\end{equation}
In particular, the ground state is indeed annihilated by $L_{-1}$ and $\overbar L_{-1}$, a satisfactory situation physically. Supporting numerical evidence for Conjecture \ref{loop_0} is found in Section \ref{W0_numerics}.

For the modules $\mathscr W_{j1}$ with $j > 0$, the numerical results in Section \ref{Wj_numerics} indicate that the highest-weight states with conformal weight $h_{ej}$ and $e > 0$ are \emph{never} annihilated by the corresponding operators $A_{ej}$, whether in the holomorphic or antiholomorphic sector. It would be tempting to conclude that the modules are now systematically of Verma type, but this not possible. Indeed, recall that for generic $\mathfrak q$, the $T^{\text a}_N(m)$-modules $\mathscr W_{j1}$ are irreducible and thus self-dual. The Virasoro generators, being obtained as continuum limits of $T^{\text a}_N(m)$ generators, should also obey this self-duality (cf.\ \textcite[Section 4.3]{Gainutdinov2015}). Verma modules clearly do not, as their structure is not invariant under reversal of the $\mathrm{Vir}\oplus\overbar{\mathrm{Vir}}$ action. To understand what might happen, let us discuss in more detail, as an example, the case $j = 2$. The generating function of levels shows a pair of primary fields
\begin{subequations}
\begin{gather}
\Phi_{12} \equiv \phi_{12} \otimes \phi_{1,-2}, \\
\overbar\Phi_{12} \equiv \phi_{1,-2} \otimes \phi_{12},
\end{gather}
\end{subequations}
with conformal weights $(h_{12}, h_{1,-2})$ and $(h_{1,-2}, h_{12})$. Note here that by $\phi_{rs}$ we simply mean a chiral primary field with conformal weight $h_{rs}$: the structure of the associated Virasoro module will be discussed below. This means in particular that $\phi_{rs} = \phi_{-r,-s}$.

By expanding the factor $1/P(q)P(\overbar q)$ in the spectrum generating functions, we see that the model also has four descendants at level two---i.e., with conformal weights $(h_{1,-2}, h_{1,-2})$, where we have used that $h_{1,-2} = h_{12} + 2$. Now, if the modules generated by $\Phi_{12}$ and $\overbar\Phi_{12}$ in the continuum limit were a product of two Verma modules, these four descendants would be the two independent fields $L_{-2}\Phi_{12}$ and $L_{-1}^2\Phi_{12}$, as well as the two fields obtained by swapping chiral and antichiral components, $\overbar L_{-2}\Phi_{12}$ and $\overbar L_{-1}^2\Phi_{12}$. The chiral/antichiral symmetry corresponds to exchanging right and left (i.e., exchanging momentum $p$ for momentum $-p$) and is present on the lattice as well, by reflecting the site index $i \to N + 1 - i$. This means one would expect to observe, in the finite-size transfer matrix, two eigenvalues, both converging (once properly scaled) to $h_{1,-2} = h_{12} + 2$, and corresponding to two linear combinations of $L_{-2}\Phi_{12}$ and $L_{-1}^2\Phi_{12}$ and their conjugates---hence both appearing in the form of doublets. This is \emph{not}, however, what is observed numerically (see Section \ref{singlet_states}). Instead, we see one doublet and two singlets, which means that the module in the continuum limit and at level two does not have, as a basis, a pair of independent states and their chiral/antichiral conjugates.

Introducing
\begin{subequations}
\begin{gather}
A_{12} = L_{-2} - \frac{3}{2 + 4h_{12}}L_{-1}^2, \\
\overbar A_{12} = \overbar L_{-2} - \frac{3}{2 + 4h_{12}}\overbar L_{-1}^2,
\end{gather}
\end{subequations}
we now claim that, in the continuum limit, the identity
\begin{equation}
A_{12}\Phi_{12} = \overbar A_{12}\overbar\Phi_{12}
\end{equation}
is satisfied. Note that both sides of the equation are primary fields---i.e., they are annihilated by $\mathrm{Vir}\oplus\overbar{\mathrm{Vir}}$ generators $L_n$, $\overbar L_n$ with $n > 0$. They are also of vanishing norm-square $\ip*{\cdot}$. Corresponding numerical results appear in Section \ref{Wj_numerics}.

We have therefore identified part of the module as a quotient of $(V_{12}^{(\text d)} \otimes\overbar V_{1,-2})\oplus (V_{1,-2}\otimes\overbar V_{12}^{(\text d)})$, corresponding to the following diagram for the degenerate fields:
\begin{equation}
\begin{tikzcd}
\Phi_{12} = \phi_{12}\otimes\overbar\phi_{1,-2} \arrow[dr,"A_{12}"'] & & \overbar\Phi_{12} = \phi_{1,-2}\otimes\overbar\phi_{12} \arrow[dl,"\overbar A_{12}"] \\
& A_{12}\Phi_{12} = \overbar A_{12}\overbar\Phi_{12}
\end{tikzcd}
\end{equation}

Note we have the quotient modules (obtained by taking the quotient by the submodule generated by the bottom field) $X_{12}\otimes\overbar V_{1,-2}$ and $V_{1,-2}\otimes\overbar X_{12}$ and with generating functions $(q^{h_{1,-2}-c/24}/P(q))\times\overbar K_{12}$ and $K_{12}\times(\overbar q^{h_{1,-2}-c/24}/P(\overbar q))$. The bottom field generates a product of Verma modules $V_{1,-2}\otimes\overbar V_{1,-2}$ with generating function $(q^{h_{1,-2}-c/24}/P(q))\times(\overbar q^{h_{1,-2}-c/24}/P(\overbar q))$.

However, this cannot be the end of the story, since the quotient identified so far is not self-dual---nor does it account for the proper multiplicity of fields. Invariance of the diagram under reversal of the arrows demands that there exists a field ``on top,'' with a quotient that is also a product of Verma modules $V_{1,-2}\otimes\overbar V_{1,-2}$. This should give rise, in terms of fields, to the diagram
\begin{equation} \label{eq:indecomposable_12}
\begin{tikzcd}
& \tilde\Psi_{12} \arrow[dr,"\overbar A^\dagger_{12}"]\arrow[dl,"A^\dagger_{12}"']\arrow[to=3-2,"L_0 - h_{1,-2}"] & \\
\Phi_{12} \arrow[dr,"A_{12}"'] & & \overbar\Phi_{12} \arrow[dl,"\overbar A_{12}"] \\
& \Psi_{12} \equiv A_{12}\Phi_{12} = \overbar A_{12}\overbar\Phi_{12}
\end{tikzcd}
\end{equation}
with $\tilde\Psi_{12}$ a field to be determined (Section \ref{OPE}).

The same construction seems to apply to all cases in the characters $F_{j1}$. The simplest example occurs, in fact, in $\mathscr W_{11}$---even though this module does not appear in the Potts model, as discussed after Eq.~\eqref{eq:torus_Z1}---with $\Phi_{11} \equiv \phi_{11} \otimes \overbar\phi_{1,-1}$ and $\overbar\Phi_{11} \equiv \phi_{1,-1} \otimes \overbar\phi_{11}$. In this case, the quotient is simply given by $L_{-1}\Phi_{11} = \overbar L_{-1}\overbar\Phi_{11}$.

The indecomposable structure for arbitrary positive integer values of $e$ and $j$ can then be conjectured to be
\begin{equation} \label{eq:indecomposable_diamond}
\begin{tikzcd}
& \tilde\Psi_{ej} \arrow[dr,"\overbar A^\dagger_{ej}"]\arrow[dl,"A^\dagger_{ej}"']\arrow[to=3-2,"L_0 - h_{e,-j}"] & \\
\Phi_{ej} = \phi_{ej}\otimes\overbar\phi_{e,-j} \arrow[dr,"A_{ej}"'] & & \overbar\Phi_{ej} = \phi_{e,-j}\otimes\overbar\phi_{ej} \arrow[dl,"\overbar A_{ej}"] \\
& \Psi_{ej} \equiv A_{ej}\Phi_{ej} = \overbar A_{ej}\overbar\Phi_{ej}
\end{tikzcd}
\end{equation}
The validity of Eq.~\eqref{eq:indecomposable_diamond} in general comes from strong numerical evidence for small values of $e$ and $j$. It is also the simplest structure we can imagine solving the problems of poles in the OPEs, based on our independent knowledge of the spectrum of the theory. More complete evidence should come from the construction of four-point functions using the corresponding regularized conformal blocks \cite{NivesvivatRibault2021}.

The corresponding structure of Virasoro modules defines the quotient modules $\mathscr L_{ej}$:
\begin{equation}
\begin{tikzcd}[column sep=small]
& & V_{e,-j}\otimes\overbar V_{e,-j} \arrow[dr]\arrow[dl] & \\
\mathscr L_{ej} \equiv \mathscr Q[(V_{ej}^{(\text d)} \otimes\overbar V_{e,-j})\oplus (V_{e,-j}\otimes\overbar V_{ej}^{(\text d)})]: & X_{ej}\otimes\overbar V_{e,-j} \arrow[dr] & & V_{e,-j}\otimes\overbar X_{ej} \arrow[dl] \\
& & V_{e,-j}\otimes\overbar V_{e,-j}
\end{tikzcd}
\end{equation}
Accordingly, we are led to our next result.
\begin{conjecture}[Loop-model modules with through-lines] \label{loop_j}
For $j > 0$ and $2j$ through lines we have the scaling limit
\begin{equation}
\mathscr W_{j1} \rightsquigarrow (V_{0,-j} \otimes \overbar V_{0j}) \oplus \bigoplus_{e=1}^\infty \mathscr L_{ej}.
\end{equation}
\end{conjecture}

As mentioned, an important piece of evidence for the correctness of the structure in Eq.~\eqref{eq:indecomposable_diamond} is based on the numerical observation of a pair of singlet states in the transfer matrix spectrum. In Section \ref{singlet_states} we identify this pair of singlets precisely in the cases $(e,j) = (1,1)$, $(2,1)$, $(1,2)$ and $(1,3)$. These observations in turn lend credence to the more general Conjecture \ref{loop_j}.

\section{Evidence from the lattice via Koo--Saleur generators} \label{evidence_lattice}

Except where necessary, in this section we suppress notating the explicit dependence on lattice size $N$.

Within this section we provide evidence for the main results given in Propositions \ref{loop_0} and \ref{loop_j}, by acting directly with the Koo--Saleur generators of Eq.~\eqref{eq:KS_generators} on eigenstates of the lattice Hamiltonian in Eq.~\eqref{eq:Potts_Hamiltonian}. In these numerical studies we partition our state space at each system size $N$ into eigenspaces of the translation operator, with eigenvalues $\{\e^{2\pi\i p/N}|0\le p\le N-1\}$. As the Hamiltonian is manifestly invariant under translation we may diagonalize it independently within each such sector. The Koo--Saleur generators exactly reproduce the fact that the action of $L_n$ ($\overbar L_n$) on a state of momentum $p$ produces a state of momentum $p- n$ ($p+n$), at finite size. For a state with an eigenvalue $\epsilon$ of the Hamiltonian at a given system size $N$, we consider its effective conformal weights, which we also denote $(h,\overbar h)$, defined as the solutions to
\begin{subequations}
\begin{gather}
\epsilon = \frac{2\pi}{N}\left(h + \overbar h - \frac{c}{12}\right), \\
p = h - \overbar h.
\end{gather}
\end{subequations}
By ``following'' a state (say, the lowest-energy state within a given sector of lattice momentum) as $N$ increases, and extrapolating the effective conformal weights $h$ and $\overbar h$, we can identify the conformal weights in the continuum limit. This process is carried out in Section \ref{singlet_states}. We will omit the qualifier ``effective'' and simply refer to ``conformal weights'' when the context makes it clear that the term is being applied to lattice quantities. Similarly, we will frequently assign conformal weights $h_{rs}$ given by the Kac formula to finite-size states---by this we mean that the effective conformal weights of a state for increasing $N$ converge to $h = h_{rs}$. In practice, we will only be able to access small values of the Kac labels $r$ and $s$, since larger system sizes are needed to accommodate a larger lattice momentum (which governs $r$) and a larger number of through-lines (which governs $s$).

Before discussing details of the numerics we must eliminate an ambiguity that may arise in the results due to phase degrees of freedom. In the following sections we will discuss quantities of the form $\|Z - \overbar Z\|_2$, where $Z$ and $\overbar Z$ are (descendants of) eigenstates of the Hamiltonian (e.g., $Z = L_{-1}\Phi_{11}$ and $\overbar Z = \overbar L_{-1}\overbar\Phi_{11}$), and $\|\cdot\|^2_2 = (\cdot|\cdot)$ is the norm induced by the native positive-definite scalar product. $Z$ and $\overbar Z$ need not be simply related (aside from the fact that they are predicted to be equal in a certain limit), as the notation might suggest. The discussion applies to any pair of expressions that may suffer from this phase ambiguity. Nevertheless, for the situations in this work where this discussion applies, $Z$ and $\overbar Z$ represent expressions that are related by the exchange of holomorphic and antiholomorphic components. In quantum mechanics the overall phase of a vector or wave function has no observable consequences and $\e^{\i\alpha}Z$ for any real $\alpha$ would serve just as well in computations of observables. Typically one chooses the phase of a state such that its components in some basis are entirely real, where possible. In the situation at hand, the eigenvectors of the Hamiltonian are generically complex---no choice of phase can make all of the components real---and there is no canonical way to fix the relative phase between eigenvectors. The measurement of $\|Z - \e^{\i\alpha}\overbar Z\|_2$ thus takes on a continuum of values. Where this ambiguity occurs, we fix the relative phase by choosing the value of $\alpha$ that minimizes this quantity:
\begin{equation}
\|Z - \overbar Z\|_{\munderbar 2} \equiv \inf_\alpha\|Z - \e^{\i\alpha}\overbar Z\|_2.
\end{equation}
This optimization is succinctly denoted by the underlined 2 in the notation $\|Z - \overbar Z\|_{\munderbar 2}$.

Our main goal is to establish certain identities by observing whether deviations from these identities at finite size decay to zero with increasing system size. Let us give two examples. In order to provide evidence for Conjecture \ref{loop_0} we will check that $L_{-1}I \to 0$, with $I$ the identity state with conformal weights $(0,0)$. Meanwhile, to provide evidence for Conjecture \ref{loop_j} we would like to establish that $L_{-1}\Phi_{11} \to \overbar L_{-1}\overbar\Phi_{11}$, or that $L_{-1}\Phi_{11} - \overbar L_{-1}\overbar\Phi_{11} \to 0$ as $N \to \infty$. Using the positive-definite norm, we examine equivalently whether $\|L_{-1}I\|_2 \to 0$ and $\|L_{-1}\Phi_{11} - \overbar L_{-1}\overbar\Phi_{11}\|_{\munderbar 2} \to 0$.

As will be seen in the tables below, this simple measurement fails to furnish the evidence we seek. Indeed, as $N$ increases, the values observed actually grow in magnitude in most cases. An interpretation of this observation is the fact that, since the finite-size Koo--Saleur generators do not yet furnish a representation of the Virasoro algebra, the action of $L_{-1}$ on $\Phi_{11}$, for instance, produces a state with nonzero components even in highly excited eigenstates of the Hamiltonian. While each such component would tend to zero on its own, the number of these so-called ``parasitic couplings'' grows rapidly, yielding a non-decaying contribution in total.

To avoid the issue of this rapid growth we project to the $d$ lowest-energy states within the relevant sector of lattice momentum, keeping $d$ fixed as $N \to \infty$.

\subsection{Scaling-weak convergence} \label{scaling_weak}

For the following discussion we consider a concrete example, the fields $L_{-1}\Phi_{11}$ and $\overbar L_{-1}\overbar\Phi_{11}$ in the loop model. In the continuum limit, these fields have conformal weights $(1,1)$. Their lattice analogues both belong to the sector of lattice momentum $p = N/2$. By following the energies $\epsilon$ of states within this sector for increasing lattice sizes $N$, we find that the two lowest-energy states correspond to these conformal weights.

Let us write schematically
\begin{equation}
L_{-1}\Phi_{11} = u + v,
\end{equation}
where $u$ is a linear combination of these two lowest states and $v$ represents all other states in the sector $p = N/2$. In order to exclude the effects of the parasitic couplings, we build a projection operator $\Pi^{(2)}$ such that $\Pi^{(2)} L_{-1}\Phi_{11} = u$. The operator $\Pi^{(2)}$ we seek is precisely the sum of the biorthogonal projection operators for the two lowest states, $\Pi^{(2)} = \Pi_1 + \Pi_2$, described in Chapter \ref{biorthogonal}. Since $\overbar L_{-1}\overbar\Phi_{11}$ has conformal weights $(1,1)$ as well, the projector $\Pi^{(2)}$ also truncates the corresponding lattice quantity to the same two states.

It is not necessary to restrict to only the components in $u$ (given by the projection to the two lowest states in the example at hand)---one could also include higher energy states. As long as the rank of the projection operator is fixed, we expect the influence of parasitic couplings within the image of the projection operator to vanish as $N \to \infty$. We call convergence of values in the context of this procedure \emph{scaling-weak convergence}. To demonstrate this type of convergence, we apply projectors of increasing rank $d$ to $L_{-1}\Phi_{11} - \overbar L_{-1}\overbar\Phi_{11}$ before measuring its norm. We expect that for any fixed projector rank $d$ (independent of $N$), so long as $\Pi^{(d)}$ eventually includes all scaling states as $d \to \infty$,
\begin{equation} \label{eq:scaling_weak_11}
\forall d \in \mathbb N,\quad \lim_{N\to\infty} \|\Pi^{(d)}(L_{-1}\Phi_{11} - \overbar L_{-1}\overbar\Phi_{11})\|_{\munderbar 2} = 0;
\end{equation}
i.e., scaling-weak convergence of the lattice quantities towards the continuum identity $L_{-1}\Phi_{11} = \overbar L_{-1}\overbar\Phi_{11}$.

The notion of scaling-weak convergence is defined and discussed in greater detail by \textcite{GSJS2021}, where it is shown that a crucial difference compared to weak convergence is that limits of products of Koo--Saleur generators are in certain cases different compared to the corresponding products of limits, necessitating the insertion of projectors. This difference is found to affect the products with dual operators that are induced by the positive-definite inner product, as in $\|L_{-1}\Phi_{11}\|_2^2 = (\Phi_{11}|L^*_{-1}L_{-1}|\Phi_{11})$, but not the product $L_{-1}^2$ inside the singular vector operator $A_{12}$.

In general, for any of the fields $Z$ relevant below, we say that its lattice analogue scaling-weakly converges to zero if
\begin{equation}
\forall d\in \mathbb N,\quad \lim_{N\to\infty} \|\Pi^{(d)}Z(N)\| = 0,
\end{equation}
with $\|\cdot\|$ some positive-definite norm, and where $Z(N)$ is the lattice analogue of $Z$ at lattice size $N$. The meaning of $\Pi^{(d)}$ is context-dependent, but should be built in such a way that $\lim_{d\to\infty}\Pi^{(d)}$ effectively functions as the identity operator:
\begin{equation} \label{eq:projector_identity}
\forall N \in 2\mathbb N, \quad \lim_{d\to\infty}\Pi^{(d)}Z(N) = Z(N).
\end{equation}
We say ``effectively,'' since $\lim_{d\to\infty}\Pi^{(d)}$ does not necessarily have to equal the identity operator. For instance, in the discussion of scaling-weak convergence of $L_{-1}\Phi_{11} - \overbar L_{-1}\overbar\Phi_{11}$ to zero, $\Pi^{(d)}$ is built from states of lattice momentum $p = N/2$. Thus $\lim_{d\to\infty}\Pi^{(d)}$ is the identity operator in the subspace of momentum $N/2$ and zero elsewhere. However, $L_{-1}\Phi_{11} - \overbar L_{-1}\overbar\Phi_{11}$ is zero in all momentum sectors save for $N/2$. Thus $\lim_{d\to\infty}\Pi^{(d)}$ effectively functions as the identity in this measurement. It is also possible to construct $\Pi^{(d)}$ using the $d$ lowest states of the entire Hamiltonian, regardless of momentum. This does not affect the limit in Eq.~\eqref{eq:projector_identity}, but merely the rate of convergence. In this case $\lim_{d\to\infty}\Pi^{(d)}$ becomes the identity operator.

An analogous discussion applies to the demonstration of the identity $A_{12}\Phi_{12} = \overbar A_{12}\overbar\Phi_{12}$, mutatis mutandis. We present numerical evidence that $A_{12}\Phi_{12} - \overbar A_{12}\overbar\Phi_{12}$ scaling-weakly converges to zero.

\begin{figure}[h]
\centering
\includegraphics[width=0.8\textwidth]{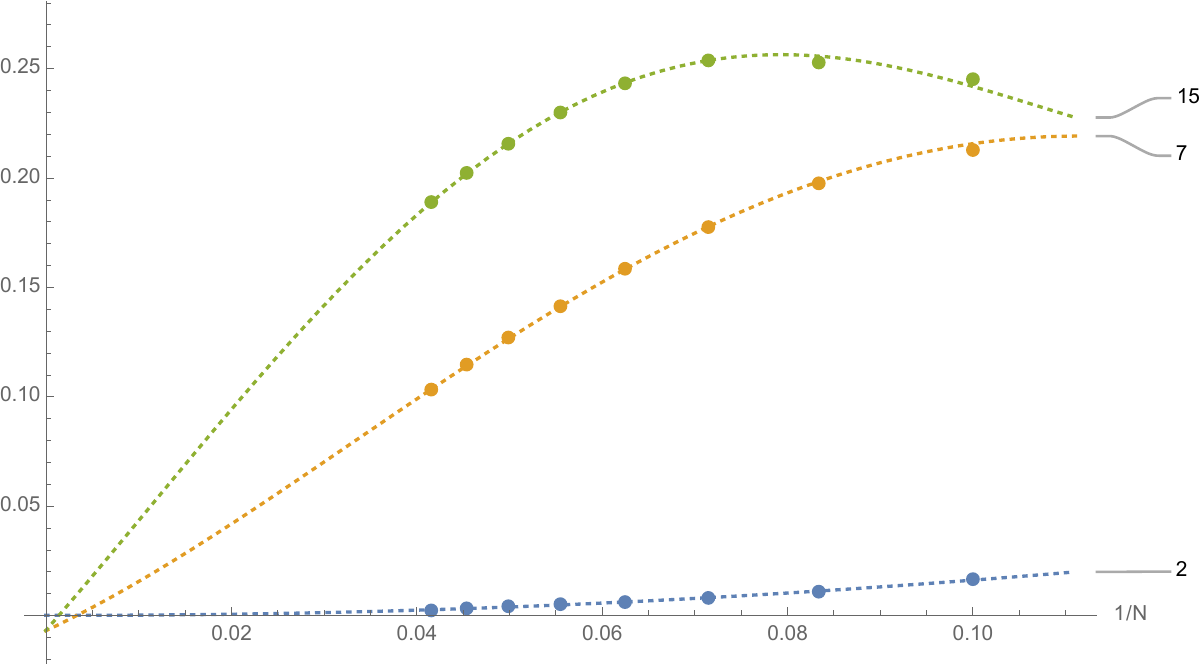}
\caption{Comparison of lattice results using projectors of different of rank, illustrating the concept of scaling-weak convergence for $j = 1$. The horizontal axis is $1/N$. The vertical axis is $\|\Pi^{(d)}(L_{-1}\Phi_{11} - \overbar L_{-1}\overbar\Phi_{11})\|_{\munderbar 2}/\|\Pi^{(d)}L_{-1}\Phi_{11}\|_2$. The tags on the graphs indicate the rank $d$ of the projector $\Pi^{(d)}$. The dotted lines are fourth-order polynomial fits (in $1/N$) to the five leftmost data points.}
\label{x1_norm_v_proj}
\end{figure}

\begin{figure}[h]
\centering
\includegraphics[width=0.8\textwidth]{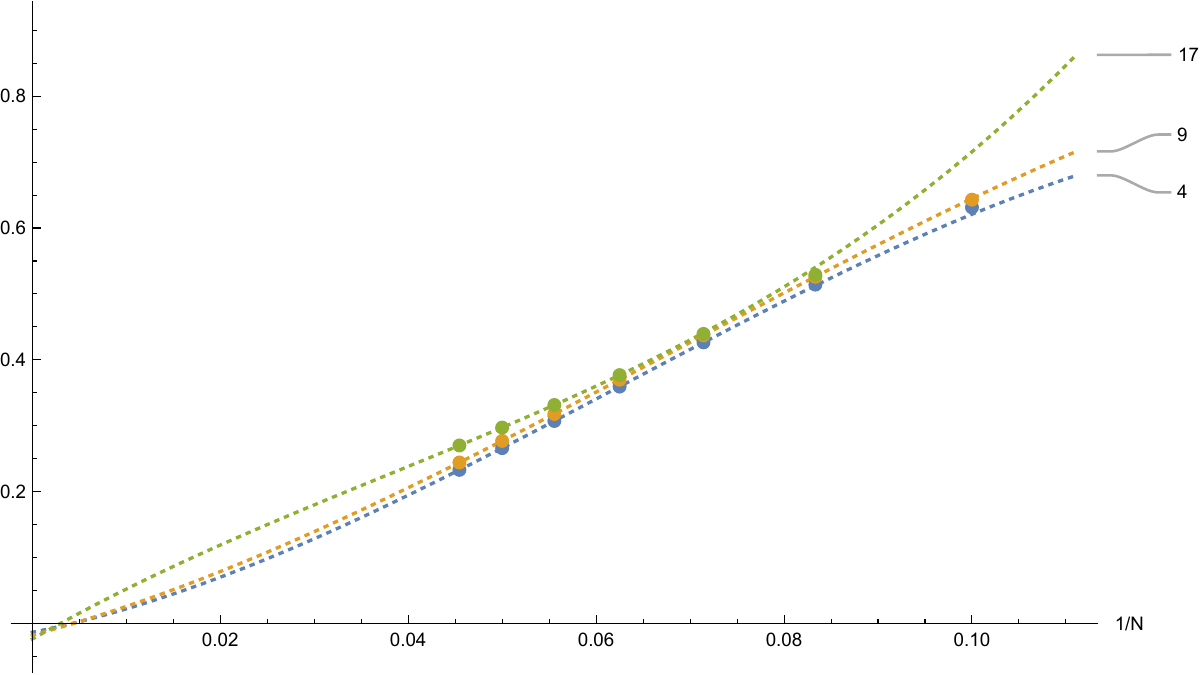}
\caption{Comparison of lattice results using projectors of different of rank, illustrating the concept of scaling-weak convergence for $j = 1$. The horizontal axis is $1/N$. The vertical axis is $\|\Pi^{(d)}(A_{12}\Phi_{12} - \overbar A_{12}\overbar\Phi_{12})\|_{\munderbar 2}/\|\Pi^{(d)}A_{12}\Phi_{12}\|_2$. The tags on the graphs indicate the rank $d$ of the projector $\Pi^{(d)}$. The dotted lines are third-order polynomial fits (in $1/N$) to the four leftmost data points.}
\label{x2_norm_v_proj}
\end{figure}

We show in Figures \ref{x1_norm_v_proj} and \ref{x2_norm_v_proj} that when applying projectors of different rank $d$, the numerical results extrapolate to almost the same values. We expect that the difference in the extrapolations can be made arbitrarily small by including data points for large enough system sizes, though they are not numerically accessible at the time of writing.


\subsection{Numerical results for $\overbar{\mathscr W}_{\!\!0,\mathfrak q^{\pm 2}}$} \label{W0_numerics}

Within the module $\mathscr W_{0,\mathfrak q^{\pm 2}}$, the link states corresponding to primary fields with degenerate conformal weights are never annihilated by the singular vector operators $A_{n1}$ or $\overbar A_{n1}$. However, the module of interest for the study of the loop model is rather the quotient module $\overbar{\mathscr W}_{\!\!0,\mathfrak q^{\pm 2}}$. In this module we consider in particular the lowest-energy link state of lattice momentum $p = 0$, which in the continuum limit will correspond to the identity state $I$ with conformal weights $(h,\overbar h) = (0,0)$. We act on this state with the Koo--Saleur generator $L_{-1}$. The norm of the resulting state $L_{-1}I$ is shown in Table \ref{Wbar_norm}. (The norm of $\overbar L_{-1}$ is the same by symmetry.) In these tables, the peculiar value $x = \pi/\sec^{-1}(2 \sqrt{2}) -1$ corresponds to $Q = 1/2$, which is further studied in Section \ref{singlet_states}.

\begin{table}[h]
\centering
\begin{tabular}{clllll}
\toprule
$N$ & \multicolumn{5}{c}{$x$} \\
\cmidrule(l){2-6}
 & \multicolumn{1}{c}{$\pi/3$} & \multicolumn{1}{c}{$\pi/2$} & \multicolumn{1}{c}{$\pi/\sec^{-1}(2 \sqrt{2}) -1$} & \multicolumn{1}{c}{$\e$} & \multicolumn{1}{c}{$\pi$} \\
\midrule
$8$ & $0.00105459$ & $0.0134764$ & $0.0140696$ & $0.0319876$ & $0.0360179$ \\
$10$ & $0.00151863$ & $0.0183461$ & $0.0191326$ & $0.0430288$ & $0.0484952$ \\
$12$ & $0.00180350$ & $0.0212243$ & $0.0221219$ & $0.0495035$ & $0.0558428$ \\
$14$ & $0.00200139$ & $0.0231978$ & $0.0241704$ & $0.0538990$ & $0.0608373$ \\
$16$ & $0.00215397$ & $0.0247167$ & $0.0257464$ & $0.0572352$ & $0.0646218$ \\
$18$ & $0.00228117$ & $0.0259884$ & $0.0270657$ & $0.0599884$ & $0.0677341$ \\
$20$ & $0.00239306$ & $0.0271153$ & $0.0282345$ & $0.0623992$ & $0.0704482$ \\
$22$ & $0.00249505$ & $0.0281508$ & $0.0293087$ & $0.0645961$ & $0.0729122$ \\
$24$ & $0.00259016$ & $0.0291244$ & $0.0303188$ & $0.0666511$ & $0.0752099$ \\
\bottomrule
\end{tabular}
\caption{The values of $\|L_{-1}I\|_2$ for various lengths $N$ and parameters $x$. $I$ is the field in the $j = 0$ sector with conformal weights $(h_{11}, h_{11}) = (0,0)$.}
\label{Wbar_norm}
\end{table}

Within this module there is no state to project to that we expect to give a nonzero contribution in the limit $N \to \infty$, the only state with the proper conformal weights having been excluded by the quotient. Projecting on the lowest-energy state still remaining in the sector of the appropriate lattice momentum we therefore expect the result to approach zero as $N \to \infty$ (Table \ref{Wbar_proj}).

\begin{table}[h]
\centering
\begin{tabular}{clllll}
\toprule
$N$ & \multicolumn{5}{c}{$x$} \\
\cmidrule(l){2-6}
 & \multicolumn{1}{c}{$\pi/3$} & \multicolumn{1}{c}{$\pi/2$} & \multicolumn{1}{c}{$\pi/\sec^{-1}(2 \sqrt{2}) -1$} & \multicolumn{1}{c}{$\e$} & \multicolumn{1}{c}{$\pi$} \\
\midrule
$8$ & $0.00105459$ & $0.0134764$ & $0.0140696$ & $0.0319876$ & $\phantom{-}0.0360179$ \\
$10$ & $0.00154453$ & $0.0201482$ & $0.0209567$ & $0.0434477$ & $\phantom{-}0.0485887$ \\
$12$ & $0.00140952$ & $0.0207429$ & $0.0216739$ & $0.0447805$ & $\phantom{-}0.0501347$ \\
$14$ & $0.00121929$ & $0.0192739$ & $0.0202699$ & $0.0427614$ & $\phantom{-}0.0480059$ \\
$16$ & $0.00103467$ & $0.0170396$ & $0.0180351$ & $0.0394988$ & $\phantom{-}0.0445480$ \\
$18$ & $0.000875168$ & $0.0147437$ & $0.0156912$ & $0.0359407$ & $\phantom{-}0.0407787$ \\
$20$ & $0.000742847$ & $0.0126649$ & $0.0135396$ & $0.0325055$ & $\phantom{-}0.0371365$ \\
$22$ & $0.000634490$ & $0.0108785$ & $0.0116714$ & $0.0293585$ & $\phantom{-}0.0337921$ \\
$24$ & $0.000545883$ & $0.0093767$ & $0.0100887$ & $0.0265463$ & $\phantom{-}0.0307936$ \\
\midrule
extrapolation & $0.0000850643$ & $0.00442133$ & $0.00495163$ & $0.000157526$ & $-0.000258504$ \\
\bottomrule
\end{tabular}
\caption{The values of $\|\Pi^{(1)}L_{-1}I\|_2$ for various lengths $N$ and parameters $x$. $\Pi^{(1)}$ is a projection to the state of lowest energy within the $j = 0$, $p = 1$ sector. This is a state that has conformal weights $(h_{1,-1}, h_{11}) = (1,0)$. The extrapolation is obtained by fitting the last five data points to a fourth-order polynomial in $1/N$.}
\label{Wbar_proj}
\end{table}

\subsection{Numerical results for $\mathscr W_{j1}$} \label{Wj_numerics}
In this section we numerically illustrate our conjectured identities $L_{-1}\Phi_{11} = \overbar L_{-1}\overbar\Phi_{11}$ and $A_{12}\Phi_{12} = \overbar A_{12}\overbar\Phi_{12}$, from Section \ref{vir_degenerate}, in the scaling-weak sense.

\sloppy For $j = 1$, we find $\Phi_{11}$ and $\overbar\Phi_{11}$ in the momentum sectors $p = N/2 - 1$ and $N/2 + 1$, and their respective descendants $L_{-1}\Phi_{11}$ and $\overbar L_{-1}\overbar\Phi_{11}$ in the sector with $p = N/2$. We show the values of $\|L_{-1}\Phi_{11} - \overbar L_{-1}\overbar\Phi_{11}\|_{\munderbar 2}/\|L_{-1}\Phi_{11}\|_2$ and see that they do not decay (Table \ref{level1}), then the values of $\|\Pi^{(2)}(L_{-1}\Phi_{11} - \overbar L_{-1}\overbar\Phi_{11})\|_{\munderbar 2}/\|\Pi^{(2)}L_{-1}\Phi_{11}\|_2$ and see that they do (Table \ref{level1proj}). Additional values of interest---$\|L_{-1}\Phi_{11}\|_2$ and $\|\Pi^{(2)}L_{-1}\Phi_{11}\|_2$---may be found in Appendix \ref{more_sw_tables}. While we numerically do observe scaling-weak convergence of $L_{-1}\Phi_{11} - \overbar L_{-1}\overbar\Phi_{11}$ to zero in the sense of Eq.~\eqref{eq:scaling_weak_11}, here we report the values of $\|L_{-1}\Phi_{11} - \overbar L_{-1}\overbar\Phi_{11}\|_{\munderbar 2}/\|L_{-1}\Phi_{11}\|_2$ and $\|\Pi^{(2)}(L_{-1}\Phi_{11} - \overbar L_{-1}\overbar\Phi_{11})\|_{\munderbar 2}/\|\Pi^{(2)}L_{-1}\Phi_{11}\|_2$ to give a measure of relative deviation from zero. The decay of the latter quantity to zero implies the scaling-weak convergence of $L_{-1}\Phi_{11} - \overbar L_{-1}\overbar\Phi_{11}$ to zero so long as the norm $\|\Pi^{(2)}L_{-1}\Phi_{11}\|_2$ does not grow too quickly. That this is the case can be seen in Table \ref{normPiLX}, in Appendix \ref{more_sw_tables}. A similar discussion applies to the scaling-weak convergence of $A_{12}\Phi_{12} - \overbar A_{12}\overbar\Phi_{12}$ to zero.

\begin{table}[h]
\centering
\begin{tabular}{clllll}
\toprule
$N$ & \multicolumn{5}{c}{$x$} \\
\cmidrule(l){2-6}
 & \multicolumn{1}{c}{$\pi/3$} & \multicolumn{1}{c}{$\pi/2$} & \multicolumn{1}{c}{$\pi/\sec^{-1}(2 \sqrt{2}) -1$} & \multicolumn{1}{c}{$\e$} & \multicolumn{1}{c}{$\pi$} \\
\midrule
 $8$ & $0.0106586$ & $0.104747$ & $0.108207$ & $0.187030$ & $0.199715$ \\
 $10$ & $0.0114921$ & $0.118893$ & $0.123110$ & $0.224566$ & $0.241811$ \\
 $12$ & $0.0117599$ & $0.124101$ & $0.128674$ & $0.243359$ & $0.263743$ \\
 $14$ & $0.0119545$ & $0.127043$ & $0.131823$ & $0.255388$ & $0.278124$ \\
 $16$ & $0.0121676$ & $0.129476$ & $0.134403$ & $0.264557$ & $0.289152$ \\
 $18$ & $0.0124147$ & $0.131931$ & $0.136979$ & $0.272425$ & $0.298551$ \\
 $20$ & $0.0126930$ & $0.134557$ & $0.139716$ & $0.279679$ & $0.307105$ \\
 $22$ & $0.0129960$ & $0.137370$ & $0.142638$ & $0.286636$ & $0.315194$ \\
\bottomrule
\end{tabular}
\caption{The values of $\|L_{-1}\Phi_{11} - \overbar L_{-1}\overbar\Phi_{11}\|_{\munderbar 2}/\|L_{-1}\Phi_{11}\|_2$ for various lengths $N$ and parameters $x$.}
\label{level1}
\end{table}
\begin{table}[h]
\centering
\begin{tabular}{clllll}
\toprule
$N$ & \multicolumn{5}{c}{$x$} \\
\cmidrule(l){2-6}
 & \multicolumn{1}{c}{$\pi/3$} & \multicolumn{1}{c}{$\pi/2$} & \multicolumn{1}{c}{$\pi/\sec^{-1}(2 \sqrt{2}) -1$} & \multicolumn{1}{c}{$\e$} & \multicolumn{1}{c}{$\pi$} \\
\midrule
 $8$ & $0.0000527492$ & $\phantom{-}0.00584360$ & $\phantom{-}0.00627048$ & $0.0215309$ & $0.0251946$ \\
 $10$ & $0.0000191654$ & $\phantom{-}0.00297587$ & $\phantom{-}0.00322369$ & $0.0133976$ & $0.0161156$ \\
 $12$ & $8.26633\times 10^{-6}$ & $\phantom{-}0.00167699$ & $\phantom{-}0.00183148$ & $0.00901364$ & $0.0111168$ \\
 $14$ & $4.03988\times 10^{-6}$ & $\phantom{-}0.00102009$ & $\phantom{-}0.00112185$ & $0.00641199$ & $0.00809058$ \\
 $16$ & $2.16718\times 10^{-6}$ & $\phantom{-}0.000658400$ & $\phantom{-}0.000728471$ & $0.00475643$ & $0.00612835$ \\
 $18$ & $1.25094\times 10^{-6}$ & $\phantom{-}0.000445437$ & $\phantom{-}0.000495467$ & $0.00364534$ & $0.00478799$ \\
 $20$ & $7.62288\times 10^{-7}$ & $\phantom{-}0.000313082$ & $\phantom{-}0.000349894$ & $0.00286778$ & $0.00383424$ \\
 $22$ & $4.87306\times 10^{-7}$ & $\phantom{-}0.000227096$ & $\phantom{-}0.000254879$ & $0.00230490$ & $0.00313292$ \\
\midrule
extrapolation & $4.38043\times 10^{-7}$ & $-0.0000700002$ & $-0.0000675678$ & $0.000161454$ & $0.0000896441$ \\
\bottomrule
\end{tabular}
\caption{The values of $\|\Pi^{(2)}(L_{-1}\Phi_{11} - \overbar L_{-1}\overbar\Phi_{11})\|_{\munderbar 2}/\|\Pi^{(2)}L_{-1}\Phi_{11}\|_2$ for various lengths $N$ and parameters $x$.}
\label{level1proj}
\end{table}

Similarly, for $j = 2$ we find $\Phi_{12}$ and $\overbar\Phi_{12}$ in the momentum sectors $p = N/2 - 2$ and $N/2 + 2$, and their respective descendants $A_{12}\Phi_{12}$ and $\overbar A_{12}\overbar\Phi_{12}$ in the sector with $p = N/2$. As before, we show the values $\|A_{12}\Phi_{12} - \overbar A_{12}\overbar\Phi_{12}\|_{\munderbar 2}/\|A_{12}\Phi_{12}\|_2$ and see that they do not decay (Table \ref{level2}), and that acting with a projector $\Pi^{(4)}$ does show a decay for $\|\Pi^{(4)}(A_{12}\Phi_{12} - \overbar A_{12}\overbar\Phi_{12})\|_{\munderbar 2}/\|\Pi^{(4)}A_{12}\Phi_{12}\|_2$ (Table \ref{level2proj}). Additional data for $\|A_{12}\Phi_{12}\|_2$ and $\|\Pi^{(4)}A_{12}\Phi_{12}\|_2$ are similarly found in Appendix \ref{more_sw_tables}.

\begin{table}[h]
\centering
\begin{tabular}{clllll}
\toprule
$N$ & \multicolumn{5}{c}{$x$} \\
\cmidrule(l){2-6}
 & \multicolumn{1}{c}{$\pi/3$} & \multicolumn{1}{c}{$\pi/2$} & \multicolumn{1}{c}{$\pi/\sec^{-1}(2 \sqrt{2}) -1$} & \multicolumn{1}{c}{$\e$} & \multicolumn{1}{c}{$\pi$} \\
\midrule
 $10$ & $0.211762$ & $0.451277$ & $0.458346$ & $0.621868$ & $0.653152$ \\
 $12$ & $0.151572$ & $0.359414$ & $0.365882$ & $0.523169$ & $0.555244$ \\
 $14$ & $0.115725$ & $0.304912$ & $0.310899$ & $0.459319$ & $0.490658$ \\
 $16$ & $0.0930500$ & $0.275545$ & $0.281251$ & $0.420970$ & $0.450592$ \\
 $18$ & $0.0783010$ & $0.264041$ & $0.269673$ & $0.402367$ & $0.429735$ \\
 $20$ & $0.0688143$ & $0.265678$ & $0.271421$ & $0.399548$ & $0.424477$ \\
 $22$ & $0.0631192$ & $0.277084$ & $0.283096$ & $0.409437$ & $0.432035$ \\
 \bottomrule
\end{tabular}
\caption{The values of $\|A_{12}\Phi_{12} - \overbar A_{12}\overbar\Phi_{12}\|_{\munderbar 2}/\|A_{12}\Phi_{12}\|_2$ for various lengths $N$ and parameters $x$.}
\label{level2}
\end{table}
\begin{table}[h]
\centering
\begin{tabular}{clllll}
\toprule
$N$ & \multicolumn{5}{c}{$x$} \\
\cmidrule(l){2-6}
 & \multicolumn{1}{c}{$\pi/3$} & \multicolumn{1}{c}{$\pi/2$} & \multicolumn{1}{c}{$\pi/\sec^{-1}(2 \sqrt{2}) -1$} & \multicolumn{1}{c}{$\e$} & \multicolumn{1}{c}{$\pi$} \\
\midrule
 $10$ & $\phantom{-}0.210763$ & $\phantom{-}0.439128$ & $0.445749$ & $\phantom{-}0.600649$ & $0.631230$ \\
 $12$ & $\phantom{-}0.148284$ & $\phantom{-}0.332095$ & $0.337881$ & $\phantom{-}0.482952$ & $0.514155$ \\
 $14$ & $\phantom{-}0.110152$ & $\phantom{-}0.257777$ & $0.262750$ & $\phantom{-}0.395760$ & $0.426568$ \\
 $16$ & $\phantom{-}0.0852125$ & $\phantom{-}0.204972$ & $0.209224$ & $\phantom{-}0.329727$ & $0.359515$ \\
 $18$ & $\phantom{-}0.0679876$ & $\phantom{-}0.166510$ & $0.170153$ & $\phantom{-}0.278915$ & $0.307392$ \\
 $20$ & $\phantom{-}0.0555742$ & $\phantom{-}0.137790$ & $0.140930$ & $\phantom{-}0.239167$ & $0.266240$ \\
 $22$ & $\phantom{-}0.0463202$ & $\phantom{-}0.115847$ & $0.118574$ & $\phantom{-}0.207563$ & $0.233243$ \\
\midrule
extrapolation & $-0.0015914$ & $-0.000137804$ & $0.0000985782$ & $-0.000155346$ & $0.000403050$ \\
\bottomrule
\end{tabular}
\caption{The values of $\|\Pi^{(4)}(A_{12}\Phi_{12} - \overbar A_{12}\overbar\Phi_{12})\|_{\munderbar 2}/\|\Pi^{(4)}A_{12}\Phi_{12}\|_2$ for various lengths $N$ and parameters $x$.}
\label{level2proj}
\end{table}

Both for $j = 1$ and $j = 2$, we find that the results gathered in Tables \ref{level1proj} and \ref{level2proj} support the proposed identities $L_{-1}\Phi_{11} = \overbar L_{-1}\overbar\Phi_{11}$ and $A_{12}\Phi_{12} = \overbar A_{12}\overbar\Phi_{12}$, when we use projection operators of ranks 2 and 4, respectively, before taking norms. In both cases we have used the lowest rank such that all states with the same total conformal dimension as the descendant fields in the proposed identities, and those with lower dimensions, have been included in the projectors. As in the discussion surrounding our definition of scaling-weak convergence, and as illustrated by Figures \ref{x1_norm_v_proj} and \ref{x2_norm_v_proj}, we expect that the result in the limit $N \to \infty$ will remain the same for higher rank projectors. However, we do not expect the result to remain the same when no projector is applied. Indeed, the values in Tables \ref{level1} and \ref{level2} do not tend to zero with increasing $N$. The numerical proximity of $\|L_{-1}\Phi_{11}\|_2$ to $\|\Pi^{(2)}L_{-1}\Phi_{11}\|_2$ and of $\|A_{12}\Phi_{12}\|_2$ to $\|\Pi^{(4)}A_{12}\Phi_{12}\|_2$ indicates that the lack of convergence comes from deviations that can be attributed to parasitic couplings to higher states, however small these couplings may be.

\section{Observation of singlet states in the loop model} \label{singlet_states}

In this section we report transfer matrix computations that support some of our main results.

We consider the transfer matrix of the loop model on $N = 2L$ sites, with $2j$ through-lines, in the geometry that corresponds to a Potts model on a square lattice (see Section \ref{Potts_description}), with $L$ spins in each row and periodic boundary conditions. The corresponding loop model then lives on a tilted square lattice (the medial lattice of the original, axially oriented square lattice). The particular method of diagonalization used here has been explained in much detail by \textcite[Appendix A]{JacobsenSaleur2019}, but it may also be carried out using the algorithms in Chapter \ref{computational} and Appendix \ref{DMA}.

We focus here on one well-chosen value, $Q = 1/2$, which can be considered representative for the case of generic values of $Q$. For each size $L = 5,6,\ldots,13$, we compute the first several hundred eigenvalues in each module $\mathscr W_{j,z^2}$ with $j = 1,2,3$ and $z^{2j} = 1$, extracting the multiplicity, finite-size scaling dimension, and lattice momentum of each eigenvalue. The multiplicities are always found to be either 1 or 2, and we pay special attention to the singlets. The lattice momentum $p$ can be identified only up to a sign, and it coincides with the conformal spin modulo $L$.

One major difficulty in the study is that the $i$th largest eigenvalue in the finite-size spectrum corresponds to the $i$th lowest-lying scaling state only for $L$ sufficiently large, and for all but the smallest few values of $i$ this simple situation is reached only when $L$ is much larger than the attainable system size. To study the scaling states numerically nevertheless, one therefore has to identify the sequences $(i_5,i_6,\ldots,i_{13})$ that correspond to any desired scaling field, using a general methodology that requires a lot of patience [\emph{Id.}\ at Appendix A.5]. Polynomial extrapolations of the finite-size scaling dimensions are then possible, most often using the data for all sizes (and only occasionally excluding the first few sizes), leading to quite accurate estimates of the conformal scaling dimension $\Delta = h + \overbar h$. Moreover, comparing the values of $p$ for several different $L$ in the sequence will permit us to lift the ``modulo $L$'' qualifier and determine $s$ (again up to a sign).

The values of $\Delta$ and $s$ allow us to identify the corresponding scaling field, up to a few ambiguities. To be precise, we are able to identify the corresponding primary field and the descendant level on both the chiral and antichiral sides, up to a possible overall exchange of chiral and antichiral components (again, because $s$ is determined only up to a sign). Our notation below implicitly incorporates this ambiguity. For instance, a field that is descendant at level $(3,2)$ of a primary field $\Phi$ will be denoted $L_{-3}\overbar L_{-2}\Phi$, although it might in fact be any linear combination of the form $(L_{-3} + \alpha L_{-1}L_{-2} + \beta L_{-1}^3)(\overbar L_{-2} + \gamma \overbar L_{-1}^2)\Phi$ for some unknown coefficients $\alpha$, $\beta$, and $\gamma$---or indeed the same field with chiral and antichiral components being exchanged.

\subsection{$j = 1$}
Results for the module $\mathscr W_{11}$ are shown in Table \ref{W11_singlets}. We have, in fact, identified the scaling fields for all lines with $i_{13} \le 60$ in this case, but to keep the table concise we show only the first 10 fields, along with several other fields that are either a primary or a singlet (or both). The ranks $i_L$ (meaning their position within the ordered spectrum of the transfer matrix, sorted by decreasing eigenvalue without repetition) of singlet fields are shown in italics, while those of the doublets are in regular type. Small numbers refer to finite-size levels for which the lattice momentum $p$ differs from the conformal spin $s$ by a nontrivial multiple of $L$ (for more details, see \emph{id.}). The table shows all singlets with $i_{12} \le 200$ (for the last seven lines the diagonalization for $L = 13$ was numerically too demanding). The extrapolation of the scaling dimension is shown to about the number of significant digits to which it agrees with the exact result.

\begin{table}[h]
\centering
\begin{tabular}{cccccccccclll}
\toprule
 $|s|$ & \multicolumn{9}{c}{$(i_5,i_6,\ldots,i_{13})$} & \multicolumn{2}{c}{$\Delta$} & \multirow{2}{2.3cm}{identification \\ of scaling field} \\
 \cmidrule(lr){2-10} \cmidrule(lr){11-12}
 & 5 & 6 & 7 & 8 & 9 & 10 & 11 & 12 & 13 & \multicolumn{1}{c}{numerics} & \multicolumn{1}{c}{exact} & \\
 \midrule
 0 & \emph{1} & \emph{1} & \emph{1} & \emph{1} & \emph{1} & \emph{1} & \emph{1} & \emph{1} & \emph{1} & 0.187014 & 0.187027 & $\Phi_{01}$ \\
 1 & 2 & 2 & 2 & 2 & 2 & 2 & 2 & 2 & 2 & 1.000003 & 1 & $\Phi_{11}$ \\
 1 & 3 & 3 & 3 & 3 & 3 & 3 & 3 & 3 & 3 & 1.187040 & 1.187027 & $L_{-1} \Phi_{01}$ \\
 2 & 4 & 4 & 4 & 4 & 4 & 4 & 4 & 4 & 4 & 1.9998 & 2 & $\overbar{L}_{-1} \Phi_{11}$ \\
 0 & \emph{6} & \emph{6} & \emph{6} & \emph{5} & \emph{5} & \emph{5} & \emph{5} & \emph{5} & \emph{5} & 2.0016 & 2 & $L_{-1} \Phi_{11}$ \\
 0 & \emph{9} & \emph{8} & \emph{7} & \emph{7} & \emph{7} & \emph{7} & \emph{7} & \emph{7} & \emph{6} & 1.9985 & 2 & $L_{-1} \Phi_{11}$ \\
 2 & 5 & 5 & 5 & 6 & 6 & 6 & 6 & 6 & 7 & 2.1882 & 2.1870 & $L_{-2} \Phi_{01}$ \\
 0 & \emph{7} & \emph{9} & \emph{8} & \emph{8} & \emph{8} & \emph{8} & \emph{8} & \emph{8} & \emph{8} & 2.18708 & 2.18703 & $L_{-1} \overbar{L}_{-1} \Phi_{01}$ \\
 2 & 8 & 10 & 10 & 9 & 9 & 9 & 9 & 9 & 9 & 2.18710 & 2.18703 & $L_{-2} \Phi_{01}$ \\
 3 & {\tiny 5} & 7 & 9 & 10 & 10 & 10 & 10 & 10 & 10 & 2.992 & 3 & $\overbar{L}_{-2} \Phi_{11}$ \\
 \midrule
 2 & 19 & 22 & 24 & 23 & 23 & 23 & 24 & 22 & 22 & 3.43895 & 3.43892 & $\Phi_{21}$ \\
 0 & \emph{18} & \emph{23} & \emph{27} & \emph{28} & \emph{27} & \emph{27} & \emph{28} & \emph{28} & \emph{27} & 4.002 & 4 & $L_{-2} \overbar{L}_{-1} \Phi_{11}$ \\
 0 & \emph{23} & \emph{30} & \emph{32} & \emph{37} & \emph{40} & \emph{38} & \emph{37} & \emph{38} & \emph{38} & 4.00016 & 4 & $L_{-2} \overbar{L}_{-1} \Phi_{11}$ \\
 0 & \emph{20} & \emph{24} & \emph{29} & \emph{34} & \emph{36} & \emph{34} & \emph{35} & \emph{37} & \emph{39} & 4.1864 & 4.1870 & $L_{-2} \overbar{L}_{-2} \Phi_{01}$ \\
 0 & \emph{33} & \emph{45} & \emph{47} & \emph{47} & \emph{49} & \emph{52} & \emph{49} & \emph{46} & \emph{46} & 4.18707 & 4.18703 & $L_{-2} \overbar{L}_{-2} \Phi_{11}$ \\
 0 & \emph{\tiny 25} & \emph{43} & \emph{56} & \emph{71} & \emph{82} & \emph{88} & \emph{99} & \emph{103} & & 6.03 & 6 & $L_{-3} \overbar{L}_{-2} \Phi_{11}$ \\
 0 & \emph{51} & \emph{78} & \emph{95} & \emph{111} & \emph{114} & \emph{119} & \emph{118} & \emph{115} & & 5.423 & 5.439 & $L_{-2} \Phi_{21}$ \\
 0 & \emph{53} & \emph{79} & \emph{100} & \emph{116} & \emph{120} & \emph{123} & \emph{121} & \emph{117} & & 5.433 & 5.439 & $L_{-2} \Phi_{21}$ \\
 0 & --- & --- & \emph{57} & \emph{76} & \emph{94} & \emph{112} & \emph{117} & \emph{120} & & 6.178 & 6.187 & $L_{-3} \overbar{L}_{-3} \Phi_{01}$ \\
 0 & --- & --- & \emph{65} & \emph{88} & \emph{103} & \emph{116} & \emph{129} & \emph{132} & & 5.989 & 6 & $L_{-3} \overbar{L}_{-2} \Phi_{11}$ \\
 0 & \emph{55} & \emph{86} & \emph{115} & \emph{133} & \emph{144} & \emph{145} & \emph{151} & \emph{155} & & 5.994 & 6 & $L_{-3} \overbar{L}_{-2} \Phi_{11}$ \\
 0 & \emph{52} & \emph{84} & \emph{119} & \emph{144} & \emph{166} & \emph{174} & \emph{175} & \emph{182} & & 6.32 & 6.19 & $L_{-3} \overbar{L}_{-3} \Phi_{01}$ \\
\bottomrule
\end{tabular}
\caption{Conformal spectrum in the sector $\mathscr W_{11}$ for $Q=1/2$.}
\label{W11_singlets}
\end{table}

The primaries $\Phi_{ej} = \phi_{ej}\otimes\overbar\phi_{e,-j}$ and $\overbar\Phi_{ej} = \phi_{e,-j}\otimes\overbar\phi_{ej}$ with $j = 1$ can be seen in the table for $e = 0,1,2$, corresponding to the lines with $i_{13} = 1,2,22$. The latter two are doublets, while the first one is a singlet, as $\Phi_{01} = \overbar\Phi_{01}$ because of the identification $\phi_{rs} = \phi_{-r,-s}$. For the same reason, we find several of the spinless descendants of $\Phi_{01}$ to be singlets (e.g., $i_{13} = 8,39$ and $i_{12} = 120,182$).

A more remarkable finding is the singlet nature of the pair of lines with $i_{13} = 5,6$. If we had been dealing with a product of Verma modules, these would have formed a degenerate doublet. Instead, we see here a manifestation of the duality $L_{-1}\Phi_{11} = \overbar L_{-1}\overbar\Phi_{11}$.

Much lower in the spectrum, we similarly draw attention to the singlet nature of the lines with $i_{12} = 115, 117$. They show the duality $A_{21}\Phi_{21} = \overbar A_{21}\overbar\Phi_{21}$.

\subsection{$j = 2$}

In the same way, we show results for the modules $\mathscr W_{2,z^2}$ in Table \ref{W2_singlets}. Note that the results for all permissible cases, $z^{2j} = 1$, are shown in the same table and the corresponding scaling levels are marked by the additional label $k = 0,1$ for $z^2 = \e^{2\pi\i k/j} = (-1)^k$. Furthermore, primary fields with fractional Kac labels appear; they are still given by $\Phi_{ej} = \phi_{ej}\otimes\overbar\phi_{e,-j}$ and $\overbar\Phi_{ej} = \phi_{e,-j}\otimes\overbar\phi_{ej}$, where $e$ is not required to be an integer. We do not study these fields further, though they appear in the table to show that they can be identified.

\begin{table}[h]
\centering
\begin{tabular}{ccccccccccclll}
\toprule
 $k$ & $|s|$ & \multicolumn{9}{c}{$(i_5,i_6,\ldots,i_{13})$} & \multicolumn{2}{c}{$\Delta$} & \multirow{2}{2.3cm}{identification \\ of scaling field} \\
 \cmidrule(lr){3-11} \cmidrule(lr){12-13}
& & 5 & 6 & 7 & 8 & 9 & 10 & 11 & 12 & 13 & \multicolumn{1}{c}{numerics} & \multicolumn{1}{c}{exact} & \\
 \midrule
 0 & 0 & \emph{1} & \emph{1} & \emph{1} & \emph{1} & \emph{1} & \emph{1} & \emph{1} & \emph{1} & \emph{1} & 1.1095698 & 1.1095673 & $\Phi_{02}$ \\
 1 & 1 & 2 & 2 & 2 & 2 & 2 & 2 & 2 & 2 & 2 & 1.312824 & 1.312810 & $\Phi_{1/2,2}$ \\
 0 & 2 & 3 & 3 & 3 & 3 & 3 & 3 & 3 & 3 & 3 & 1.92264 & 1.92254 & $\Phi_{12}$ \\
 0 & 1 & 5 & 4 & 4 & 4 & 4 & 4 & 4 & 4 & 4 & 2.1099 & 2.1096 & $\overbar{L}_{-1} \Phi_{02}$ \\
 1 & 2 & 4 & 5 & 5 & 5 & 5 & 5 & 5 & 5 & 5 & 2.31304 & 2.31281 & $\overbar{L}_{-1} \Phi_{1/2,2}$ \\
 1 & 0 & 7 & 7 & 6 & 6 & 6 & 6 & 6 & 6 & 6 & 2.31297 & 2.31281 & $L_{-1} \Phi_{1/2,2}$ \\
 0 & 3 & {\tiny 6} & 8 & 8 & 7 & 7 & 7 & 7 & 7 & 7 & 2.9230 & 2.9225 & $\overbar{L}_{-1} \Phi_{12}$ \\
 0 & 2 & 6 & 9 & 9 & 9 & 9 & 8 & 8 & 8 & 8 & 3.1075 & 3.1096 & $\overbar{L}_{-2} \Phi_{02}$ \\
 1 & 3 & {\tiny 8} &10 & 11 & 11 & 10 & 10 & 10 & 9 & 9 & 2.9393 & 2.9388 & $\Phi_{3/2,2}$ \\
 0 & 1 & 13 & 13 & 13 & 13 & 11 & 11 & 11 & 10 & 10 & 2.9229 & 2.9225 & $L_{-1} \Phi_{12}$ \\
 \midrule
 0 & 0 & \emph{9} & \emph{11} & \emph{12} & \emph{15} & \emph{12} & \emph{12} & \emph{12} & \emph{12} & \emph{12} & 3.1101 & 3.1096 & $L_{-1} \overbar{L}_{-1} \Phi_{02}$ \\
 0 & 0 & \emph{19} & \emph{24} & \emph{28} & \emph{30} & \emph{28} & \emph{29} & \emph{27} & \emph{25} & \emph{24} & 3.9244 & 3.9225 & $L_{-2} \Phi_{12}$ \\
 0 & 0 & 21 & 27 & 31 & 32 & 32 & 33 & 31 & 30 & 25 & 3.9231 & 3.9225 & $L_{-2} \Phi_{12}$ \\
 0 & 0 & \emph{31} & \emph{40} & \emph{40} & \emph{42} & \emph{41} & \emph{40} & \emph{36} & \emph{37} & \emph{35} & 3.9202 & 3.9225 & $L_{-2} \Phi_{12}$ \\
 0 & 4 & {\tiny 23} & {\tiny 34} & {\tiny 37} & 43 & 47 & 48 & 45 & 46 & & 4.356 & 4.361 & $\Phi_{22}$ \\
\bottomrule
\end{tabular}
\caption{Conformal spectrum in the sector $\mathscr W_{2,z^2}$ for $Q=1/2$. The label $k$ corresponds to $z^2 = (-1)^k$.}
\label{W2_singlets}
\end{table}

The primary $\Phi_{02}$ on the line with $i_{13} = 1$, and its descendant at level $(1,1)$ on the line with $i_{13} = 12$, are both singlets due to self-duality. But more importantly, we find a pair of singlets on the lines with $i_{12} = 24,35$. They show the duality $A_{12}\Phi_{12} = \overbar A_{12}\overbar\Phi_{12}$. By contrast, the remaining two states with conformal weights $(h_{12} + 2, h_{12} + 2)$ can be identified as the doublet on the line with $i_{13} = 25$. They correspond to a descendant of $\Phi_{12}$ at level $(2,0)$ and a descendant of $\overbar\Phi_{12}$ at level $(0,2)$, with coefficients that are unknown but different from those of the operators $A_{12}$ and $\overbar A_{12}$, respectively.

\subsection{$j = 3$}

Finally, the results for the modules $\mathscr W_{3,z^2}$ are given in Table \ref{W3_singlets}. The label $k$ now takes on three possible values, $0$, $1$, and $2$, again signifying $z^2 = \e^{2\pi\i k/j} = \e^{2\pi\i k/3}$.

\begin{table}[h]
\centering
\begin{tabular}{cccccccccclll}
\toprule
 $k$ & $|s|$ & \multicolumn{8}{c}{$(i_5,i_6,\ldots,i_{12})$} & \multicolumn{2}{c}{$\Delta$} & \multirow{2}{2.3cm}{identification \\ of scaling field} \\
 \cmidrule(lr){3-10} \cmidrule(lr){11-12}
& & 5 & 6 & 7 & 8 & 9 & 10 & 11 & 12 & \multicolumn{1}{c}{numerics} & \multicolumn{1}{c}{exact} & \\
 \midrule
0 & 0 & \emph{1} & \emph{1} & \emph{1} & \emph{1} & \emph{1} & \emph{1} & \emph{1} & \emph{1} & 2.64732 & 2.64713 & $\Phi_{03}$ \\
 1 & 1 & 2 & 2 & 2 & 2 & 2 & 2 & 2 & 2 & 2.73773 & 2.73746 & $\Phi_{1/3,3}$ \\
 2 & 2 & 3 & 3 & 3 & 3 & 3 & 3 & 3 & 3 & 3.00867 & 3.00846 & $\Phi_{2/3,3}$ \\
 0 & 3 & 5 & 5 & 4 & 4 & 4 & 4 & 4 & 4 & 3.463 & 3.46011 & $\Phi_{13}$ \\
 0 & 0 & \emph{10} & \emph{13} & \emph{16} & \emph{17} & \emph{19} & \emph{17} & \emph{17} & \emph{17} & 4.637 & 4.647 & $L_{-1} \overbar{L}_{-1} \Phi_{03}$ \\
 2 & 0 & --- & --- & \emph{36} & \emph{48} & \emph{61} & \emph{74} & \emph{84} & \emph{84} & 6.640 & 6.647 & $L_{-2} \overbar{L}_{-2} \Phi_{03}$ \\
 0 & 0 & \emph{31} & \emph{55} & \emph{79} & \emph{100} & \emph{107} & \emph{109} & \emph{111} & \emph{108} & 6.454 & 6.460 & $L_{-3} \Phi_{13}$ \\
 0 & 0 & \emph{39} & \emph{66} & \emph{98} & \emph{122} & \emph{140} & \emph{149} & \emph{144} & & 6.6478 & 6.6471 & $L_{-2} \overbar{L}_{-2} \Phi_{03}$ \\
 0 & 0 & \emph{43} & \emph{85} & \emph{126} & \emph{155} & \emph{163} & \emph{175} & \emph{169} & & 6.468 & 6.460 & $L_{-3} \Phi_{13}$ \\
 0 & 0 & \emph{39} & \emph{91} & \emph{153} & \emph{216} & \emph{280} & \emph{327} & \emph{367} & & 8.70 & 8.64 & $L_{-3} \overbar{L}_{-3} \Phi_{03}$ \\
 0 & 0 & \emph{31} & \emph{91} & \emph{169} & \emph{251} & \emph{332} & \emph{402} & \emph{456} & & 8.85 & 8.64 & $L_{-3} \overbar{L}_{-3} \Phi_{03}$ \\
\bottomrule
\end{tabular}
\caption{Conformal spectrum in the sector $\mathscr W_{3,z^2}$ for $Q=1/2$. The label $k$ corresponds to $z^2 = \e^{2\pi\i k/3}$.}
\label{W3_singlets}
\end{table}

We observe here a pair of singlets on the lines with $i_{11} = 111,169$. They show the duality $A_{13}\Phi_{13} = \overbar A_{13}\overbar\Phi_{13}$.

\sectionbreak

Summarizing, through the particular cases $(e,j) = (1,1),(2,1),(1,2)$ and $(1,3)$, we have identified pairs of singlets to give evidence of the structure proposed in Conjecture \ref{loop_j}. While the mere presence of singlets is insufficient to claim definitively that the structure must be exactly as described, the possibility of a product of Verma modules has been ruled out, and as discussed, the diamond structure is the simplest alternative consistent with self-duality.

\section{Parity and the structure of modules} \label{parity}

In Section \ref{singlet_states}, we showed that the presence of certain singlets ruled out the possibility of the structure of the continuum field theory being a product of Verma modules. Though the diamond structure proposed is consistent with these observations, the number of singlets or doublets observed cannot directly yield insight into the structure of the continuum limit Virasoro modules. That this is so can be seen by comparison with a study of the XXZ spin chain \cite{GSJS2021}, where a similar investigation is performed. There, one obtains the same spectrum of the Hamiltonian, and thus one must have the same numbers of singlets and doublets, while the numerical results from using the Koo--Saleur generators indicate that the structures of the continuum limit Virasoro modules are not the same.

To distinguish between the two types of modules that appear for loop models and the XXZ spin chain, we must instead think more carefully about symmetry under parity, under which chiral and antichiral sectors are mapped to each other. On the lattice this corresponds to a mapping of site $j$ to $-j$, which by Eq.~\eqref{eq:KS_generators} maps the Koo--Saleur generators $L_n$ and $\overbar L_n$ to each other. The parity operation is an involution, so we can distinguish the states by its eigenvalues, $P = \pm 1$. Let us consider the two states $\Phi_{12} = \phi_{12}\otimes\overbar\phi_{1,-2}$ and $\overbar\Phi_{12} = \phi_{1,-2}\otimes\overbar\phi_{12}$, which are mapped to each other under parity. Different situations can occur for their descendants $L_{-1}^2\Phi_{12}$, $L_{-2}\Phi_{12}$, $\overbar L_{-1}^2\overbar\Phi_{12}$, and $\overbar L_{-2}\overbar\Phi_{12}$. On the lattice, the four scaling states that have the correct lattice momenta and energies to be identified with these descendants form two singlets and one doublet. The states $v_1$ and $v_2$ in the doublet are mapped to each other by parity and can thus be combined into one $P = 1$ state $v_1+v_2$ and one $P = -1$ state $v_1 - v_2$. We now distinguish between the types of modules using the parity of the singlets.

If the four descendants are independent, we can form four linear combinations that are eigenstates of the parity operator, $L_{-1}^2\Phi_{12} \pm \overbar L_{-1}^2\overbar\Phi_{12}$ and $L_{-2}\Phi_{12} \pm \overbar L_{-2}\overbar\Phi_{12}$. Now consider instead the states depicted in the following diagram (the same as that in Eq.~\eqref{eq:indecomposable_12}):
\begin{equation}
\begin{tikzcd}
& \tilde\Psi_{12} \arrow[dr,"\overbar A^\dagger_{12}"]\arrow[dl,"A^\dagger_{12}"']\arrow[to=3-2,"L_0 - h_{-1,2}"] & \\
\Phi_{12} \arrow[dr,"A_{12}"'] & & \overbar\Phi_{12} \arrow[dl,"\overbar A_{12}"] \\
& A_{12}\Phi_{12} = \overbar A_{12}\overbar\Phi_{12}
\end{tikzcd}
\end{equation}
The four descendants are no longer independent. The bottom field $A_{12}\Phi_{12} = \overbar A_{12}\overbar\Phi_{12}
$ is clearly invariant under parity. Meanwhile, the top field satisfies $A_{12}A^\dagger_{12}\tilde\Psi_{12} = \overbar A_{12}\overbar A^\dagger_{12}\tilde\Psi_{12}$ and therefore also has $P = 1$. These two fields should both appear as singlets, while the doublet would correspond to the two linear combinations that can be formed with what remains (not shown in the diagram).

The exposition of the preceding argument relies on the continuum formulation. In order to validate our approach of inferring properties of the continuum theory from lattice discretizations, a similar scenario had better hold on the lattice as well. We therefore return to the finite-size numerics to seek the verdict. When acting with $P: j\mapsto -j$ on the two singlets, we find that the results depend on the representation. In the loop model, both singlets have $P = 1$, corresponding to the situation in the diagram, while in the XXZ spin chain we find that one has $P = 1$ and the other has $P = -1$, so that we rather have the parities expected from four linearly independent descendants. The two lattice discretizations thus confirm the general argument, and we find that only the loop model has the Jordan-block structure of Eq.~\eqref{eq:indecomposable_12} with the dependence $A_{12}\Phi_{12} = \overbar A_{12}\overbar\Phi_{12}$, which is one of the main points of this study.

\section{Summary}

One of the lessons of this chapter is that Jordan blocks for $L_0$ and $\overbar L_0$ are expected to appear in the continuum limit of the $Q$-state Potts model and the loop models, even though there are no such Jordan blocks in the finite-size lattice model. This possibility was already mentioned by \textcite{Gainutdinov2015} in the particular case $c = 0$, but occurs quite generically, whenever fields with degenerate conformal weights $h_{rs}$ (with $r,s\in\mathbb N^*$) appear in the spectrum. It is, in fact, a logical consequence of the self-duality of the modules $\mathscr W_{j1}$, and thus can be argued on very general grounds. At the time the paper on which this chapter is based was written, we stated that \cite[footnote 11]{Grans-Samuelsson2020}
\begin{quote}
The absence of Jordan [blocks] on the lattice makes measuring the logarithmic couplings $b_{rs}$ [see Section \ref{OPE}] appearing in the indecomposable [structures of Eq.~\eqref{eq:indecomposable_diamond}] quite difficult, as there seems to be no simple way of normalizing the lattice version of the field $\tilde\Psi_{rs}$.
\end{quote}
We will see in Section \ref{b_emerging} how the concept of emerging Jordan blocks, introduced in Chapter \ref{emerging}, surmounts this difficulty and allows for direct measurement of the logarithmic couplings.

The CFT for the XXZ spin chain seems well described by the somewhat mundane Dotsenko--Fateev twisted boson theory \cite{GSJS2021}. By contrast, the $Q$-state Potts model or loop model CFTs appear to be new objects, related to but not identical with the $c < 1$ Liouville theory \cite{DelfinoViti2011,PSVD2013,IJS2016}, and are slowly getting under control thanks to this and other recent work. A possible direction for future progress in understanding these CFTs better would be to revisit the bootstrap approach of \textcite{He2020} by taking into account properly regularized conformal blocks \cite{NivesvivatRibault2021}. More pressing qualitative questions, perhaps, include a better understanding of the OPEs: in particular, the OPEs for the hull operators, which should have some interesting geometrical \cite{DuplantierLudwig1991} and algebraic \cite{GJS2018} meanings, or the OPEs of the currents, where logarithmic features should explain why there are much fewer than $D^2_{\text{adj}}$ fields with weights $(1,1)$---or the behavior when $\mathfrak q$ approaches a root of unity, and more Jordan blocks appear, probably of rank higher than two. This last question seems particularly suited for a treatment via the concept of emerging Jordan blocks.

\chapter{Numerical study of Jordan blocks in the dense loop model CFT} \label{Jordan_loop}

In Chapter \ref{Virasoro}, we argued for the general structure of the Virasoro modules that arise in the continuum limit of the dense loop model CFT, and gave evidence that essentially ruled out simpler possiblities. In this chapter, we turn to more direct measurable consequences of such indecomposable structures.

This chapter is organized as follows. After some brief remarks to put our findings in context, Section \ref{amplitudes} describes an indirect probe of the Jordan-block structure via measurements of four-point functions, building upon some of the findings of the previous chapter. We turn back to the continuum field theory in Section \ref{OPE} to assign quantifiable parameters to the diamond structures of Conjecture \ref{loop_j}. Finally, in Sections \ref{b_emerging} and \ref{c_0}, we bring in the new machinery of emerging Jordan blocks to observe directly the gradual formation of Jordan blocks, and also to measure directly the quantities described in Section \ref{OPE}.

\section{Overview}

The bootstrap approach and the detailed study of four-point correlation functions have helped clarify the logarithmic properties of the $Q$-state Potts-model and $O(n)$-model CFTs for generic values of the parameters $Q$ and $n$. Many of the results in this field, however, rely on self-consistency arguments, and have not been checked directly, in particular via numerical calculations. A noticeable exception is the $c = 0$ case, where the Jordan block mixing the stress--energy tensor $T$ and its logarithmic partner $t$ was first observed by \textcite{DJS2010c}, leading to measurements of the famous ``$b$-numbers'' (the logarithmic couplings), and the solution of several paradoxes involving the polymer/percolation or the bulk/boundary differences \cite{VJS2011,VGJS2012}. This progress at $c = 0$ was rendered considerably easier by the presence, in finite size, of Jordan blocks for the Hamiltonian going over the CFT Jordan blocks for $L_0$ and $\overbar L_0$ in the continuum limit.

By contrast, despite the conjectured appearance of Jordan blocks in $L_0$ and $\overbar L_0$ for the Potts model in the continuum limit (Conjecture \ref{loop_j}), their known lattice versions have Hamiltonians that, although non-hermitian, are fully diagonalizable for generic values of $Q$, making the confirmation of their logarithmic structure---let alone the measurement of the corresponding logarithmic couplings---significantly more difficult.

We will see in this chapter how the framework of emerging Jordan blocks is naturally suited to overcoming these difficulties and thus strengthen, in particular, the evidence supporting Conjecture \ref{loop_j}. While we focus here on a specific lattice model---the dense loop model---we believe our technique is completely general.

\section{Numerical amplitudes and Jordan blocks} \label{amplitudes}

In Section \ref{singlet_states} we have identified some singlet levels in the transfer matrix of the loop model that support the existence of the indecomposable structure of Conjecture \ref{loop_j}, in the sense that the simple product of Verma modules has been ruled out. To go further and find numerical evidence for the existence of the expected Jordan blocks for $L_0$ and $\overbar L_0$ is more subtle, since it turns out that the Hamiltonian and transfer matrices of the Potts model for generic $Q$ remain, for the levels we are interested in, completely diagonalizable in finite size. In other words, the Jordan blocks appear only in the continuum limit. While this possibility was foreseen by \textcite{Gainutdinov2015}, it makes the problem quite different from the one studied by \textcite{VJS2011,DJS2010c}, where Jordan blocks were present for finite systems as a result of Temperley--Lieb representation theory, with the indecomposable structures in the continuum limit being identical to those observed in the lattice model. We will see that, even before invoking the formalism of emerging Jordan blocks, it is nonetheless possible for the case at hand to observe the build-up of Jordan blocks in the lattice model indirectly.

To this end, we now return to the four-point functions of the order operator in the Potts model. In lattice terms, they are of the form $P_{a_1a_2a_3a_4}$, where a label $a_i$ is associated to each of the four insertion points $z_i$ (with $i = 1,2,3,4$), the convention being that points are required to belong to the same FK cluster if and only if their corresponding labels are identical. For instance, $P_{abab}$ denotes the four-point function in which $z_1$ and $z_3$ belong to the same cluster, while $z_2$ and $z_4$ belong to a different cluster (see \textcite[Figure 2]{JacobsenSaleur2019}). To study such correlation functions on the lattice by the transfer matrix technique, it is convenient to place points $z_1$ and $z_2$ on the same time slice (i.e., lattice row) and points $z_3$ and $z_4$ on a different, distant slice [\emph{Id.}, Figure 1]. This geometric arrangement amounts to performing the $s$-channel expansion of the correlation function \cite{He2020,JacobsenSaleur2019,HGSJS2020}. The simplest example of the structure conjectured in Eq.~\eqref{eq:indecomposable_diamond} involves the fields $\Phi_{ej}$ and $\overbar\Phi_{ej}$ from the standard module $\mathscr W_{j,z^2}$ with $j = 1$, but we have seen in Eq.~\eqref{eq:torus_Z1} and the ensuing discussion that these fields decouple from the Potts-model partition function, and the results of \citeauthor{JacobsenSaleur2019} show that they also decouple from the correlation functions of the order parameter.

It is therefore natural to turn to the next available case, $j = 2$, and thus the representation $\mathscr W_{2,z^2}$. $P_{abab}$ and $P_{abba}$ both have the property of coupling $\mathscr W_{21}$ and $\mathscr W_{2,-1}$ in their $s$-channel expansions, and they are the only four-point functions that contain these two representations as their \emph{leading} contributions (other correlation functions couple to $\overbar{\mathscr W}_{\!\!0,\mathfrak q^{\pm 2}}$ and/or $\mathscr W_{0,-1}$ as well) [\emph{Id.}]. Moreover, the symmetric combination
\begin{equation}
P_{\text S} = P_{abab} + P_{abba}
\end{equation}
decouples from $\mathscr W_{2,-1}$ for symmetry reasons, and since $\mathscr W_{21}$ contains the fields $\Phi_{e2}$ and $\overbar\Phi_{e2}$ for integers $e \ge 0$, it transpires that $P_{\text S}$ is the most convenient correlation function to investigate in the present context. Finally, the lowest-lying levels that can give rise to the indecomposable structure correspond to the case $e = 1$. For all these reasons we henceforth focus on the case $(e,j) = (1,2)$.

Denoting the separation between the two groups of points $(z_1, z_2)$ and $(z_3, z_4)$ along the imaginary time direction by $l$, the correlation function in the cylinder geometry generically takes the form
\begin{equation}
P_{\text S} = \sum_i A_i \left(\frac{\Lambda_i}{\Lambda_0}\right)^l,
\end{equation}
where the sum is over the contributing eigenvalues $\Lambda_i$ (with $\Lambda_0$ referring to the ground state), and $A_i$ are the corresponding amplitudes. A shift between the two groups of points along the spacelike direction was shown to be irrelevant [\emph{Id.}]. In the notations of Figure 1 therein, one can therefore consider the two groups to be aligned---i.e., with a shift $x = 0$. A rank-2 Jordan block for the transfer matrix on the lattice manifests itself by a ``generalized amplitude,'' with $A_i$ of the form $a_i + l b_i$. This structure can be observed in many cases when $\mathfrak q$ is a root of unity \cite{GSJLS}. In our problem, however, the Jordan blocks are not present for finite $L$, and only expected to appear in the limit $L \to \infty$. A natural scenario for how this might happen is as follows: we should have two eigenvalues which become close as $L \to\infty$, with divergent and opposite amplitudes. Assuming that $\Lambda_1 = \Lambda(1 + \epsilon a)$ and $\Lambda_2 = \Lambda(1 - \epsilon a)$ appear with respective amplitudes $A_1 = A + b/\epsilon$ and $A_2 = A - b/\epsilon$, where the small parameter $\epsilon \to 0$ as $L \to \infty$, we have then
\begin{equation} \label{eq:amplitudes}
\begin{aligned}
A_1\left(\frac{\Lambda_1}{\Lambda_0}\right)^l + A_2 \left(\frac{\Lambda_2}{\Lambda_0}\right)^l &\approx \left(A + \frac{b}{\epsilon}\right)\left(\frac{\Lambda}{\Lambda_0}\right)^l(1+\epsilon a l) + \left(A - \frac{b}{\epsilon}\right)\left(\frac{\Lambda}{\Lambda_0}\right)^l (1 - \epsilon a l) \\
&= (2A + 2abl) \left(\frac{\Lambda}{\Lambda_0}\right)^l,
\end{aligned}
\end{equation}
reproducing as $L \to\infty$ the behavior expected from the presence of a Jordan block for the continuum-limit Hamiltonian.

The method best adapted to identifying the scenario of Eq.~\eqref{eq:amplitudes} is based on scalar products, as discussed by \textcite[Section 4.3.2]{JacobsenSaleur2019}. Notice that although this method measures the amplitudes $A_i$ directly in the $l\to\infty$ limit, the hypotheses leading to the scaling form can still be tested, and in particular the scaling of the amplitudes under the approach to the thermodynamic limit $L \to \infty$.

We now investigate this issue in the context of the structure containing $\Phi_{12}$ and $\overbar\Phi_{12}$, which is numerically the most accessible case for the reasons given above.

The finite-size level corresponding to the pair of fields $(\Phi_{12}, \overbar\Phi_{12})$ has been identified in Section \ref{singlet_states} as the line with $i_{13} = 3$ in Table \ref{W2_singlets}. Note that this is a twice-degenerate level (a doublet) in the transfer matrix spectrum, because the fields $\Phi_{12}$ and $\overbar\Phi_{12}$ are related by the exchange of chiral and antichiral components. The corresponding combined amplitude (i.e., summed over the doublet) for the contribution of this level to $P_{\text S}$ is shown in the first line of Table \ref{tab:amp}. The amplitudes are normalized by that of the leading contribution to $P_{\text S}$---namely the amplitude of the line with $i_{13} = 1$ in Table \ref{W2_singlets}. To be precise, the table shows the amplitudes for cylinders of circumference $L = 5,6,\ldots,10$, and in all cases the distance $d$ between the two points in each group $(z_1, z_2)$ and $(z_3, z_4)$ is taken the largest possible: $d = L/2$ for even $L$, and $d = (L-1)/2$ for odd $L$. This choice (which was also used in similar numerical work \cite{He2020,JacobsenSaleur2019}) corresponds to a fixed, finite distance between the two points in the continuum limit. Unfortunately, it also leads to parity effects in $L$, which are clearly visible in Table \ref{tab:amp}. It is nevertheless clear that the amplitude of the line with $i_{13} = 3$ converges to a finite constant, as expected for this non-logarithmic pair of fields, and this can be confirmed by independent fits of even and odd sizes. Regrettably, the situation for the remaining lines of the table is less clear. Na\"ively, the amplitude for each one of the last three lines appears to grow with $L$, but our attempts to quantify this have not been very compelling, due to the fact that we only have three sizes of each parity at our disposal.

\begin{table}[h]
\centering
\begin{tabular}{cllllll}
\toprule
 $i_{13}$ & \multicolumn{6}{c}{$L$} \\ 
 \cmidrule(l){2-7}
 & \multicolumn{1}{c}5 & \multicolumn{1}{c}6 & \multicolumn{1}{c}7 & \multicolumn{1}{c}8 & \multicolumn{1}{c}9 & \multicolumn{1}{c}{10} \\
 \midrule
 3 & $\phantom{-}0.515846$ & $\phantom{-}0.537395$ & $\phantom{-}0.514353$ & $\phantom{-}0.534697$ & $\phantom{-}0.524267$ & $\phantom{-}0.539497$ \\
 \emph{24} & $-0.00418366$ & $-0.0124738$ & $-0.0186079$ & $-0.0326019$ & $-0.0417739$ & $-0.0596335$ \\
 25 & $-0.0118071$ & $ -0.0252680$ & $-0.0241138$ & $-0.0342282$ & $-0.0331532$ & $-0.0404785$ \\
 \emph{35} & $\phantom{-}0.0230056$ & $\phantom{-}0.0532078$ & $\phantom{-}0.0610276$ & $\phantom{-}0.0930659$ & $\phantom{-}0.102977$ & $\phantom{-}0.134391$ \\
\bottomrule
\end{tabular}
\caption{Amplitudes $A_i$ of the correlation function $P_{\text S}$ corresponding to selected fields within $\mathscr W_{21}$, in finite size $L$. The distance between the two points within each group is taken as $d=\lfloor L/2 \rfloor$. The lines of the table are labeled, as in Table \ref{W2_singlets}, by the index $i_{13}$. These values are given to 8 significant figures in \textcite{Grans-Samuelsson2020}.}
 \label{tab:amp}
\end{table}

We therefore turn to another strategy, in which the same amplitudes are measured with the smallest possible distance $d = 1$ between the two points in each group. This will eliminate the parity effects, so that more reliable fits can be studied. Note that the choice $d = 1$ corresponds to a vanishing distance in the continuum limit, so one might expect the finite-size amplitudes to pick up an extra factor of $1/L$. In particular, the amplitude of a generic, non-logarithmic field contributing to $P_{\text S}$ is the expected to vanish as $L^{-1}$ in the $L\to\infty$ limit. Indeed, the amplitude of the line with $i_{13} = 3$ in Table \ref{tab:amp1} fits very nicely to $c_0 + c_1 L^{-1} + c_2 L^{-2} + \cdots$, and the absolute value of the constant term $c_0$ can be determined to be at least 80 times smaller than the data point with $L = 10$. We therefore conjecture that, in this case, $c_0 = 0$ indeed.

\begin{table}[h]
\centering
{\small\begin{tabular}{cllllll}
\toprule
 $i_{13}$ & \multicolumn{6}{c}{$L$} \\ 
 \cmidrule(l){2-7}
 & \multicolumn{1}{c}5 & \multicolumn{1}{c}6 & \multicolumn{1}{c}7 & \multicolumn{1}{c}8 & \multicolumn{1}{c}9 & \multicolumn{1}{c}{10} \\
 \midrule
3 & $\phantom{-}0.296317$ & $\phantom{-}0.233276$ & $\phantom{-}0.190424$ & $\phantom{-}0.159168$ & $\phantom{-}0.135394$ & $\phantom{-}0.116790$ \\
 \emph{24} & $-0.0000022603$ & $-0.0000057466$ & $-0.0000088156$ & $-0.0000107666$ & $-0.0000116358$ & $-0.0000117097$ \\
 25 & $-0.00194026$ & $-0.00264918$ & $-0.00297698$ & $-0.00305503$ & $-0.00298975$ & $-0.00285018$ \\
 \emph{35} & $\phantom{-}0.0000380855$ & $\phantom{-}0.0000508765$ & $\phantom{-}0.0000532218$ & $\phantom{-}0.0000497383$ & $\phantom{-}0.0000439510$ & $\phantom{-}0.0000377665$ \\
 \bottomrule
\end{tabular}}
\caption{Amplitudes $A_i$ of the correlation function $P_{\text S}$ corresponding to selected fields within $\mathscr W_{21}$, in finite size $L$. The distance between the two points within each group is now chosen the smallest possible, $d=1$. These values are given to 8 significant figures in \textcite{Grans-Samuelsson2020}.}
 \label{tab:amp1}
\end{table}

For the line with $i_{13} = 24$ (a singlet level) we attempt a fit of the form $c_0 + c_1 L^{-\delta} + c_2 L^{-2\delta} + c_3 L^{-3\delta}$. This matches the data nicely with $\delta \approx 1.005$, indicating that $\delta = 1$ might be the exact value of the exponent. But we find now that the absolute value of the constant term $c_0$ is about 3 times \emph{larger} than the data point with $L = 10$, which is strongly indicative of $c_0$ being nonzero in this case. We therefore conjecture that this line should be identified with one of the top or bottom fields in the Jordan block of Eq.~\eqref{eq:indecomposable_12}.

The same type of fit for the line with $i_{13} = 35$ (the other singlet level) yields $\delta \approx 2.05$ and a constant term $c_0$ which is about 4 times \emph{smaller} than the $L = 10$ data point. Finally, the line with $i_{13} = 25$ (a doublet) matches the fit with $\delta \approx 1.3$ and $c_0$ about 3 times smaller than the data point with $L = 10$. Seen in isolation, these fits do not permit us to convincingly conclude whether the value of $c_0$ is finite or zero for those two lines. However, structural considerations provide more compelling evidence. According to the argument given in Eq.~\eqref{eq:amplitudes}, the logarithmic singlet with $i_{13} = 24$ needs to be accompanied by another singlet field with an opposite and diverging (for finite conformal distance) amplitude. Being a singlet, the line with $i_{13} = 35$ is the only possible candidate for such a logarithmic partner.

\begin{figure}[h]
\centering
\includegraphics[width=0.8\textwidth]{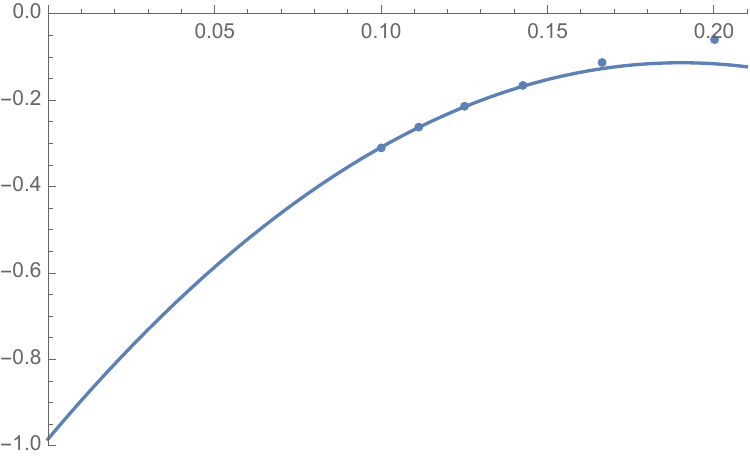}
\caption{Ratio $A_1/A_2$ between the amplitudes of the two singlet fields (see Table \ref{tab:amp1}), corresponding to the lines with $i_{13} = 24$ and $i_{13}=35$ (see Table \ref{W2_singlets}), plotted against $1/L$. The curve is a second-order polynomial fit to the last three data points.}
\label{logfits}
\end{figure}

As a decisive test, we therefore plot in Figure \ref{logfits} the ratio between the amplitudes of the two singlets. A second-order polynomial in $1/L$ fits the data nicely and gives an extrapolated value of the ratio of $-0.985$, very close to the exact ratio of $-1$ expected from Eq.~\eqref{eq:amplitudes}. We believe that this shows that the two singlets correspond to the conformal fields $A_{12}\Phi_{12} + \overbar A_{12}\overbar\Phi_{12}$ and $\tilde\Psi_{12}$, and that the indecomposable structure of Eq.~\eqref{eq:indecomposable_diamond} builds up only in the $L \to\infty$ limit (recall that $A_{12}\Phi_{12} = \overbar A_{12}\overbar\Phi_{12}$ only holds in the limit; the symmetrized combination is the natural guess to consider here). On the other hand, Figure \ref{logfits} vividly illustrates that a maximum size of $L = 10$ is still quite far from the thermodynamic limit, and with hindsight it is therefore hardly surprising that only a combination of arguments can reveal the detailed nature of the four fields from Table \ref{W2_singlets} having conformal weights $(h_{12} + 2, h_{12} + 2)$.

\section{Modules for the loop model in the degenerate case: the OPE point of view} \label{OPE}

As in the early works on logarithmic CFTs \cite{VJS2011,GurarieLudwig2002}, it is possible to understand the appearance of indecomposable modules in the continuum limit of $\mathscr W_{j1}$ by carefully examining the OPEs and their potential divergences when one of the fields in the $s$-channel has a degenerate conformal weight.

To start, imagine that we have some OPE of a field of dimension $\Delta$ with itself where a field with conformal weights $(h_{12}, h_{12})$ appears. In ordinary CFT, the descendants of this field at level two in the chiral and in the antichiral sector would not be independent: this fact is crucial to cancel the divergence arising in the OPE coefficients from the fact that $h_{12}$ is in the Kac table, resulting in a finite OPE such as the ones arising in the minimal-model CFTs \cite{BPZ1984}. Let us now see what happens if the null descendants are not zero, and the divergences potentially remain. To proceed, we factor out the $(z\overbar z)^{-2\Delta}$ term, with $\Delta = \overbar\Delta$ denoting the conformal weight of the fields being fused, and analyze the potential divergences by slightly shifting the conformal weights of the field on the right hand side of the OPE:
\begin{equation} \label{eq:OPE}
\begin{aligned}
\Phi_{h\overbar h}(z)\Phi_{h\overbar h}(0) &\sim \frac{C(\epsilon)}{(z\overbar z)^{2\Delta}} \left\{(z\overbar z)^{h_{1+\epsilon,2}}\left[\left(X_\epsilon + \frac z2 \partial X_\epsilon + \alpha^{(-2)}(\Delta, h_{1+\epsilon,2})z^2 L_{-2}X_\epsilon\right.\right.\right. \\
&\hphantom{\frac{C(\epsilon)}{(z\overbar z)^{2\Delta}} \left\{(z\overbar z)^{h_{1+\epsilon,2}}\left[\left(\right.\right.\right.} \left.\left.\left.\qquad{}+ \alpha^{(-1,-1)}(\Delta, h_{1+\epsilon,2})z^2 L_{-1}^2 X_\epsilon\right) \otimes \text{h.\ c.}\right] + \cdots \right\},
\end{aligned}
\end{equation}
where $C(\epsilon)$ is a number to be determined, the dots stand for other fields, and we have used the shorthand notations
\begin{subequations}
\begin{gather}
X_\epsilon = \phi_{1+\epsilon,2} \qq{and} \\
\overbar X_\epsilon = \overbar\phi_{1+\epsilon,2}.
\end{gather}
\end{subequations}
Fields in OPEs without position arguments are understood to be evaluated at $z = 0$ ($\overbar z = 0$ for antiholomorphic fields). The coefficients $\alpha$ in Eq.~\eqref{eq:OPE} are fully determined by conformal invariance:
\begin{subequations}
\begin{gather}
\alpha^{(-2)}(\Delta,h) = \frac{(h-1)h + 2\Delta(1+2h)}{16(h-h_{12})(h-h_{21})}, \label{eq:alpha_2} \\
\alpha^{(-1,-1)}(\Delta,h) = \frac{(1+h)(c+8h) - 12(\Delta+h)}{64(h-h_{12})(h-h_{21})},
\end{gather}
\end{subequations}
and note that we have
\begin{equation} \label{eq:alpha_identity}
\alpha^{(-1,-1)}(\Delta,h)L_{-1}^2 + \alpha^{(-2)}(\Delta,h) L_{-2} = \alpha^{(-2)}(\Delta,h)A(h) + \alpha^{(-1,-1)}_0(h)L_{-1}^2,
\end{equation}
where
\begin{subequations}
\begin{gather}
A(h) \equiv L_{-2} - \frac{3}{2+4h}L_{-1}^2, \qq{and} \\
\alpha^{(-1,-1)}_0(h) = \frac{1+h}{4(1+2h)}.
\end{gather}
\end{subequations}
It is important to notice that in writing Eq.~\eqref{eq:alpha_identity}, the dependence on the external field $\Delta$ only appears in the coefficient $\alpha^{(-2)}$---i.e., the operator $A$ that will turn out to give rise to the Jordan-block structure is independent of the external field. This point will become more clear in the following discussion.

With $\epsilon \to 0$, we maintain all calculations only to the lowest necessary orders in $\epsilon$. Going back to $h = h_{1+\epsilon,2}$, and writing $A_\epsilon \equiv A(h_{1+\epsilon,2})$, it is convenient to define
\begin{equation} \label{eq:gamma_null}
\gamma(\epsilon) \equiv \mel*{X_\epsilon}{A^\dagger_\epsilon A_\epsilon}{X_\epsilon} = \frac{8(h-h_{12})(h-h_{21})}{1+2h} = \nu\epsilon,
\end{equation}
with
\begin{equation} \label{eq:nu_beta}
\nu = -\frac{2(1 - 2\beta^2 - \beta^4 + 2\beta^6)}{\beta^6},
\end{equation}
where we have used the parametrization $\beta^2 = x/(x+1)$. On the other hand, notice that as $\epsilon\to 0$, the coefficient $\alpha^{(-2)}(\Delta,h_{1+\epsilon,2})$ has a simple pole, since the denominator is proportional to the Kac determinant, as is clear from Eq.~\eqref{eq:alpha_2}. This means that the OPE potentially presents singularities, which must be properly canceled by the contribution of other fields with the proper dimensions---a point well understood since earlier studies on indecomposability in logarithmic CFT \cite{VJS2011,DJS2010c,VGJS2012,GurarieLudwig2002}. The leading singularity in the OPE is a second-order pole coming from the descendants at level 2 of $X_\epsilon\otimes\overbar X_\epsilon$. Keeping in mind that $h_{12} + 2 = h_{-1,2}$, and of course $h_{rs} = h_{-r,-s}$, we therefore introduce the other fields
\begin{subequations}
\begin{gather}
Y_\epsilon = \phi_{-1+\epsilon,2} \qq{and} \\
\overbar Y_\epsilon = \overbar\phi_{-1+\epsilon,2}
\end{gather}
\end{subequations}
in order to cancel such singularities, and we complete the OPE as follows:
\begin{equation} \label{eq:OPE_complete}
\begin{aligned}
\Phi_{h\overbar h}(z)\Phi_{h\overbar h}(0) &\sim \frac{C(\epsilon)}{(z\overbar z)^{2\Delta}} \left\{(z\overbar z)^{h_{1+\epsilon,2}}\left[\left(X_\epsilon + \frac z2 \partial X_\epsilon + \alpha^{(-1,-1)}_0(\epsilon)z^2 L_{-1}^2 X_\epsilon+ \alpha^{(-2)}(\epsilon)z^2 A_\epsilon X_\epsilon\right) \otimes \text{h.\ c.}\right]\right. \\
&\hphantom{\frac{C(\epsilon)}{(z\overbar z)^{2\Delta}}} \left.\qquad{}+ (z\overbar z)^{h_{-1+\epsilon,2}}a(\epsilon)Y_\epsilon\otimes\overbar Y_\epsilon\right\},
\end{aligned}
\end{equation}
where we have adopted the shorthand notations $\alpha^{(-2)}(\epsilon) \equiv \alpha^{(-2)}(\Delta, h_{1+\epsilon,2})$, $\alpha^{(-1,-1)}_0(\epsilon) \equiv \alpha^{(-1,-1)}_0(h_{1+\epsilon,2})$, and the new coefficient $a(\epsilon)$ is yet to be determined.

To study the necessary cancelation of singularities, we focus on the most divergent term at level 2:
\begin{equation} \label{eq:level_2_divergence}
\begin{aligned}
(z\overbar z)^{h_{1+\epsilon,2}+2}[\alpha^{(-2)}(\epsilon)]^2A_\epsilon X_\epsilon\otimes\overbar A_\epsilon\overbar X_\epsilon &+ a(\epsilon)(z\overbar z)^{h_{-1+\epsilon,2}+2}Y_\epsilon\otimes\overbar Y_\epsilon \\
&= \epsilon\kappa\log(z\overbar z)(z\overbar z)^{h_{-1,2}}[\alpha^{(-2)}(\epsilon)]^2A_\epsilon X_\epsilon\otimes\overbar A_\epsilon\overbar X_\epsilon + \frac{1}{\sqrt\epsilon}(z\overbar z)^{h_{-1+\epsilon,2}}\Phi_\epsilon,
\end{aligned}
\end{equation}
where we have defined
\begin{equation} \label{eq:kappa_beta}
\kappa\equiv\frac{h_{1+\epsilon,2} + 2 - h_{-1+\epsilon,2}}{\epsilon} = \frac{1}{\beta^2}
\end{equation}
and introduced the new field
\begin{equation}
\Phi_\epsilon \equiv \sqrt\epsilon\{[\alpha^{(-2)}(\epsilon)]^2A_\epsilon X_\epsilon\otimes\overbar A_\epsilon\overbar X_\epsilon + a(\epsilon)Y_\epsilon\otimes\overbar Y_\epsilon\}.
\end{equation}
The two-point function of this field is given by
\begin{equation} \label{eq:phi_epsilon_2_point}
\ev*{\Phi_\epsilon(z,\overbar z)\Phi_\epsilon(0,0)} = \epsilon\{[\alpha^{(-2)}(\epsilon)]^4\gamma^2(\epsilon)(z\overbar z)^{-2h_{1+\epsilon,2}-4} + a^2(\epsilon)(z\overbar z)^{-2h_{-1+\epsilon,2}}\}.
\end{equation}
Recall Eq.~\eqref{eq:gamma_null} and that $\alpha^{(-2)}(\epsilon)$ has a simple pole in $\epsilon$. One can write
\begin{equation} \label{eq:alpha_pole}
[\alpha^{(-2)}(\Delta, h_{1+\epsilon,2})]^2\gamma(\epsilon) \equiv \frac{r}{\epsilon} + s + O(\epsilon).
\end{equation}
It is then clear that the coefficient of the first term in Eq.~\eqref{eq:phi_epsilon_2_point} has a double pole that must be canceled by the divergence from the second term. This requires $a^2(\epsilon)$ to be of the form
\begin{equation}
a^2(\epsilon) = \frac{\lambda}{\epsilon^2} + \frac{\mu}{\epsilon} + O(1).
\end{equation}
Such behavior can in fact be established using that $\phi_{21}$ is degenerate in the theory \cite[Section 7]{Grans-Samuelsson2020}. The singularity cancelation condition then reads
\begin{equation}
\lambda = -r^2,
\end{equation}
and the two-point function of Eq.~\eqref{eq:phi_epsilon_2_point} becomes as $\epsilon\to 0$
\begin{equation}
\ev*{\Phi(z,\overbar z)\Phi(0,0)} = \frac{-2\kappa r^2\log(z\overbar z) + 2rs + \mu}{(z\overbar z)^{2h_{-1,2}}}.
\end{equation}
Taking into account the $1/\sqrt\epsilon$ factor in Eq.~\eqref{eq:level_2_divergence}, we must therefore take $C(\epsilon) = \sqrt\epsilon$, so that the contribution of $\Phi_\epsilon$ in the OPE is of order unity.

At this point, it is natural to introduce the normalized field
\begin{equation} \label{eq:X_normalized}
\hat X_\epsilon \equiv \frac{1}{\sqrt\gamma}A_\epsilon X_\epsilon,
\end{equation}
and identify it as another copy of $Y_\epsilon$ in the limit $\epsilon\to 0$, since both then have dimension $h_{-1,2}$ and are annihilated by $L_1$ and $L_2$. The first term on the right hand side of Eq.~\eqref{eq:level_2_divergence}, upon multiplication by $C(\epsilon)$, is then given by
\begin{equation}
\frac{\kappa r}{2\sqrt\nu}(z\overbar z)^{h_{-1,2}}\log(z\overbar z)(AX\otimes\overbar Y + Y\otimes\overbar A\overbar X).
\end{equation}
Combining with the remaining terms in the OPE of Eq.~\eqref{eq:OPE_complete},
\begin{equation}
\sqrt\epsilon(z\overbar z)^{h_{1+\epsilon,2}}\left[\left(X_\epsilon + \frac z2\partial X_\epsilon + \alpha^{(-1,-1)}_0(\epsilon) z^2 L_{-1}^2 X_\epsilon\right)\otimes\alpha^{(-2)}(\epsilon)\sqrt{\gamma}\overbar z^2 \hat{\overbar X}_\epsilon + \text{h.\ c.}\right],
\end{equation}
and recalling Eq.~\eqref{eq:alpha_pole}, we then have the full OPE as $\epsilon \to 0$, after factoring out a global factor of $\sqrt r$:
\begin{equation}
\left[z^{h_{12}}\overbar z^{h_{-1,2}}\left(X + \frac z2\partial X + \alpha^{(-1,-1)}_0 z^2 L_{-1}^2 X\right)\otimes\overbar Y+ \text{h.\ c.} \right] + \frac{(z\overbar z)^{h_{-1,2}}\kappa r}{2\sqrt\nu}\left[\log(z\overbar z)(AX\otimes\overbar Y + Y\otimes\overbar A\overbar X) + \frac{2\sqrt\nu}{\kappa r}\Phi\right].
\end{equation}
Setting
\begin{equation}
AX\otimes\overbar Y = \sqrt\gamma\hat X\otimes\overbar Y = \sqrt\gamma Y\otimes\hat{\overbar X} = Y\otimes\overbar A\overbar X,
\end{equation}
using the identifications of $\hat X$ and $\hat{\overbar X}$ with $Y$ and $\overbar Y$ in the $\epsilon \to 0$ limit, the OPE becomes
\begin{equation} \label{eq:OPE_final}
\left[z^{h_{12}}\overbar z^{h_{-1,2}}\left(X + \frac z2\partial X + \alpha^{(-1,-1)}_0 z^2 L_{-1}^2 X\right)\otimes\overbar Y+ \text{h.\ c.} \right] + (z\overbar z)^{h_{-1,2}}\frac{\kappa r}{\sqrt\nu}\left[\log(z\overbar z)AX\otimes\overbar Y + \frac{\sqrt\nu}{\kappa r}\Phi\right].
\end{equation}
As will become obvious below, this has the interpretation that $L_0 - \overbar L_0$ is diagonalizable.

We are interested in the logarithmic mixing at level 2---i.e., the second term of the OPE. Inspecting the terms, it is natural to redefine the field
\begin{equation}
\tilde\Psi \equiv \frac{\sqrt\nu}{\kappa r}\Phi,
\end{equation}
which, as we will see, becomes the logarithmic partner of $AX\otimes\overbar Y = Y\otimes\overbar A\overbar X$. It is a simple exercise to calculate their two-point functions and one arrives at
\begin{subequations}
\begin{gather}
\ev*{(AX\otimes\overbar Y)(z,\overbar z)(AX\otimes\overbar Y)(0,0)} = 0, \\
\ev*{\tilde\Psi(z,\overbar z)(AX\otimes\overbar Y)(0,0)} = \frac{\kappa^{-1}\nu}{(z\overbar z)^{2h_{-1,2}}}, \\
\ev*{\tilde\Psi(z,\overbar z)\tilde\Psi(0,0)} = \frac{-2\kappa^{-1}\nu\log(z\overbar z) + (\kappa r)^{-2}\nu(2rs+\mu)}{(z\overbar z)^{2h_{-1,2}}}. \label{eq:OPE_log}
\end{gather}
\end{subequations}
We recognize the usual logarithmic structure of a rank-2 Jordan block \cite{Gurarie1993}.

As a final step, we compute the action of the Virasoro generator $L_0$ on the pair $(AX\otimes\overbar Y,\tilde\Psi)$:
\begin{subequations}
\begin{gather}
L_0AX\otimes\overbar Y = h_{-1,2}AX\otimes\overbar Y, \\
\begin{aligned}
L_0\tilde\Psi &= h_{-1,2}\tilde\Psi + \frac{\sqrt{\nu\epsilon}}{\kappa r}[\alpha^{(-2)}(\epsilon)]^2(h_{1+\epsilon,2} + 2 - h_{-1+\epsilon,2})A_\epsilon X_\epsilon\otimes\overbar A_\epsilon\overbar X_\epsilon \\
&= h_{-1,2}\tilde\Psi + AX\otimes\overbar Y,
\end{aligned}
\end{gather}
\end{subequations}
and similarly for $\overbar L_0$. Therefore we see that in the basis $(AX\otimes\overbar Y, \tilde\Psi) = (Y\otimes\overbar A\overbar X,\tilde\Psi)$, we have
\begin{equation} \label{eq:L_0_matrix}
L_0 = \begin{pmatrix}
h_{-1,2} & 1 \\ 0 & h_{-1,2} \end{pmatrix} = \overbar L_0,
\end{equation}
forming a rank-2 Jordan block. In addition, we find
\begin{equation}
A^\dagger\tilde\Psi = \frac{\sqrt{\nu\epsilon}}{\kappa r}[\alpha^{(-2)}(\epsilon)]^2\gamma X\otimes\overbar A\overbar X = \kappa^{-1}\nu X\otimes\overbar Y,
\end{equation}
where we have used Eqs.~\eqref{eq:gamma_null}, \eqref{eq:alpha_pole}, and \eqref{eq:X_normalized}. Note also that $L_1\tilde\Psi = 0$. Hence, the module is depicted as
\begin{equation} \label{eq:indecomposable_XY}
\begin{tikzcd}
& \tilde\Psi \arrow[dr,"\overbar A^\dagger/\kappa^{-1}\nu"]\arrow[dl,"A^\dagger/\kappa^{-1}\nu"']\arrow[to=3-2,"L_0 - h_{-1,2}"] & \\
X\otimes\overbar Y \arrow[dr,"A"'] & & Y\otimes\overbar X \arrow[dl,"\overbar A"] \\
& AX\otimes\overbar Y = Y\otimes\overbar A\overbar X
\end{tikzcd}
\end{equation}
This structure coincides with that of Eq.~\eqref{eq:indecomposable_12}.

As we have briefly commented, the logarithmic coupling $\kappa^{-1}\nu$ in Eq.~\eqref{eq:OPE_log}, which characterizes the Jordan-block structure, does not depend on the dimension $\Delta$ of the external fields. More explicitly, from Eqs.~\eqref{eq:nu_beta} and \eqref{eq:kappa_beta}, we have
\begin{equation} \label{eq:kappa_nu}
\kappa^{-1}\nu = -\frac{2(1 - 2\beta^2 - \beta^4 + 2\beta^6)}{\beta^4},
\end{equation}
which is entirely determined by the Kac formula and the Kac determinant. By contrast, the coefficient $\kappa\sqrt r/\sqrt\nu$ in the OPE of Eq.~\eqref{eq:OPE_final} does depend on $\Delta$ through $r$, due to Eq.~\eqref{eq:alpha_pole}. Similarly, the constant in the two-point function of Eq.~\eqref{eq:OPE_log} also depends on $\delta$. This is, however, compatible with the Jordan-block structure, since the field $\tilde\Psi$ always admits a shift by a multiple of the null field \cite{Gurarie1993},
\begin{equation}
\tilde\Psi \to \tilde\Psi + \theta AX\otimes \overbar Y = \tilde\Psi + \theta Y\otimes\overbar A\overbar X, \qquad (\theta\in\mathbb C)
\end{equation}
which does not change Eq.~\eqref{eq:L_0_matrix}.

The construction also generalizes to the case of operators $\phi_{rs}$ and $\phi_{r,-s}$. In general, the module has the structure in Eq.~\eqref{eq:indecomposable_XY} with $X \to \phi_{rs}$, $Y \to \phi_{r,-s}$, and $A$ replaced by the proper combination of Virasoro generators $A_{rs}$. Setting
\begin{equation}
\ev*{A^\dagger_{rs}A_{rs}}{\phi_{r+\epsilon_s}} = \nu_{rs}\epsilon,
\end{equation}
and observing that
\begin{gather}
h_{r+\epsilon,s} + rs - h_{-r+\epsilon,s} = \kappa_{rs}\epsilon, \qq{with} \\
\kappa_{rs} = \frac{r}{\beta^2},
\end{gather}
we find that the free parameter of the module (the so-called logarithmic coupling, or indecomposability parameter) is
\begin{equation}
b_{rs} = \kappa_{rs}^{-1}\nu_{rs},
\end{equation}
so that
\begin{subequations}
\begin{gather}
(L_0 - h_{-r,s})\tilde\Psi_{rs} = (\overbar L_0 - h_{r,-s})\tilde\Psi_{rs} = A_{rs}\phi_{rs}\otimes\overbar\phi_{r,-s} = \phi_{r,-s}\otimes\overbar A_{rs}\overbar\phi_{rs}, \\
A^\dagger_{rs}\tilde\Psi_{rs} = b_{rs}\phi_{rs}\otimes\overbar\phi_{r,-s}, \\
\overbar A^\dagger_{rs} \tilde\Psi_{rs} = b_{rs}\phi_{r,-s}\otimes\overbar\phi_{rs}
\end{gather}
\end{subequations}
with the structure
\begin{equation} \label{eq:indecomposable_rs}
\begin{tikzcd}
& \tilde\Psi_{rs} \arrow[dr,"\overbar A^\dagger_{rs}/b_{rs}"]\arrow[dl,"A^\dagger_{rs}/b_{rs}"']\arrow[to=3-2,"L_0 - h_{-r,s}"] & \\
\phi_{rs}\otimes\overbar\phi_{r,-s} \arrow[dr,"A_{rs}"'] & & \phi_{r,-s}\otimes\overbar\phi_{rs} \arrow[dl,"\overbar A_{rs}"] \\
& A_{rs}\phi_{rs}\otimes\overbar\phi_{r,-s} = \phi_{r,-s}\otimes\overbar A_{rs}\overbar\phi_{rs}
\end{tikzcd}
\end{equation}
This structure again coincides with that of Eq.~\eqref{eq:indecomposable_diamond}.

For the special case $r = s = 1$, for instance, we find that $\nu_{11} = -1 + 1/\beta^2$ and therefore
\begin{equation} \label{eq:b_11}
b_{11} = 1-\beta^2.
\end{equation}
Similarly, the calculation of Eq.~\eqref{eq:kappa_nu} yields, using these notations,
\begin{equation} \label{eq:b_12}
b_{12} = \kappa_{12}^{-1}\nu_{12} = -\frac{2(1-2\beta^2-\beta^4+2\beta^6)}{\beta^4}.
\end{equation}

\section{Measurement of indecomposability parameters via emerging Jordan blocks} \label{b_emerging}
As a result of indecomposable operator product expansions in a logarithmic CFT, a parameter called $b$ was first introduced by \textcite{GurarieLudwig2002} to quantify this indecomposability, in terms of the stress--energy tensor $T$ and its logarithmic partner $t$. The indecomposability manifests as non-diagonalizability of $L_0$ and $\overbar L_0$:
\begin{subequations} \label{eq:Tt_Jordan}
\begin{gather}
L_0T = 2T, \\
L_0t = 2t+T, \\
\overbar L_0T = 0, \qq{and} \\
\overbar L_0t = T.
\end{gather}
\end{subequations}
For the two-point functions involving $T$ and $t$, we have (at $c = 0$)
\begin{subequations}
\begin{gather}
\ev*{T(z)T(0)} = 0, \\
\ev*{T(z)t(0)} = \frac{b}{z^4}, \qq{and} \\
\ev*{t(z)t(0)} = \frac{-2b\log z+\theta}{z^4},
\end{gather}
\end{subequations}
where $\theta$ is an arbitrary constant that can be adjusted using the freedom $t\to t+\lambda T$, with $\lambda$ an arbitrary scalar. Recalling the Virasoro inner product (for holomorphic fields) \cite{DFMS1997}
\begin{equation}
\braket*{\Phi}{\Psi} = z^{2h_\Phi}\lim_{z\to\infty}\ev*{\Phi(z)\Psi(0)},
\end{equation}
we have, in the CFT, $\braket*{T} = 0$, $\braket*{t} = \infty$, and $\braket*{T}{t} = b$.

The measurement of $b$ on the lattice would thus seem straightforward---identify $T$ and $t$ on the lattice and evaluate their conformal scalar product. The difficulty with this approach is the otherwise-mundane matter of normalization. Because $\ip*{T} = 0$ in the CFT, also observed in finite size on the lattice, the value of $\ip*{T}{t}$ on the lattice is completely dependent on the way one normalizes $T$. The solution is to write an expression for $b$ that is independent of the normalization of $T$. In the CFT, $T = L_{-2}I$, with $I$ the identity field. Being primary, we may normalize its lattice analogue to unity: $\ip*{I} = 1$. Then
\begin{equation}
b = \frac{b^2}{b} = \frac{|\mel*{t}{L_{-2}}{I}|^2}{\ip*{t}{T}}
\end{equation}
is independent of the normalization of $T$, since the normalization of $t$ is linked to that of $T$ via the coefficients in \eqref{eq:Tt_Jordan}.

The method of measuring $b$ between $T$ and $t$ generalizes to any logarithmic pair, especially regarding the fact that $T$ is a null state. In general, whenever a Jordan block forms, there is a top and a bottom state. If this Jordan block comes from a Hamiltonian describing a logarithmic CFT, from general arguments involving the action of the Virasoro algebra, it is expected that the bottom state in such a Jordan block has zero conformal norm square. In the CFT, calling $\tilde\Psi,\Psi$ the top and bottom state, the result follows from the fact that we have 
\begin{subequations}
\begin{gather}
L_0\tilde\Psi = h\tilde\Psi + \Psi \qq{and}\\
L_0\Psi = h\Psi.
\end{gather}
\end{subequations}
Hence we have 
\begin{equation}
\ip*{\Psi}=\ip*{\Psi}{(L_0-h)\tilde\Psi}=\ip*{(L_0-h)\Psi}{\tilde\Psi}=0
\end{equation}
where in the last equation we used the fact that $L_n^\dagger=L_{-n}$ (i.e., $L_n$ and $L_{-n}$ are conjugate for the Virasoro norm).

On the lattice, recall that we observe pairs of singlets (Section \ref{singlet_states}) that have close eigenvalues at finite size that approach each other in the limit $L\to\infty$. Nevertheless, both singlets are still proper eigenvectors, and we expect that they both converge to the bottom state $\Psi$ of an emerging Jordan block. They should thus have a loop norm going to zero in the limit $L\to\infty$. This will be checked in cases where we find emerging Jordan blocks using the $J$ measure (Chapter \ref{emerging}).

The method for calculating $b$ is summarized as follows:
\begin{enumerate}
\item Identify an emerging Jordan block by finding pairs of vectors $\Psi$, $\Psi'$ on the lattice corresponding to given conformal fields such that $\lim J(\Psi,\Psi') = 1$.

\item Construct the emerging Jordan vector $\tilde\Psi$ by orthogonalizing $\Psi'$ against $\Psi$ with respect to the standard inner product $(\cdot|\cdot)$. Normalize $\tilde\Psi$ so that $(\Psi|H_0|\tilde\Psi) = 2$ (recall generally that $H_n = L_n + \overbar L_{-n}$).

\item Using the CFT, express $\Psi = A\Phi$ as a descendant of a primary field $\Phi$, with $A$ the corresponding null vector operator. Identify the lattice analogue of $\Phi$, and normalize it to $\ip*{\Phi} = 1$.

\item Finally,
\begin{equation} \label{eq:b_measurement}
b = \frac{|\mel*{\tilde\Psi}{A}{\Phi}|^2}{\ip*{\tilde\Psi}{\Psi}},
\end{equation}
where, optionally, $L_{-n} \to H_{-n}$ and $\overbar L_{-n} \to H_n$ in the expression for $A$.
\end{enumerate}
There are many possibilities for the unspecified limit in step 1. In this work we focus on the Jordan blocks that form as $L \to \infty$ and $c \to 0$, by studying the emerging Jordan blocks at finite $L$ and generic $c$, respectively. Additionally, the procedure gives two possible measurements of $b$. Since $\Psi$ and $\Psi'$ both converge to the same vector, they should be treated on an equal footing, and by switching the roles of $\Psi$ and $\Psi'$ we obtain a second value for $b$. Generically, this value will be different from the first one, but should become the same when the limit is taken. In practice, we may have reason to favor one measurement over the other.

\subsection{Emerging Jordan blocks in $\mathscr W_{11}$ at generic $c$}

\begin{figure}
\centering
\includegraphics[width=0.8\textwidth]{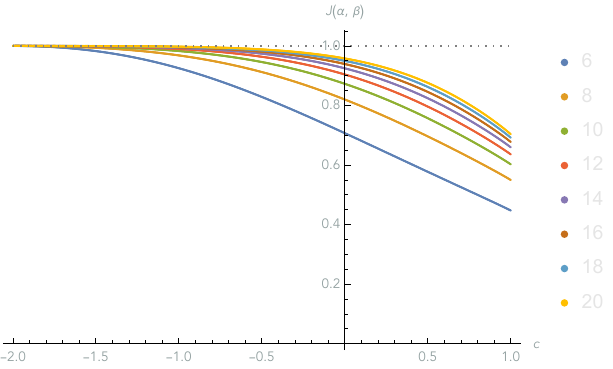}
\caption{The formation of a Jordan block between two singlets in $\mathscr W_{11}$. The legend indicates the value of $N = 2L$.}
\label{W11J2}
\end{figure}

In $\mathscr W_{11}$ there are two eigenstates at zero momentum whose conformal weights go to $h + \overbar h = 2$ as $L \to \infty$ (Table \ref{W11_singlets}, $i_{13} = 5$ and $6$). For generic $c$ and finite $L$, these eigenstates are non-degenerate, and we refer to them by the labels $\alpha$ and $\beta$. In practice, $\alpha$ is the first excitation (above the ground state) in the zero-momentum sector of $\mathscr W_{11}$. $\beta$ is the second or third excitation, depending on the values of $L$ and $c$. At values of $c$ close to 1 and small values of $L$, it is the third excitation, but crosses over to the second excitation at large values of $L$. The emerging Jordan block involving $\alpha$ and $\beta$, based on the measure $J(\alpha,\beta)$, can be seen in Figure \ref{W11J2}. The plot suggests that there is a Jordan block at finite size for $c = -2$, and this is indeed the case \cite{GRS2013a}.

\begin{figure}
\centering
\includegraphics[width=0.8\textwidth]{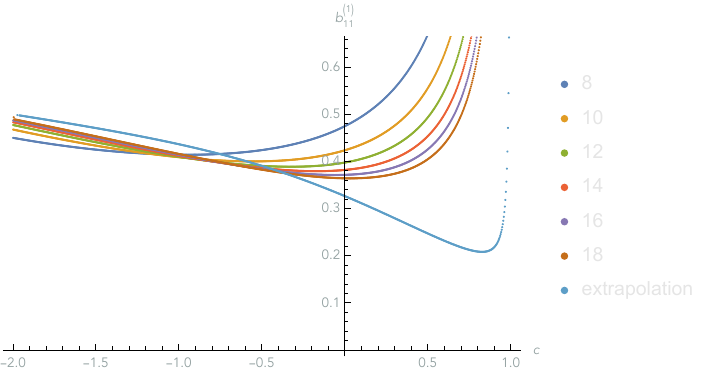}
\caption{The measurement of $b_{11}^{(1)}$ between $\alpha$ and $\beta$.}
\label{W11b1}
\end{figure}

\begin{figure}
\centering
\includegraphics[width=0.8\textwidth]{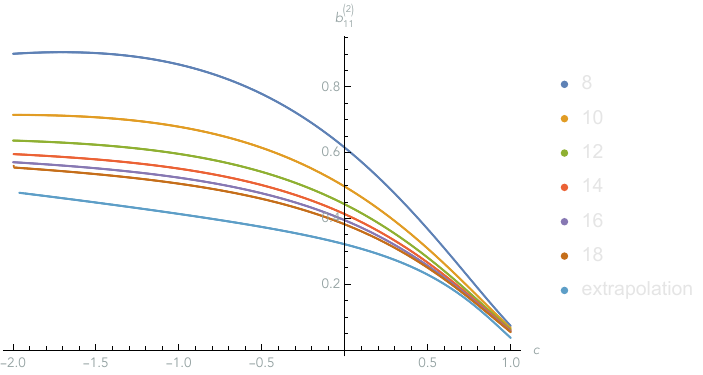}
\caption{The measurement of $b_{11}^{(2)}$ between $\alpha$ and $\beta$.}
\label{W11b2}
\end{figure}

Following our procedure, we calculate the parameter $b_{11}$ that describes this Jordan block at finite size. Both $\alpha$ and $\beta$ converge to the same vector as $L\to\infty$, the primary field $\Psi_{11} = L_{-1}\Phi_{11} = \overbar L_{-1}\overbar\Phi_{11}$ of Eq.~\eqref{eq:indecomposable_diamond} (again, the equality holds in the continuum limit). Take, for the lattice approximation of $\Phi_{11}$, the lowest field of momentum 1 ($i_{13} = 2$ in Table \ref{W11_singlets}) and normalize it to $\ip*{\Phi_{11}} = 1$. Then, orthogonalize $\alpha$ and $\beta$ by subtracting from $\beta$ its component along $\alpha$, resulting in $\tilde\beta$, and normalize it such that $(\alpha|H_0|\tilde\beta) = 2$. Then, we obtain a lattice measurement of the logarithmic coupling by calculating
\begin{equation}
b^{(1)}_{11}(L,c) = \frac{|\mel*{\tilde\beta}{H_{-1}}{\Phi_{11}}|^2}{\ip*{\tilde\beta}{\alpha}}.
\end{equation}
As $\alpha$ and $\beta$ should be treated on an equal footing, we similarly define $b^{(2)}_{11}$ by exchanging the roles of $\alpha$ and $\beta$ in this procedure. $b^{(2)}_{11}$ will generically be different from $b^{(1)}_{11}$, but when a genuine Jordan block forms, they should be equal, as we see in the analytic examples. These two measurements of $b_{11}$ are given graphically in Figures \ref{W11b1} and \ref{W11b2}.


\begin{figure}
\centering
\includegraphics[width=0.8\textwidth]{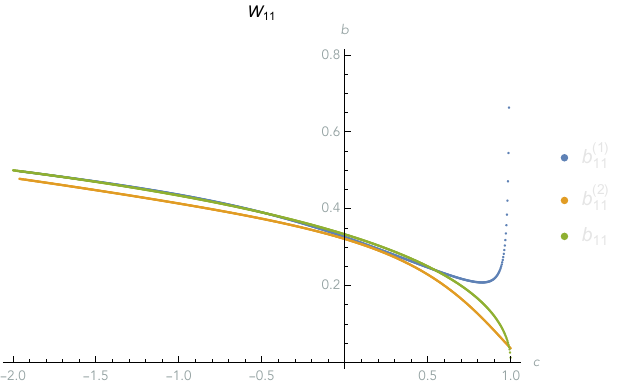}
\caption{The two measurements of $b_{11}$ compared with the expected value.}
\label{W11bs}
\end{figure}

We predicted in Eq.~\eqref{eq:b_11} that
\begin{equation}
b_{11} = \frac{1}{x+1},
\end{equation}
now written in terms of $x$. The comparison of this value with the extrapolations is given in Figure \ref{W11bs}.

\subsection{Emerging Jordan blocks in $\mathscr W_{21}$ at generic $c$}
In $\mathscr W_{21}$ there are four fields that have $h + \overbar h = 2h_{1,-2}$. For generic $c$ and finite $L$, these fields correspond to two non-degenerate eigenvalues (singlets; lines $i_{13} = 24$ and $35$ of Table \ref{W2_singlets}), which we denote by $\mu$ and $\nu$, and one doubly degenerate eigenvalue (line $i_{13} = 25$ of Table \ref{W2_singlets}). At generic $c$, the two singlets are expected to form a Jordan block in the $L\to\infty$ limit. The doublet occurs because of a left-right symmetry (Section \ref{parity}), and is not expected to be involved in the Jordan-block structure.

\begin{figure}
\centering
\includegraphics[width=0.8\textwidth]{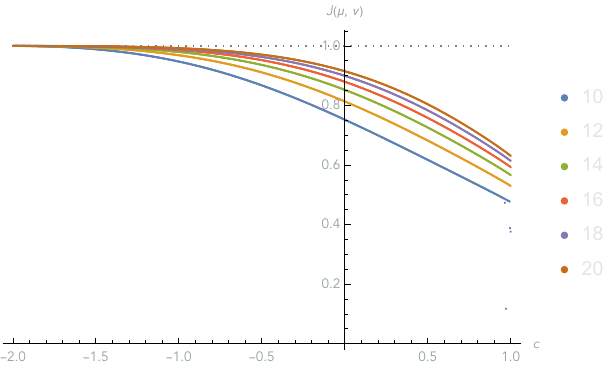}
\caption{The formation of a Jordan block between two singlets in $\mathscr W_{21}$. The legend indicates the value of $N = 2L$.}
\label{W21J4}
\end{figure}

Using the $J$ measure, we can directly observe the emerging Jordan block involving $\mu$ and $\nu$ (Figure \ref{W21J4}). As with the fields $\alpha$ and $\beta$ in $\mathscr W_{11}$, the plot correctly suggests that there is a Jordan block at finite size for $c = -2$.

\begin{figure}
\centering
\includegraphics[width=0.8\textwidth]{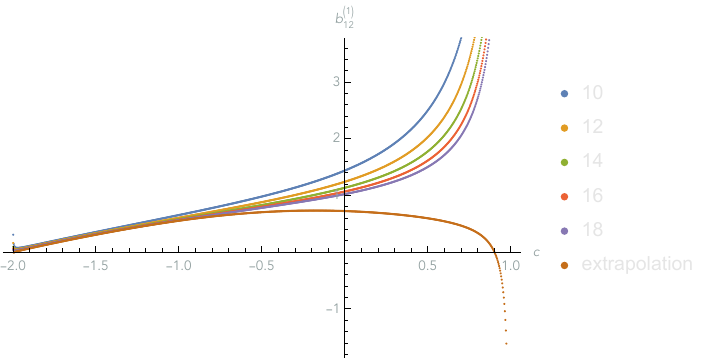}
\caption{The measurement of $b_{12}^{(1)}$ between $\mu$ and $\nu$.}
\label{W21b1}
\end{figure}

\begin{figure}
\centering
\includegraphics[width=0.8\textwidth]{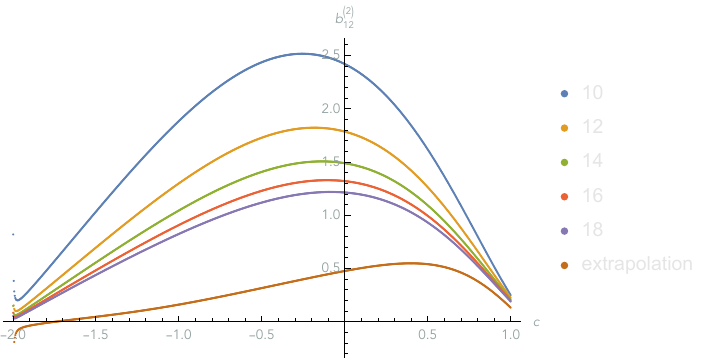}
\caption{The measurement of $b_{12}^{(2)}$ between $\mu$ and $\nu$.}
\label{W21b2}
\end{figure}

We again obtain estimates for the parameter $b_{12}$ that characterizes the Jordan block involving $\mu$ and $\nu$. Both $\mu$ and $\nu$ converge to the same vector as $L\to\infty$, the field $\Psi_{12}$ in Eq.~\eqref{eq:indecomposable_12}. The field $\Phi_{12}$ is realized on the lattice as the lowest field of momentum $2$ (line $i_{13} = 3$ in Table \ref{W2_singlets}), which we normalize to $\ip*{\Phi_{12}} = 1$. Orthogonalize $\mu$ and $\nu$ by subtracting from $\nu$ its component along $\mu$, resulting in $\tilde\nu$, and normalize it so that $(\mu|H_0|\tilde\nu) = 2$. Then we have
\begin{equation}
b^{(1)}_{12}(L,c) = \frac{|\mel*{\tilde\nu}{A}{\Phi_{12}}|^2}{\ip*{\tilde\nu}{\mu}},
\end{equation}
where
\begin{equation}
A = H_{-2} - \frac{3}{2(2h_{12} + 1)}H_{-1}^2.
\end{equation}
We similarly define $b^{(2)}_{12}$ by exchanging the roles of $\mu$ and $\nu$ in this procedure. The two measurements of $b_{12}$ are given graphically in Figures \ref{W21b1} and \ref{W21b2}.


\begin{figure}
\centering
\includegraphics[width=0.8\textwidth]{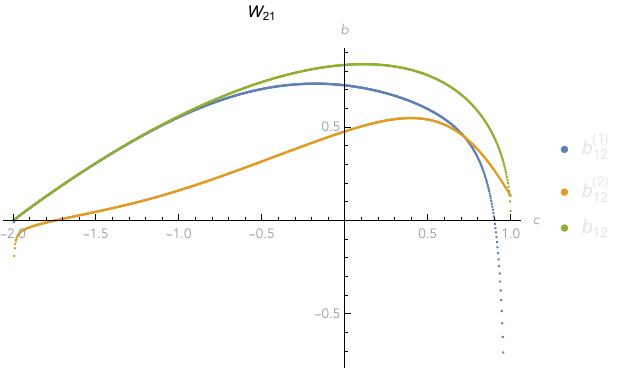}
\caption{The two measurements of $b_{12}$ compared with the expected value.}
\label{W21bs}
\end{figure}

From Eq.~\eqref{eq:b_12} we have
\begin{equation}
b_{12} = \frac{4}{x+1} - \frac{2}{x^2}.
\end{equation}
The comparison of this value with the two extrapolated measurements is shown in Figure \ref{W21bs}.

\sectionbreak

Particularly with the measured values of $b_{11}^{(1)}$ and $b_{12}^{(1)}$, we see excellent agreement with the theoretical $b$ values in Eqs.~\eqref{eq:b_11} and \eqref{eq:b_12}. As these values of $b$ were derived from the diamond structure of Eq.~\eqref{eq:indecomposable_diamond} in Section \ref{OPE}, these measurements almost certainly establish the correctness of Conjecture \ref{loop_j}. Admittedly, the measurement is not as compelling near $c = 1$, and the values of $b_{11}^{(1)}$ and $b_{12}^{(1)}$ differ significantly from $b_{11}^{(2)}$ and $b_{12}^{(2)}$, even though they should give the same values in the limit. On the former point, the emergence of the Jordan block is much slower near $c = 1$, as seen in Figures \ref{W11J2} and \ref{W21J4}, and this may be remedied once we have access to larger lattice sizes. On the latter point, the equivalence of the two measurements followed from the fact that in the limit $L \to \infty$, the two eigenvectors eventually became equal. But at finite size, one of the two may be more equal than the other, as we now describe.

In Section \ref{further_study}, it was remarked that in each example with a continuous parameter, one of the two vectors that formed an emerging Jordan block was already equal to the proper eigenvector, even before the limit was taken. In the case at hand, we studied the limit $L\to\infty$ and there is no sense in which an eigenvector can be considered independent of $L$, as the dimension of the vector strongly depends on $L$. But if there is a sense in which a similar idea can be made precise---that $\alpha$ and $\mu$ are less strongly dependent on $L$ than $\beta$ and $\nu$---it would favor the former fields over the latter and break the symmetry of treating each pair on an equal footing. The ``strength'' of the dependence on $L$ could very well vary with $c$, a concept we explore in another context in Chapter \ref{mixing}. This could explain why $b_{11}^{(1)}$ and $b_{12}^{(1)}$ are more successful measurements of $b_{11}$ and $b_{12}$ than $b_{11}^{(2)}$ and $b_{12}^{(2)}$ for a wide range of $c$ values, with their degrees of success trading places near $c = 1$.

\section{The limit $c \to 0$: Jordan blocks between different modules} \label{c_0}
\subsection{Glued cell representations}

We previously described cell representations of the Jones--Temperley--Lieb algebra, the standard modules $\mathscr W_{j,z^2}$ and $\overbar{\mathscr W}_{\!\!0,\mathfrak q^{\pm 2}}$. The salient feature of the modules $\mathscr W_{j,z^2}$ was that the number of through-lines was kept fixed: whenever the action of a generator $e_i$ resulted in a state with fewer through-lines, the result was annihilated. One may equally well consider a representation consisting of a collection of standard modules with a range of consecutive values of $j$. The underlying vector space is simply the vector-space direct sum of the underlying vector spaces of the involved modules, and the basis is simply the union of their bases. The generators $e_i$ act naturally on the basis states, and the number of through-lines is allowed to decrease (except possibly for the module with the lowest value of $j$, which is usually 0). The new representation is thus not a module direct sum, so it is indecomposable. But it is not irreducible, since the standard module with the lowest value of $j$ is an invariant submodule. In this section we consider the representation $\overbar{\mathscr W}_{\!\!0,\mathfrak q^{\pm 2}} + \mathscr W_{11} + \mathscr W_{21}$---the primary ``glued'' module or representation of this section.

Recall the rules previously set out for determining the loop scalar product between link states:
\begin{quote}
First, unless all through-lines connect from top to bottom the result is zero. We also take into account the weight of straightening the connected through-lines, using the idea of the graduated phase: a through-line that has moved to the right (left) is assigned the weight $\e^{\i\phi/2N}$ ($\e^{-\i\phi/2N}$) for each step. Each contractible loop carries the weight $m = \mathfrak q + \mathfrak q^{-1}$, while each non-contractible loop carries the weight $\e^{\i\phi/2} + \e^{-\i\phi/2}$.
\end{quote}

In the glued representations, we define an inner product that is the same as in the standard-module case, except we drop the requirement that through-lines connect. Importantly, the ex ante inner product of any two states, even those with different numbers of through-lines, is nonzero---if the diagram produced by gluing two link states produces no loops and movement of through-lines, its inner product is $m^0 (\e^{\i\phi/2N})^0 = 1$. Under these rules the inner product of the first example $\braket*{(12)(3)(4)}{(1)(2)(34)}$ in Eq.~\eqref{eq:ip_examples} likewise becomes $1$ instead of $0$. The JTL generators $e_i$ remain self-adjoint with respect to this new inner product.

\subsection{Absence or presence of Jordan blocks at finite size} \label{finite_Jordan}

Strictly at $c = 0$, it so happens that there are no Jordan blocks on the lattice for the dense loop model Hamiltonian at $m = 1$, even in the glued module $\overbar{\mathscr W}_{\!\!0,\mathfrak q^{\pm 2}} + \mathscr W_{11} + \mathscr W_{21}$. This is a somewhat surprising fact since we know from representation theory that the JTL algebra is not semisimple at that point, and that the modules $\overbar{\mathscr W}_{\!\!0,\mathfrak q^{\pm 2}}$ are glued by the action of generic elements of the algebra (Section \ref{root_of_unity}). Furthermore, other representations, such as the $s\ell(2|1)$ spin chain, do exhibit the expected Jordan blocks. Nonetheless, the fact remains that the Hamiltonian itself is fully diagonalizable, and that, for instance, the eigenvalues corresponding to the conformal states $T$ and its logarithmic partner $t$ are degenerate without being part of a Jordan block. This fact was actually observed previously by \textcite{DJS2010c} for the case of open boundary conditions, though it was not clear at the time whether this was a bizarre effect due to the small sizes studied. However, the situation does not seem to change when exploring larger sizes, and it seems to be a definite feature of the model.

In the preceding work as well as in subsequent studies, this problem was circumvented by introducing a parameter $y$ that did not change the (generalized) eigenvalues of the Hamiltonian but allowed the Jordan blocks at $c = 0$ to ``reappear'' somewhat miraculously (Appendix \ref{y_parameter}). The meaning of this parameter was not very clear, except for the fact that it broke the symmetry of translation by one site---a somewhat pleasant feature, since the dense loop model is naturally described using an underlying oriented lattice that is only compatible with translation invariance by two sites. With $y \ne 1$ for open boundary conditions ($y \ne \pm 1$ for periodic boundary conditions) the measured values of the logarithmic couplings known at the time of those studies for $c = 0$ were found to be independent of $y$ and in excellent agreement with theoretical expectations. Some illustrative calculations are carried out analytically, in Appendix \ref{y_calculations}, to show that the absence of a Jordan block for $y = \pm 1$ is no accident.

The parameter $y$ only affects matrix elements between different modules. As long as one restricts to a single standard module (e.g., $\mathscr W_{11}$ or $\mathscr W_{21}$), the values of the overlaps $J$ or the parameters $b_{11}$ and $b_{12}$ studied in this chapter for generic $c$ are unaffected by the introduction of $y$---and this is true even at $c = 0$. However, it turns out that, if we measure the same quantities by embedding these in the glued module, the values of $J$, $b_{11}$, and $b_{12}$ depend on $y$ in a nontrivial, most unsatisfactory way: they encounter seemingly-random divergences and sign changes when considered as functions of $c$, and these divergences and sign flips change positions as $y$ varies. This occurs despite the fact that the JTL algebra is simple for $c \ne 0$, so a change of basis provides modules isomorphic to $\mathscr W_{11}$ or $\mathscr W_{21}$. The only strategy to recover results which are compatible with conformal invariance seems to be to consider the limit $y \to 1$ at finite $L$, and then extrapolate to the thermodynamic limit, if it can be done. It is fair to ask why one would have to go through these gymnastics, and what they mean.

We now observe that the JTL algebra has a symmetry under $e_j \to -e_j$ and $m\to -m$. We can therefore study the loop model with opposite signs for the Hamiltonian and loop fugacity (as well as the ground-state energy density $e_\infty$). While it is widely expected that these two descriptions are equivalent, they do differ right at $m = 1$ ($c = 0$). To see this, consider the simple example of $L = 2$ ($N = 4$). Using the basis $\{(12)(34),(23)(14),(2)(3)(41),(3)(4)(12),(1)(4)(23),(1)(2)(34),(1)(2)(3)(4)\}$ (the notation is defined in Section \ref{standard}), we have, excluding the prefactor $v_{\text F}$ in Eq.~\eqref{eq:Potts_Hamiltonian},
\begin{equation}
H(m,e_\infty) = \begin{pmatrix} \label{eq:loop_H_4}
 4 e_\infty-2m & -2 & 0 & -1 & 0 & -1 & 0 \\
 -2 & 4 e_\infty-2m & -1 & 0 & -1 & 0 & 0 \\
 0 & 0 & 4 e_\infty-m & -1 & 0 & -1 & -1 \\
 0 & 0 & -1 & 4 e_\infty-m & -1 & 0 & -1 \\
 0 & 0 & 0 & -1 & 4 e_\infty-m & -1 & -1 \\
 0 & 0 & -1 & 0 & -1 & 4 e_\infty-m & -1 \\
 0 & 0 & 0 & 0 & 0 & 0 & 4 e_\infty
\end{pmatrix}.
\end{equation}
Both 
\begin{equation}
H(1,1) = \begin{pmatrix}
 2 & -2 & 0 & -1 & 0 & -1 & 0 \\
 -2 & 2 & -1 & 0 & -1 & 0 & 0 \\
 0 & 0 & 3 & -1 & 0 & -1 & -1 \\
 0 & 0 & -1 & 3 & -1 & 0 & -1 \\
 0 & 0 & 0 & -1 & 3 & -1 & -1 \\
 0 & 0 & -1 & 0 & -1 & 3 & -1 \\
 0 & 0 & 0 & 0 & 0 & 0 & 4
\end{pmatrix}
\end{equation} and 
\begin{equation}
-H(-1,-1) = \begin{pmatrix}
 2 & 2 & 0 & 1 & 0 & 1 & 0 \\
 2 & 2 & 1 & 0 & 1 & 0 & 0 \\
 0 & 0 & 3 & 1 & 0 & 1 & 1 \\
 0 & 0 & 1 & 3 & 1 & 0 & 1 \\
 0 & 0 & 0 & 1 & 3 & 1 & 1 \\
 0 & 0 & 1 & 0 & 1 & 3 & 1 \\
 0 & 0 & 0 & 0 & 0 & 0 & 4 
\end{pmatrix}
\end{equation} have eigenvalues $\{\lambda_1,\ldots,\lambda_7\} = \{0,1,3,3,4,4,5\}$. It turns out however that $H(1,1)$ is diagonalizable with the 7 corresponding geometric eigenvectors
\begin{subequations}
\begin{gather}
u_1 = (1, 1, 0, 0, 0, 0, 0), \\
u_2 = (2, 2, -1, -1, -1, -1, 0), \\
u_3 = (0, 0, 1, 0, -1, 0, 0), \\
u_4 = (0, 0, 0, 1, 0, -1, 0), \\
u_5 = (1, -1, 0, 0, 0, 0, 0), \\
u_6 = (1, 1, -2, -2, -2, -2, 6), \\
u_7 = (2, -2, 1, -1, 1, -1, 0),
\end{gather}
\end{subequations}
but $-H(-1,-1)$ is not. It has an identical spectrum, but its generalized eigenbasis in Jordan canonical form is
\begin{subequations} \label{eq:eigenvectors_m_-1}
\begin{gather}
v_1 = (1, -1, 0, 0, 0, 0, 0), \\
v_2 = (2, -2, -1, 1, -1, 1, 0), \\
v_3 = (0, 0, 0, 1, 0, -1, 0), \\
v_4 = (0, 0, 1, 0, -1, 0, 0), \\
v_5 = (1, 1, 0, 0, 0, 0, 0), \\
\tilde v_6 = \left(0, 0, \frac{1}{2}, \frac{1}{2}, \frac{1}{2}, \frac{1}{2}, -\frac{1}{2}\right), \\
v_7 = (2, 2, 1, 1, 1, 1, 0),
\end{gather}
\end{subequations}
where the tilde on $\tilde v_6$ indicates it is not a proper eigenvector.

The reader may be unwilling to believe the astonishing fact that a symmetry of the algebra as a whole leads to such profoundly different behavior. My collaborators and I are also displeased with this state of affairs, although repeated calculation with varied parameters continues to yield this conclusion. On the plus side, we have now recovered nontrivial Jordan blocks without invoking the exogenous parameter $y$. Let us reinforce this observation using the formalism of emerging Jordan blocks. With $m$ left unspecified, we have more generally the corresponding eigenvectors (of $H(m,e_\infty)$ and $-H(-m,-e_\infty)$, where $m > 0$)
\begin{subequations} \label{eq:eigenvectors_generic_m}
\begin{gather}
u_5 = (1, -1, 0, 0, 0, 0, 0), \\
u_6 = (1, 1, -(m+1), -(m+1), -(m+1), -(m+1), (m+1)(m+2)), \\
v_5 = (1, 1, 0, 0, 0, 0, 0), \\
v_6 = (1, 1, m-1, m-1, m-1, m-1, (m-1)(m-2)).
\end{gather}
\end{subequations}
One can immediately see that $J(u_5, u_6) = 0$, while $\lim_{m\to 1}J(v_5, v_6) = 1$. More generally,
\begin{equation}
J(v_5, v_6) = \sqrt{\frac{2}{2 + 4(m-1)^2 + (m-1)^2(m-2)^2}}.
\end{equation}
Furthermore, $\tilde v_6$ is obtained by orthogonalizing $v_6$ against $v_5$, then rescaling the result in the limit $m \to 1$. Finally, we illustrate the concept of the connection between the rate of convergence and the Jordan coupling. For generic $m$, the eigenvalues are $\{\lambda_5,\lambda_6\} = \{4e_\infty, 4e_\infty + 2 - 2m\}$. Their difference is $\lambda_6 - \lambda_5 = 2 - 2m$. If we \emph{orthonormalize} $v_6$ against $v_5$ instead, we obtain
\begin{subequations}
\begin{gather}
\hat v_5 = \frac{1}{\sqrt 2}(1,1,0,0,0,0,0), \\
v_6' = \frac{\sgn(m-1)}{\sqrt{m^2 - 4m + 8}}(0,0,1,1,1,1,m-2).
\end{gather}
\end{subequations}
We then have
\begin{equation}
\lim_{m\to 1}v_6' = \pm\frac{1}{\sqrt 5}(0,0,1,1,1,1,-1) = \pm\frac{1}{\sqrt 2}\times\sqrt{\frac85}\,\tilde v_6.
\end{equation}
The first prefactor $1/\sqrt 2$ comes from the normalization of $v_5$. Otherwise, $v_6'$ differs from $\tilde v_6$ by a constant, which must be the Jordan coupling. As in Eq.~\eqref{eq:rank_2_coupling}, one may compute
\begin{equation} \label{eq:Jordan_coupling_56}
\left|\lim_{m\to 1}\frac{(\lambda_6 - \lambda_5)J(v_5,v_6)}{\sqrt{1 - J(v_5,v_6)^2}}\right| = \sqrt{\frac85}
\end{equation}
(the absolute value is taken to remove the sign ambiguity present also in Section \ref{coupling}). Therefore, if we scale $v_6'$ down by this factor, adjust its sign if necessary, and then rescale both $\hat v_5$ and $v_6'$ by $1/\sqrt 2$, we obtain $v_5$ and $\tilde v_6$---the basis of the Jordan canonical form.

\subsection{Measurement of $b$}

Although we now have real, rather than emerging, Jordan blocks, it is useful to consider the measurement of $b$ using emerging Jordan blocks anyway, because we already have measurements of $b$ itself using the Jordan blocks \cite{VGJS2012}, and we would like to see agreement between the two methods. The procedure of Section \ref{b_emerging} reads as follows. For fixed $L$, identify an eigenvector of $H$ in the representation $\overbar{\mathscr W}_{\!\!0,\mathfrak q^{\pm 2}} + \mathscr W_{11} + \mathscr W_{21}$ corresponding to the stress--energy tensor $T$. (Actually, it is known that $T$ lives in $\overbar{\mathscr W}_{\!\!0,\mathfrak q^{\pm 2}}$, and since the action of JTL can only decrease the first index $j$, it suffices to find $T$ in $\overbar{\mathscr W}_{\!\!0,\mathfrak q^{\pm 2}}$ and embed it in the full glued module.) Find an eigenvector $T'$ that becomes degenerate and parallel with $T$ at $c = 0$, but is otherwise distinct---it has nonzero components in $\mathscr W_{21}$. Define $t$ via the Gram--Schmidt process and normalize it as specified. Then
\begin{equation}
b(L) = \lim_{c\to 0} \frac{|\mel*{t}{H_{-2}}{I}|^2}{\ip*{t}{T}\ip*{I}{I}},
\end{equation}
where $I$ is the ground state (fully contained in $\overbar{\mathscr W}_{\!\!0,\mathfrak q^{\pm 2}}$) and $A = L_{-2}$ as $T = L_{-2}I$.

The last expression requires some explanation. The modifications $e_j \to -e_j$, $m \to -m$, and $e_\infty \to -e_\infty$ must be propagated throughout the entire process. This includes not only the Hamiltonian but also the inner product and Koo--Saleur generators. When we do this, it is found that $\ip*{I} = (-1)^L$ upon normalization---i.e., it is negative for odd $L$. The sign of the norm square is invariant and cannot be changed by rescaling, even by a complex constant. However, because CFT demands that primary fields have unit norm square, particularly the ground state, in these cases we must enforce this by manually adding in the sign, and declaring this to be the correct inner product. This is the meaning of the appearance of $\ip*{I}$ in the denominator.

The results of this process are shown in Table \ref{b_W012}.
\begin{table}[h]
\centering
\begin{tabular}{cc}
\toprule
$N = 2L$ & $b$ \\
\midrule
4 & $-1.96028$ \\
6 & $-3.24046$ \\
8 & $-3.94952$ \\
10 & $-4.33295$ \\
12 & $-4.55079$ \\
14 & $-4.68234$ \\
16 & $-4.76633$ \\
18 & $-4.82253$ \\
\midrule
conjectured & $-5$ \\
\bottomrule
\end{tabular}
\caption{Different measurements for $b = \ip*{t}{T}$ in $\overbar{\mathscr W}_{\!\!0,\mathfrak q^{\pm 2}} + \mathscr W_{11} + \mathscr W_{21}$.}
\label{b_W012}
\end{table}

In this case, when the roles of $T$ and $T'$ are exchanged, the same value of $b(L)$ obtains, unlike what is observed in Section \ref{b_emerging}. The two measurements match here because the limit $c \to 0$ is actually reached, rather than extrapolated as must be the case for the limit $L \to \infty$. Furthermore, the measurements match those reported in \textcite{VGJS2012}. The crucial difference is that we were able to compute the same quantities without ever calculating a Jordan form, a demanding task (Chapter \ref{computational}). We needed only to diagonalize the Hamiltonian and take limits; the emerging Jordan block found itself for us.

Because exact diagonalization is also algebraically much less daunting a task than finding a Jordan canonical form, we are able to calculate closed form expressions for $b(L)$ for sufficiently small $L$. For $L = 2$, $T$ and $T'$ are $v_5$ and $v_6$ of Eq.~\eqref{eq:eigenvectors_generic_m}, and $I$ is $v_1$ of Eq.~\eqref{eq:eigenvectors_m_-1}, but normalized to $\ip*{I} = 1$. This leads to two measurements of $b$:
\begin{subequations}
\begin{gather}
b^{(1)} = -\frac{8(m+1)}{\pi v_{\text F}}, \qq{and} \\
b^{(2)} = \frac{4m(m+1)(m^4 + 2m^3 + m^2 - 12m + 10)^2}{\pi v_{\text F} (m^3 + 7m^2 + 6m - 16)}.
\end{gather}
\end{subequations}
When $m \to 1$, $v_{\text F} \to 3\sqrt 3/2$, and $b^{(1)}, b^{(2)} \to -32\sqrt 3/9\pi \approx -1.96028$, precisely the measured value. An analogous calculation is possible for $L = 3$. One value of $b$ is
\begin{equation}
b^{(1)} = -\frac{3}{\pi v_{\text F}}\frac{4 m^2 - \frac{2m (2 m^2-9)}{\sqrt{m^2+48}}-6}{m-1}.
\end{equation}
The expression for $b^{(2)}$ is much too complicated to be worth showing here, and is not particularly illuminating. (A related convoluted expression does appear in Appendix \ref{y_calculations}.) However, for $m \to 1$ both tend to $b = -288\sqrt 3/49\pi \approx -3.24026$, again the measured value. (Despite the denominator in the expression for $b^{(1)}$, it is a removable singularity, and yields a finite value as $m \to 1$.)

It seems plausible that the finite-size $b$ numbers have the form $b(L) = -q(L)\sqrt 3/\pi$, where $q(L) \in \mathbb Q$. If this sequence could be found, one could rigorously extract the conjectured limit $b = -5$. This would also further validate the lattice approach.

\chapter{The periodic alternating $s\ell(2|1)$ superspin chain and its scaling-weak convergence to a logarithmic conformal field theory} \label{sl21_chapter}

The $s\ell(2|1)$ superspin chain (Section \ref{sl21_section}) is only defined for the JTL parameter $m = 1$. There is no known modification that allows for representations for other values of $m$. In particular, we cannot use the framework of emerging Jordan blocks to study the Jordan blocks observed at finite size, since there is no way to take the limit $m \to 1$. However, precisely because Jordan blocks proliferate in this model even at finite size, this model will serve well as a testing ground for the dual Jordan projection operators of Chapter \ref{biorthogonal}.

\section{Identification of fields and Jordan structure at finite size}

The identification of fields proceeds as described in Section \ref{field_identification}. This task may seem to be much more difficult, because the Hilbert space of the $s\ell(2|1)$ chain at any given size $L$ is much larger than that of the corresponding loop model. However, since the $s\ell(2|1)$ representation and the loop model representation share the JTL standard modules $\mathscr W_{j,\e^{\i\phi}}$ (for $m = 1$, $\mathfrak q = \e^{\i\pi/3}$), we may transfer the field identifications from the loop model to $s\ell(2|1)$ by matching the spectrum. However, we must use the $s\ell(2|1)$ model itself to determine the Jordan block structure. Examples of this field and structure identification for $N = 10$ sites is shown in Table \ref{sl21_table_10}. (The identification of the continuum scaling fields was performed by my collaborator Jesper Jacobsen, following \textcite[Appendix A.5]{JacobsenSaleur2019}.) The Jordan form of the Hamiltonian was obtained using the K\aa gstr\"om--Ruhe algorithm (Section \ref{jordan_algorithm}).
\begin{table}
\centering
{\small\begin{tabular}{cccccccc}
\toprule 
$i$ & $\epsilon$ & $n_i$ & $d_i$ & \makecell{Jordan \\ structure} & module(s) & $h + \overbar h$ & \makecell{scaling field(s)} \\
\midrule
1 & $0$ & 1 & 1 & & $\overbar{\mathscr W}_{\!\!0,\mathfrak q^{\pm 2}}$ & $0$ & $I$ \\ 
2 & $0.248568$ & 2 & 3 & & $\mathscr W_{11}$ & $1/4,5$ & $\Phi_{01},(0,4)\Phi_{11},(4,0)\Phi_{-1,1}$ \\ 
\cmidrule(lr){1-8}
3 & $1.18586$ & 4 & 7 & $2$ & $\overbar{\mathscr W}_{\!\!0,\mathfrak q^{\pm 2}},\mathscr W_{21}$ & $5/4,5$ & \makecell{$\phi_{21}\otimes\overbar\phi_{21},(5,0)I,(0,5)I,$ \\ $\Phi_{02},(0,3)\Phi_{12},(3,0)\Phi_{-1,2}$} \\ 
\cmidrule(lr){1-8}
4 & $1.69315$ & 2 & 9 & & $\mathscr W_{11}$ & $2,21/4$ & $(1,0)\Phi_{11},(5,0)\Phi_{01},(0,5)\Phi_{01}$ \\ 
5 & $2.00969$ & 2 & 11 & & $\mathscr W_{11}$ & $2$ & $(0,1)\Phi_{-1,1}$ \\ 
6 & $2.01198$ & 2 & 13 & & $\mathscr W_{11}$ & $9/4,5$ & $(1,1)\Phi_{01},(0,4)\Phi_{11},(4,0)\Phi_{-1,1}$ \\ 
\cmidrule(lr){1-8}
7 & $2.16868$ & 4 & 17 & & $\mathscr W_{21}$ & $39/16,79/16$ & \makecell{$(1,0)\Phi_{1/2,2},(0,1)\Phi_{-1/2,2},$ \\ $(0,2)\Phi_{3/2,2},(2,0)\Phi_{-3/2,2}$} \\ 
\cmidrule(lr){1-8}
8 & $2.58777$ & 8 & 25 & & $\mathscr W_{3,\mathfrak q^{\pm 2}}$ & $35/12$ & $\Phi_{03}$ \\ 
\cmidrule(lr){1-8}
9 & $2.63744$ & 4 & 29 & $2$ & $\overbar{\mathscr W}_{\!\!0,\mathfrak q^{\pm 2}},\mathscr W_{21}$ & $13/4,5$ & \makecell{$(1,1)\phi_{21}\otimes\overbar\phi_{21},$ \\ $(5,0)I,(0,5)I,(1,1)\Phi_{02}$} \\ 
\cmidrule(lr){1-8}
10 & $2.91833$ & 2 & 31 & & $\mathscr W_{11}$ & $4$ & \\ 
11 & $3.05347$ & 2 & 33 & & $\mathscr W_{11}$ & $17/4$ & $(2,2)\Phi_{01}$ \\ 
\cmidrule(lr){1-8}
12 & $3.17526$ & 4 & 37 & $2$ & $\overbar{\mathscr W}_{\!\!0,\mathfrak q^{\pm 2}},\mathscr W_{21}$ & $4,21/4$ & \makecell{$\phi_{31}\otimes\overbar\phi_{31},(2,2)\phi_{21}\otimes\overbar\phi_{21},$ \\ $(2,0)\Phi_{12},(0,2)\Phi_{-1,2}$} \\ 
\cmidrule(lr){1-8}
13 & $3.19435$ & 4 & 41 & & $\mathscr W_{21}$ & $71/16$ & $(2,1)\Phi_{1/2,2},(1,2)\Phi_{-1/2,2}$ \\ 
14 & $3.23433$ & 2 & 43 & & $\mathscr W_{11}$ & $4$ & \\ 
\cmidrule(lr){1-8}
15 & $3.24694$ & 24 & 67 & $8\times2$ & $\mathscr W_{11},\mathscr W_{21},\mathscr W_{3,\mathfrak q^{\pm 2}}$ & $4,21/4$ & \makecell{$(2,1)\Phi_{11},(1,2)\Phi_{-1,1},$ \\ $(2,0)\Phi_{12},(0,2)\Phi_{-1,2},$ \\ $(1,0)\Phi_{1/3,3},(0,1)\Phi_{-1/3,3},$ \\ $(0,1)\Phi_{4/3,3},(1,0)\Phi_{-4/3,3}$} \\ 
\cmidrule(lr){1-8}
16 & $3.29693$ & 4 & 71 & & $\mathscr W_{11}$ & $17/4$ & $(2,2)\Phi_{01}$ \\ 
17 & $3.50028$ & 2 & 73 & & $\mathscr W_{11}$ & $17/4$ & $(2,2)\Phi_{01}$ \\ 
18 & $3.52212$ & 4 & 77 & & $\mathscr W_{21}$ & $71/16$ & $(2,1)\Phi_{1/2,2},(1,2)\Phi_{-1/2,2}$ \\ 
19 & $3.59093$ & 8 & 85 & & $\mathscr W_{3,\mathfrak q^{\pm 2}}$ & $59/12$ & $(1,1)\Phi_{03}$ \\ 
20 & $3.67553$ & 4 & 89 & $2$ & $\overbar{\mathscr W}_{\!\!0,\mathfrak q^{\pm 2}},\mathscr W_{21}$ & $4$ & $(2,2)I,(2,0)\Phi_{12},(0,2)\Phi_{-1,2}$ \\ 
21 & $3.74941$ & 24 & 113 & $8\times2$ & $\mathscr W_{11},\mathscr W_{21},\mathscr W_{3,\mathfrak q^{\pm 2}}$ & $5,21/4$ & $\Phi_{-5/3,3},\Phi_{5/3,3}$ \\ 
22 & $3.90951$ & 4 & 117 & & $\mathscr W_{11}$ & & \\ 
23 & $4.00967$ & 4 & 121 & $2$ & $\overbar{\mathscr W}_{\!\!0,\mathfrak q^{\pm 2}},\mathscr W_{21}$ & $6$ & $(3,3)I$ \\ 
24 & $4.05177$ & 4 & 125 & & $\mathscr W_{21}$ & & \\
\bottomrule
\end{tabular}}
\caption{Jordan structure of the lowest 125 of 711 eigenvalues of $H_0$ on $N = 10$ sites, in the vacuum sector at momentum 0. $n_i$ is the algebraic multiplicity of the eigenvalue on line $i$. $d_i = \sum_{j=1}^i n_j$ is the running dimension. In the ``Jordan structure'' column, $m\times n$ means $m$ rank-$n$ Jordan blocks appear for that eigenvalue, and $n \equiv 1\times n$. $(m,n)\Phi$ means a level-$(m,n)$ descendant of $\Phi$.} 
\label{sl21_table_10}
\end{table}

The analysis of the $c = 0$ logarithmic CFT makes a clear prediction that the stress--energy tensor $T$ and its logarithmic partner $t$ are found in a diamond structure much like Eq.~\eqref{eq:indecomposable_rs}, with an associated $b = -5$. Using the field--vector correspondence, we are able to identify the lattice analogue of the $T$ and $t$. We may obtain a measurement of $b$ following the procedure of Section \ref{b_emerging}, except we can start from step 2 as there is already a Jordan block. The results are again identical to those in Table \ref{b_W012}. We then renormalize $T$ and $t$ by jointly rescaling them by the same constant---$T\to zT$ and $t\to zT$---so that $\braket*{t}{T}$ matches the finite-size measured value of $b$. From the diamond structure, one then has \cite{VasseurOPE}
\begin{equation}
\overbar{A}^\dagger \overbar{A} t = bT,
\end{equation}
with
\begin{subequations}
\begin{gather}
\overbar{A} = \overbar L_{-2} - \frac{3}{2} \overbar L_{-1}^2, \\
\overbar{A}^\dagger = \overbar L_{2} - \frac{3}{2} \overbar L_{1}^2.
\end{gather}
\end{subequations}
It follows that
\begin{subequations}
\begin{gather}
\mel*{T}{\overbar{A}^\dagger \overbar{A}}{t} = 0, \\
\mel*{t}{\overbar{A}^\dagger \overbar{A}}{t} = b^2.
\end{gather}
\end{subequations}
These values on some lattice sizes are reported in Table \ref{b_test_1}.

\begin{table}
\centering
\begin{tabular}{cll}
\toprule
$L$ & $(\mel*{t}{\overbar{A}^\dagger \overbar{A}}{t})^{1/2}$ & $\mel*{T}{\overbar{A}^\dagger \overbar{A}}{t}$ \\
\midrule
10 & $3.94536$ & $-0.00856632$ \\
12 & $4.24481$ & $-0.0237170$ \\
14 & $4.43279$ & $-0.0275028$ \\
16 & $4.55716$ & $-0.0262987$ \\
\midrule
conjectured & 5 & 0 \\
\bottomrule
\end{tabular}
\caption{Tests of $\mel*{T}{\overbar{A}^\dagger \overbar{A}}{t} = 0$ and $\mel*{t}{\overbar{A}^\dagger \overbar{A}}{t} = b^2$.}
\label{b_test_1}
\end{table}

Another check comes from \textcite{Ridout2012}. Since
\begin{equation}
\overbar L_1 \overbar L_{-1} t = 2\overbar L_0 t = 2T,
\end{equation}
it follows that
\begin{subequations}
\begin{gather}
\mel*{T}{\overbar L_1 \overbar L_{-1}}{t} = 0, \\
\mel*{t}{\overbar L_1 \overbar L_{-1}}{t} = 2b.
\end{gather}
\end{subequations}
The validation of these identities on the lattice are reported in Table \ref{b_test_2}.

\begin{table}
\centering
\begin{tabular}{cll}
\toprule
$L$ & $\mel*{t}{\overbar L_1 \overbar L_{-1}}{t}$ & $\mel*{T}{\overbar L_1 \overbar L_{-1}}{t}$ \\
\midrule
10 & $-6.34905$ & $0.00287950$ \\
12 & $-7.25285$ & $0.00285605$ \\
14 & $-7.88091$ & $0.00261639$ \\
16 & $-8.32548$ & $0.00233339$ \\
\midrule
conjectured & $-10$ & 0 \\
\bottomrule
\end{tabular}
\caption{Tests of $\mel*{T}{\overbar L_1 \overbar L_{-1}}{t} = 0$ and $\mel*{t}{\overbar L_1 \overbar L_{-1}}{t} = 2b$.}
\label{b_test_2}
\end{table}

One reasonable objection to the preceding tests is that measurements involving the supersymmetric inner product don't necessarily imply convergence or equalities of the underlying fields; since the inner product is indefinite, lots of independent random fluctuations could be hidden by the fact that they tend to cancel out. So, for instance, to test $\overbar{A}^\dagger \overbar{A} t = bT$, we should see if $\|\overbar{A}^\dagger \overbar{A} t - bT\| = 0$, with $\|\cdot\|$ some positive definite norm. It will become apparent that the measurements using positive definite norms don't converge as nicely as those using the supersymmetric inner product, if at all. A possible solution to this is to find restricted subspaces of interest of some fixed dimension $d$ (for instance, by working in a subspace with certain conformal dimensions) while increasing the lattice size and measuring quantities there and seeing if there is convergence for every $d$. This would be a notion of convergence even weaker than ``weak convergence''---it is precisely the notion of scaling-weak convergence introduced in Section \ref{scaling_weak}.

\section{Scaling-weak convergence of conformal identities} \label{sl21_projections}

A detailed analysis of the scaling limit of tilting modules, composed of the standard modules, predicts that in the full logarithmic CFT, $A\overbar T = \overbar A T$ and $A\overbar t = \overbar A t$, where $\overbar T$ and $\overbar t$ are the antiholomorphic counterparts to $T$ and $t$ \cite[Figure 2 and Section 7]{Gainutdinov2015}. Because we are able to identify these four fields on the lattice and construct the singular vector operators, we study these identities in the context of scaling-weak convergence, now using the dual Jordan projection operators of Section \ref{dual_Jordan}.

These identities involve descendant fields with zero spin (i.e., $h = \overbar h$). We identify $T$ and $t$ ($\overbar T$ and $\overbar t$) as eigenvectors of the Hamiltonian with momentum $2$ and conformal weights $(2,0)$ (momentum $-2$ and conformal weights $(0,2)$), and bring it to the momentum $0$ sector by applying $\overbar A$ ($A$). To begin applying the projectors, we first find a reduced Hamiltonian in the sector of momentum $0$ at the lowest energy levels, using the implicitly restarted Arnoldi method (Section \ref{Arnoldi}) with dynamic multiplicity adjustment (Appendix \ref{DMA}), which greatly improves convergence. Next, we find a dual Jordan basis of the reduced Hamiltonian, using the K\aa gstr\"om--Ruhe algorithm. We partition the two bases by eigenvalue, and follow the procedure of Section \ref{dual_Jordan} to construct the resolution of the identity in this reduced subspace. In order to apply these projection operators as constructed, one must express the descendant field in the basis of the reduced Hamiltonian, using the partial change of basis found by the implicitly restarted Arnoldi method. We subtract the two and take the 2-norm, with the phase optimization described in Section \ref{evidence_lattice}. Finally, with an exact Jordan block, another complication is that $t$ is not uniquely defined either: there is a degree of freedom $t \to t + \theta T$ for $\theta \in \mathbb C$ (and similarly $\overbar t \to \overbar t + \theta^*\overbar T$). We must therefore measure instead
\begin{equation}
\|A\overbar t - \overbar A t\|_{\munderbar 2} \equiv \inf_{\alpha,\theta} \|A(\overbar t + \theta^*\overbar T) - \e^{\i\alpha}\overbar A(t + \theta T)\|_2 \label{t_optimize}
\end{equation}
to obtain a well-defined measurement (there are probably others but this seems to be the simplest).

Numerical results from this procedure are found in Tables \ref{table_AT}--\ref{table_At_norm}. The second of each pair of tables differs in that the normed difference is divided by the norm of one of the two vectors, to give a relative measure of the deviation from zero. Since $\|\Pi^{(d)}A\overbar T\|_2 = \|\Pi^{(d)}\overbar A T\|_2$ and $\|\Pi^{(d)}A\overbar t\|_2 = \|\Pi^{(d)}\overbar A t\|_2$, the choice is immaterial. For a fixed value of $d$, the values in Tables \ref{table_AT} and \ref{table_AT_norm} appear to decay to zero with increasing $L$ quite convincingly. We thus say the identity $\overbar A T = A \overbar T$ holds in the scaling-weak sense.

\begin{table}
\centering
\begin{tabular}{cccc}
\toprule
$d$ & \multicolumn{3}{c}{$N$} \\
\cmidrule(l){2-4}
 & $10$ & 12 & 14 \\
\midrule
1 & 0* & 0* & 0* \\
3 & 0* & 0* & 0* \\
7 & 0* & 0* & 0* \\
9 & 0* & 0* & 0* \\
11 & 0* & 0* & 0* \\
13 & 0* & 0* & 0* \\
17 & 0* & 0* & 0* \\
25 & 0* & 0* & 0* \\
29 & 0* & 0* & 0* \\
31 & 0* & 0* & 0* \\
33 & 0* & --- & --- \\
35 & --- & 0* & 0* \\
37 & 0* & 0* & --- \\
41 & 0* & --- & --- \\
43 & 0* & --- & --- \\
59 & --- & --- & 0* \\
61 & --- & 0* & 0* \\
63 & --- & 0* & 0* \\
67 & 0* & 0* & 0* \\
71 & 0* & 0* & 0* \\
73 & 0* & 0* & 0* \\
77 & 0* & 0* & $0.129267$ \\
81 & --- & 0* & $0.129267$ \\
85 & 0* & $0.140133$ & $0.129267$ \\
87 & --- & $0.140133$ & --- \\
89 & $0.153817$ & --- & --- \\
93 & --- & --- & $0.129267$ \\
95 & --- & $0.140133$ & $0.129267$ \\
\midrule
no projector & $0.153611$ & $0.181995$ & $0.320142$ \\
\bottomrule
\end{tabular}
\caption{The norm $\|\Pi^{(d)}(A\overbar T - \overbar A T)\|_{\munderbar 2}$ for various projector ranks $d$ and system lengths $N$. In this table ``0*'' means a number that is less than about $2\times 10^{-6}$. I estimate the uncertainty in the nonzero numbers to be about $10^{-6}$. ``{---}'' means that the projector of a given rank $d$ at length $N$ is ill-defined due to degeneracies.}
\label{table_AT}
\end{table}

\begin{table}
\centering
\begin{tabular}{cccc}
\toprule
$d$ & \multicolumn{3}{c}{$N$} \\
\cmidrule(l){2-4}
 & $10$ & 12 & 14 \\
\midrule
1 & * & * & * \\
3 & 0* & 0* & * \\
7 & 0* & 0* & * \\
9 & 0* & 0* & * \\
11 & 0* & 0* & * \\
13 & * & * & * \\
17 & * & * & * \\
25 & * & * & * \\
29 & * & * & * \\
31 & * & * & * \\
33 & * & --- & --- \\
35 & --- & 0* & 0* \\
37 & 0* & 0* & --- \\
41 & 0* & --- & --- \\
43 & 0* & --- & --- \\
59 & --- & --- & 0* \\
61 & --- & 0* & 0* \\
63 & --- & 0* & 0* \\
67 & 0* & 0* & 0* \\
71 & 0* & 0* & 0* \\
73 & 0* & 0* & 0* \\
77 & 0* & 0* & $0.015756$ \\
81 & --- & 0* & $0.015756$ \\
85 & 0* & $0.0220225$ & $0.015756$ \\
87 & --- & $0.0220225$ & --- \\
89 & $0.0315451$ & --- & --- \\
93 & --- & --- & $0.015756$ \\
95 & --- & $0.0220225$ & $0.015756$ \\
\midrule
no projector & $0.0317082$ & $0.0288243$ & $0.0393106$ \\
\bottomrule
\end{tabular}
\caption{The norm $\|\Pi^{(d)}(A\overbar T - \overbar A T)\|_{\munderbar 2}/\|\Pi^{(d)}A\overbar T\|_2$ for various projector ranks $d$ and system lengths $N$. In this table ``*'' means a highly variable number of order 1. They are likely the result of a 0/0 since the corresponding values in Table \ref{table_AT} are so small. ``0*'' means a number that is less than $10^{-6}$. I estimate the uncertainty in the nonzero numbers to be about $10^{-6}$. ``{---}'' means that the projector of a given rank $d$ at length $N$ is ill-defined due to degeneracies.}
\label{table_AT_norm}
\end{table}

\begin{table}
\centering
\begin{tabular}{cccc}
\toprule
$d$ & \multicolumn{3}{c}{$N$} \\
\cmidrule(l){2-4}
 & $10$ & 12 & 14 \\
\midrule
1 & 0* & 0* & 0* \\
3 & 0* & 0* & 0* \\
7 & 0* & 0* & 0* \\
9 & 0* & 0* & 0* \\
11 & 0* & 0* & 0* \\
13 & 0* & 0* & 0* \\
17 & 0* & 0* & 0* \\
25 & 0* & 0* & 0* \\
29 & 0* & 0* & 0* \\
31 & 0* & 0* & 0* \\
33 & 0* & --- & --- \\
35 & --- & $11.052$ & $12.3243$ \\
37 & $9.97002$ & $10.6092$ & --- \\
41 & $9.97002$ & --- & --- \\
43 & $9.97002$ & --- & --- \\
59 & --- & --- & $12.8422$ \\
61 & --- & $11.136$ & $12.8422$ \\
63 & --- & $11.0192$ & $12.8422$ \\
67 & $10.8326$ & $11.4142$ & $12.8422$ \\
71 & $10.8326$ & $11.2607$ & $12.8422$ \\
73 & $10.8326$ & $11.2429$ & $12.8422$ \\
77 & $10.8326$ & $11.4929$ & $12.5025$ \\
81 & --- & $11.4057$ & $12.5025$ \\
85 & $10.8326$ & $11.5618$ & $12.5025$ \\
87 & --- & $11.5618$ & --- \\
89 & $10.8411$ & --- & --- \\
93 & --- & --- & $12.5025$ \\
95 & --- & $11.5618$ & $12.5025$ \\
\midrule
no projector & $11.0886$ & $12.0692$ & $13.2406$ \\
\bottomrule
\end{tabular}
\caption{The norm $\|\Pi^{(d)}(A\overbar t - \overbar A t)\|_{\munderbar 2}$ for various projector ranks $d$ and system lengths $N$. In this table ``0*'' means a number that is less than $10^{-5}$. I estimate the uncertainty in the nonzero numbers to be about $10^{-4}$. ``{---}'' means that the projector of a given rank $d$ at length $N$ is ill-defined due to degeneracies.}
\label{table_At}
\end{table}

\begin{table}
\centering
\begin{tabular}{cccc}
\toprule
$d$ & \multicolumn{3}{c}{$N$} \\
\cmidrule(l){2-4}
 & $10$ & 12 & 14 \\
\midrule
1 & 0* & 0* & 0* \\
3 & 0* & 0* & 0* \\
7 & 0* & 0* & 0* \\
9 & 0* & 0* & 0* \\
11 & 0* & 0* & 0* \\
13 & 0* & 0* & 0* \\
17 & 0* & 0* & 0* \\
25 & 0* & 0* & 0* \\
29 & 0* & 0* & 0* \\
31 & 0* & 0* & 0* \\
33 & 0* & --- & --- \\
35 & --- & $0.163094$ & $0.120853$ \\
37 & $0.230968$ & 0* & --- \\
41 & $0.230968$ & --- & --- \\
43 & $0.230968$ & --- & --- \\
59 & --- & --- & $0.123166$ \\
61 & --- & 0* & $0.123166$ \\
63 & --- & 0* & $0.123166$ \\
67 & $0.24383$ & 0* & $0.123166$ \\
71 & $0.24383$ & 0* & $0.123166$ \\
73 & $0.24383$ & 0* & $0.123166$ \\
77 & $0.24383$ & 0* & $0.0452794$ \\
81 & --- & 0* & $0.0452796$ \\
85 & $0.24383$ & $0.0674564$ & $0.0452796$ \\
87 & --- & $0.0674562$ & --- \\
89 & $0.0998496$ & --- & --- \\
93 & --- & --- & $0.0452792$ \\
95 & --- & $0.0674561$ & $0.0452794$ \\
\midrule
no projector & $0.147268$ & $0.318627$ & $0.299784$ \\
\bottomrule
\end{tabular}
\caption{The norm $\|\Pi^{(d)}(A\overbar t - \overbar A t)\|_{\munderbar 2}/\|\Pi^{(d)}A\overbar t\|_2$ for various projector ranks $d$ and system lengths $N$. In this table ``0*'' means a number that is less than $10^{-6}$. I estimate the uncertainty in the nonzero numbers to be about $10^{-5}$. ``{---}'' means that the projector of a given rank $d$ at length $N$ is ill-defined due to degeneracies.}
\label{table_At_norm}
\end{table}

The situation with $\overbar A t = A \overbar t$ is not so clear, and even a bit deceptive. The relevant results are in Tables \ref{table_At} and \ref{table_At_norm}. In Table \ref{table_At} I see no clear pattern, except for the ``no projector'' row increasing monotonically. The $d = 73$ row in Table \ref{table_At_norm} is monotonically decreasing for those three values. Near the $d = 77$ row the values drop suddenly for each column (although the row is different for different columns). I believe this drop is deceptive. Recall from Eq. \eqref{t_optimize} that $\|A\overbar t - \overbar A t\|_{\munderbar 2}$ is really shorthand for $\inf_{\alpha,\theta} \|A(\overbar t + \theta^*\overbar T) - \e^{\i\alpha}\overbar A(t + \theta T)\|_2$. Thus $\|\Pi^{(d)}(A\overbar t - \overbar A t)\|_{\munderbar 2}/\|\Pi^{(d)}A\overbar t\|_2$ is shorthand for
\begin{equation}
\left.\frac{\|\Pi^{(d)}[A(\overbar t + \theta^*\overbar T) - \e^{\i\alpha}\overbar A(t + \theta T)]\|_2}{\|\Pi^{(d)}A(\overbar t + \theta^*\overbar T)\|_2}\right\rvert_{\alpha,\theta},
\end{equation}
where the entire expression is evaluated with the parameter values $\alpha$ and $\theta$ that minimize the numerator alone. The minimal values of the denominators are reported in Table \ref{table_At}. However, the value of the denominator can be unbounded if the value of $\theta$ that happens to minimize the numerator is very large. I believe this is responsible for the sudden drop that I mentioned. For instance, in the $L = 16$, $d = 73$ entry the parameter $\theta$ that minimizes the difference is $\theta = -22.3744 + 14.6383\i$. But the value that minimizes the $L = 16$, $d = 77$ entry is $\theta = 27.5284 + 154.983\i$. For $L = 16$, $\|T\|_2 \approx 8.54$ and $\|t\|_2 \approx 29.4$ as they come out of the diagonalization process ($t$ will vary because it has the $\theta T$ degree of freedom, but these numbers all refer to the same evaluation). Furthermore $\|\overbar A T\|_2 \approx 2.08$ and $\|\overbar A t\|_2 \approx 52.4$. Hence the small values in Table \ref{table_At_norm} are probably controlled by a large $\|\theta^*A\overbar T\|$ in the denominator. Thus I cannot say how meaningful Table \ref{table_At_norm} really is since $\theta$ can vary so much.

Note that the ranks of the projectors $d$ should be chosen so that they completely include all of the eigenspaces they encompass, otherwise they are ill-defined. As $N$ increases, level crossings cause these ranks to change, but they should stabilize for large enough $N$.

The full interpretation of these results in terms of conformal fields requires a further set of tables. One of them is Table \ref{sl21_table_10}, and the others are in Appendix \ref{sl21_tables}, for which the analysis is analogous.

Consider the $L = 10$ column of Table \ref{table_AT}. This table measures the deviation from the identity $A\overbar T = \overbar A T$ by projecting the difference $A\overbar T - \overbar A T$ to an increasing chain of subspaces of given ranks $d$ and taking the 2-norm. According to the table this measurement is essentially zero until the rank of the projector surpasses 85 and becomes 89. Thus the first substantial deviation of this identity is associated to the eigenspace whose eigenvalues are associated with positions 86--89 when sorted in increasing order. According to Table \ref{sl21_table_10}, this eigenvalue is $3.67553$, in row $i = 20$. The eigenvalues of row 20 extrapolate to scaling fields with $(h,\overbar h) = (2,2)$, and the four possibilities are given in the scaling fields column there. (The smaller values of order $10^{-6}$ in these tables are not very large, but they are not negligibly small either. They change every time I run the code, but always remain at about the same order of magnitude. It is possible that they should really be numerically zero, and are just the result of compounded numerical error.)

From this observation we could make the deduction that the identity $A\overbar T = \overbar A T$ holds at finite size in the subspace of states that extrapolate to a total conformal dimension of $h + \overbar h < 4$. In fact it holds for almost all of the states that do extrapolate to $h + \overbar h = 4$ as well, except for row 12. There are a few good reasons for the deviation here. At finite size the Koo--Saleur operators we use for $L_n$ have not yet converged well to the Virasoro operators. Similarly, the states also have not yet converged well to their continuum limits. By way of example, the stress-energy tensor $T$ is found with $h = 1.79319$ at $L = 10$, when it should be $h = 2$ in the continuum limit. Nevertheless, the very small measurements found in the rows with $d = 1$ to 85 are a very good indication of the internal consistency of the algebraic analysis of the $s\ell(2|1)$ chain, the extrapolation of the finite-size data, and the construction of the projection operators when the Hamiltonian is not diagonalizable, all put together. This result furthermore suggests that $A\overbar T$ and $\overbar AT$ should be linear combinations of $L_{-2}\overbar L_{-2}I = T\overbar T$, $L_{-2}\phi_1'$, and $\overbar L_{-2}\Phi_{-1,2}$ in the continuum limit. The prediction is indeed that $A\overbar T = \overbar AT = T\overbar T$.

The analysis proceeds similarly for the $L = 12$ column, referencing the appropriate table in Appendix \ref{sl21_tables}. For Table \ref{table_At}, $L = 10$, we find jumps in the measurement at $d = 37$, $67$, and $89$. Turning again to Table \ref{sl21_table_10}, these are rows $i = 12$, $15$, and $20$, which again correspond to fields that extrapolate to $h + \overbar h = 4$, another good sign.

One might notice in Tables \ref{table_AT}--\ref{table_At_norm} that for fixed $N$, the numbers do not always behave monotonically with increasing $d$. Recall that if $\{u_i\}_{i=1}^{\dim(N)}$ is a (generalized) right eigenbasis of the hamiltonian $H$, sorted in any order (though practically by increasing eigenvalue), then for an arbitrary vector 
\begin{gather}
u = \sum_{i=1}^{\dim(N)} c_i u_i, \\
\Pi^{(d)}u = \sum_{i=1}^d c_i u_i
\end{gather}
is its restriction to the first $d$ states. The (squared) 2-norm of this vector is given by
\begin{equation}
\|\Pi^{(d)}u\|_2^2 = \sum_{i=1}^d\sum_{j=1}^d c^*_i c_j u^*_i u_j.
\end{equation}
In hermitian quantum mechanics we are guaranteed orthonormality of the basis (or it can be made so by Gram--Schmidt) and $u^*_i u_j = \delta_{ij}$. The above sum collapses to $\sum_{i=1}^d |c_i|^2$ and is monotonically increasing in $d$. In dual Jordan quantum mechanics we do not have such a nice property of the basis, and, as far as I know currently, the double sum cannot be simplified. As a consequence, non-monotonic behavior is not ruled out. We could consider defining a new norm $\|\cdot\|$ that would be monotonic in $d$:
\begin{equation}
\|u\|^2 \equiv \sum_{i=1}^{\dim(N)} \|c_i u_i\|_2^2,
\end{equation}
where $\{c_i\}$ are the coefficients of $u$ in the basis $\{u_i\}$. On the one hand, this reduces to the standard 2-norm when applied to hermitian quantum mechanics, so it is a reasonable guess. On the other hand, it is not obvious to me yet that this is a norm, and it is not obvious that it is induced by an inner product, which seems to be an important property for quantum mechanics. One might guess the inner product
\begin{equation}
(v,u) \equiv \sum_{i=1}^{\dim(N)} c_{v,i}^* c_{u,i} \|u_i\|_2^2
\end{equation}
(with obvious notation) induces the norm but it is not obvious whether this is an inner product. This norm might not even be meaningful here since it uses the 2-norm, which is induced by the conjugate-transpose inner product, and that is obviously meaningful in hermitian quantum mechanics, thus giving the 2-norm meaning in that context. We do have the notion of inner product and hermitian conjugation from the CFT, but it is generally not positive-definite and not as convincing to say ``$X = 0$ because $\|X\| = 0$'' when the norm is indefinite.

We may be able to get better results by carefully choosing our projection steps. By carefully following states as the lattice size increases and tracking various properties, one can associate a pair of conformal weights $(h,\overbar h)$ to each state on the lattice. The Koo--Saleur generators generally do a good job of coupling vectors to those expected in the continuum theory. For example, one expects $\overbar At$ to have $(h,\overbar h) = (4,4)$ in the limit, and on the lattice we generally find the strongest components of $\overbar At$ in this sector. Nevertheless we occasionally find some unexpected contributions. For instance, on the lattice I observe that $\overbar At$ has a significant component along the ground state. Fortunately, this ground state component tends to decrease with $N$.

With this in mind, there are a few options we could try. First is to pick our lattice projection operator $\Pi$ to restrict $A\overbar t - \overbar A t$ (as well as the $T$ identity) to states whose conformal weights have been identified as $(2,2)$. This projector $\Pi$ is distinct from $\Pi^{(d)}$ as $\Pi^{(d)}$ projects to the lowest $d$ states by energy, which is closely related to (but not quite the same as, due to finite-size effects) the conformal weights. Another possibility is to apply a projection operator after every action of a Koo--Saleur operator. Thus in a term like $L_{-1}^2$ we consider $\Pi L_{-1}\Pi L_{-1}$ where (abusing notation) each $\Pi$ is the correct projection operator for the context. I will use square brackets $\Pi[\cdot]$ to denote the above described procedure of applying projection operators after each Koo--Saleur operator.

Thus we will be interested in $\|\Pi(A\overbar T - \overbar AT)\|_{\munderbar 2}$ and $\|\Pi[A\overbar T - \overbar AT]\|_{\munderbar 2}$, as well as the $t$ versions and the normed versions. In each of these measurements the $\theta$ and $\alpha$ optimizations remain implicit. Thus the notation $\|\Pi[A\overbar t - \overbar At]\|_{\munderbar 2}$ is hiding the complexity of a rather long expression:
\begin{align}
\|\Pi[A\overbar t - \overbar At]\|_{\munderbar 2} = \inf_{\alpha,\theta}{}&\left\|\Pi^{(h+\overbar h=4)}L_{-2}(\overbar t + \theta^*\overbar T) - \frac{3}{2} \Pi^{(h+\overbar h=4)}L_{-1}\Pi^{(h+\overbar h=3)}L_{-1}(\overbar t + \theta^*\overbar T) \right. \nonumber \\
&\qquad \left.{}- \e^{\i\alpha}\left[\Pi^{(h+\overbar h=4)}\overbar L_{-2}(t + \theta T) - \frac{3}{2} \Pi^{(h+\overbar h=4)}\overbar L_{-1}\Pi^{(h+\overbar h=3)}\overbar L_{-1}(t + \theta T)\right]\right\|_2.
\end{align}
So far, I have performed these measurements for $N = 10, 12,$ and $14$, and they do not shed much light nor do they show much structure. It is plausible that larger lattice sizes will reveal same patterns. But for now, the data has been omitted.

I note that the same measurements can be performed using one of the modified loop models, described in Section \ref{finite_Jordan}. Though the numbers differ, the conclusions are very similar. Tables for these measurements can be found in Appendix \ref{sl21_tables}.

\section{Action of the $s\ell(2|1)$ superalgebra on singlet states}

The $s\ell(2|1)$ algebra is spanned by 8 generators whose commutation relations are standard \cite{GQS2007}. For the fundamental representation on even sites $i$ I use
\begin{subequations}
\begin{gather}
B_i = f_i^\dagger f_i + \frac12 (b_{i1}^\dagger b_{i1} + b_{i2}^\dagger b_{i2}) = f_i^\dagger f_i + \frac{(-1)^i}{2} (b_{i1}^\dagger b_{i1} + b_{i2}^\dagger b_{i2}), \\
Q_i^z = \frac12(b_{i1}^\dagger b_{i1} - b_{i2}^\dagger b_{i2}) = \frac{(-1)^i}{2}(b_{i1}^\dagger b_{i1} - b_{i2}^\dagger b_{i2}), \\
Q_i^+ = b_{i1}^\dagger b_{i2} = (-1)^i b_{i1}^\dagger b_{i2}, \\
Q_i^- = b_{i2}^\dagger b_{i1} = (-1)^i b_{i2}^\dagger b_{i1}, \\
F_i^+ = f_i^\dagger b_{i2} = (-1)^i f_i^\dagger b_{i2}, \\
F_i^- = f_i^\dagger b_{i1} = (-1)^i f_i^\dagger b_{i1}, \\
\overbar F_i^+ = b_{i1}^\dagger f_i = (-1)^i b_{i1}^\dagger f_i, \\
\overbar F_i^- = b_{i2}^\dagger f_i = (-1)^i b_{i2}^\dagger f_i,
\end{gather}
\end{subequations}
and on odd sites $i$ I use
\begin{subequations}
\begin{gather}
B_i = f_i^\dagger f_i - \frac12 (b_{i1}^\dagger b_{i1} + b_{i2}^\dagger b_{i2}) = f_i^\dagger f_i + \frac{(-1)^i}{2} (b_{i1}^\dagger b_{i1} + b_{i2}^\dagger b_{i2}), \\
Q_i^z = -\frac12(b_{i1}^\dagger b_{i1} - b_{i2}^\dagger b_{i2}) = \frac{(-1)^i}{2}(b_{i1}^\dagger b_{i1} - b_{i2}^\dagger b_{i2}), \\
Q_i^+ = -b_{i2}^\dagger b_{i1} = (-1)^i b_{i2}^\dagger b_{i1}, \\
Q_i^- = -b_{i1}^\dagger b_{i2} = (-1)^i b_{i1}^\dagger b_{i2}, \\
F_i^+ = -b_{i2}^\dagger f_i = (-1)^i b_{i2}^\dagger f_i, \\
F_i^- = -b_{i1}^\dagger f_i = (-1)^i b_{i1}^\dagger f_i, \\
\overbar F_i^+ = -f_i^\dagger b_{i1} = (-1)^i f_i^\dagger b_{i1}, \\
\overbar F_i^- = -f_i^\dagger b_{i2} = (-1)^i f_i^\dagger b_{i2}.
\end{gather}
\end{subequations}
The signs $(-1)^i$ are added to emphasize where relative signs are important. Note that I have used different forms of the generators compared to \textcite[Appendix A]{Gainutdinov2015}. But I have checked carefully that my generators do produce the $s\ell(2|1)$ relations, and deviations here represent corrections to errors in the reference.

Also note the proliferation of signs one must keep track of carefully, in addition to the signs in the odd $i$ generators. First, when applying an operator to a state, the super product of elements in a graded tensor product is used. For an operator $A = A_1\otimes\cdots\otimes A_L$ with $A_i$ built up from $\{b_{i1}, b_{i2}, f_i, b_{i1}^\dagger, b_{i2}^\dagger, f_i^\dagger\}$ and a basis state $\ket*{\psi} = \ket*{n_1\cdots n_L} = \ket*{n_1}\otimes\cdots\otimes\ket*{n_L} \in \bigotimes_{i=1}^L \{b_{i1}^\dagger\ket*{0}, b_{i2}^\dagger\ket*{0}, f_i^\dagger\ket*{0}\}$,
\begin{equation}
A\ket*{\psi} = (A_1\otimes\cdots\otimes A_L)(\ket*{n_1}\otimes\cdots\otimes\ket*{n_L}) = (-1)^C A_1\ket*{n_1}\otimes\cdots\otimes A_L\ket*{n_L},
\end{equation}
where $C$ is the number of fermionic exchanges needed to bring the operators to the corresponding states. Second, $f_i\ket*{2}_i = f_i f_i^\dagger\ket*{0} = ((-1)^i - f_i^\dagger f_i)\ket*{0} = (-1)^i\ket*{0}$, not just $\ket*{0}$. This consideration is important for operators such as $F_i^\pm$ for odd $i$ and $\overbar F_i^\pm$ for even $i$. These signs were not an issue previously since $e_i$ (and consequently the Koo--Saleur operators) is built of pairs of neighboring fermionic operators and has a nonzero action only on states containing adjacent pairs of fermions. Similarly, the $B = Q^z = 0$ Hilbert space was built by iterating the procedure of replacing pairs with pairs (see the discussion at the end of Section \ref{sl21_section}).

From the $s\ell(2|1)$ commutation relations (or just by looking at the operators) it is clear that $Q^\pm$, $F^\pm$, and $\overbar F^\pm$ will take us out of the $B = Q^z = 0$ sector. However, since we are most interested in the action of $s\ell(2|1)$ generators on states from that sector, such as the ground state, we can first identify all the states of interest by working in the $B = Q^z = 0$ sector and then embedding them in the whole $3^L$-dimensional space to compute the action of the generators.

As a first check, the ground state, which should belong to an $s\ell(2|1)$ singlet representation, is annihilated by all generators $T_a$ where as usual
\begin{equation}
T_a = \sum_{i=1}^L T_{a,i} = \sum_{i=1}^L \underset{\substack{\uparrow\\ 1}}{I_3} \otimes \cdots \otimes I_3 \otimes \underset{\substack{\uparrow\\ i}}{T_{a,i}} \otimes I_3 \otimes \cdots \otimes \underset{\substack{\uparrow\\ L}}{I_3}.
\end{equation}
This is identically true in finite size. (It is known for this Hamiltonian that the vector components of the ground state are all integers before normalization. Therefore after numerically finding the ground state I can rationalize it by dividing through by the smallest nonzero element and rounding all the results, setting them to integers. Then I divide through by the $s\ell(2|1)$ norm, which ends up being an integer. Results of this type are generally known as Razumov--Stroganov conjectures and have been proven in some cases. The takeaway is that the ground state can be treated exactly, to infinite precision.)

Similarly, the stress-energy tensor $T$ should also belong to a singlet. While this does not happen exactly, for $L = 10$, typical values are $\|Q^\pm T\|_2 \simeq 3\times 10^{-12}$ and $\|F^\pm T\|_2 \simeq \|\overbar F^\pm T\|_2 \simeq 3 \times 10^{-7}$, and for $L = 12$, typical values are $\|Q^\pm T\|_2 \simeq 7\times 10^{-12}$ and $\|F^\pm T\|_2 \simeq \|\overbar F^\pm T\|_2 \simeq 3 \times 10^{-6}$. The value of $\|F^\pm T\|_2$ is of the same order as $d^{1/2}\epsilon$, where $d$ is the dimension of the vacuum sector and $\epsilon$ is the precision of the eigenvalues. So these numbers can probably be improved by determining $T$ to higher numerical precision. Nevertheless the evidence is in favor of $T$ transforming trivially under the action of the generators.

\chapter{Mixing of conformal fields at finite size} \label{mixing}

Much of the success of the lattice approach to conformal field theory came from the identification of lattice eigenvectors as analogues of continuum fields. Indeed, we have made the innocuous assumption, justified by most of the data, that each vector at finite size could cleanly be associated to a single conformal field, even if we did not know precisely what that field was. However, two lattice excitations in $\overbar{\mathscr W}_{\!\!0,\mathfrak q^{\pm 2}}$ pose a challenge to this assumption, and it appears that they each represent mixtures of the same two conformal fields.

\emph{N.\ B.: In this chapter $\Phi_{rs}$ refers to the \emph{diagonal} field $\phi_{rs}\otimes\overbar\phi_{rs}$, not $\phi_{rs}\otimes\overbar\phi_{r,-s}$ as in Eq.~\eqref{eq:indecomposable_diamond}.}

\section{Mixing of $T\overbar T$ and $\Phi_{31}$}

Begin with the loop model in the standard module $\overbar{\mathscr W}_{\!\!0,\mathfrak q^{\pm 2}}$, with the lattice size denoted by $N$. Consider the lowest three right eigenstates of the Hamiltonian in the sector of momentum zero, labeled $\ket*{0}$, $\ket*{1}$, and $\ket*{2}$. $\ket*{0}$ is the ground state. Various calculations show that for $c > 0$, $\ket*{1}$ has the expected properties of $\ket*{\Phi_{31}}$ and, likewise, $\ket*{2}$ tends to behave like $\ket*{T\overbar T}$. This behavior is ``sharper'' the closer one gets to $c = 1$. It is also observed that for $c < 0$, the roles are exchanged; $\ket*{1}$ has the properties of $\ket*{T\overbar T}$ and $\ket*{2}$ has the properties of $\ket*{\Phi_{31}}$, with such behavior most clear approaching $c = -2$. Close to $c = 0$, one observes that both states $\ket*{1}$ and $\ket*{2}$ ``contain'' a fairly even mixture of the two. The identification of $\ket*{1}$ or $\ket*{2}$ with $\ket*{T\overbar T}$ or $\ket*{\Phi_{31}}$ also sharpens with increasing $N$, except at $c = 0$.

An ansatz that would account for these observations is
\begin{subequations}\label{eq:exponents}
\begin{gather}
\ket*{2} = A_2(N,c)\left(\ket*{T\overbar T} + \frac{B_2(c)}{N^{\epsilon_2(c)}}\ket*{\Phi_{31}}\right), \\
\ket*{1} = A_1(N,c)\left(\ket*{\Phi_{31}} + \frac{B_1(c)}{N^{\epsilon_1(c)}}\ket*{T\overbar T}\right),
\end{gather}
\end{subequations}
where $A_1$ and $A_2$ are overall normalization constants, and $B_1$ and $B_2$ represent a relative normalization between the two terms. This may be written more symmetrically as
\begin{subequations}
\begin{gather}
\ket*{2} = A'_2(N,c)\left(N^{\epsilon_2(c)/2}\ket*{T\overbar T} + \frac{B_2(c)}{N^{\epsilon_2(c)/2}}\ket*{\Phi_{31}}\right), \\
\ket*{1} = A'_1(N,c)\left(N^{\epsilon_1(c)/2}\ket*{\Phi_{31}} + \frac{B_1(c)}{N^{\epsilon_1(c)/2}}\ket*{T\overbar T}\right),
\end{gather}
\end{subequations}
with $A'_i(N,c) = A_i(N,c) N^{-\epsilon_i(c)/2}$. Let us suppose that the exponents $\epsilon_i(c)$ have the property that their sign is that of $c$: $\sgn\epsilon_i(c) = \sgn c$. This needs to be demonstrated, either numerically or analytically. However, assuming this to be the case, the ansatz displays exactly the exchange of states observed numerically.

We turn now to numerical calculation of the exponents $\epsilon_i$. Hereafter, as much as possible, we will suppress notation of functional dependence on $N$ and $c$. Write
\begin{subequations}
\label{eq:expansion}
\begin{gather}
\ket*{2} = \alpha\ket*{T\overbar T} + \beta\ket*{\Phi_{31}}, \\
\ket*{1} = \gamma\ket*{T\overbar T} + \delta\ket*{\Phi_{31}}.
\end{gather}
\end{subequations}
We also define a normalized version,
\begin{subequations}
\label{eq:norm_expansion}
\begin{gather}
\ket*{\tilde 2} = \tilde\alpha\ket*{T\overbar T} + \tilde\beta\ket*{\Phi_{31}}, \\
\ket*{\tilde 1} = \tilde\gamma\ket*{T\overbar T} + \tilde\delta\ket*{\Phi_{31}}
\end{gather}
\end{subequations}
such that $\ip*{\tilde2} = \ip*{\tilde1} = \pm 1$. This is not possible exactly at $c = 0$ since $\ip*{1} = \ip*{2} = 0$ exactly. Up to phases we must have $\ket*{\tilde 1} = \ket*{1}/\sqrt{\ip*{1}}$ and $\ket*{\tilde 2} = \ket*{2}/\sqrt{\ip*{2}}$. Furthermore, $\ket*{\tilde 1}$ and $\ket*{\tilde 2}$ are orthogonal with respect to the loop scalar product.

It turns out we must partition the range of $c$, $(-2,1)$, into three regions. These will be $R_0 = (-2,-3/5)\cup(1/2,1)$, $R_2 = (-3/5,0)$, and $R_1 = (0,1/2)$. The boundaries at $c = -2, -3/5, 0, 1/2, 1$ correspond to $x = 1,3/2,2,3,\infty$. The regions are characterized by the loop norms of the states above. Let $\sigma_i = \ip*{\tilde i}$. In $R_0$ we have $\sigma_1 = \sigma_2 = 1$. In $R_2$ we have $\sigma_1 = 1$, $\sigma_2 = -1$. In $R_1$ we have $\sigma_1 = -1$, $\sigma_2 = 1$. The labeling is such that $\sigma_i = -1$ in $R_i$.

We have the following expansions of a generic field $\ket*{X}$:
\begin{subequations}
\begin{align}
\ket*{X} &= c_0\ket*{0} + c_1\ket*{1} + c_2\ket*{2} + \cdots \\
&= \tilde c_0\ket*{\tilde 0} + \tilde c_1\ket*{\tilde 1} + \tilde c_2\ket*{\tilde 2} + \cdots.
\end{align}
\end{subequations}
Combining the last equation above with \eqref{eq:norm_expansion} we have
\begin{equation}
\ket*{X} = c_0\ket*{0} + (\tilde c_1\tilde\gamma + \tilde c_2\tilde\alpha)\ket*{T\overbar T} + (\tilde c_1\tilde\delta + \tilde c_2\tilde\beta)\ket*{\Phi_{31}} + \cdots.
\label{eq:field_expansion}
\end{equation}
$c_0 = \tilde c_0$ and $\ket*{0} = \ket*{\tilde 0}$ since $\ip*{0} = 1$ already. The question then turns to the extraction of the coefficients $\tilde c_i$ and the parameters $\tilde\alpha$, $\tilde\beta$, $\tilde\gamma$, and $\tilde\delta$. To be clear, the $\tilde c_i$ are properties of the expansion, and so depend on the field $\ket*{X}$. $\tilde\alpha$, $\tilde\beta$, $\tilde\gamma$, and $\tilde\delta$ are properties of the guess \eqref{eq:norm_expansion} and are fixed once determined. All of $\tilde c_i$ and $\tilde\alpha$, $\tilde\beta$, $\tilde\gamma$, and $\tilde\delta$ depend on $N$ and $c$.

It is rather straightforward to get $\tilde c_i$. The Hamiltonian $H$ is self-adjoint with respect to the loop inner product, and thus $\braket*{\tilde i}{\tilde j} = \delta_{ij}\sigma_i$. Thus $\tilde c_i = \ip*{\tilde i}{X}$.

Let us concretely try to choose $\ket*{X} = \ket*{T\overbar T}$. On the lattice, we do not have direct access to this quantity. The closest we can get is to choose a definition for $\ket*{T\overbar T}$ that is hopefully close to its ``correct'' identification. We thus let $\ket*{T\overbar T} = L_{-2}\overbar L_{-2}\ket*{0}$, or, on the lattice, a symmetrized combination $(L_{-2}\overbar L_{-2} + \overbar L_{-2}L_{-2})\ket*{0}/2$, which I will leave implicit. We should thus have a unit coefficient for $\ket*{T\overbar T}$ and zero for $\ket*{\Phi_{31}}$. Let $\tau = \ip*{T\overbar T}$, which is numerically calculated and thus given. Finally, let $\psi = \ip*{T\overbar T}{\Phi_{31}}$, which is left as an unknown, as it may be nonzero on the lattice. 

The expected coefficients of the expansion, along with orthonormality of the states $\ket*{\tilde 1}$ and $\ket*{\tilde 2}$, gives us 5 equations in 5 unknowns:
\begin{subequations}
\begin{gather}
\tilde c_1\tilde\gamma + \tilde c_2\tilde\alpha = 1, \label{eq:a} \\
\tilde c_1\tilde\delta + \tilde c_2\tilde\beta = 0, \label{eq:b} \\
\tau\tilde\alpha^*\tilde\alpha + \psi\tilde\alpha^*\tilde\beta + \psi^*\tilde\beta^*\tilde\alpha + \tilde\beta^*\tilde\beta = \sigma_2, \label{eq:c} \\
\tau\tilde\gamma^*\tilde\gamma + \psi\tilde\gamma^*\tilde\delta + \psi^*\tilde\delta^*\tilde\gamma + \tilde\delta^*\tilde\delta = \sigma_1, \label{eq:d} \\
\tau\tilde\gamma^*\tilde\alpha + \psi\tilde\gamma^*\tilde\beta + \psi^*\tilde\delta^*\tilde\alpha + \tilde\delta^*\tilde\beta = 0. \label{eq:e} 
\end{gather}
\end{subequations}
The inputs to this equation are $\sigma_1, \sigma_2 \in \{1,-1\}$, $\tilde c_1, \tilde c_2 \in \mathbf C$, and $\tau\in\mathbf R$. The unknowns are $\psi$, $\tilde\alpha$, $\tilde\beta$, $\tilde\gamma$, and $\tilde\delta$.

Eq.~\eqref{eq:b} gives 
\begin{equation}
\tilde\beta = -\frac{\tilde c_1\tilde\delta}{\tilde c_2}.
\end{equation}
Eq.~\eqref{eq:e} gives \begin{equation}
\tilde\alpha = -\frac{\tilde\beta(\psi\tilde\gamma^*+\tilde\delta^*)}{\tau\tilde\gamma^*+\psi^*\tilde\delta^*} = \frac{\tilde c_1\tilde\delta(\psi\tilde\gamma^*+\tilde\delta^*)}{\tilde c_2(\tau\tilde\gamma^*+\psi^*\tilde\delta^*)}.
\end{equation}
Putting this into Eq.~\eqref{eq:a} and rearranging gives
\begin{equation}
\tilde c_1(\tau\tilde\gamma^*\tilde\gamma + \psi\tilde\gamma^*\tilde\delta + \psi^*\tilde\delta^*\tilde\gamma + \tilde\delta^*\tilde\delta) = \tau\tilde\gamma^* + \psi^*\tilde\delta^*.
\end{equation}
Because of Eq.~\eqref{eq:d}, the left hand side is $\sigma_1\tilde c_1$. Thus
\begin{equation}
\psi = \frac{\sigma_1\tilde c_1^* - \tau\tilde\gamma}{\tilde\delta},
\end{equation}
and, in turn,
\begin{equation}
\tilde\alpha = \frac{\sigma_1\tilde c_1^*\tilde\gamma^* + \tilde\delta^*\tilde\delta - \tau\tilde\gamma^*\tilde\gamma}{\sigma_1\tilde c_2}.
\end{equation}
Putting this back into Eq.~\eqref{eq:a}, one finds that
\begin{equation}
\tilde\delta^*\tilde\delta = \tau\tilde\gamma^*\tilde\gamma + \sigma_1(1 - \tilde c_1\tilde\gamma - \tilde c_1^*\tilde\gamma^*).
\end{equation}
Using this previous equality and using Eqs.~\eqref{eq:a} and \eqref{eq:b} to write $\tilde\alpha$ and $\tilde\beta$ in terms of $\tilde\gamma$ and $\tilde\delta$, Eq.~\eqref{eq:c} gives us
\begin{equation}
\sigma_1\tilde c_1^*\tilde c_1 + \sigma_2\tilde c_2^*\tilde c_2 = \tau.
\label{TT_norm_square_equation}
\end{equation}
Since this is a relation among given parameters and does not contain the unknowns, it would appear that one of the equations is redundant. At the same time, we are short one to determine a value for $\tilde\gamma$. Numerically, Eq.~\eqref{TT_norm_square_equation} is not satisfied exactly, though it is close. However, the discrepancy seems to grow with $N$. The appearance of this relation can likely be traced back to the assumptions encoded in Eqs.~\eqref{eq:a} and \eqref{eq:b} that $L_{-2}\overbar L_{-2}I$ can be identified exactly with $T\overbar T$ with no component along $\Phi_{31}$ or any other field (other calculations indicate that there might be a component along the ground state). Another way to put it is that, with $\ket*{X} = \ket*{T\overbar T}$, the $\ket*{T\overbar T}$ on the left hand side of Eq.~\eqref{eq:field_expansion} is not the same as the $\ket*{T\overbar T}$ on the right hand side, because the left hand side is defined and in general this definition is inconsistent with its correct identification. Despite the preceding, numerical evidence suggests that the equality appears to be satisfied exactly for $c = -3/5$, $c = 0$, and $c = 1/2$. The ratio of the left hand side to the right hand side approaches unity for $c = -3/5$ and $c = 1/2$ and all values of $N$, though it diverges at $c = 0$ (where both quantities approach 0, though clearly at different rates).

As a first approximation, let's suppose $\psi = 0$ (or that it is the least significant component to the above system). Then we have a solution where
\begin{subequations}
\begin{gather}
\tilde\delta^*\tilde\delta = \frac{\sigma_1\sigma_2\tilde c_2^*\tilde c_2}{\tau}, \\
\tilde\alpha = \frac{\sigma_2\tilde c_2}{\tau}, \\
\tilde\beta = -\frac{\tilde c_1\tilde\delta}{\tilde c_2}, \\
\tilde\gamma = \frac{1}{\tilde c_1}\left(1 - \frac{\sigma_2\tilde c_2^* c_2}{\tau}\right).
\end{gather}
\end{subequations}
Again, these are based on the approximate equality above. We may simply choose $\tilde\delta$ to be real, which amounts to a choice of phase for $\Phi_{31}$. Since $\tilde\beta$ is proportional to $\tilde\delta$, this does not lead to an inconsistency. Thus
\begin{equation}
\tilde\delta = \left\lvert{\sqrt{\frac{\sigma_1\sigma_2\tilde c_2^*\tilde c_2}{\tau}}}\right\rvert.
\end{equation}

Let us now reconcile Eqs.~\eqref{eq:exponents}, \eqref{eq:expansion}, and \eqref{eq:norm_expansion}. We may trivially pull out various factors:
\begin{subequations}
\begin{gather}
\ket*{2} = \alpha\left(\ket*{T\overbar T} + \frac\beta\alpha\ket*{\Phi_{31}}\right), \\
\ket*{1} = \delta\left(\ket*{\Phi_{31}} + \frac\gamma\delta\ket*{T\overbar T}\right), \\
\ket*{\tilde 2} = \tilde\alpha\left(\ket*{T\overbar T} + \frac{\tilde\beta}{\tilde\alpha}\ket*{\Phi_{31}}\right), \\
\ket*{\tilde1} = \tilde\delta\left(\ket*{\Phi_{31}} + \frac{\tilde\gamma}{\tilde\delta}\ket*{T\overbar T}\right).
\end{gather}
\end{subequations}
We thus have $\alpha = A_2$, $\delta = A_1$. However, the overall normalization is not particularly meaningful at the moment, and besides, we do not have $\alpha$ and $\delta$. What is consistent, though, are the relative normalizations between the two states, regardless of the overall normalizations. Thus we should have
\begin{subequations}
\begin{gather}
\frac{B_2}{N^{\epsilon_2}} = \frac\beta\alpha = \frac{\tilde\beta}{\tilde\alpha}, \\
\frac{B_1}{N^{\epsilon_1}} = \frac\gamma\delta = \frac{\tilde\gamma}{\tilde\delta}.
\end{gather}
\end{subequations}
For a fixed value of $c$, we can thus measure $\tilde c_1$ and $\tilde c_2$ for various values of $N$, and compute the values of the ratios $\tilde\beta/\tilde\alpha$ and $\tilde\gamma/\tilde\delta$ using the relations derived above. We can then fit the data to a curve of the form $\tilde\beta/\tilde\alpha= CN^{-\epsilon}$ and let the parameter values which give the best fit define $\epsilon_2 = \epsilon$, and similarly for $\epsilon_1$.

We can perform the preceding procedure for both definitions of $\ket*{T\overbar T}$. To distinguish between the cases where $\ket*{T\overbar T} = L_{-2}\overbar L_{-2}\ket*{0}$ and $(L_{-2}\overbar L_{-2} + \overbar L_{-2}L_{-2})\ket*{0}/2$, we will append a subscript ``s'' to the exponents corresponding to the symmetric combination. (When the symmetry of the definition is not important to the discussion I will simply refer to the exponents as $\epsilon_i$ with no subscript. The subscripts appear primarily in the plots.) Plots of the exponents $\epsilon_i(c)$ are given in Figures \ref{e1s} and \ref{e2s}. (For all plots, the nonsymmetrized versions appear in Appendix \ref{mixing_plots}.)
\begin{figure}
\centering
\includegraphics[width=\textwidth]{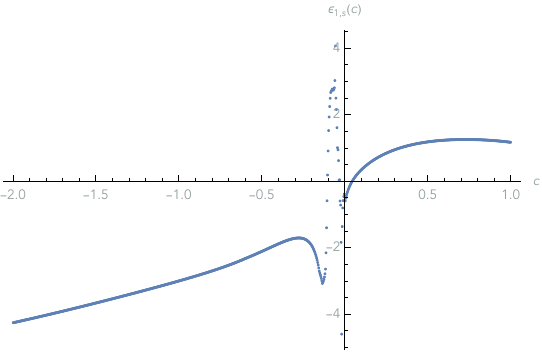}
\caption{Plot of the exponent $\epsilon_{1,\text s}$ as a function of $c$.}
\label{e1s}
\end{figure}
\begin{figure}
\centering
\includegraphics[width=\textwidth]{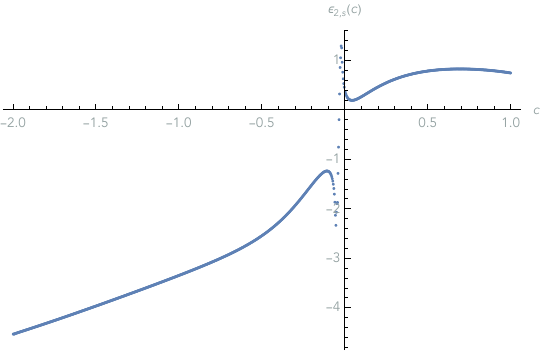}
\caption{Plot of the exponent $\epsilon_{2,\text s}$ as a function of $c$.}
\label{e2s}
\end{figure}

\begin{figure}
\centering
\includegraphics[width=\textwidth]{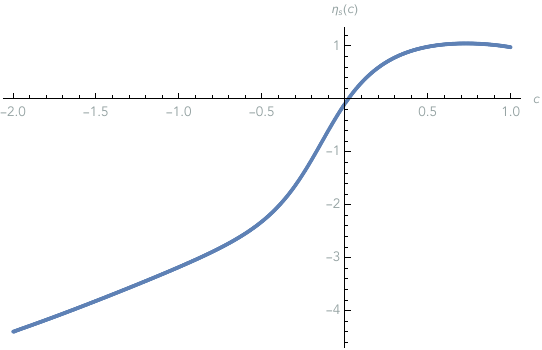}
\caption{Plot of the exponent $\eta_{\text s}$ as a function of $c$.}
\label{etas}
\end{figure}

\begin{figure}
\centering
\includegraphics[width=\textwidth]{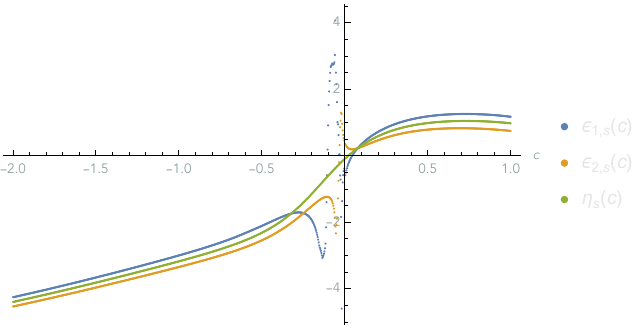}
\caption{Plots of the exponents $\epsilon_{1,\text s}$, $\epsilon_{2,\text s}$, and $\eta_{\text s}$ as functions of $c$.}
\label{alls}
\end{figure}

Another quantity that is consistent regardless of the normalization of the states is $\tilde c_1/\tilde c_2$. Let us also guess that for fixed $c$, this quantity also obeys a power law (to be justified later) so that $\tilde c_1/\tilde c_2 = C/N^{\eta(c)}$. We may perform the procedure described above in order to get a numerical measurement for $\eta$. The results of this calculation are given in Figures \ref{etas} and \ref{eta}. Finally, all three exponents are shown together in Figures \ref{alls} and \ref{all}.

From Figures \ref{alls} and \ref{all}, the relation $\epsilon_1(c) = \epsilon_2(c) = \eta(c)$ obviously suggests itself. Aside from some artifacts near $c = 0$, it seems quite plausible that the three exponents should, in fact, be equal. Assuming this to be the case, the deviations from equality would come from finite-size effects and the curve-fitting method, which currently is a least-squares optimization---it minimizes the sum of squares of the absolute deviations. Since the data values can become quite large and indicate divergences, larger values of the data have greater weight in the fit. Thus using a fitting algorithm that minimizes the sum of squares of the relative deviations may yield closer convergence of the curves.

I now give an analytical argument that implies the equality of all three, at least for the region $R_0$. Take $N$ large enough such that on the lattice, $\ip*{T\overbar T}$ deviates negligibly from $c^2/4$ and $\ip*{T\overbar T}{\Phi_{31}}$ is essentially zero. The loop-orthonormality of $\ket*{\tilde 1}$ and $\ket*{\tilde 2}$ implies that
\begin{equation}
R = \begin{pmatrix}
\tilde\alpha c/2 & \tilde\beta \\
\tilde\gamma c/2 & \tilde\delta
\end{pmatrix}
\end{equation}
is a unitary matrix. Fix the phases of $\ket*{T\overbar T}$ and $\ket*{\Phi_{31}}$ such that the diagonal elements of $R$ are real. Then the most general form of $R$ is
\begin{equation}
R = \begin{pmatrix}
\cos\theta & \e^{\i\phi}\sin\theta \\
-\e^{-\i\phi}\sin\theta & \cos\theta
\end{pmatrix},
\end{equation}
with real parameters $\theta$ and $\phi$, both possibly dependent on $N$ and $c$. We thus have
\begin{subequations}
\begin{gather}
\frac\beta\alpha = \frac{\tilde\beta}{\tilde\alpha} = \frac{c}{2}\e^{\i\phi}\tan\theta, \\
\frac\gamma\delta = \frac{\tilde\gamma}{\tilde\delta} = -\frac{2}{c}\e^{-\i\phi}\tan\theta.
\end{gather}
\end{subequations}
Since $\e^{\i\phi}$ is a number of unit modulus, aside from a normalizing factor $(c/2)^{\pm 1}$, the exponential dependence on $N$ of both ratios is entirely contained in $\tan\theta$, which is common to both. Thus $\epsilon_1 = \epsilon_2$. From this we also see that $B_1(c)B_2(c) = -1$. Similarly,
\begin{equation}
\frac{\tilde c_1}{\tilde c_2} = -\frac{\tilde\beta}{\tilde\delta} = -\e^{\i\phi}\tan\theta,
\end{equation}
and thus $\eta = \epsilon_1 = \epsilon_2$. The plots indicate that the preceding equality is also true for parts of $R_2$ and $R_1$, so it remains to be seen how $\sigma_1$ and $\sigma_2$ figure into this argument. (It seems to me that instead of $R\in U(2)$, we would have $R\in U(1|1)$ instead. I have not worked out the form of $R$ in this case.)

\begin{figure}
\centering
\includegraphics[width=\textwidth]{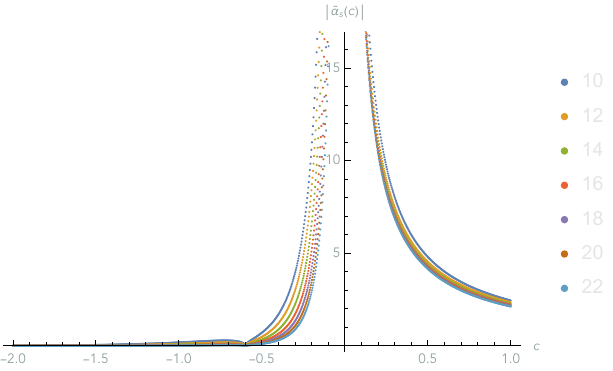}
\caption{}
\label{alphas}
\end{figure}

\begin{figure}
\centering
\includegraphics[width=\textwidth]{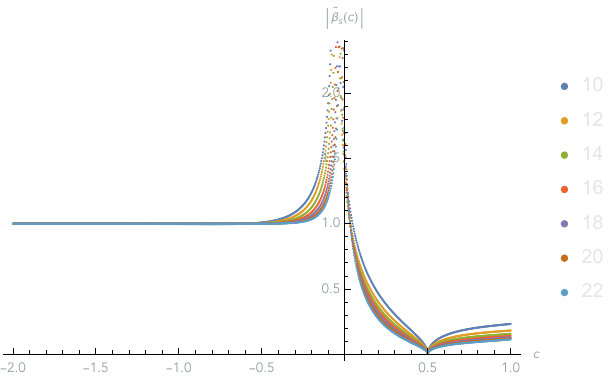}
\caption{}
\label{betas}
\end{figure}

\begin{figure}
\centering
\includegraphics[width=\textwidth]{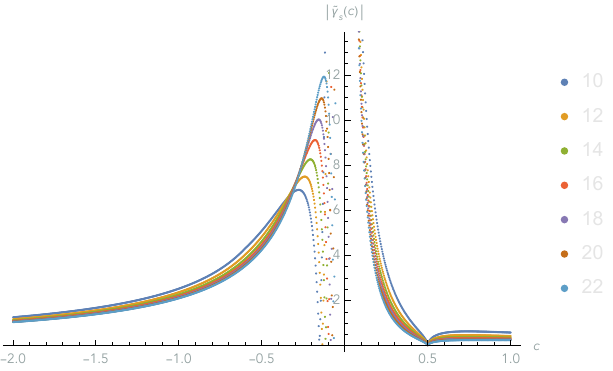}
\caption{}
\label{gammas}
\end{figure}

\begin{figure}
\centering
\includegraphics[width=\textwidth]{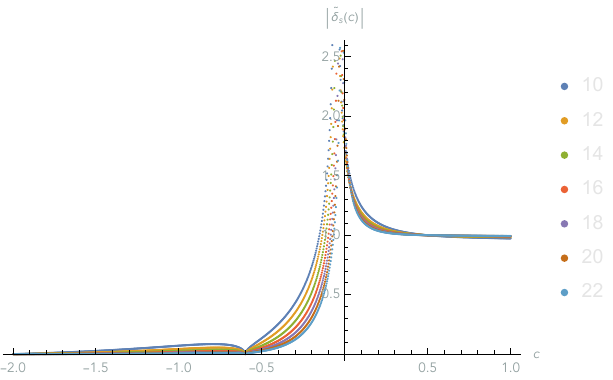}
\caption{}
\label{deltas}
\end{figure}

\begin{figure}
\centering
\includegraphics[width=\textwidth]{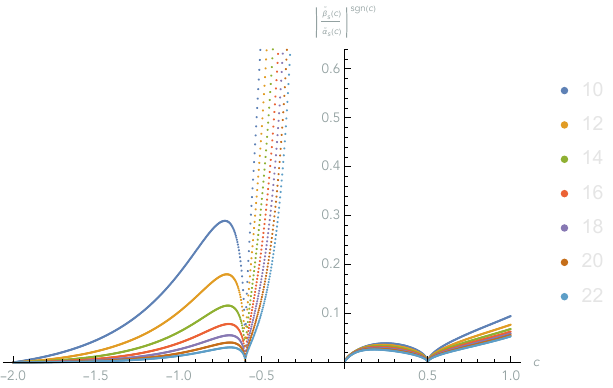}
\caption{}
\label{betaalphas}
\end{figure}

\begin{figure}
\centering
\includegraphics[width=\textwidth]{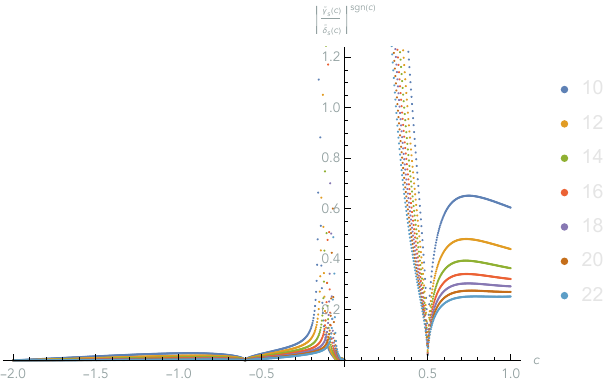}
\caption{}
\label{gammadeltas}
\end{figure}

\begin{figure}
\centering
\includegraphics[width=\textwidth]{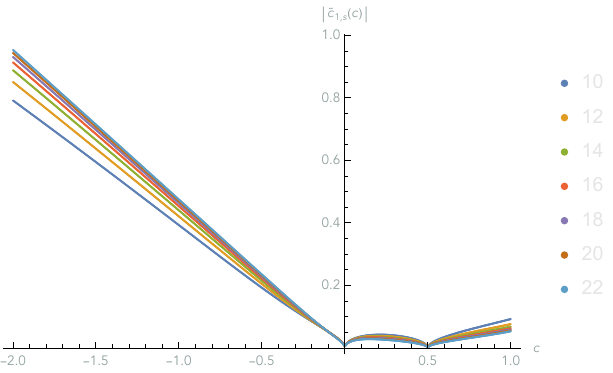}
\caption{}
\label{c1s}
\end{figure}

\begin{figure}
\centering
\includegraphics[width=\textwidth]{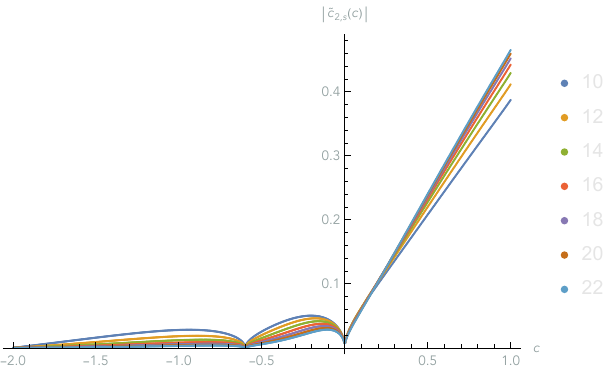}
\caption{}
\label{c2s}
\end{figure}

\begin{figure}
\centering
\includegraphics[width=\textwidth]{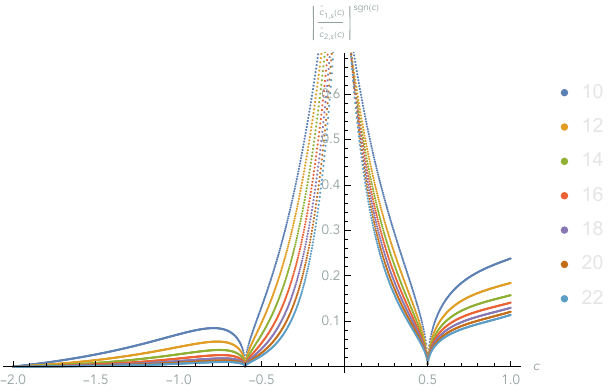}
\caption{}
\label{c1c2s}
\end{figure}

Finally, there seem to be significant deviations from $\eta = \epsilon_1 = \epsilon_2$ near $c = 0$. A closer look at the data explains why these deviations occur, and how the data might be refined to yield closer agreement. Figures \ref{alphas}--\ref{c1c2s} show the values of the individual parameters and some of their ratios. They may help to explain some features that we see in the plots of the exponents. For each of these I have only given the absolute value of each, rather than the real and imaginary parts separately. Not much information is lost here since for a given value of $N$ and $c$ (more broadly, for a given region $R_i$ of the $c$ axis), all the values plotted tend to be close to purely real or purely imaginary. This information is summarized in Table \ref{real_or_complex}. For certain ratios such as $\tilde\beta/\tilde\alpha$, I have instead plotted $(\tilde\beta(c)/\tilde\alpha(c))^{\sgn c} = (\tilde\beta(c)/\tilde\alpha(c))^{\pm 1}$ so that it is easier to view on the same plot---for $c < 0$ one thus sees $\tilde\alpha(c)/\tilde\beta(c)$.

\begin{table}
\centering
\begin{tabular}{c|c|c|c|c|c|c|c|c|c|c|c|c}
region & $\tilde c_1$ & $\tilde c_2$ & $\tilde\alpha$ & $\tilde\beta$ & $\tilde\gamma$ & $\tilde\delta$ & $\tilde c_{1,\text s}$ & $\tilde c_{2,\text s}$ & $\tilde\alpha_{\text s}$ & $\tilde\beta_{\text s}$ & $\tilde\gamma_{\text s}$ & $\tilde\delta_{\text s}$ \\
\hline
$R_{0,-}$ & $\mathbb C_{\mathbb R^*}$ & $\mathbb C_{\mathbb R}$ & $\mathbb C_{\mathbb R}$ & $\mathbb C_{\mathbb R}$ & $\mathbb C_{\mathbb R^*}$ & $\mathbb R$ & $\mathbb R$ & $\mathbb R$ & $\mathbb R$ & $\mathbb R$ & $\mathbb C_{\mathbb R^*}$ & $\mathbb R$ \\
\hline
$R_2$ & $\mathbb C_{\mathbb R}$ & $\mathbb C_{\i\mathbb R}$ & $\mathbb C_{\i\mathbb R}$ & $\mathbb C_{\i\mathbb R}$ & $\mathbb C_{\mathbb R}$ & $\mathbb R$ & $\mathbb R$ & $\i\mathbb R$ & $\i\mathbb R$ & $\i\mathbb R$ & $\mathbb C_{\mathbb R^*}$ & $\mathbb R$ \\
\hline
$R_1$ & $\mathbb C_{\i\mathbb R}$ & $\mathbb C_{\mathbb R}$ & $\mathbb C_{\mathbb R}$ & $\mathbb C_{\i\mathbb R}$ & $\mathbb C_{\i\mathbb R}$ & $\mathbb R$ & $\i\mathbb R$ & $\mathbb R$ & $\mathbb R$ & $\i\mathbb R$ & $\mathbb C_{\i\mathbb R}$ & $\mathbb R$ \\
\hline
$R_{0,+}$ & $\mathbb C_{\mathbb R}$ & $\mathbb C_{\mathbb R}$ & $\mathbb C_{\mathbb R}$ & $\mathbb C_{\mathbb R}$ & $\mathbb C_{\mathbb R}$ & $\mathbb R$ & $\mathbb R$ & $\mathbb R$ & $\mathbb R$ & $\mathbb R$ & $\mathbb C_{\mathbb R}$ & $\mathbb R$
\end{tabular}
\caption{The domains in which the values of the given parameters lie. $\i\mathbf R$ means purely imaginary. $\mathbb C_{\mathbb R}$ means that $10 < \re x/\im x < 1000$, $\mathbb C_{\mathbb R^*}$ means $\re x/\im x \ge 1000$, $\mathbb C_{\i\mathbb R}$ means $10 < \im x/\re x < 1000$, and $\mathbb C_{\i\mathbb R^*}$ means $\im x/\re x \ge 1000$. In no case did I observe the real and imaginary parts to be the same order of magnitude.}
\label{real_or_complex}
\end{table}

One noticeable feature is that for any given $N$, there is a value $c<0$, $c\in R_2$ where $\tilde\gamma(c)$ passes through zero. This value of $c$ increases (i.e., moves closer to 0) with $N$. This means that there will be zeros in the ratio $\gamma/\delta$, and these zeros are located in different places with different $N$. Clearly such a behavior cannot be modeled with a pure power law. The way out of this may be to wait for large enough $N$ such that we are past the zero and fit only those points to the power law, instead of including all data points for small and large $N$. Of course, this would mean the validity of our data and fits are reduced from all of $R_2 = (-3/5,0)$ to approximately $(-3/5,c(N))$, where $c(N)$ is such that $\tilde\gamma(c(N)) = 0$. This would explain the departure of $\epsilon_1(c)$ in Figures \ref{all} and \ref{alls} from a smooth curve at small negative values of $c$. 

It appears to me that the locations of the zeros may correspond to the same places where the measured of $\tau = \ip*{T\overbar T}$ is zero, but $\sigma_1\tilde c_1^*\tilde c_1 + \sigma_2\tilde c_2^*\tilde c_2$ is close to $c^2/4$. Thus the equality \eqref{TT_norm_square_equation} fails badly and the expressions for the coefficients are not valid. (The positions of the zeros are not quite the same [in fact for $N \ge 20$ they move to positive values] but I am still inclined to think that this is related.)

Along the same lines, we see a divergence in $\tilde\beta$ for some $c\in R_2$ that is moving towards $c = 0$ (equivalently, the discussion for $\tilde\gamma$ applies to $1/\tilde\beta$ with its moving zeros). Thus the result is the deviation of $\epsilon_2(c)$ from a smooth curve close to $c = 0$. It seems that the defects in these graphs do not apply to $\eta(c)$, since the plots of $\tilde c_1$ and $\tilde c_2$ do not show zeros except possibly at the boundaries of the regions $R_i$. Along these lines the direct measurement of $\tilde c_1$ and $\tilde c_2$ are not subject to approximations such as those used in the computation of $\tilde\alpha$, $\tilde\beta$, $\tilde\gamma$, and $\tilde\delta$. We might be tempted to conclude that $\eta$ is the ``true'' value of the exponent, and most accurate of the three at finite size.


\section{The region $R_1$}
\begin{figure}
\centering
\includegraphics[width=0.45\textwidth]{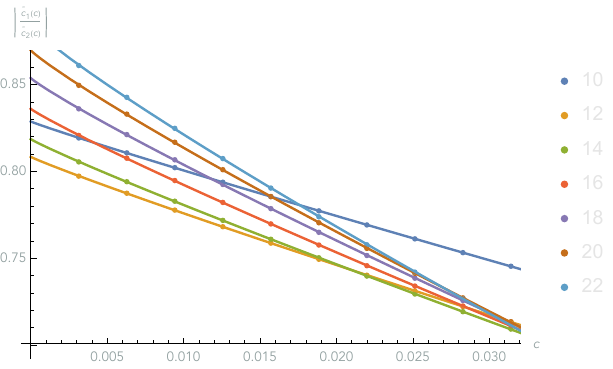}
\includegraphics[width=0.45\textwidth]{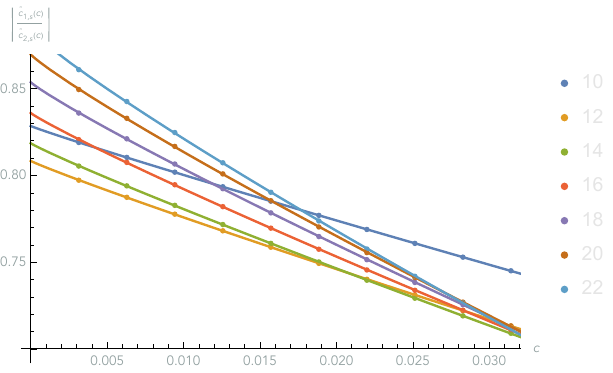}
\caption{The ratio $|\tilde c_1/\tilde c_2|$ close to $c = 0$, along with curves from Table \ref{c_ratio_parameters}.}
\label{c_ratio_parameter_plots}
\end{figure}

Focusing on $R_1$, particularly close to $c = 0$, we seek more detailed information on the behavior of the ratios $\beta/\alpha$, $\gamma/\delta$, and $\tilde c_1/\tilde c_2$, as they depend on $c$ rather than $N$. From theoretical arguments given above, we expect that
\begin{subequations}
\begin{gather}
\frac\beta\alpha = -\frac{c}{2}\frac{\tilde c_1}{\tilde c_2}, \\
\frac\gamma\delta = \frac{2}{c}\frac{\tilde c_1^*}{\tilde c_2^*},
\end{gather}
\end{subequations}
with the caveat that these relations were derived for $R_0$. Plots of $\tilde c_1/\tilde c_2$ indicate that it approaches a finite value as $c \to 0$, and decreases thereafter. For any given value of $N$, we thus use the ansatz
\begin{equation}
\left|\frac{\tilde c_1}{\tilde c_2}\right| = C_0 - A_0c^\lambda,
\end{equation}
valid close to $c = 0$. The results are given in Table \ref{c_ratio_parameters} by fitting the first 10 data points to the given curve. A comparison is given in Figure \ref{c_ratio_parameter_plots}.
\begin{table}
\centering
\subfloat[$|\tilde c_1/\tilde c_2|$]{
\begin{tabular}{cccc}
\toprule
$N$ & $C_0$ & $A_0$ & $\lambda$ \\
\midrule
$10$ & $0.828808$ & $2.21892$ & $0.947905$ \\
$12$ & $0.808445$ & $2.43479$ & $0.936954$ \\
$14$ & $0.818670$ & $2.70472$ & $0.926392$ \\
$16$ & $0.836131$ & $2.97936$ & $0.916099$ \\
$18$ & $0.853852$ & $3.24119$ & $0.906020$ \\
$20$ & $0.869871$ & $3.48412$ & $0.896121$ \\
$22$ & $0.883846$ & $3.70654$ & $0.886379$ \\
\bottomrule
\end{tabular}}
\qquad
\subfloat[$|\tilde c_{1,\text s}/\tilde c_{2,\text s}|$]{
\begin{tabular}{cccc}
\toprule
$N$ & $C_0$ & $A_0$ & $\lambda$ \\
\midrule
$10$ & $0.828551$ & $2.21891$ & $0.947901$ \\
$12$ & $0.808445$ & $2.43479$ & $0.936954$ \\
$14$ & $0.818627$ & $2.70474$ & $0.926391$ \\
$16$ & $0.836056$ & $2.97942$ & $0.916098$ \\
$18$ & $0.853772$ & $3.24127$ & $0.906018$ \\
$20$ & $0.869801$ & $3.48421$ & $0.896119$ \\
$22$ & $0.883788$ & $3.70662$ & $0.886377$ \\
\bottomrule
\end{tabular}}
\caption{Parameters in the guess $|\tilde c_1/\tilde c_2| = C_0 - A_0c^\lambda$.}
\label{c_ratio_parameters}
\end{table}

\begin{figure}
\centering
\includegraphics[width=0.45\textwidth]{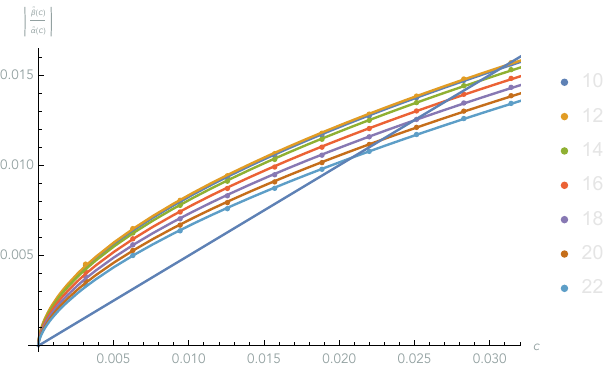}
\includegraphics[width=0.45\textwidth]{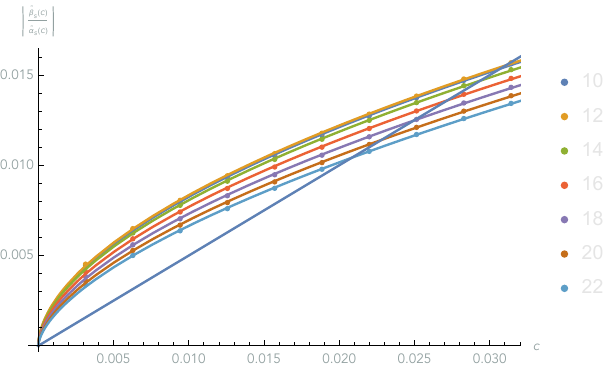}
\caption{The ratio $|\beta/\alpha|$ close to $c = 0$, along with curves from Table \ref{ba_ratio_parameters}. For comparison is the straight line $c/2$.}
\label{ba_ratio_parameter_plots}
\end{figure}

Plots of $\beta/\alpha = \tilde\beta/\tilde\alpha$ suggest a zero at $c = 0$. As before, we fit the first 10 data points to
\begin{equation}
\left|\frac\beta\alpha\right| = A_2c^{\mu_2}.
\end{equation}
These are in Table \ref{ba_ratio_parameters} and Figure \ref{ba_ratio_parameter_plots}. Based on theoretical arguments, we expect
\begin{equation}
\left|\frac\beta\alpha\right| = \frac{c}{2}\left|\frac{\tilde c_1}{\tilde c_2}\right| = \frac{C_0 c}{2} + \frac{A_0 c^{\lambda+1}}{2}.
\end{equation}
As $c \to 0$ this would suggest $\mu_2 = 1$ if $\lambda > 0$. However, what is observed is that $\mu_2 < 1$. While the convergence to $\mu_2 = 1$ seems plausible, the consequence of this is an infinite slope approaching $c = 0$ rather than linear behavior for finite $N$.
\begin{table}
\centering
\subfloat[$|\tilde\beta/\tilde\alpha|$]{
\begin{tabular}{ccc}
\toprule
$N$ & $A_2$ & $\mu_2$ \\
\midrule
$10$ & $0.105648$ & $0.553664$ \\
$12$ & $0.104027$ & $0.547802$ \\
$14$ & $0.104449$ & $0.556083$ \\
$16$ & $0.105804$ & $0.569144$ \\
$18$ & $0.107676$ & $0.583945$ \\
$20$ & $0.109775$ & $0.598920$ \\
$22$ & $0.111860$ & $0.613150$ \\
\bottomrule
\end{tabular}}
\qquad
\subfloat[$|\tilde\beta_{\text s}/\tilde\alpha_{\text s}|$]{
\begin{tabular}{ccc}
\toprule
$N$ & $A_2$ & $\mu_2$ \\
\midrule
$10$ & $0.105610$ & $0.553536$ \\
$12$ & $0.104027$ & $0.547802$ \\
$14$ & $0.104440$ & $0.556060$ \\
$16$ & $0.105785$ & $0.569095$ \\
$18$ & $0.107652$ & $0.583882$ \\
$20$ & $0.109750$ & $0.598855$ \\
$22$ & $0.111836$ & $0.613089$ \\
\bottomrule
\end{tabular}}
\caption{Parameters in the guess $|\beta/\alpha| = A_2c^{\mu_2}$.}
\label{ba_ratio_parameters}
\end{table}

\begin{figure}
\centering
\includegraphics[width=0.45\textwidth]{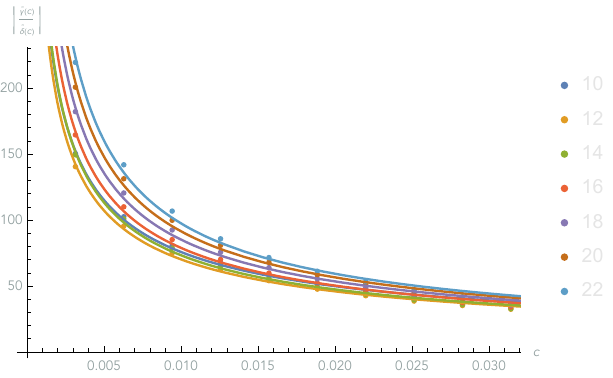}
\includegraphics[width=0.45\textwidth]{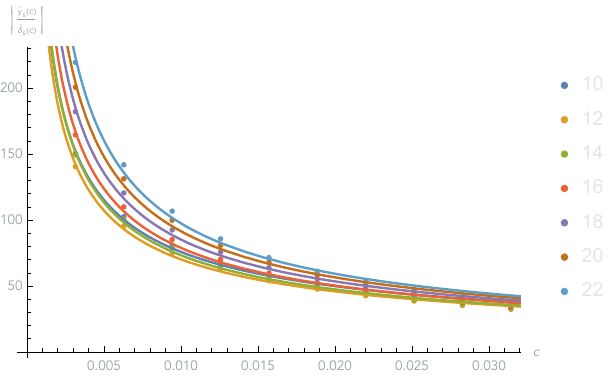}
\caption{The ratio $|\gamma/\delta|$ close to $c = 0$, along with curves from Table \ref{gd_ratio_parameters}.}
\label{gd_ratio_parameter_plots}
\end{figure}

Similarly, due to an apparent divergence, we guess
\begin{equation}
\left|\frac\gamma\delta\right| = A_1c^{\mu_1}.
\end{equation}
These are in Table \ref{gd_ratio_parameters} and Figure \ref{gd_ratio_parameter_plots}. Based on theoretical arguments, we expect
\begin{equation}
\left|\frac\gamma\delta\right| = \frac{2}{c}\left|\frac{\tilde c_1}{\tilde c_2}\right| = 2C_0c^{-1} + 2A_0 c^{\lambda-1}.
\end{equation}
If $\lambda > 0$, we thus expect $\mu_1 = -1$, but this is not observed either. However, the convergence seems plausible.

\begin{table}
\centering
\subfloat[$|\tilde\gamma/\tilde\delta|$]{
\begin{tabular}{ccc}
\toprule
$N$ & $A_1$ & $\mu_1$ \\
\midrule
$10$ & $4.72444$ & $-0.602725$ \\
$12$ & $4.21977$ & $-0.610617$ \\
$14$ & $4.06902$ & $-0.627908$ \\
$16$ & $3.98118$ & $-0.648606$ \\
$18$ & $3.87961$ & $-0.670884$ \\
$20$ & $3.74892$ & $-0.693805$ \\
$22$ & $3.59302$ & $-0.716786$ \\
\bottomrule
\end{tabular}}
\qquad
\subfloat[$|\tilde\gamma_{\text s}/\tilde\delta_{\text s}|$]{
\begin{tabular}{ccc}
\toprule
$N$ & $A_1$ & $\mu_1$ \\
\midrule
$10$ & $4.72675$ & $-0.602800$ \\
$12$ & $4.21977$ & $-0.610617$ \\
$14$ & $4.06924$ & $-0.627923$ \\
$16$ & $3.98144$ & $-0.648639$ \\
$18$ & $3.87975$ & $-0.670928$ \\
$20$ & $3.74892$ & $-0.693853$ \\
$22$ & $3.59292$ & $-0.716833$ \\
\bottomrule
\end{tabular}}
\caption{Parameters in the guess $|\gamma/\delta| = A_1c^{\mu_1}$.}
\label{gd_ratio_parameters}
\end{table}

\section{The field $\Phi_{21}$}

\begin{figure}
\centering
\includegraphics[width=\textwidth]{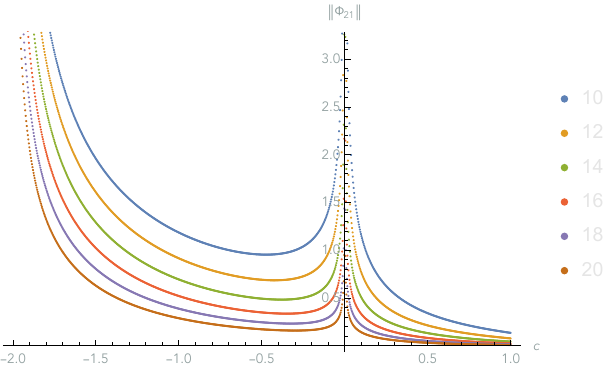}
\caption{}
\label{norm21}
\end{figure}

\begin{figure}
\centering
\includegraphics[width=\textwidth]{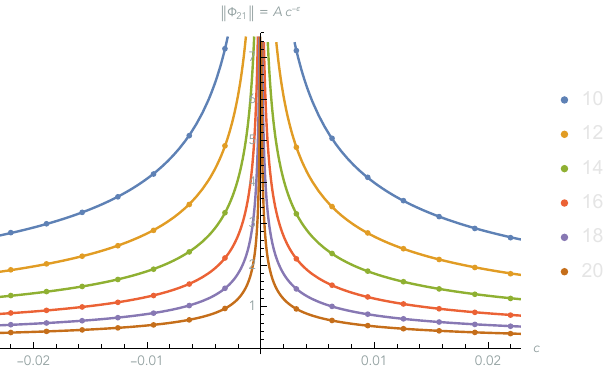}
\caption{}
\label{norm21_fit}
\end{figure}

To test some of our conjectures, we look at a situation where there is presumably no mixing---the field $\Phi_{21}$ in the momentum $N/2$ sector. When normalized to $\braket*{\Phi_{21}} = 1$, its standard 2-norm behaves as shown in Figure \ref{norm21}. Interestingly, $\Phi_{21}$ does not seem to ``know'' that other quantities behave unusually for $c = -3/5$ or $c = 1/2$. Close to $c = 0$, our analysis leads us to guess that $\|\Phi_{21}\|_2 \sim 1/\sqrt{|c|}$. By fitting the points closest to $c = 0$ ($|c| < 0.025$) to a power law, we can see if this holds. The results are given in Table \ref{norm21_power}. The quality of the fit can be inspected in Figure \ref{norm21_fit}.
\begin{table}
\centering
\subfloat[$c > 0$]{
\begin{tabular}{cc}
\toprule
$N$ & $\epsilon_+$ \\
\midrule
10 & $0.508331$ \\
12 & $0.509521$ \\
14 & $0.510640$ \\
16 & $0.511718$ \\
18 & $0.512768$ \\
20 & $0.513801$ \\
\bottomrule
\end{tabular}}
\qquad
\subfloat[$c < 0$]{
\begin{tabular}{cc}
\toprule
$N$ & $\epsilon_-$ \\
\midrule
10 & $0.491626$ \\
12 & $0.490427$ \\
14 & $0.489299$ \\
16 & $0.488212$ \\
18 & $0.487150$ \\
20 & $0.486105$ \\
\bottomrule
\end{tabular}}
\caption{Best fit parameters for $\|\Phi_{21}\|_2 = A|c|^{-\epsilon}$.}
\label{norm21_power}
\end{table}

\begin{figure}
\centering
\includegraphics[width=\textwidth]{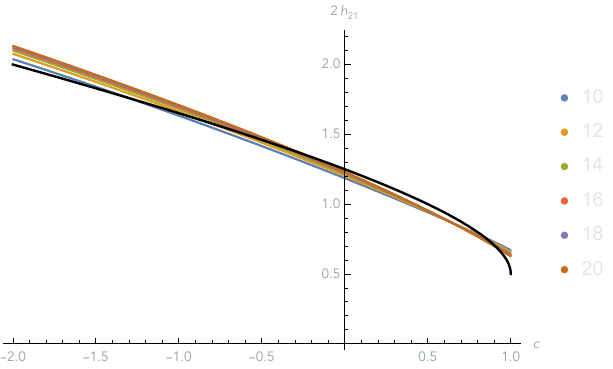}
\caption{Measurements of the eigenvalue of $\Phi_{21}$, giving the total conformal dimension. The black line is the exact value.}
\label{eig21}
\end{figure}

The eigenvalue of the diagonal field $\Phi_{21} = \phi_{21}\otimes\overbar\phi_{21}$ is expected to approach $2h_{21}$. However, inspection of Figure \ref{eig21} indicates that in some regions the data is repelled from the curve $2h_{21}$. This might signal that at finite size, the lowest eigenstate is still a mixture of $\Phi_{21}$ with some higher excitation.

\chapter*{Conclusion and outlook}
\addcontentsline{toc}{chapter}{Conclusion and outlook}

Much of the story of studying logarithmic conformal field theory has involved the use of standard methods in studying its problems, only for those methods to be thwarted by null results, followed by the development of increasingly elaborate methods. For instance, progress in the numerical observation of a number of theoretical predictions stalled for a number of years due to the typically-uninteresting issue of normalization: fields of primary importance, such as the stress--energy tensor, had zero conformal norms, and studies using the standard norm were initially not very promising. The resolution of these---the quotient expression of Eq.~\eqref{eq:b_measurement} that circumvents the issue of absolute normalization and the introduction of the concept of scaling-weak convergence---restarted progress, though each fix felt like an ad hoc approach that was limited in applicability to the problem at hand. Against this background---the search for a systematic understanding of logarithmic CFTs---the ideas of emerging Jordan blocks began to, well, emerge.

An immediate continuation of the present work is to search for an emerging rank-3 Jordan block in the continuum limits of our lattice models, whose existence has been argued for by \textcite{Gainutdinov2015} and \textcite{HeSaleur2022}. At present, the primary obstacle seems to be that the bottom field, $T\overbar T$, is difficult to identify---it is buried in a mixture with $\Phi_{31}$ (Chapter \ref{mixing}). Disentangling this mixture would lead to further understanding of the representations of the various Temperley--Lieb algebras, why this mixture occurs in $\overbar{\mathscr W}_{\!\!0,\mathfrak q^{\pm 2}}$, but not in the other standard modules (as far as we have observed).

For the field of logarithmic CFTs as a whole, \textcite{VGJS2012} suggest that it ``may be solved by a careful exercise in the representation theory of the (associative) algebras satisfied by the local energy terms (such as the Temperley--Lieb algebra).'' This exercise could likely involve the use of emerging Jordan blocks to get a gradual handle on emerging indecomposability. The fact that logarithms appear in the correlation functions of logarithmic conformal field theory is well-connected to the fact that the dilation operator $L_0$ is not diagonalizable. These logarithmic terms were initially found by subtracting divergences and taking limits, a procedure that probably also has a parallel to subtracting components along earlier vectors in the Gram--Schmidt process. Furthermore, there exist formulas for the indecomposability parameters $b$ in terms of how quickly the eigenvalues converge---a fact almost certainly related to the connection between the Jordan coupling and the rates of convergence.

I have mentioned that using the dual Jordan quantum formalism for a density matrix formulation of non-hermitian quantum spin chains could lead to the adaptation of the density matrix renormalization group technique to these problems. The completion of this undertaking would yield a wide range of useful numerical results for long spin chains, whose absence is preventing us from getting closer to the continuum limit.

The new frameworks of Jordan forms and dual Jordan quantum physics seem to be very general, and this is a very exciting prospect. Regarding the material of Part \ref{applications}, we may be able to resolve the remaining outstanding problems in a way that feels more intellectually complete. This will be no small task. We see bits and pieces of these frameworks in the ad hoc studies, and so they will need to be incorporated systematically.

If the intuition is correct that emerging Jordan blocks are lurking in such a well-studied branch of mathematics as linear differential equations (Section \ref{emerging_applications}), then the emerging Jordan block could indeed be a far-reaching concept.

\uspunctuation
\emergencystretch=1em
\printbibliography[heading=bibintoc]

\appendix
\clearpage
\phantomsection
\addcontentsline{toc}{part}{Appendices}
\part*{Appendices}
\chapter{Dynamic multiplicity adjustment in the implicitly restarted Arnoldi method} \label{DMA}

The implicitly restarted Arnoldi method is the primary algorithm in use to partially diagonalize a sparse matrix when it suffices to find only a part of the spectrum. This use case is the most commonly encountered one---indeed, in the applications described here, and in quantum physics more generally, the ground state, and a few excited states above that, are of primary interest. The implicitly restarted Arnoldi algorithm avoids the inefficiency of diagonalizing the full matrix and throwing out the unneeded eigenstates, instead directly producing the desired eigenstates and eigenvalues with minimal overhead.

The algorithm is described in Chapter \ref{computational} (Algorithm \ref{IRAM}). Intuitively, the procedure can be thought of as follows. Suppose one wants $k$ eigenvalues of an $n\times n$ matrix. For concreteness, assume that the desired eigenvalues are the lowest ones in magnitude---many criteria are possible, of course. In order to determine these lowest $k$ eigenvalues, the algorithm first determines $k+p$ approximate eigenvalues, then throws out the $p$ largest values determined in this iteration, while projecting the information about the eigenvectors associated to these (approximate) eigenvalues out of the Arnoldi basis $V_k$ (to use the notation of Algorithm \ref{IRAM}). The algorithm then calculates $p$ more approximate eigenvalues, aggregates them with the $k$ eigenvalues kept from the first step (which become refined), then throws out the $p$ largest ones from this set, and this process is repeated until convergence.

While performing some of the numerical calculations in Part \ref{applications}, the starting point was naturally the (partial) diagonalization of some Hamiltonian or transfer matrix. I began by using some black-box software, though the results were inconsistent, and on occasion, clearly incorrect. I then wrote my own implementation of the implicitly restarted Arnoldi method. On some occasions, this algorithm would work very well, and on some occasions, it would reach the maximal number of iterations without converging. Paradoxically, I found in these latter cases that increasing $k$ to a larger value would sometimes lead to convergence, even though this would seem to imply higher computational demands. 

Experimenting further, I found that asking for some values of $k$ would lead to rapid convergence, and others that would lead to non-convergence. It turned out that whenever $\lambda_k = \lambda_{k+1}$, the algorithm would not converge. Turning back to the intuitive description, this makes sense---whenever information about $\lambda_{k+1}$ is thrown out upon application of the filter, information about $\lambda_k$ is thrown out with it. Since $\lambda_k$ is one of the desired eigenvalues, it must eventually end up in the $k$ kept eigenvalues. At the same time, $\lambda_{k+1}$ is almost sure to end up in the spectrum of $H_m$, and it will get thrown out repeatedly, taking $\lambda_k$ along with it.

Most matrices naturally occurring in applications are completely non-degenerate. Indeed, the set of square matrices of a given dimension that have a repeated eigenvalue are a subset of measure zero of the set of all square matrices. Thus the implicitly restarted Arnoldi method works very well in a large majority of cases. In the problems of Part \ref{applications}, however, symmetries of the lattice models lead to patterns of degeneracies. It is similar to how rotational invariance in three dimensions leads to independence of the magnetic quantum number $m$, with a $(2j+1)$-fold degeneracy when the total angular momentum is $j$, and furthermore for the basic hydrogen atom with a Coulomb potential, conservation of the Runge--Lenz vector implies that all $n^2$ orbital states with principal quantum number $n$ are degenerate.

It would seem that in order to get the algorithm to converge for highly-degenerate matrices, one would have to specify a proper value of $k$ that does not ``split'' the eigenspace associated to a particular repeated eigenvalue. However, knowing exactly the pattern of degeneracies absent other input would essentially require the spectrum to be known, which makes the question moot. The solution is to have the algorithm dynamically adjust $k$ if it encounters a situation of multiplicity.

As mentioned, this situation can be encountered when any criterion is used to decide which eigenvalues to keep. The discussion assumed that one was looking to compute the $k$ lowest eigenvalues, but to generalize the discussion, I use a linear ordering $\succeq$ to denote the criterion determining that the (approximate) eigenvalue $\mu_1$ is more important than $\mu_2$ if and only if $\mu_1\succeq\mu_2$. The improved algorithm is given in Algorithm \ref{IRAMDMA}.

\begin{algorithm}
\KwData{$n\times n$ matrix $A$, $n$-dimensional column vector $v$, positive integers $k,p$ such that $k+p\le n$}
\KwResult{$V_k$, $H_k$, $f_k$ as in Eq.~\eqref{eq:arnoldi_k}}
$m = k + p$\;
compute a $k$-step Arnoldi factorization using Algorithm \ref{arnoldi_k}\;
\Repeat{convergence}{
beginning with the $k$-step Arnoldi factorization $(V_k, H_k, f_k)$, apply $p$ additional steps of the Arnoldi process to obtain an $m$-step Arnoldi factorization $(V_m, H_m, f_m)$ [i.e., run Algorithm \ref{arnoldi_k}, lines 2--9, with line 2 replaced by ``for $j = k$ to $k + p - 1$ do'']\;
compute $\sigma(H_m)$ and sort using $\succeq$, obtaining $\{\lambda_1,\ldots,\lambda_m\}$\;
\While{$\lambda_k = \lambda_{k+1}$}{
$k \leftarrow k+1$; $p \leftarrow p-1$\;
}
select a set of $p$ shifts $\mu_1$, $\mu_2$, $\ldots$, $\mu_p$ based on $\sigma(H_m)$ or other information\;
$Q = I_m$\;
\For{$j = 1$ \KwTo $p$}{
$(Q_j, R_j) = \text{QR}(H_m - \mu_j I_m)$\;
$H_m \leftarrow Q_j^* H_m Q_j$; $Q \leftarrow QQ_j$\;
}
$V_m \leftarrow V_mQ$; $v = V_m e_{k+1}$\;
$f_k \leftarrow v(e_{k+1}^TH_m e_k) + f_m (e_{k+p}^TQe_k)$; $V_k \leftarrow V_m(1:n,1:k)$; $H_k \leftarrow H_m(1:k,1:k)$\;
if $p$ was modified in line 7 it may optionally be reset to its original value\;
}
\caption{Implicitly restarted Arnoldi method with dynamic multiplicity adjustment}
\label{IRAMDMA}
\end{algorithm}

\chapter{Efficient Mathematica implementation of lattice field theories} \label{mathematica}
In this appendix I describe how I have implemented the loop model and the $s\ell(2|1)$ superspin chain to give efficient computations, with an eye towards generating minimal overhead. Each description is followed by sample code that can be run in Mathematica, although the descriptions should suffice to allow the reader to create analogous code in other suitable computing languages. The beginning of my implementation of these lattice models are representations of the basis in a particular sector under study. Next is a representation of the translation operator, and the construction of its eigenstates. This is followed by the construction of the Temperley--Lieb generators $e_j$, from which the Hamiltonian follows. The diagonalization of this Hamiltonian is accomplished using the methods of Chapter \ref{computational} and Appendix \ref{DMA}. Finally, I describe considerations for the efficient computation of conformal inner products, which are not needed in every calculation.

Note: in the remainder of the text the total number of sites is usually denoted by $N = 2L$, so that $N$ is even. However, since \texttt{N} is a protected symbol in Mathematica, for this appendix I replace $N$ by $L$, so that $L$ is an even number and the total length. The loop model is assumed to be periodic---it should not be terribly difficult to adapt the following considerations to the open case.

\section{Common definitions}
The user first sets values for \texttt{x} and \texttt{L}, corresponding to the parameters that determine the central charge and the total length of the spin chain (see preceding note).
\begin{Verbatim}
c=1-6/(x(x+1));
\[Gamma]=\[Pi]/(x+1);
M=-2Cos[\[Gamma]];
e\[Infinity]=
	-Sin[\[Gamma]]NIntegrate[Sinh[(\[Pi]-\[Gamma])t]/(Sinh[\[Pi] t]Cosh[\[Gamma] t]),
		{t,-\[Infinity],\[Infinity]}];
vF=\[Pi] Sin[\[Gamma]]/\[Gamma];
h[r_,s_]:=((r(x+1)-s x)^2-1)/(4x(x+1));
\end{Verbatim}
(The appearance of many of the code listings here should simplify substantially when copied and pasted into Mathematica, since symbols denoted in text by escaped brackets will condense to a single character.)

\section{Loop model}

\subsection{Construction of the basis}

There are two representations I use for the link states. First is the pairing notation described in Section \ref{standard}: $(ij)(kl)\cdots(mn)$. Second is the identification of each of these with a binary string. The correspondence is as follows: in a length $L$ binary string, whenever an index appears as the first of a pair $(ij)$ or indicates a singleton $(i)$, then the binary string has a 1 in the corresponding position. In terms of diagrams, the ones represent through-lines or the left side of a half-loop where it ``opens,'' and zeros represent the right side of a half-loop where it ``closes.'' To use the examples of Section \ref{standard}, we have
\begin{subequations}
\begin{gather}
\vcenter{\hbox{\begin{tikzpicture}
\newcommand{\dist}{0.2}
\draw[thick] (0,0) arc (-180:0:\dist);
\draw[thick] (4*\dist,0) -- (4*\dist,-2*\dist);
\draw[thick] (6*\dist,0) -- (6*\dist,-2*\dist);
\draw[thick, dotted] ($(current bounding box.north east) + (0.05+\dist,0.05)$) rectangle ($(current bounding box.south west)+ (-0.05-\dist,-0.05)$);
\end{tikzpicture}}} \leftrightarrow (12) \leftrightarrow 1011 \\
\vcenter{\hbox{\begin{tikzpicture}
\newcommand{\dist}{0.2}
\draw[thick] (4*\dist,0) arc (-180:0:\dist);
\draw[thick] (2*\dist,0) -- (2*\dist,-2*\dist);
\draw[thick] (0,0) -- (0,-2*\dist);
\draw[thick, dotted] ($(current bounding box.north east) + (0.05+\dist,0.05)$) rectangle ($(current bounding box.south west)+ (-0.05-\dist,-0.05)$);
\end{tikzpicture}}} \leftrightarrow (34) \leftrightarrow 1110 \\
\vcenter{\hbox{\begin{tikzpicture}
\newcommand{\dist}{0.2}
\draw[thick] (\dist,-\dist) arc (-90:0:\dist);
\draw[thick] (4*\dist,0) arc (-180:0:\dist);
\draw[thick] (9*\dist,-\dist) arc (-90:-180:\dist);
\draw[thick, dotted] ($(current bounding box.north east) + (0.05,0.05)$) rectangle ($(current bounding box.south west)+ (-0.05,-0.05-1*\dist)$);
\end{tikzpicture}}} \leftrightarrow (23)(41) \leftrightarrow 0101 \\
\vcenter{\hbox{\begin{tikzpicture}
\newcommand{\dist}{0.2}
\draw[thick] (2*\dist,0) arc (-180:0:\dist);
\draw[thick] (0,0) arc (-180:0:3*\dist);
\draw[thick, dotted] ($(current bounding box.north east) + (0.05+\dist,0.05)$) rectangle ($(current bounding box.south west)+ (-0.05-\dist,-0.05)$);
\end{tikzpicture}}} \leftrightarrow (14)(23) \leftrightarrow 1100 \\
\vcenter{\hbox{\begin{tikzpicture}
\newcommand{\dist}{0.2}
\draw[thick] (2*\dist,0) arc (-180:0:\dist);
\draw[thick] (0,0) arc (-180:0:3*\dist);
\draw[thick] (8*\dist,0) -- (8*\dist,-3*\dist);
\draw[thick] (10*\dist,0) -- (10*\dist,-3*\dist);
\draw[thick, dotted] ($(current bounding box.north east) + (0.05+\dist,0.05)$) rectangle ($(current bounding box.south west)+ (-0.05-\dist,-0.05)$);
\end{tikzpicture}}} \leftrightarrow (14)(23) \leftrightarrow 110011 \\
\vcenter{\hbox{\begin{tikzpicture}
\newcommand{\dist}{0.2}
\draw[thick] (2*\dist,0) arc (-180:0:\dist);
\draw[thick] (6*\dist,0) arc (-180:0:\dist);
\draw[thick] (0,0) -- (0,-2*\dist);
\draw[thick] (10*\dist,0) -- (10*\dist,-2*\dist);
\draw[thick, dotted] ($(current bounding box.north east) + (0.05+\dist,0.05)$) rectangle ($(current bounding box.south west)+ (-0.05-\dist,-0.05)$);
\end{tikzpicture}}} \leftrightarrow (23)(45) \leftrightarrow 110101
\end{gather}
\end{subequations}
The binary string can be more compactly represented using its base-10 equivalent (offset by $1$; thus note that $00\cdots00$ corresponds to $1$, a choice that can easily be reversed by the reader's own implementation), an integer between $1$ and $2^L$, and I do so interchangeably.

To convert from the binary representation to the pairing representation, I index the sites, and iteratively replace an appearance of ``10'' in the binary string with its corresponding pair using the labels, until only ones are left. This is best illustrated by example:
\begin{equation}
11001101 \to \begin{pmatrix}
11001101 \\
12345678
\end{pmatrix} \to (23)(67) \begin{pmatrix}
1011 \\
1458
\end{pmatrix} \to (23)(67)(14) \begin{pmatrix}
11 \\
58
\end{pmatrix} \to (14)(23)(67)(5)(8)
\end{equation}
If a half-loop crosses the periodic boundary, I translate the state to the left until the diagram is planar again (to ensure that the binary string will never start with a zero), determine the pairings, then add the number of sites back to the indices in the pairings. This entire procedure is accomplished by the following code in Mathematica:
\small
\begin{Verbatim}
pairings[n_]:=pairings[n]=Block[{bin=IntegerDigits[n-1,2,L],rot=0,l,m,r,s},
	If[Total[bin]<L/2,Abort[]];
While[Min[Accumulate[2bin-1]]<0,rot++;
	bin=RotateLeft[bin]];
bin=MapIndexed[{#2,#1}&,bin]//.{l___,{{r_},1},{{s_},0},m___}->{l,m,{r,s}}/.{{l_},1}->{l};
Mod[bin+rot,L,1]];
\end{Verbatim}
\normalsize
This function takes the base-10 representation of a binary string and outputs the pairing representation of that binary string. It exits if the sum of digits of the binary string is less than $L/2$ since such a binary string cannot describe a valid link state.

To construct the basis in the sector with $2j$ through lines, one simply runs the following.
\begin{verbatim}
jbasis[j_]:=Select[Range[2^L],Total[IntegerDigits[#-1,2]]==j+L/2&];
\end{verbatim}
The output of this function on a value of $j$ is a set of base-10 numbers that can be converted into the binary representations---valid link states with $2j$ more ones than zeros.

The case of $\mathscr W_{0,\mathfrak q^{\pm 2}}$ versus $\overbar{\mathscr W}_{\!\!0,\mathfrak q^{\pm 2}}$ must be handled separately. One sets a boolean variable \texttt{identify} equal to \texttt{True} if one only wants to consider states in $\overbar{\mathscr W}_{\!\!0,\mathfrak q^{\pm 2}}$, and \texttt{False} to construct the entire sector $\mathscr W_{0,\mathfrak q^{\pm 2}}$. If \texttt{identify} is \texttt{True}, then we must eliminate the extra states:
\small
\begin{verbatim}
If[identify,jbasis[0]=Select[jbasis[0],Min[Accumulate[2IntegerDigits[#-1,2,L]-1]]==0&]];
\end{verbatim}
\normalsize
This is equivalent to keeping states whose pairing representations have only increasing pairs $(ij)$ with $i < j$, but this implementation is somewhat faster since it avoids calling \texttt{pairings}.

Next, it is useful, given an integer representing a state, to determine how many through-lines it has. This is most easily achieved by taking the sum of digits of the binary string and subtracting $L/2$:
\begin{verbatim}
jval[n_]:=Total[IntegerDigits[n-1,2,L]]-L/2;
\end{verbatim}
This function takes an integer and outputs $j$, where the associated link state has $2j$ through-lines. This function only makes sense when the integer represents a valid binary string, and we will see that it is only called in those instances where we are sure that we have one.

The user must indicate which modules are of interest, in terms of the values of $j$. This is stored in a user-specified variable \texttt{jvals}, which is a set of nonnegative integers. To study the module $\overbar{\mathscr W}_{\!\!0,\mathfrak q^{\pm 2}} + \mathscr W_{11} + \mathscr W_{21}$, for instance, the value of \texttt{jvals} is set to \texttt{\{0,1,2\}} (and \texttt{identify} to \texttt{True}). If one is interested in a single module $\mathscr W_{j1}$, then \texttt{jvals} is just \texttt{\{j\}}, with a numerical value for \texttt{j}. Then the basis consists of all integers that are the base-10 representations of the appropriate binary strings:
\begin{verbatim}
basis=Join@@jbasis/@jvals;
dim=Length[basis];
dim0=If[MemberQ[0,jvals],Length[jbasis[0]],0];
Do[pos[basis[[n]]] = n, {n, dim}];
\end{verbatim}
I have additionally defined \texttt{dim} and \texttt{dim0}, the total dimension and the dimension of the $j = 0$ sector, the latter defined to be 0 if $j = 0$ is not part of \texttt{jvals}. Finally, the basis integers are given a position index from 1 to \texttt{dim}.

Finally, two more functions are useful:
\begin{verbatim}
toBinary[diagram_]:=Block[{blank=ConstantArray[0,L]},blank[[First/@diagram]]=1;blank];
projPairs[n_]:=projPairs[n]=FromDigits[toBinary[Sort/@pairings[n]],2]+1;
\end{verbatim}
The first, \texttt{toBinary}, is simply the conversion of a pairing diagram $(ij)(kl)\cdots(mn)$ to its binary representation, which starts with an array of zeros and replaces the zeros corresponding to the first indices of pairings and singletons with ones. The second, \texttt{projPairs}, is simply the map $\mathscr W_{0,\mathfrak q^{\pm 2}} \to \overbar{\mathscr W}_{\!\!0,\mathfrak q^{\pm 2}}$ in the base-10 representation. Using the first of these, it is easy to construct the action of the parity operator $P$:
\begin{verbatim}
P[n_?IntegerQ]:=pos[FromDigits[toBinary[L+1-Reverse/@pairings[basis[[n]]]],2]+1];
parity=P/@Range[dim];
P[vec_?ListQ]:=vec[[parity]];
\end{verbatim}

\subsection{Translation operator and diagonalization}
The translation operator by one site simply increases all of the indices in the pairing representation by 1, modulo $L$. In the binary representation, the entire string is simply shifted periodically by 1 as well. If \texttt{identify} is \texttt{True} and we are in the module with $j = 0$, then the result must be projected back to $\overbar{\mathscr W}_{\!\!0,\mathfrak q^{\pm 2}}$. Using the base-10 representation, I have implemented this in the following, as a translation \texttt{Tnum} and its inverse \texttt{Tinvnum}.
\begin{Verbatim}
Tnum[n_]:=Tnum[n]=Block[{val=FromDigits[RotateRight[IntegerDigits[n-1,2,L]],2]+1},
	If[identify&&jval[n]==0,projPairs[val],val]];
Do[Tinvnum[Tnum[n]]=n,{n,basis}];
\end{Verbatim}

It is also useful to consider the action of the translation operator in terms of the positions that index the basis vectors:
\begin{verbatim}
shuffle=pos/@Tinvnum/@basis;
translate=pos/@Tnum/@basis;
T[i_]:=translate[[i]];
Tinv[i_]:=shuffle[[i]];
T[i_,n_]:=Which[n>0,Nest[T,i,n],n<0,Nest[Tinv,i,-n],True,i];
\end{verbatim}
If translation takes the basis vector with index $i$ to the basis vector with index $j$, then \texttt{translate} is the ordered image of the translation operator on the ordered range from 1 to \texttt{dim} and \texttt{shuffle} is its inverse. These are useful to apply the translation operator on vectors and matrices without having to construct its matrix explicitly. The function \texttt{T} is the function $i \mapsto j$ as above, and \texttt{Tinv} its inverse. Finally, \texttt{T[i,n]} is the $n$-fold application of \texttt{T} to \texttt{i}.

Because of the translation invariance of the loop model it is useful to use translation to accomplish calculations using only a subset of the basis. I first define another set \texttt{dist}, consisting of the ``distinct'' link states whose translates generate the whole basis. This set is generated by writing \texttt{translate}, a permutation, as a product of cycles, and taking an element corresponding to one index of each cycle. For each distinct state, its order \texttt{ord} is also recorded, defined as the smallest positive integer such that translating this state this many times gives back the original state. Clearly this number divides $L$.
\begin{verbatim}
With[{cycles=PermutationCycles[translate,Identity]},dist=First/@cycles;
Evaluate[orbit/@dist]=cycles;
Evaluate[ord/@dist]=Length/@cycles];
\end{verbatim}

For each link basis state it is also useful to have at hand the number of steps away that a given site is ``linked'' by the loops, keeping in mind the periodic boundary condition. For through-lines this number is zero. For a pair such as $(16)$, the number of steps at site 1 is $5$, and the number of steps $6$ is $-5$.
\begin{Verbatim}
Block[{blank,a,b},Do[blank=ConstantArray[0,L];
	Do[If[Length[link]==2,{a,b}=link;
		If[link==Sort[link],
			blank[[a]]=b-a;blank[[b]]=-blank[[a]],
			blank[[a]]=b-a+L;blank[[b]]=-blank[[a]]],
		blank[[link]]={0}];steps[i]=blank,{link,pairings[basis[[i]]]}],{i,dim}]];
\end{Verbatim}

Since we know the distinct basis elements and the order of each, it is easy to construct the basis that diagonalizes the translation operator $\tau$. The eigenvalues of $\tau$ are $\{\e^{2\pi\i k /L}\}$, where $k$ is any consecutive range of integers of length $L$. For convenience, each of these eigenvalues will be referenced just using the value of $k$ in the exponent. To construct the basis diagonalizing $\tau$, we begin by taking the distinct vectors and taking linear combinations of their translates, multiplied by phases. For example, to construct $\ket{i,k}$, the eigenvector of momentum $k$ built upon basis vector $i$,
\begin{equation}
\ket{i,k} \sim \sum_{m=0}^{L-1} \e^{-2\pi\i k m/L}\tau^m\ket{i}.
\end{equation}
However, it must be checked that this vector actually belongs to the space associated to the eigenvalue $k$. It turns out that this is the case if and only if $\mathrm{ord}(i) k/L \in \mathbb Z$. When this is the case, the normalized eigenvector is given by
\begin{equation}
\ket{i,k} = \frac{1}{\sqrt{\mathrm{ord}(i)}}\sum_{m=0}^{\mathrm{ord(i)-1}} \e^{-2\pi\i k m/L}\tau^m\ket{i}.
\end{equation}
This basis is accomplished with the following code. \texttt{kbasis} determines which elements of $\texttt{dist}$ give rise to a nonzero vector given a value of $k$, and \texttt{kvec} constructs the vector above. Finally, \texttt{kvecs} creates a matrix whose \emph{rows} are the constructed eigenvectors.
\begin{Verbatim}
kbasis[k_]:={#,k}&/@Select[dist,IntegerQ[ord[#]k/L]&];
kvec[{n_,k_}]:=kvec[{n,k}]=Block[{rules},
	rules=Table[{orbit[n][[m+1]]}->E^(-2\[Pi] I k m/L),{m,0,ord[n]-1}];
	1/Sqrt[ord[n]]SparseArray[rules,dim]];
kvecs[k_]:=SparseArray[kvec/@kbasis[k]];
\end{Verbatim}

\subsection{Construction of the Temperley--Lieb generators}
The construction of the operators $e_j$ is most easily done using the pairing representation. The action of $e_j$ on a basis state yields the pair $(j,j+1)$ in the result. Then an examination of cases in which loops can possibly connect to sites $j$ and $j+1$ gives the rest of the state. The user must decide whether to allow contraction of through-lines, depending on the module under study. If it is one of the standard modules, then a boolean variable named \texttt{irreducible} should be set to \texttt{True}, and if it is a glued module, then it should be \texttt{False}. There is a closed loop formed upon the action of $e_j$ when either $(j,j+1)$ or $(j+1,j)$ is present in the pairing representation of the basis state. The preceding considerations suffices to construct all of the generators $e_j$. To save computation time, however, I construct only $e_1$ and use translation to generate the other generators.
\begin{Verbatim}
eTL[1,m_]:=eTL[1,m]=Block[{step1=steps[m][[1]],step2=steps[m][[2]],val,
	return,n=basis[[m]],bin},
	bin=IntegerDigits[basis[[m]]-1,2,L];
	bin[[{1,2}]]={1,0};
	return=Mod[Which[step1==0&&step2==0,{},step1==0,{2+step2},
		step2==0,{1+step1},step2-step1+1>0,{1+step1,2+step2},
		True,{2+step2,1+step1}],L,1];
	bin[[return]]=Take[{1,0},Length[return]];
	val=FromDigits[bin,2]+1;
	If[irreducible&&jval[val]!=jval[n],0,
		If[identify&&jval[val]==0,pos[projPairs[val]],pos[val]]]];
loopfactor[j_,m_]:=loopfactor[j,m]=Boole[steps[m][[j]]==1||steps[m][[j]]==1-L];
e[1]=SparseArray[Table[If[eTL[1,i]!=0,{eTL[1,i],i}->M^loopfactor[1,i],
	Nothing],{i,dim}],{dim,dim}];
e[j_]:=e[j]=e[j-1][[shuffle,shuffle]];
\end{Verbatim}

From these the Hamiltonian, the Hamiltonian when restricted to the sector of momentum $k$, and Koo--Saleur generators follow:
\begin{Verbatim}
Htl=(Sum[e[i],{i,L}]-L e\[Infinity] IdentityMatrix[dim,SparseArray])L/(2\[Pi] vF);
Hk[k_]:=Hk[k]=Simplify[Conjugate[kvecs[k]].Htl.Transpose[kvecs[k]]];
Hv[n_Integer,vec_List]:=
	-L/(2\[Pi] vF)Sum[(e[j].vec-e\[Infinity] vec)E^(2\[Pi] I n j/L),{j,L}]
		+c/12 vec KroneckerDelta[n,0];
\end{Verbatim}

\subsection{Inner products}

I begin with some practical aspects to consider when computing the loop inner product. As mentioned in Section \ref{c_0}, almost all inner products are nonzero, in contrast with, say, the standard inner product in an orthonormal basis. This makes it very time-consuming to compute inner products between all pairs.

The standard inner product is easy to calculate on a computer, since it involves only multiplication and addition of complex numbers, with $d$ total terms, $d$ being the dimension of the space (\texttt{dim} in the code listings). The loop inner product is much more computationally expensive, since not only are there $d^2$ terms, but for each pair of basis vectors one must do a ``geometric'' computation to count the number of loops (and also the number of steps the defect lines move in the case where $z^2$ is different from $1$). It is therefore desirable to cut down on as many of these geometric computations as possible.

Hereafter, I will abbreviate $\tau v_i = v_j$ to $\tau i = j$; this means that if $v_1,\ldots,v_d$ is an enumeration of the basis vectors, then translating vector $i$ results in vector $j$. This is helpful because on linear combinations, $(\tau\psi)_i = \psi_{\tau^{-1}i}$.

One helpful observation is that the inner product between basis vectors is obviously symmetric: $\ip*{i}{j} = \ip*{j}{i}$. This cuts down the number of computations by about a factor of 2. But we can do better.

The inner product is invariant under translation: $\ip*{\tau i}{\tau j} = \ip*{i}{j}$. We have $S$, a generating set for the basis under $\tau$ (this is not quite the same as \texttt{dist} in the code listings, but rather \texttt{pos/@dist}). That is,
\begin{equation}
\bigcup_{n=0}^{L-1}\tau^nS = \{1,2,\ldots,d\}.
\end{equation}
For each $i\in S$, we have already computed $\mathrm{ord}(i)$, the order of $i$ under translation. (Here I am abusing notation by identifying the position index of a basis vector with its base-10 representation.) We then have
\begin{equation}
\sum_{i=1}^d f(i) = \sum_{i\in S}\sum_{n=0}^{\mathrm{ord}(i)-1}f(\tau^n i).
\end{equation}
Applying this to the loop inner product of vectors $\phi$ and $\psi$,
\begin{equation}
\begin{aligned}
\ip*{\phi}{\psi} &= \sum_{i=1}^d\sum_{j=1}^d \phi_i^*\psi_j\ip*{i}{j} = \sum_{i\in S}\sum_{n=0}^{\mathrm{ord}(i)-1}\sum_{j=1}^d \phi_{\tau^n i}^*\psi_j\ip*{\tau^n i}{j} \\
&= \sum_{i\in S}\sum_{n=0}^{\mathrm{ord}(i)-1}\sum_{j=1}^d \phi_{\tau^n i}^*\psi_j\ip*{i}{\tau^{-n}j} = \sum_{i\in S}\sum_{n=0}^{\mathrm{ord}(i)-1}\sum_{j=1}^d \phi_{\tau^n i}^*\psi_{\tau^n j}\ip*{i}{j}.
\end{aligned}
\end{equation}
It is therefore only necessary to compute those basis inner products in which the first state belongs to $S$. In practice, $|S|$ is slightly larger than $d/L$ (which is not necessarily an integer), reflecting the fact that most states in $S$ have order $L$. Thus we have reduced the number of basis inner product computations by a factor of about $L$.

The following function counts the number of loops when two link states are sandwiched together.
\begin{Verbatim}
ipLoops[i_,j_]:=Block[{f,sign,start,startsign,a,b,k,loop=0,points=Range[L]},
	f[-1]=steps[i];
	f[1]=steps[j];
	While[points!={},a=First[points];
	sign=1;
	start=a;
	points=Rest[points];
	Do[b=Mod[a+f[sign][[a]],L,1];
	points=Complement[points,{b}];
	If[b==a,Break[]];
	If[b==start&&sign==-1,loop++;Break[]];
	a=b;
	sign=-sign,L]];loop];
\end{Verbatim}

Then, the computation of the inner product, using the reduction in the number of calculations above, is given by the following, when contraction of through-lines is allowed and $z^2 = 1$.
\begin{Verbatim}
distLoopAssociation=Association[ParallelTable[i->(ipLoops[i,#]&/@Range[dim]),
	{i,Union[dist,Range[dim0]]}]];
ip[u_,v_]:=Block[{tot=0,u2=Conjugate[u]},
Do[Block[{vec=M^distLoopAssociation[i],v2},
v2=v;
Do[v2=v2[[translate]];
tot+=u2[[T[i,m]]]vec.v2,{m,ord[i]}]],{i,dist}];
tot];
\end{Verbatim}

In the glued modules, when $c = 0$ and $z^2 = 1$ things would seem to be easy. $m = 1$ so that $\ip*{i}{j} = 1$ for all $i$ and $j$. Thus
\begin{equation}
\ip*{\phi}{\psi} = \sum_{i=1}^d \sum_{j=1}^d \phi_i^*\psi_j \ip*{i}{j} = \Big(\sum_{i=1}^d\phi_i^*\Big)\Big(\sum_{j=1}^d \psi_j\Big).
\end{equation}
Here one simply adds up the components of each vector and multiplies the two sums together. 

This simple picture changes once one introduces a parameter $y$ (usually when $c = 0$, as that is the only case where the introduction of $y$ controls the appearance of Jordan blocks) that arises each time two strings are contracted, with the first one living on an even site and the second one on an odd site (Appendix \ref{y_parameter}). This breaks the symmetry of the original problem in two ways: first is that the system is no longer translationally invariant, and second is that a chirality is introduced into the system by giving it a preferred direction.

\subsection{The contraction parameter $y$}

The following algorithms allow one to study the glued modules with a non-unity value of $y$.

Importantly, since the system is no longer invariant under translations by one site, but only translations by two sites, the translation operator must be modified appropriately. I retain \texttt{T} to effect translations by one site, and use \texttt{U} as the translation operator by two sites. \texttt{U} is most easily defined by the action of \texttt{T} twice.
\begin{verbatim}
Unum[n_]:=Tnum[Tnum[n]];
Uinvnum[n_]:=Tinvnum[Tinvnum[n]];
shuffle=pos/@Uinvnum/@basis;
translate=pos/@Unum/@basis;
U[i_]:=translate[[i]];
Uinv[i_]:=shuffle[[i]];
U[i_,n_]:=Which[n>0,Nest[U,i,n],n<0,Nest[Uinv,i,-n],True,i];
\end{verbatim}

There are some changes in constructing the basis that diagonalizes \texttt{U}.
\begin{Verbatim}
kbasis[k_]:={#,k}&/@Select[dist,IntegerQ[2ord[#]k/L]&];
kvec[{n_,k_}]:=kvec[{n,k}]=Block[{rules},
	rules=Table[{orbit[n][[m+1]]}->E^(-4\[Pi] I k m/L),{m,0,ord[n]-1}];
	1/Sqrt[ord[n]]SparseArray[rules,dim]];
\end{Verbatim}

The introduction of $y$ also implies a modification to the even generators $e_j$. It becomes necessary to construct both $e_1$ and $e_2$, and the rest can be generated again by translation.
\begin{Verbatim}
yfactor[j_/;OddQ[j],m_]:=0;
yfactor[j_,m_]:=yfactor[j,m]=Boole[steps[m][[j]]==0&&steps[m][[Mod[j+1,L,1]]]==0];
e[1]=SparseArray[Table[If[eTL[1,i]!=0,{eTL[1,i],i}->M^loopfactor[1,i],
	Nothing],{i,dim}],{dim,dim}];
e[2]=SparseArray[Table[If[eTL[1,Tinv[i]]!=0,
	{T[eTL[1,Tinv[i]]],i}->M^loopfactor[1,Tinv[i]] y^yfactor[2,i],
	Nothing],{i,dim}],{dim,dim}];
e[j_]:=e[j]=e[j-2][[shuffle,shuffle]];
\end{Verbatim}

Finally, I return to some considerations for computing the inner product, now with a non-unity value of $y$. In $\overbar{\mathscr W}_{\!\!0,\mathfrak q^{\pm 2}}$ there are no defect lines so that the calculation of $\ip*{i}{j}$ remains the same. At $c = 0$, $\ip*{i}{j} = 1$.

Let $U = \tau^2$ denote translations to the right by two sites. It is easily seen that $\ip*{Ui}{Uj} = \ip*{i}{j}$ when both $i$ and $j$ are in $\overbar{\mathscr W}_{\!\!0,\mathfrak q^{\pm 2}}$, or when both $i$ and $j$ are in $\mathscr W_{j1}$ (henceforth collectively abbreviated to $\mathscr W = \bigcup_j \mathscr W_{j1}$, where the union is over all $j$ values of interest [typically $j = 1$ and $2$]). Simple counterexamples show that the factor $y$ and the projection $\mathscr W_{0,\mathfrak q^{\pm 2}} \to \overbar{\mathscr W}_{\!\!0,\mathfrak q^{\pm 2}}$ are responsible for the fact that $\ip*{Ui}{Uj}$ need not equal $\ip*{i}{j}$ when $i\in\overbar{\mathscr W}_{\!\!0,\mathfrak q^{\pm 2}}$ and $j\in\mathscr W$.

For an inner product $\ip*{\phi}{\psi}$, break up the sum into four pieces:
\begin{equation}
\begin{aligned}
\ip*{\phi}{\psi} &= \sum_{i=1}^d\sum_{j=1}^d \phi_i^* \psi_j \ip*{i}{j} = \Big(\sum_{i\in\overbar{\mathscr W}_{\!\!0,\mathfrak q^{\pm 2}}} + \sum_{i\in\mathscr W}\Big)\Big(\sum_{j\in\overbar{\mathscr W}_{\!\!0,\mathfrak q^{\pm 2}}} + \sum_{j\in\mathscr W}\Big)\phi_i^* \psi_j \ip*{i}{j} \\
&= \sum_{i\in\overbar{\mathscr W}_{\!\!0,\mathfrak q^{\pm 2}}}\sum_{j\in\overbar{\mathscr W}_{\!\!0,\mathfrak q^{\pm 2}}}\phi_i^* \psi_j \ip*{i}{j} + \sum_{i\in\overbar{\mathscr W}_{\!\!0,\mathfrak q^{\pm 2}}}\sum_{j\in\mathscr W}\phi_i^* \psi_j \ip*{i}{j} + \sum_{i\in\mathscr W}\sum_{j\in\overbar{\mathscr W}_{\!\!0,\mathfrak q^{\pm 2}}}\phi_i^* \psi_j \ip*{i}{j} + \sum_{i\in\mathscr W}\sum_{j\in\mathscr W}\phi_i^* \psi_j \ip*{i}{j} \\
&= \sum_{i\in\overbar{\mathscr W}_{\!\!0,\mathfrak q^{\pm 2}}}\phi_i^* \sum_{j\in\overbar{\mathscr W}_{\!\!0,\mathfrak q^{\pm 2}}} \psi_j + \sum_{i\in\overbar{\mathscr W}_{\!\!0,\mathfrak q^{\pm 2}}}\sum_{j\in\mathscr W}(\phi_i^* \psi_j + \phi_j^* \psi_i) \ip*{i}{j} + \sum_{i\in\mathscr W}\sum_{j\in\mathscr W}\phi_i^* \psi_j \ip*{i}{j}.
\end{aligned}
\end{equation}
The last equality holds only at $c = 0$: the first term factors into two pieces if $\ip*{i}{j} = 1$. The two ``off-diagonal'' terms are reindexed and combined into one sum, using the symmetry $\ip*{i}{j} = \ip*{j}{i}$. Finally, we can handle the last piece as before, with some modifications. Let $S$ be a generating set for $\mathscr W$, except now under $U$ instead of $\tau$. Similarly, $\mathrm{ord}(i)$ for $i\in S$ is the smallest integer $n$ for which $U^ni = i$. Thus
\begin{equation}
\ip*{\phi}{\psi} = \sum_{i\in\overbar{\mathscr W}_{\!\!0,\mathfrak q^{\pm 2}}}\phi_i^* \sum_{j\in\overbar{\mathscr W}_{\!\!0,\mathfrak q^{\pm 2}}} \psi_j + \sum_{i\in\overbar{\mathscr W}_{\!\!0,\mathfrak q^{\pm 2}}}\sum_{j\in\mathscr W}(\phi_i^* \psi_j + \phi_j^* \psi_i) \ip*{i}{j} + \sum_{i\in S}\sum_{n=0}^{\mathrm{ord}(i)-1}\sum_{j\in\mathscr W} \phi_{U^n i}^*\psi_{U^n j}\ip*{i}{j}.
\end{equation}
Thus the only inner products that need to be computed are those between $\overbar{\mathscr W}_{\!\!0,\mathfrak q^{\pm 2}}$ and $\mathscr W$, and those between $S$ and $\mathscr W$. The reduction is not as great as in the $y = 1$ case since $U$ has smaller orbits than $u$---generically they are half the size.

The following function counts the number of times a factor of $y$ must be introduced when two link states are sandwiched together to compute their inner product.
\begin{verbatim}
evens=Range[2,L,2];
evenLines[j_]:=evenLines[j]=Pick[evens,steps[j][[evens]],0];
evenLines/@Range[dim];
ipyfactor[i_,j_]:=Block[{f,sign,shift,start,startsign,a,b,m,n,count=0},
{n,m}=Sort[{i,j}];
f[-1]=steps[m];
f[1]=steps[n];
Do[b=start=k;
shift=0;
sign=startsign=1;
Do[a=f[sign][[b]];
If[a==0,If[sign==-startsign,If[shift>0,count+=1]];Break[]];
b=Mod[b+a,L,1];
shift+=a;
sign=-sign,L],{k,Complement[evenLines[m],evenLines[n]]}];
Do[b=start=k;
shift=0;
sign=startsign=-1;
Do[a=f[sign][[b]];
If[a==0,If[sign==-startsign,If[shift>0,count+=1]];Break[]];
b=Mod[b+a,L,1];
shift+=a;
sign=-sign,L],{k,Complement[evenLines[n],evenLines[m]]}];
count];
\end{verbatim}

Finally, the computation of the inner product, following the considerations above, is as follows. Note that an additional optimization has been made to put the vector with fewer nonzero components on the left, so that fewer computations are needed.
\begin{Verbatim}
distyAssociation=Association[ParallelTable[i->(ipyfactor[i,#]&/@Range[dim]),
	{i,Union[dist,Range[dim0]]}]];
dist0=Intersection[dist,Range[dim0]];
dist12=Complement[dist,Range[dim0]];
ip[u_,v_]:=Block[{upos=Intersection[Union[Flatten[NestList[U,Flatten[
	SparseArray[u]["NonzeroPositions"]],L/2-1]]],
	Union[Range[dim0],dist12]],vpos=Intersection[Union[Flatten[NestList[U,Flatten[
	SparseArray[v]["NonzeroPositions"]],L/2-1]]],
	Union[Range[dim0],dist12]],short,long,shortpos,longpos,u0,u1,v0,v1,tot=0,
	translate0=Take[translate,dim0],translate1=Drop[translate,dim0]-dim0,v2,a},
	If[Length[upos]<Length[vpos],{short,shortpos,long,longpos}={u,upos,v,vpos},
	{short,shortpos,long,longpos}={Conjugate[v],vpos,Conjugate[u],upos}];
	u0=Conjugate[Take[short,dim0]];u1=Conjugate[Drop[short,dim0]];
	v0=Take[long,dim0];v1=Drop[long,dim0];
	Do[Block[{vec},
		vec=M^Take[distLoopAssociation[i],dim0] y^Take[distyAssociation[i],dim0];
	v2=v0;
	Do[v2=v2[[translate0]];
	tot+=u0[[U[i,m]]]vec.v2,{m,ord[i]}]],{i,dist0}];
	tot+=Sum[u0[[i]]((M^Drop[distLoopAssociation[i],dim0] 
		y^Drop[distyAssociation[i],dim0]).v1),{i,Intersection[shortpos,Range[dim0]]}];
	tot+=Sum[v0[[i]]((M^Drop[distLoopAssociation[i],dim0] 
		y^Drop[distyAssociation[i],dim0]).u1),{i,Intersection[longpos,Range[dim0]]}];
	Block[{vec},
	Do[vec=M^Drop[distLoopAssociation[i],dim0] y^Drop[distyAssociation[i],dim0];
	v2=v1;
	Do[v2=v2[[translate1]];
	tot+=u1[[U[i,m]-dim0]]vec.v2,{m,ord[i]}],{i,Intersection[shortpos,dist12]}]];
	tot];
\end{Verbatim}

\section{$s\ell(2|1)$ superspin chain}

The $s\ell(2|1)$ superspin chain is much simpler to implement computationally than the loop model. However, the drawback is that its dimension grows much more quickly with $L$.

\subsection{Construction of the basis}

The computational basis can easily be identified with ternary strings of length $L$. As with the loop model, these ternary strings can be identified with their base-10 equivalents, again with an offset of 1. If the user specifies $L$, $B$, and $Q^z$, then one may construct the basis of that sector as follows.
\begin{Verbatim}
Bnum[n_]:=Block[{str=IntegerDigits[n-1,3,L],k},
	Sum[If[str[[k]]==2,(-1)^k,(-1)^k/2],{k,1,L}]];
Qznum[n_]:=Block[{str=IntegerDigits[n-1,3,L],k},
	Sum[If[str[[k]]==2,0,(1/2-str[[k]])(-1)^k],{k,1,L}]];
sector[B_,Qz_]:=Select[Range[3^L],Bnum[#]==B&&Qznum[#]==Qz&];
basis=sector[B,Qz];
Do[pos[j basis[[i]]]=j i,{i,1,Length[basis]},{j,{-1,1}}];
dim=Length[basis];
\end{Verbatim}
With an eye on what follows, the basis has been stored as a positively indexed version and a negatively indexed version, with a corresponding sign for the basis element. For instance, if \texttt{pos[m]} gives \texttt{n} then \texttt{pos[-m]} gives \texttt{-n}.

One may also construct the basis corresponding to an invariant submodule of the $B = Q^z = 0$ sector, as described at the end of Section \ref{sl21_section}.
\begin{Verbatim}
Block[{old={},new=FromDigits[#,3]&/@Table[ConstantArray[i,L],{i,0,2}]+1,mid,str,vec},
	While[old!=new,
	mid=Complement[new,old];
	old=new;
	new=Reap[Do[str=IntegerDigits[j-1,3,L];
	Do[If[str[[i]]==str[[i+1]],
	vec=str;
	Do[vec[[i]]=b;
	vec[[i+1]]=b;
	Sow[FromDigits[vec,3]+1];,{b,Complement[{0,1,2},{str[[i]]}]}];],{i,1,L-1}];
	If[str[[L]]==str[[1]],vec=str;
	Do[vec[[L]]=b;
	vec[[1]]=b;
	Sow[FromDigits[vec,3]+1];,{b,Complement[{0,1,2},{str[[L]]}]}];],{j,mid}];][[2,1]];
	new=Union[new,old]];
	basis=new];
\end{Verbatim}

\subsection{Translation operator and diagonalization}
Because of the alternating representations, the fundamental translation operator for this system is the translation by two sites, $U = \tau^2$. If $c_{ij}^\dagger$ is any of the bosonic or fermionic creation operators for site $i$, then
\begin{equation}
Uc^\dagger_{1j_1} c^\dagger_{2j_2}\cdots c^\dagger_{Lj_L}\ket*{0} = Uc^\dagger_{1j_1} c^\dagger_{2j_2}\cdots c^\dagger_{Lj_L}U^{-1}U\ket*{0} = c^\dagger_{3j_1} c^\dagger_{4j_2}\cdots c^\dagger_{Lj_{L-2}} c^\dagger_{1j_{L-1}} c^\dagger_{2j_L} \ket*{0}
\end{equation}
defines the action of $U$ on the computational basis. That is, $U$ shifts all of the site indices on the creation operators by 2, but does not change the type of particle created. Note, however, that the expression on the right side of the equation isn't necessarily equal to an element of the computational basis. To make this expression comparable to the computational basis, one must use the commutation relations to move $c^\dagger_{1j_{L-1}} c^\dagger_{2j_L}$ all the way to the left, so that the sites are in order again. If either of these are fermionic operators, then there are potential signs that arise. Thus, the action of $U$ on a computational basis state is again another computational basis state, up to a sign. This is represented schematically as $Ui = \pm j$, where the sign is to be determined. In Mathematica, this is accomplished with the following function.
\begin{Verbatim}
Unum[0]=0;
Unum[n_]:=Unum[n]=Block[{bas=IntegerDigits[basis[[Abs[n]]]-1,3,L],
	fcount1,fcount2,rot},
	fcount1=Count[Drop[bas,-2],2];
	fcount2=Count[Take[bas,-2],2];
	rot=FromDigits[RotateRight[bas,2],3]+1;
	Sign[n](-1)^(fcount1 fcount2) pos[rot]];
	Do[Uinvnum[Unum[n]]=n,{n,-dim,dim}];
	U[vec_]:=Block[{new=ConstantArray[0,dim]},
	Do[new[[Abs[Unum[n]]]]=Sign[Unum[n]]vec[[n]],{n,1,dim}];
	new];
\end{Verbatim}
One may check that $U$ applied $L/2$ times gives the identity operator.

As before, it is useful to find the distinct elements that generate the whole basis under translation, and to use them to construct the basis that diagonalizes $U$.
\begin{Verbatim}
split=GatherBy[(IntegerDigits[#-1,3,L]&/@basis),Sort[Tally[#]]&];
split=Flatten[GatherBy[#,Sort[Tally[ListConvolve[{1,-1},#,1]]]&]&/@split,1];
split=Flatten[GatherBy[#,Sort[Tally[ListConvolve[{1,1},#,1]]]&]&/@split,1];
Do[splitBasis[i]=1+FromDigits[#,3]&/@split[[i]],{i,1,Length[split]}];
dist=Reap[Do[Block[{track=splitBasis[i]},
	While[Length[track]>0,
	Block[{n,abs,signs},
	n=pos[track[[1]]];
	Sow[n];
	orbit[n]=NestList[Unum,n,L/2-1];
	track=Complement[track,basis[[Abs[orbit[n]]]]];
	orbit[n]=DeleteDuplicatesBy[orbit[n],Abs];
	ord[n]=Length[orbit[n]];
	torsion[n]=Unum[orbit[n][[ord[n]]]]/n;]]],{i,1,Length[split]}]][[2,1]];
dist=Sort[dist];
kbasis[k_]:=kbasis[k]=Reap[Do[
	If[(IntegerQ[2ord[n]k/L]&&torsion[n]==1)||((OddQ[L/(2ord[n])]||OddQ[4ord[n]k/L])
	&&torsion[n]==-1),Sow[{n,k}]],{n,dist}]][[2,1]];
kvec[{n_,k_}]:=kvec[{n,k}]=Block[{rules},
	rules=Table[{Abs[orbit[n][[m+1]]]}->Sign[orbit[n][[m+1]]]E^(-4\[Pi] I k m/L),
	{m,0,ord[n]-1}];
	1/Sqrt[ord[n]]SparseArray[rules,dim]];
kvecs[k_]:=kvecs[k]=SparseArray[kvec/@kbasis[k]];
\end{Verbatim}

\subsection{Construction of the Temperley--Lieb generators}

Following the explicit expressions of the generators $e_j$, the fact that each $e_j$ replaces pairs of identical neighboring particles with linear combinations of the same indicates that it is useful to work in the ternary representation. The following code creates Temperley--Lieb generators $e_1$, $e_2$, and $e_L$ (denoted as \texttt{h} in the following), and uses translation by two sites to generate the remainder.

\begin{Verbatim}
Do[h[i]=Block[{rules,fac},
	rules={};
	rules=Reap[Do[str=IntegerDigits[basis[[j]]-1,3,L];
	If[str[[i]]==str[[i+1]],
	bos1=str;
	bos1[[i]]=0;
	bos1[[i+1]]=0;
	bos2=str;
	bos2[[i]]=1;
	bos2[[i+1]]=1;
	fer=str;
	fer[[i]]=2;
	fer[[i+1]]=2;
	bos1=FromDigits[bos1,3]+1;
	bos2=FromDigits[bos2,3]+1;
	fer=FromDigits[fer,3]+1;
	bos1=pos[bos1];
	bos2=pos[bos2];
	fer=pos[fer];
	If[str[[i]]==2,fac=(-1)^(i+1),fac=1];
	Sow[{{bos1,j}->fac,{bos2,j}->fac,{fer,j}->(-1)^i fac}];];,{j,1,dim}]][[2]];
	rules=Flatten[rules];
	SparseArray[rules,dim]];,{i,{1,2}}];
h[L]=Block[{rules,fac,f},
	rules={};
	rules=Reap[Do[str=IntegerDigits[basis[[j]]-1,3,L];
	If[str[[L]]==str[[1]],
	bos1=str;
	bos1[[L]]=0;
	bos1[[1]]=0;
	bos2=str;
	bos2[[L]]=1;
	bos2[[1]]=1;
	fer=str;
	fer[[L]]=2;
	fer[[1]]=2;
	bos1=FromDigits[bos1,3]+1;
	bos2=FromDigits[bos2,3]+1;
	fer=FromDigits[fer,3]+1;
	bos1=pos[bos1];
	bos2=pos[bos2];
	fer=pos[fer];
	f=Count[Rest[Most[str]],2];
	If[str[[L]]==2,fac=(-1)^f,fac=1];
	Sow[{{bos1,j}->fac,{bos2,j}->fac,{fer,j}->(-1)^(f+1) fac}];];,{j,1,dim}]][[2]];
	rules=Flatten[rules];
	SparseArray[rules,dim]];
shuffle=Abs[Uinvnum/@Range[dim]];
Do[h[i]=h[i-2][[shuffle,shuffle]],{i,3,L-1}];
h[i_]:=h[Mod[i,L,1]];
Htl=-Sum[h[j],{j,1,L}]+L IdentityMatrix[dim,SparseArray];
Hv[n_,vec_]:=-L/(2\[Pi] vF)Sum[(h[j].vec-vec)E^(2\[Pi] I n j/L),{j,1,L}];
\end{Verbatim}

\subsection{Inner products}

The $s\ell(2|1)$ inner product is computationally very simple, since different basis states are orthogonal and the inner product of a basis state with itself is $\pm 1$. Its calculation is described in Section \ref{sl21_section}, and implemented as follows.
\begin{verbatim}
ipbas[i_]:=ipbas[i]=(-1)^Total[Flatten[Position[IntegerDigits[basis[[i]]-1,3,L],2]]];
ips=ipbas/@Range[dim];
ip[v_,w_]:=Conjugate[v].(ips w);
\end{verbatim}

\chapter{Direct calculation of the subquotient structure of Temperley--Lieb algebra representations} \label{subquotient}

In \textcite{Gainutdinov2015} the subquotient structure of $JTL(1)$-modules for various numbers of sites $N$ was presented based on algebraic arguments. In this appendix I describe a method to compute this subquotient structure directly and thus corroborate their conclusions from a different approach.

The identity module containing the ground state can be constructed following an algorithm presented in Section \ref{sl21_section}, where starting from a computational basis state along which the ground state is expected to have a nonzero component, act with the $JTL(1)$ generators $e_i$ repeatedly until no new states are generated. Then the ground state can be found by constructing the Hamiltonian within this subspace.

We may generalize this idea to determine the $JTL(1)$ structure of various indecomposable modules. Start with the Hamiltonian restricted to a particular module. Its generalized right eigenstates are denoted by column vectors $v_i$. For each $i$, find the $JTL(1)$ submodule generated by $v_i$. Explicitly, start with the vector $v_i$, and apply to it all the $e_j$ operators and $u^2$, yielding vectors $v_i,e_1v_i, e_2v_i,\ldots,e_L v_i, u^2v_i$. Gather them all into a matrix $[v_i;e_1v_i; e_2v_i;\cdots;e_L v_i;u^2v_i]$ and column-reduce it. Keep the nonzero columns and repeat the procedure, applying $e_j$ and $u^2$ to each column, then gathering them all and column-reducing, until the number of vectors stabilizes. This is a basis of the $JTL(1)$ submodule generated by $v_i$. Thus to each $v_i$ we associate a set of vectors $\{v_i^j\}_{j=1}^{d_j}$ (or its span $V_i$), which may or may not contain $v_i$ itself (but $v_i \in V_i$ either way). Of course many of these modules will coincide for different indices.

Then with these submodules generated by eigenvectors in hand we can observe the subquotient structure. Say $d_i \le d_j$. Take the vectors $\{v_i^k\}$ and $\{v_j^l\}$ and form one big matrix $M_{ij} = [v_i^1;\cdots;v_i^{d_i};v_j^1;\cdots;v_j^{d_j}]$ and column-reduce it. If $\rank M_{ij} = d_j$ then that implies $V_i\subset V_j$. We then have the subquotient structure $(d_j - d_i) \to d_i$, where $d_i$ labels the dimension of a factor. By doing this for every pair of submodules one should be able to infer the full subquotient structure, particularly for small lattice sizes which can be handled exactly.

\chapter{Corrections to some tables in \citeauthor{KooSaleur1994} (\citeyear{KooSaleur1994})} \label{corrections}
\emph{N.\ B.: Because it pertains only to the paper under discussion, this appendix will use notation consistent with that of Koo and Saleur }\cite{KooSaleur1994}\emph{. In particular, the Koo--Saleur generators as elements of a lattice Virasoro algebra will be denoted by $l_n$ and $\overbar l_n$.}

There exist several processes that serve as guardrails to maintain the integrity of published scientific research. One of these is peer review, in which experts in the same or a related field of study determine whether a paper under consideration merits publication. Because of timeline constraints, the review process typically addresses methodology and logic rather than the finer details of the study.

Equally important or more is post review, during which attempts are made to replicate published results, sometimes using the same methodology described in an original publication. Rarely is this done to ``find the fraud'' in the literature. I would venture to guess that most of the time replicating established results serves as a prequel to follow-up studies that extend the methodology or apply it to different situations. It is in this context that I find myself proposing these corrections, thus engaging with the literature and participating in the scientific process.

\section*{Corrected data}

Tables \ref{corrected1} and \ref{corrected2} contain my own measurements of the quantities given in the respective captions, that correct Tables 1 and 2 in \textcite{KooSaleur1994}. The extrapolations are obtained using a third-order polynomial in $1/N$ fitted to the four last data points. Comparing with the published tables, the obvious trend to notice is that the data values are quite different but the extrapolations are very close (and by transitivity, close to the conjectured values).

\begin{table}[h]
\centering
\begin{tabular}{clllll}
\toprule
$N$ & \multicolumn{5}{c}{$x$} \\
\cmidrule(l){2-6}
 & \multicolumn{1}{c}{$3$} & \multicolumn{1}{c}{$4$} & \multicolumn{1}{c}{$5$} & \multicolumn{1}{c}{$6$} & \multicolumn{1}{c}{$7$} \\
\midrule
$4$ & $0.172749$ & $0.134770$ & $0.110859$ & $0.0942739$ & $0.0820654$ \\
$6$ & $0.190565$ & $0.148760$ & $0.122293$ & $0.103957$ & $0.0904698$ \\
$8$ & $0.196899$ & $0.153548$ & $0.126146$ & $0.107184$ & $0.0932494$ \\
$10$ & $0.199752$ & $0.155643$ & $0.127795$ & $0.108544$ & $0.0944074$ \\
$12$ & $0.201253$ & $0.156709$ & $0.128611$ & $0.109202$ & $0.0949585$ \\
$14$ & $0.202128$ & $0.157308$ & $0.129054$ & $0.109549$ & $0.0952415$ \\
$16$ & $0.202679$ & $0.157669$ & $0.129308$ & $0.109740$ & $0.0953921$ \\
$18$ & $0.203045$ & $0.157897$ & $0.129461$ & $0.109847$ & $0.0954721$ \\
$20$ & $0.203298$ & $0.158047$ & $0.129552$ & $0.109907$ & $0.0955121$ \\
\midrule
extrapolation & $0.204065$ & $0.158103$ & $0.129210$ & $0.109357$ & $0.0948614$ \\
\bottomrule
\end{tabular}
\caption{Numerical values of $|\ip*{u}{l_1v}|$ in the $S_z = 0$ sector with $\phi = 0$.}
\label{corrected1}
\end{table}
\begin{table}[h]
\centering
\begin{tabular}{clllll}
\toprule
$N$ & \multicolumn{5}{c}{$x$} \\
\cmidrule(l){2-6}
 & \multicolumn{1}{c}{$3$} & \multicolumn{1}{c}{$4$} & \multicolumn{1}{c}{$5$} & \multicolumn{1}{c}{$6$} & \multicolumn{1}{c}{$7$} \\
\midrule
$4$ & $0.331475$ & $0.262672$ & $0.217753$ & $0.186055$ & $0.162462$ \\
$6$ & $0.373264$ & $0.293407$ & $0.242177$ & $0.206384$ & $0.179910$ \\
$8$ & $0.388620$ & $0.304308$ & $0.250630$ & $0.213303$ & $0.185777$ \\
$10$ & $0.395791$ & $0.309229$ & $0.254344$ & $0.216281$ & $0.188264$ \\
$12$ & $0.399682$ & $0.311811$ & $0.256233$ & $0.217757$ & $0.189473$ \\
$14$ & $0.402018$ & $0.313308$ & $0.257290$ & $0.218557$ & $0.190110$ \\
$16$ & $0.403527$ & $0.314241$ & $0.257922$ & $0.219016$ & $0.190462$ \\
$18$ & $0.404554$ & $0.314854$ & $0.258317$ & $0.219289$ & $0.190660$ \\
$20$ & $0.405285$ & $0.315273$ & $0.258572$ & $0.219453$ & $0.190770$ \\
\midrule
extrapolation & $0.408197$ & $0.316209$ & $0.258395$ & $0.218681$ & $0.189692$ \\
\bottomrule
\end{tabular}
\caption{Numerical values of $|\ip*{u}{l_1v}|$ in the $S_z = 0$ sector with $\phi = -2\pi/(x+1)$.}
\label{corrected2}
\end{table}

Re-generating Tables 2, 3, and 4 from \textcite{KooSaleur1994} are quite similar. One only needs to change the parameter values of $S_z$ and $\phi$ and run the rest of the computations as usual. Occasionally, the phase $\phi$ I use to generate data is of the opposite sign as that quoted in the paper; whether the sign errors are in my own calculation or the paper I do not know. For example, Table 2 is supposed to be generated with $\phi = -4\pi\alpha_0\alpha_- = 2\pi/(x+1)$ (by Eq.~(4.13)), but this results in a table that is identically zero (within about $10^{-15}$), so I have used the opposite sign.

To summarize, the evidence remains firm that the Koo--Saleur generators produce many of the expected results from conformal field theory. The extrapolated data remain close to the conjectured values, and only the finite-size data differ significantly. This pattern is consistent for the other tables, and so it is only necessary to show corrections to the first two. The reader capable of replicating the data in Tables \ref{corrected1} and \ref{corrected2} likely has set up their computations correctly.

\chapter{Additional calculations} \label{more_calculations}

\section{Another loop model scalar product}

We have seen that the module $\overbar{\mathscr W}_{\!\!0,\mathfrak q^{\pm 2}} + \mathscr W_{11} + \mathscr W_{21}$ of the standard loop model is missing the expected Jordan blocks at $c = 0$. This problem was remedied in two different ways: either introduce a contraction parameter $y$, or invoke a symmetry of the TL algebra that simultaneously negates the generators $e_i$ and the loop weights $m$ (Section \ref{finite_Jordan}).

Another inner product, obtained via inspired guesswork, is as follows. Imagine that an alternating sign lives between the sites:
\begin{equation}
1+2-3+4-\cdots+N-1
\end{equation}
where I have used the fact that $L$ is even and the chain is periodic. Now, whenever in $\ip*{i}{j}$ there are two defect lines $(k)$ and $(l)$ on the same side that are eventually contracted together, starting from $(k)$ and ending at $(l)$, multiply all the signs together between $k$ and $l$. If the result is positive, do nothing. If the result is negative, call this a \emph{signed contraction}, for lack of a better term. The inner product is then
\begin{equation}
\ip*{i}{j} = y^{\tilde s(i,j)}[m^{l(i,j)}],
\end{equation}
where $\tilde s(i,j)$ counts the number of signed contractions. The factor $m^{l(i,j)}$ is optional, as $c = 0$ and thus $m = 1$, so it is placed in brackets as a reminder that it should be there otherwise. Here are two examples on 6 sites:
\begin{equation}
\ip*{(14)(23)(56)}{(23)(56)} = y,\,\ip*{(41)(23)(56)}{(23)(56)} = 1.
\end{equation}
In the first example, defect lines 1 and 4 in the right state are contracted from $1\to 4$. The overall enclosed sign is $+-+ = -$, so this is a signed contraction. There are no others. In the second example, the same lines are contracted but from $4\to 1$ (draw this out and you will see what I mean). The overall enclosed sign is $-+- = +$, so this is not a signed contraction.

It turns out that this inner product gives $y$-independent measurements for $b$ (except $y = 1$) when taken with the Hamiltonian deformation. The two deformations described also help with the comparison to $s\ell(2|1)$ since now the deformed loop model is only invariant under translations by even numbers of sites as is the $s\ell(2|1)$ chain. It turns out that the measurements of $b$ are quite off using this inner product, but the fact that the results are still $y$-independent made this inner product worth recording, because it may indicate a symmetry of the loop model that has yet to be explored.

\section{Loop model contraction parameter} \label{y_parameter}

The correct deformed inner product for the glued modules of the loop model, in fact, is much simpler than the alternating sign one. Whenever two defect lines $(k)$ and $(l)$ on the same side are contracted together, starting from $(k)$ and ending at $(l)$, and $k$ is even, introduce a factor $y$. For example, on $N = 6$ sites
\begin{equation}
\ip*{(23)(14)(56)}{(41)(56)} = \ip*{(23)}{(25)(34)} = y.
\end{equation}
In the first case $(23)$ in the left state contracts $(2)$ with $(3)$ in the right state, and since the first of these, $(2)$, is even, there results a factor $y$ (and there is one closed loop). In the second case, $(4)$ and $(5)$ are contracted though via a more complicated path, and this again gives a factor $y$. But note that
\begin{equation}
\ip*{(32)(45)(61)}{(45)(61)} = 1,
\end{equation}
since in this case the paired defect in the right state starts at $(3)$ and ends at $(2)$, and the deformation requires starting on an even site.
Then, the new inner product is
\begin{equation}
\ip*{i}{j} = y^{s(i,j)}[m^{l(i,j)}],
\end{equation}
where $s(i,j)$ counts the number of paired defects as described above. Measurements of $b$ using this inner product give $y$-independent outcomes, which match Table \ref{b_W012}. Because of the smaller dimension, larger lattice sizes are possible.

\section{Illustration of the absence of Jordan blocks for $y = \pm 1$, and the $y$-independence of the measurement of $b$ at $c = 0$} \label{y_calculations}

In the periodic loop model with deformation parameter $y$, on four sites, in the basis 
\begin{equation}
\{(12)(34),(23)(14),(2)(3)(41),(3)(4)(12),(1)(4)(23),(1)(2)(34),(1)(2)(3)(4)\},
\end{equation}
the Hamiltonian matrix is
\begin{equation}
H_0 = \frac{N}{2\pi v_{\text F}} \begin{pmatrix}
 4 e_\infty-2m & -2 & 0 & -1 & 0 & -1 & 0 \\
 -2 & 4 e_\infty-2m & -y & 0 & -y & 0 & 0 \\
 0 & 0 & 4 e_\infty-m & -1 & 0 & -1 & -y \\
 0 & 0 & -1 & 4 e_\infty-m & -1 & 0 & -1 \\
 0 & 0 & 0 & -1 & 4 e_\infty-m & -1 & -y \\
 0 & 0 & -1 & 0 & -1 & 4 e_\infty-m & -1 \\
 0 & 0 & 0 & 0 & 0 & 0 & 4 e_\infty
\end{pmatrix},
\end{equation}
where, in terms of $x$, $\gamma = \pi/(x+1)$, $v_{\text F} = \pi\sin\gamma/\gamma$, $m = 2\cos\gamma$, and
\begin{equation}
e_\infty = \sin\gamma\int_{-\infty}^\infty\!\frac{\sinh[(\pi-\gamma)t]}{\sinh\pi t\cosh\gamma t}\,\d t.
\end{equation}
Excluding the $N/2\pi v_{\text F}$ factor, the eigenvalues of this matrix are
\begin{subequations}
\begin{gather}
\lambda_1 = 4 e_\infty-2m-2, \\
\lambda_2 = 4 e_\infty-m-2, \\
\lambda_3 = 4 e_\infty-m, \\
\lambda_4 = 4 e_\infty-m, \\
\lambda_5 = 4 e_\infty, \\
\lambda_6 = 4 e_\infty-2m+2, \\
\lambda_7 = 4 e_\infty-m+2.
\end{gather}
\end{subequations}
They are ordered so that for $x = 2$, $e_\infty = 1$, $\gamma = \pi/3$, $\{\lambda_1,\lambda_2,\lambda_3,\lambda_4,\lambda_5,\lambda_6,\lambda_7\} = \{0,1,3,3,4,4,5\}$. For $x = 2$, $y\ne 1$, it is known that there is a Jordan block at $\lambda_5 = \lambda_6 = 4$. Their corresponding eigenvectors are
\begin{subequations}
\begin{gather}
v_1 = (1, 1, 0, 0, 0, 0, 0) \\
v_2 = \Big(\frac{-2 m+4 y+4}{(m-4) m},\frac{4-2 (m-2) y}{(m-4) m},1,1,1,1,0\Big) \\
v_3 = (0, 0, 0, -1, 0, 1, 0) \\
v_4 = (0, 0, -1, 0, 1, 0, 0) \\
v_5 = \Big(\frac{m^2 - my^2 + 2y(1-m)}{m^4-5 m^2+4},\frac{m^2 y^2 - m + 2y(1-m)}{m^4-5 m^2+4},\frac{2-m y}{m^2-4},\frac{m-2 y}{4-m^2},\frac{2-m y}{m^2-4},\frac{m-2 y}{4-m^2},1\Big) \label{loop_4_v5} \\
v_6 = (-1, 1, 0, 0, 0, 0, 0) \label{loop_4_v6} \\
v_7 = \Big({-\frac{2 (m+2 y+2)}{m (m+4)}},\frac{2 (m+2) y+4}{m (m+4)},-1,1,-1,1,0\Big)
\end{gather}
\end{subequations}
The inner product expression signifying the appearance of Jordan blocks is then
\begin{subequations}
\begin{equation}
\frac{(v_5|v_6)}{\sqrt{(v_5|v_5)(v_6|v_6)}} = \frac{m (m+1) (y^2-1)}{\sqrt{2P(m,y)}}
\end{equation}
where
\begin{align}
P(m,y) &= m^8+m^6 (2 y^2-8)-16 m^5 y+m^4 (y^4+4 y^2+38)+m^3 (-4 y^3-4 y^2+28 y) \nonumber \\
&\qquad{}+m^2 (y^4+8 y^3-6 y^2+8 y-53)+m (-4 y^3-16 y^2-20 y)+16 y^2+24.
\end{align}
\end{subequations}
Importantly, this expression contains a factor $y^2-1$ in the numerator. Thus Jordan blocks cannot be observed with $y = 1$, as then $v_5$ and $v_6$ are always orthogonal. This is observed to be the case when directly trying to find the Jordan canonical form of $H_0$ at $y = \pm 1$. We may take a series expansion around $m = 1$. The result is
\begin{equation}
\lim_{m\to 1}\frac{(v_5|v_6)}{\sqrt{(v_5|v_5)(v_6|v_6)}} = \frac{1-y^2}{|1-y^2|} - \frac{(m-1)^2( y^4-8 y^3+98 y^2-136 y+153 )}{8 (1-y^2) |1-y^2|}+O((m-1)^3).
\end{equation}
As long as $y \ne \pm1$, the leading term goes to $\pm 1$, as expected. In \textcite{DJS2010c}, it was found that only $y = 1$ led to diagonalizability. Here, this was due to expressions that contained $y-1$ in the denominator. However, things seem to differ with the inclusion of $y = -1$. Note that in that work open boundary conditions were used, whereas here they are periodic, which may be responsible for the additional special value $y = -1$. The $(1-y^2) |1-y^2|$ factor in the denominator of the quadratic term is also rather bothersome.

Clearly in the $(v_5, v_6)$ pair, $v_6$ is the bottom state as $v_6$ is contained in $\overbar{\mathscr W}_{\!\!0,\mathfrak q^{\pm 2}}$ while $v_5$ is in the full $\overbar{\mathscr W}_{\!\!0,\mathfrak q^{\pm 2}} + \mathscr W_{1,1} + \mathscr W_{2,1}$ (although as $m \to 1$, they converge to the same vector). The conformal norm square of $v_6$ is
\begin{equation}
\ip*{v_6}{v_6} = 2m(m-1),
\end{equation}
with the caveat that it is not normalized. But the preceding expression vanishes at $m = 1$ ($c = 0$). 

We have
\begin{equation}
\lim_{m\to 1} v_5 = \bigg(\frac{y^2-1}{6 (m-1)}+\frac{7 y^2+12 y-13}{36},\frac{1-y^2}{6 (m-1)} + \frac{-13 y^2+12 y+7}{36},\frac{y-2}{3},\frac{1-2 y}{3}, \frac{y-2}{3},\frac{1-2 y}{3},1\bigg)
\end{equation}
The two diverging components proportional to $(m-1)^{-1}$ are clearly parallel to $v_6$. Subtracting off this component (and a little more, in order to make the result orthogonal to $v_6$), we must have the Jordan partner,
\begin{equation}
\tilde v_5 = \bigg(\frac{-y^2 + 4y - 1}{12},\frac{-y^2 + 4y - 1}{12},\frac{y-2}{3},\frac{1-2 y}{3}, \frac{y-2}{3},\frac{1-2 y}{3},1\bigg).
\end{equation}
I rescale $\tilde v_5$ so that $(\hat v_6|H_0\tilde v_5) = 2$. The normalization of $\tilde v_5$ is thus tied to that of $v_6$, and we should get
\begin{equation}
b = \frac{|\mel*{\tilde v_5}{H_{-2}}{\tilde v_1}|^2}{\ip*{\tilde v_5}{\hat v_6}},
\end{equation}
where $\langle\tilde v_1,\tilde v_1\rangle = 1$. For $N = 4$, the result is $b = -32/3\sqrt 3\pi \approx -1.96028$, which is $y$-independent. Note that $y$ remained generic in this calculation, so that all $y$ terms ended up disappearing. This value is identical to the one obtained in the discussion surrounding Table \ref{b_W012}.

The preceding procedure makes sense for all values of $c$, not just $c = 0$. If we follow it and define $b$ in the same way, I find that $b = -8(m+1)/\pi v_{\text F}$, which is still independent of $y$.

We also know that $v_5$ and $v_6$ become parallel as $m \to 1$. They should thus be treated on an equal footing, and I should be able to follow the same process but with the roles of $v_5$ and $v_6$ reversed. If I do this, I get a horrendous expression for $b$:
\begin{equation}
b = \frac{16 \bigg[\substack{\displaystyle m^8+2 m^6 (y^2-4)-16 m^5 y+m^4 (y^4+4 y^2+38)-4 m^3 y (y^2+y-7) \\ \displaystyle {}+m^2 (y^4+8 y^3-6 y^2+8 y-53)-4 m y (y^2+4 y+5)+8 (2 y^2+3)}\bigg]^2}{m (m+1)^2 (y^2-1)^2\pi v_{\text F} \bigg[\substack{\displaystyle m^6-2 m^5 (y^2+1)+m^4 (y^4+8 y-8)-2 m^3 (2 y^3-6 y^2+2 y-5) \\ \displaystyle{}-2 m^2 (y^4+2 y^3-4 y^2+22 y-11)+8 m (y^3+y-1)-16 (y-1)^2}\bigg]}.
\end{equation}
But, when I set $m = 1$ in this expression, it reduces to $b = -32\sqrt 3/9\pi$, again without specifying $y$ (although there is a $(y^2 - 1)^2$ in the denominator of the full expression).

To see that this is no accident, we can do the same thing for $N = 6$. In the basis of momentum-2 states, which I will not write out, one has two eigenvectors that degenerate at $m = 1$:
\begin{align}
v_1 &= (1,0,0,0,0,0,0,0), \label{loop_6_v1} \\
v_2 &= \bigg(\frac{m (y^2-1)}{m^3-m^2-4 m+4},\frac{(i\sqrt{3}-1) (m-2 y)}{2 (m^2-4)},\frac{2-m y}{m^2-4},\frac{(1-i \sqrt{3}) (y-1)}{2 (m-2)}, \nonumber \\
&\qquad\frac{(1+i \sqrt{3}) (m y-2)}{2 (m^2-4)},-\frac{(1+i\sqrt{3}) (m-2 y)}{2 (m^2-4)},\frac{1}{2} (1-i \sqrt{3}),1\bigg).
\end{align}
I find
\begin{equation}
J(v_1, v_2) = \frac{m (y^2-1)}{\sqrt{\substack{\displaystyle 2 m^6-4 m^5+m^4 (3 y^2-2 y-11)-2 m^3 (y^2+10 y-15) \\ \displaystyle {}+m^2 (y^4+5 y^2+38 y+24)-4 m (5 y^2+2 y+21)+12 y^2-8 y+44}}}.
\end{equation}
A series expansion around $m = 1$ gives
\begin{equation}
\lim_{m\to 1}J(v_1, v_2) = \frac{1-y^2}{|1-y^2|} - \frac{(m-1)^2(19y^2 -34y + 37)}{2 (1-y^2) |1-y^2|}+O((m-1)^3).
\end{equation}
$v_1$ and $v_2$ thus become parallel, and are evidently can be orthogonalized to obtain $T$ and $t$. A value of $b$ is
\begin{equation}
b = \frac{3}{\pi v_{\text F}}\frac{4 m^2+\frac{2 (9-2 m^2) m}{\sqrt{m^2+48}}-6}{1-m}.
\end{equation}
Despite the denominator, the function is perfectly analytic at $m = 1$. We then have $b = -288 \sqrt{3}/49 \pi \approx -3.24046$.

Exchanging the roles of $v_1$ and $v_2$ leads to a $b$ number of
\begin{equation}
\begin{aligned}
b &= \frac{6 (m+1)^2 (-m^2+\sqrt{m^2+48} m+12)^2}{\pi v_{\text F} (y^2-1)^2 m} \\
&\qquad \!{}\times \frac{\bigg[\substack{\displaystyle 2 m^6-4 m^5+m^4 (3 y^2-2 y-11)-2 m^3 (y^2+10 y-15) \\ \displaystyle {}+m^2 (y^4+5 y^2+38 y+24)-4 m (5 y^2+2 y+21)+12 y^2-8 y+44}\bigg]^2}{\bigg(\substack{\displaystyle m^4+50 m^2+22 \sqrt{m^2+48} m \\ \displaystyle {}-\sqrt{m^2+48} m^3+96}\bigg) \bigg[\substack{\displaystyle m^6-2 m^5 (y^2+1)+m^4 (y^4+8 y-8)-2 m^3 (2 y^3-6 y^2+2 y-5) \\ \displaystyle {}-2 m^2 (y^4+2 y^3-4 y^2+22 y-11)+8 m (y^3+y-1)-16 (y-1)^2}\bigg]}.
\end{aligned}
\end{equation}
The $(y^2 - 1)^2$ in the denominator says that there is no $b$ when $y = \pm 1$, but when $y$ is generic and $m \to 1$, we get the same value as before, $b = -288 \sqrt{3}/49 \pi$. This value is also identical to the one obtained in the discussion surrounding Table \ref{b_W012}.

The conformal norm of $v_1$ is
\begin{equation}
\ip*{v_1}{v_1} = m(m^2 - 1).
\end{equation}

\chapter{Additional tables and figures} \label{more_tables}

\section{Supplementary data for Section \ref{scaling_weak}} \label{more_sw_tables}

\begin{table}[H]
\centering
\begin{tabular}{cccccc}
\toprule
$N$ & \multicolumn{5}{c}{$x$} \\
\cmidrule(l){2-6}
 & $\pi/3$ & $\pi/2$ & $\pi/\sec^{-1}(2 \sqrt{2}) -1$ & $\e$ & $\pi$ \\
\midrule
8 & 0.325797 & 0.350787 & 0.351607 & 0.372103 & 0.376185 \\
10 & 0.329276 & 0.352661 & 0.353516 & 0.376516 & 0.381388 \\
12 & 0.328903 & 0.351523 & 0.352390 & 0.376800 & 0.382216 \\
14 & 0.327108 & 0.349302 & 0.350173 & 0.375408 & 0.381203 \\
16 & 0.324801 & 0.346738 & 0.347609 & 0.373350 & 0.379417 \\
18 & 0.322360 & 0.344140 & 0.345010 & 0.371078 & 0.377346 \\
20 & 0.319948 & 0.341637 & 0.342507 & 0.368803 & 0.375224 \\
22 & 0.317637 & 0.339281 & 0.340151 & 0.366621 & 0.373165 \\
\bottomrule
\end{tabular}
\caption{The values of $\|L_{-1}\Phi_{11}\|_2$ for various lengths $N$ and parameters $x$.}
\label{normLX}
\end{table}

\begin{table}[H]
\centering
\begin{tabular}{cccccc}
\toprule
$N$ & \multicolumn{5}{c}{$x$} \\
\cmidrule(l){2-6}
 & $\pi/3$ & $\pi/2$ & $\pi/\sec^{-1}(2 \sqrt{2}) -1$ & $\e$ & $\pi$ \\
\midrule
 8 & 0.325786 & 0.349822 & 0.350572 & 0.368581 & 0.372080 \\
 10 & 0.329244 & 0.350622 & 0.351340 & 0.369610 & 0.373372 \\
 12 & 0.328845 & 0.348905 & 0.349594 & 0.367684 & 0.371562 \\
 14 & 0.327027 & 0.346360 & 0.347031 & 0.364836 & 0.368753 \\
 16 & 0.324704 & 0.343605 & 0.344262 & 0.361773 & 0.365685 \\
 18 & 0.322251 & 0.340885 & 0.341533 & 0.358769 & 0.362653 \\
 20 & 0.319832 & 0.338301 & 0.338943 & 0.355932 & 0.359775 \\
 22 & 0.317515 & 0.335886 & 0.336524 & 0.353297 & 0.357092 \\
\bottomrule
\end{tabular}
\caption{The values of $\|\Pi^{(2)}L_{-1}\Phi_{11}\|_2$ for various lengths $N$ and parameters $x$.}
\label{normPiLX}
\end{table}

\begin{table}[H]
\centering
\begin{tabular}{cccccc}
\toprule
$N$ & \multicolumn{5}{c}{$x$} \\
\cmidrule(l){2-6}
 & $\pi/3$ & $\pi/2$ & $\pi/\sec^{-1}(2 \sqrt{2}) -1$ & $\e$ & $\pi$ \\
\midrule
 10 & 0.724502 & 0.855299 & 0.859681 & 0.965547 & 0.985531 \\
 12 & 0.769370 & 0.879924 & 0.883733 & 0.977938 & 0.996234 \\
 14 & 0.794399 & 0.894860 & 0.898393 & 0.987034 & 1.00452 \\
 16 & 0.808303 & 0.903779 & 0.907193 & 0.993487 & 1.01062 \\
 18 & 0.815690 & 0.908984 & 0.912369 & 0.998190 & 1.01523 \\
 20 & 0.819132 & 0.912010 & 0.915425 & 1.00203 & 1.01916 \\
 22 & 0.820131 & 0.913885 & 0.917377 & 1.00577 & 1.02311 \\
 \bottomrule
\end{tabular}
\caption{The values of $\|A_{12}\Phi_{12}\|_2$ for various lengths $N$ and parameters $x$.}
\label{normAX}
\end{table}

\begin{table}[H]
\centering
\begin{tabular}{cccccc}
\toprule
$N$ & \multicolumn{5}{c}{$x$} \\
\cmidrule(l){2-6}
 & $\pi/3$ & $\pi/2$ & $\pi/\sec^{-1}(2 \sqrt{2}) -1$ & $\e$ & $\pi$ \\
\midrule
 10 & 0.724473 & 0.853535 & 0.857809 & 0.959148 & 0.977917 \\
 12 & 0.769136 & 0.874667 & 0.878230 & 0.964825 & 0.981331 \\
 14 & 0.793913 & 0.886208 & 0.889353 & 0.967146 & 0.982324 \\
 16 & 0.807582 & 0.891969 & 0.894866 & 0.967190 & 0.981519 \\
 18 & 0.814765 & 0.894091 & 0.896830 & 0.965628 & 0.979399 \\
 20 & 0.818030 & 0.893910 & 0.896546 & 0.962969 & 0.976361 \\
 22 & 0.818871 & 0.892283 & 0.894845 & 0.959593 & 0.972719 \\
 \bottomrule
\end{tabular}
\caption{The values of $\|\Pi^{(4)}A_{12}\Phi_{12}\|_2$ for various lengths $N$ and parameters $x$.}
\label{normPiAX}
\end{table}

\section{Supplementary data for Section \ref{sl21_projections}} \label{sl21_tables}

\begin{table}[H] \label{sl21_table_12}
\centering
{\small\begin{tabular}{cccccccc}
\toprule 
$i$ & $\epsilon$ & $n_i$ & $d_i$ & \makecell{Jordan \\ structure} & module(s) & $h + \overbar h$ & \makecell{scaling field(s)} \\
\midrule
1 & $0$ & 1 & 1 & & $\overbar{\mathscr W}_{\!\!0,\mathfrak q^{\pm 2}}$ & $0$ & $I$ \\ 
2 & $0.248922$ & 2 & 3 & & $\mathscr W_{11}$ & $1/4,6$ & $\Phi_{01},(0,5)\Phi_{11},(5,0)\Phi_{-1,1}$ \\ 
\cmidrule(lr){1-8}
3 & $1.20295$ & 4 & 7 & $2$ & $\overbar{\mathscr W}_{\!\!0,\mathfrak q^{\pm 2}},\mathscr W_{21}$ & $5/4,6$ & \makecell{$\phi_{21}\otimes\overbar\phi_{21},(6,0)I,(0,6)I,$ \\ $\Phi_{02},(0,4)\Phi_{12},(4,0)\Phi_{-1,2}$} \\ 
\cmidrule(lr){1-8}
4 & $1.75032$ & 2 & 9 & & $\mathscr W_{11}$ & $2$ & $(1,0)\Phi_{11}$ \\ 
5 & $2.03178$ & 2 & 11 & & $\mathscr W_{11}$ & $2$ & $(0,1)\Phi_{-1,1}$ \\ 
6 & $2.07518$ & 2 & 13 & & $\mathscr W_{11}$ & $9/4$ & $(1,1)\Phi_{01}$ \\ 
7 & $2.23935$ & 4 & 17 & & $\mathscr W_{21}$ & $39/16$ & $(1,0)\Phi_{1/2,2},(0,1)\Phi_{-1/2,2}$ \\ 
8 & $2.6721$ & 8 & 25 & & $\mathscr W_{3,\mathfrak q^{\pm 2}}$ & $35/12$ & $\Phi_{03}$ \\ 
9 & $2.79865$ & 4 & 29 & $2$ & $\overbar{\mathscr W}_{\!\!0,\mathfrak q^{\pm 2}},\mathscr W_{21}$ & $13/4,6$ & $(1,1)\phi_{21}\otimes\overbar\phi_{21},(6,0)I,(0,6)I,(1,1)\Phi_{02}$ \\ 
10 & $3.14032$ & 2 & 31 & & $\mathscr W_{11}$ & $4$ & \\ 
11 & $3.33436$ & 4 & 35 & $2$ & $\overbar{\mathscr W}_{\!\!0,\mathfrak q^{\pm 2}},\mathscr W_{21}$ & $4$ & $\phi_{31}\otimes\overbar\phi_{31},(2,0)\Phi_{12},(0,2)\Phi_{-1,2}$ \\ 
12 & $3.36307$ & 2 & 37 & & $\mathscr W_{11}$ & $17/4$ & $(2,2)\Phi_{01}$ \\ 
\cmidrule(lr){1-8}
13 & $3.43849$ & 24 & 61 & $8\times2$ & $\mathscr W_{11},\mathscr W_{21},\mathscr W_{3,\mathfrak q^{\pm 2}}$ & $4$ & \makecell{$(2,1)\Phi_{11},(1,2)\Phi_{-1,1},$ \\ $(2,0)\Phi_{12},(0,2)\Phi_{-1,2},$ \\ $(1,0)\Phi_{1/3,3},(0,1)\Phi_{-1/3,3}$} \\ 
\cmidrule(lr){1-8}
14 & $3.50199$ & 2 & 63 & & $\mathscr W_{11}$ & $4$ & \\ 
15 & $3.51279$ & 4 & 67 & & $\mathscr W_{21}$ & $71/16$ & $(2,1)\Phi_{1/2,2},(1,2)\Phi_{-1/2,2}$ \\ 
16 & $3.52945$ & 4 & 71 & & $\mathscr W_{11}$ & $17/4$ & $(2,2)\Phi_{01}$ \\ 
17 & $3.67553$ & 2 & 73 & & $\mathscr W_{11}$ & $17/4$ & $(2,2)\Phi_{01}$ \\ 
18 & $3.69711$ & 4 & 77 & $2$ & $\overbar{\mathscr W}_{\!\!0,\mathfrak q^{\pm 2}},\mathscr W_{21}$ & $21/4$ & $(2,2)\phi_{21}\otimes\overbar\phi_{21}$ \\ 
19 & $3.74793$ & 4 & 81 & & $\mathscr W_{21}$ & $71/16$ & $(2,1)\Phi_{1/2,2},(1,2)\Phi_{-1/2,2}$ \\ 
20 & $3.81032$ & 4 & 85 & $2$ & $\overbar{\mathscr W}_{\!\!0,\mathfrak q^{\pm 2}},\mathscr W_{21}$ & $4$ & $(2,2)I,(2,0)\Phi_{12},(0,2)\Phi_{-1,2}$ \\ 
21 & $3.91012$ & 2 & 87 & & $\mathscr W_{11}$ & & \\ 
22 & $3.92424$ & 8 & 95 & & $\mathscr W_{3,\mathfrak q^{\pm 2}}$ & $59/12$ & $(1,1)\Phi_{03}$ \\ 
23 & $4.11904$ & 24 & 119 & $8\times2$ & $\mathscr W_{11},\mathscr W_{21},\mathscr W_{3,\mathfrak q^{\pm 2}}$ & $21/4$ & \\ 
24 & $4.30698$ & 24 & 143 & $8\times2$ & $\mathscr W_{11},\mathscr W_{21},\mathscr W_{3,\mathfrak q^{\pm 2}}$ & & \\ 
25 & $4.32986$ & 4 & 147 & & $\mathscr W_{11}$ & & \\ 
26 & $4.33042$ & 4 & 151 & & $\mathscr W_{11}$ & & \\ 
27 & $4.41063$ & 4 & 155 & $2$ & $\overbar{\mathscr W}_{\!\!0,\mathfrak q^{\pm 2}},\mathscr W_{21}$ & $6$ & $(3,3)I$ \\
\bottomrule
\end{tabular}}
\caption{Jordan structure of the lowest 155 of 3991 eigenvalues of $H_0$ on $N = 12$ sites, in the vacuum sector at momentum 0. $n_i$ is the algebraic multiplicity of the eigenvalue on line $i$. $d_i = \sum_{j=1}^i n_j$ is the running dimension. In the ``Jordan structure'' column, $m\times n$ means $m$ rank-$n$ Jordan blocks appear for that eigenvalue, and $n \equiv 1\times n$. $(m,n)\Phi$ means a level-$(m,n)$ descendant of $\Phi$.} 
\end{table}

\begin{table}[H] \label{sl21_table_14}
\centering
{\small\begin{tabular}{cccccccc}
\toprule 
$i$ & $\epsilon$ & $n_i$ & $d_i$ & \makecell{Jordan \\ structure} & module(s) & $h + \overbar h$ & \makecell{scaling field(s)} \\
\midrule
1 & $0$ & 1 & 1 & & $\overbar{\mathscr W}_{\!\!0,\mathfrak q^{\pm 2}}$ & $0$ & $I$ \\ 
2 & $0.249157$ & 2 & 3 & & $\mathscr W_{11}$ & $1/4$ & $\Phi_{01}$ \\ 
3 & $1.21393$ & 4 & 7 & $2$ & $\overbar{\mathscr W}_{\!\!0,\mathfrak q^{\pm 2}},\mathscr W_{21}$ & $5/4,7$ & $\phi_{21}\otimes\overbar\phi_{21},(7,0)I,(0,7)I,\Phi_{02}$ \\ 
4 & $1.7908$ & 2 & 9 & & $\mathscr W_{11}$ & $2$ & $(1,0)\Phi_{11}$ \\ 
5 & $2.04203$ & 2 & 11 & & $\mathscr W_{11}$ & $2$ & $(0,1)\Phi_{-1,1}$ \\ 
6 & $2.11598$ & 2 & 13 & & $\mathscr W_{11}$ & $9/4$ & $(1,1)\Phi_{01}$ \\ 
7 & $2.28522$ & 4 & 17 & & $\mathscr W_{21}$ & $39/16$ & $(1,0)\Phi_{1/2,2},(0,1)\Phi_{-1/2,2}$ \\ 
8 & $2.72747$ & 8 & 25 & & $\mathscr W_{3,\mathfrak q^{\pm 2}}$ & $35/12$ & $\Phi_{03}$ \\ 
9 & $2.90368$ & 4 & 29 & $2$ & $\overbar{\mathscr W}_{\!\!0,\mathfrak q^{\pm 2}},\mathscr W_{21}$ & $13/4$ & $(1,1)\phi_{21}\otimes\overbar\phi_{21},(1,1)\Phi_{02}$ \\ 
10 & $3.2963$ & 2 & 31 & & $\mathscr W_{11}$ & $4$ & \\ 
11 & $3.44746$ & 4 & 35 & $2$ & $\overbar{\mathscr W}_{\!\!0,\mathfrak q^{\pm 2}},\mathscr W_{21}$ & $4$ & $\phi_{31}\otimes\overbar\phi_{31},(2,0)\Phi_{12},(0,2)\Phi_{-1,2}$ \\ 
\cmidrule(lr){1-8}
12 & $3.56541$ & 24 & 59 & $8\times2$ & $\mathscr W_{11},\mathscr W_{21},\mathscr W_{3,\mathfrak q^{\pm 2}}$ & $4$ & \makecell{$(2,1)\Phi_{11},(1,2)\Phi_{-1,1},$ \\ $(2,0)\Phi_{12},(0,2)\Phi_{-1,2},$ \\ $(1,0)\Phi_{1/3,3},(0,1)\Phi_{-1/3,3}$} \\ 
\cmidrule(lr){1-8}
13 & $3.56731$ & 2 & 61 & & $\mathscr W_{11}$ & $17/4$ & $(2,2)\Phi_{01}$ \\ 
14 & $3.6637$ & 2 & 63 & & $\mathscr W_{11}$ & $4$ & \\ 
15 & $3.68765$ & 4 & 67 & & $\mathscr W_{11}$ & $17/4$ & $(2,2)\Phi_{01}$ \\ 
16 & $3.72401$ & 4 & 71 & & $\mathscr W_{21}$ & $71/16$ & $(2,1)\Phi_{1/2,2},(1,2)\Phi_{-1/2,2}$ \\ 
17 & $3.79679$ & 2 & 73 & & $\mathscr W_{11}$ & $17/4$ & $(2,2)\Phi_{01}$ \\ 
18 & $3.89009$ & 4 & 77 & $2$ & $\overbar{\mathscr W}_{\!\!0,\mathfrak q^{\pm 2}},\mathscr W_{21}$ & $4$ & $(2,2)I,(2,0)\Phi_{12},(0,2)\Phi_{-1,2}$ \\ 
19 & $3.90027$ & 4 & 81 & & $\mathscr W_{21}$ & $71/16$ & $(2,1)\Phi_{1/2,2},(1,2)\Phi_{-1/2,2}$ \\ 
20 & $4.04741$ & 4 & 85 & $2$ & $\overbar{\mathscr W}_{\!\!0,\mathfrak q^{\pm 2}},\mathscr W_{21}$ & $21/4$ & $(2,2)\phi_{21}\otimes\overbar\phi_{21}$ \\ 
21 & $4.14726$ & 8 & 93 & & $\mathscr W_{3,\mathfrak q^{\pm 2}}$ & $59/12$ & $(1,1)\Phi_{03}$ \\ 
22 & $4.30023$ & 2 & 95 & & $\mathscr W_{11}$ & & \\ 
23 & $4.36054$ & 2 & 97 & & $\mathscr W_{11}$ & & \\ 
24 & $4.3695$ & 24 & 121 & $8\times2$ & $\mathscr W_{11},\mathscr W_{21},\mathscr W_{3,\mathfrak q^{\pm 2}}$ & $21/4$ & \\ 
25 & $4.51031$ & 4 & 125 & & $\mathscr W_{21}$ & & \\ 
26 & $4.56181$ & 2 & 127 & & $\mathscr W_{11}$ & & \\ 
27 & $4.63272$ & 4 & 131 & & $\mathscr W_{11}$ & & \\
\bottomrule
\end{tabular}}
\caption{Jordan structure of the lowest 131 of 23391 eigenvalues of $H_0$ on $N = 14$ sites, in the vacuum sector at momentum 0. $n_i$ is the algebraic multiplicity of the eigenvalue on line $i$. $d_i = \sum_{j=1}^i n_j$ is the running dimension. In the ``Jordan structure'' column, $m\times n$ means $m$ rank-$n$ Jordan blocks appear for that eigenvalue, and $n \equiv 1\times n$. $(m,n)\Phi$ means a level-$(m,n)$ descendant of $\Phi$.} 
\end{table}

\begin{table}[H] \label{sl21_table_16}
\centering
{\small\begin{tabular}{cccccccc}
\toprule 
$i$ & $\epsilon$ & $n_i$ & $d_i$ & \makecell{Jordan \\ structure} & module(s) & $h + \overbar h$ & \makecell{scaling field(s)} \\
\midrule
1 & $0$ & 1 & 1 & & $\overbar{\mathscr W}_{\!\!0,\mathfrak q^{\pm 2}}$ & $0$ & $I$ \\ 
2 & $0.249321$ & 2 & 3 & & $\mathscr W_{11}$ & $1/4$ & $\Phi_{01}$ \\ 
3 & $1.22141$ & 4 & 7 & $2$ & $\overbar{\mathscr W}_{\!\!0,\mathfrak q^{\pm 2}},\mathscr W_{21}$ & $5/4,8$ & $\phi_{21}\otimes\overbar\phi_{21},(8,0)I,(0,8)I,\Phi_{02}$ \\ 
4 & $1.82075$ & 2 & 9 & & $\mathscr W_{11}$ & $2$ & $(1,0)\Phi_{11}$ \\ 
5 & $2.04663$ & 2 & 11 & & $\mathscr W_{11}$ & $2$ & $(0,1)\Phi_{-1,1}$ \\ 
6 & $2.14386$ & 2 & 13 & & $\mathscr W_{11}$ & $9/4$ & $(1,1)\Phi_{01}$ \\ 
7 & $2.31668$ & 4 & 17 & & $\mathscr W_{21}$ & $39/16$ & $(1,0)\Phi_{1/2,2},(0,1)\Phi_{-1/2,2}$ \\ 
8 & $2.76578$ & 8 & 25 & & $\mathscr W_{3,\mathfrak q^{\pm 2}}$ & $35/12$ & $\Phi_{03}$ \\ 
9 & $2.97579$ & 4 & 29 & $2$ & $\overbar{\mathscr W}_{\!\!0,\mathfrak q^{\pm 2}},\mathscr W_{21}$ & $13/4$ & $(1,1)\phi_{21}\otimes\overbar\phi_{21},(1,1)\Phi_{02}$ \\ 
10 & $3.41025$ & 2 & 31 & & $\mathscr W_{11}$ & $4$ & \\ 
11 & $3.53106$ & 4 & 35 & $2$ & $\overbar{\mathscr W}_{\!\!0,\mathfrak q^{\pm 2}},\mathscr W_{21}$ & $4$ & $\phi_{31}\otimes\overbar\phi_{31},(2,0)\Phi_{12},(0,2)\Phi_{-1,2}$ \\ 
\midrule
12 & $3.65361$ & 24 & 59 & $8\times2$ & $\mathscr W_{11},\mathscr W_{21},\mathscr W_{3,\mathfrak q^{\pm 2}}$ & $4$ & \makecell{$(2,1)\Phi_{11},(1,2)\Phi_{-1,1},(2,0)\Phi_{12},$ \\ $(0,2)\Phi_{-1,2},(1,0)\Phi_{1/3,3},(0,1)\Phi_{-1/3,3}$} \\ 
\midrule
13 & $3.70857$ & 2 & 61 & & $\mathscr W_{11}$ & $17/4$ & $(2,2)\Phi_{01}$ \\ 
14 & $3.76720$ & 2 & 63 & & $\mathscr W_{11}$ & $4$ & \\ 
15 & $3.79947$ & 4 & 67 & & $\mathscr W_{11}$ & $17/4$ & $(2,2)\Phi_{01}$ \\ 
16 & $3.87063$ & 4 & 71 & & $\mathscr W_{21}$ & $71/16$ & $(2,1)\Phi_{1/2,2},(1,2)\Phi_{-1/2,2}$ \\ 
17 & $3.88376$ & 2 & 73 & & $\mathscr W_{11}$ & $17/4$ & $(2,2)\Phi_{01}$ \\ 
18 & $3.93996$ & 4 & 77 & $2$ & $\overbar{\mathscr W}_{\!\!0,\mathfrak q^{\pm 2}},\mathscr W_{21}$ & $4$ & $(2,2)I,(2,0)\Phi_{12},(0,2)\Phi_{-1,2}$ \\ 
19 & $4.00740$ & 4 & 81 & & $\mathscr W_{21}$ & $71/16$ & $(2,1)\Phi_{1/2,2},(1,2)\Phi_{-1/2,2}$ \\ 
20 & $4.29274$ & 4 & 85 & $2$ & $\overbar{\mathscr W}_{\!\!0,\mathfrak q^{\pm 2}},\mathscr W_{21}$ & $21/4$ & $(2,2)\phi_{21}\otimes\overbar\phi_{21}$ \\ 
21 & $4.30309$ & 8 & 93 & & $\mathscr W_{3,\mathfrak q^{\pm 2}}$ & $59/12$ & $(1,1)\Phi_{03}$ \\ 
22 & $4.54588$ & 24 & 117 & $8\times2$ & $\mathscr W_{11},\mathscr W_{21},\mathscr W_{3,\mathfrak q^{\pm 2}}$ & $21/4$ & \\ 
23 & $4.58776$ & 2 & 119 & & $\mathscr W_{11}$ & & \\
\bottomrule
\end{tabular}}
\caption{Jordan structure of the lowest 119 of 143073 eigenvalues of $H_0$ on $N = 16$ sites, in the vacuum sector at momentum 0. $n_i$ is the algebraic multiplicity of the eigenvalue on line $i$. $d_i = \sum_{j=1}^i n_j$ is the running dimension. In the ``Jordan structure'' column, $m\times n$ means $m$ rank-$n$ Jordan blocks appear for that eigenvalue, and $n \equiv 1\times n$. $(m,n)\Phi$ means a level-$(m,n)$ descendant of $\Phi$.} 
\end{table}

\begin{table}[H]
\centering
\begin{tabular}{cccc}
\toprule
$d$ & \multicolumn{3}{c}{$N$} \\
\cmidrule(l){2-4}
 & $10$ & 12 & 14 \\
\midrule
1 & 0* & 0* & 0* \\
2 & 0* & 0* & 0* \\
4 & 0* & 0* & 0* \\
5 & 0* & 0* & 0* \\
6 & 0* & 0* & 0* \\
7 & 0* & 0* & 0* \\
9 & 0* & 0* & 0* \\
10 & 0* & 0* & 0* \\
11 & 0* & --- & --- \\
12 & --- & 0* & 0* \\
13 & 0* & 0* & --- \\
14 & 0* & --- & --- \\
16 & --- & --- & 0* \\
17 & --- & 0* & 0* \\
18 & 0* & 0* & 0* \\
20 & 0* & 0* & 0* \\
21 & 0* & 0* & 0* \\
23 & $0.011488$ & 0* & $0.00296685$ \\
25 & --- & $0.00585179$ & $0.00296685$ \\
26 & --- & $0.00585179$ & $0.00296685$ \\
27 & $0.011488$ & --- & $0.00296685$ \\
29 & $0.011488$ & --- & --- \\
30 & --- & $0.00585179$ & --- \\
31 & $0.0115711$ & --- & $0.00296685$ \\
32 & --- & --- & $0.00296685$ \\
34 & $0.0115711$ & $0.00585179$ & $0.00296685$ \\
35 & $0.0115711$ & --- & --- \\
36 & $0.0115711$ & $0.00585179$ & --- \\
37 & --- & --- & $0.00296685$ \\
38 & --- & $0.00585179$ & --- \\
39 & --- & $0.00617649$ & --- \\
40 & $0.0115711$ & $0.00627736$ & $0.00296685$ \\
41 & --- & --- & $0.00317961$ \\
\midrule
no projector & $0.0115289$ & $0.0076238$ & $0.00712264$ \\
\bottomrule
\end{tabular}
\caption{The norm $\|\Pi^{(d)}(A\overbar T - \overbar A T)\|_{\munderbar 2}$ for various projector ranks $d$ and system lengths $N$, in the loop model representation. In this table ``0*'' means a number that is less than about $2\times 10^{-8}$. I estimate the uncertainty in the nonzero numbers to be about $10^{-7}$.}
\label{loop_table_AT}
\end{table}

\begin{table}[H]
\centering
\begin{tabular}{cccc}
\toprule
$d$ & \multicolumn{3}{c}{$N$} \\
\cmidrule(l){2-4}
 & $10$ & 12 & 14 \\
\midrule
1 & * & * & * \\
2 & * & * & * \\
4 & * & * & * \\
5 & * & * & * \\
6 & * & * & * \\
7 & * & * & * \\
9 & * & * & * \\
10 & * & * & * \\
11 & * & --- & --- \\
12 & --- & 0* & 0* \\
13 & 0* & 0* & --- \\
14 & 0* & --- & --- \\
16 & --- & --- & 0* \\
17 & --- & 0* & 0* \\
18 & 0* & 0* & 0* \\
20 & 0* & 0* & 0* \\
21 & 0* & 0* & 0* \\
23 & $0.035351$ & 0* & $0.0185908$ \\
25 & --- & $0.0253789$ & $0.0185908$ \\
26 & --- & $0.0253789$ & $0.0185908$ \\
27 & $0.035351$ & --- & $0.0185908$ \\
29 & $0.035351$ & --- & --- \\
30 & --- & $0.0253789$ & --- \\
31 & $0.035679$ & --- & $0.0185908$ \\
32 & --- & --- & $0.0185908$ \\
34 & $0.035679$ & $0.0253789$ & $0.0185908$ \\
35 & $0.035679$ & --- & --- \\
36 & $0.035679$ & $0.0253789$ & --- \\
37 & --- & --- & $0.0185908$ \\
38 & --- & $0.0253789$ & --- \\
39 & --- & $0.0268784$ & --- \\
40 & $0.035679$ & $0.0272667$ & $0.0185908$ \\
41 & --- & --- & $0.0199863$ \\
\midrule
no projector & $0.0355239$ & $0.0330649$ & $0.044643$ \\
\bottomrule
\end{tabular}
\caption{The norm $\|\Pi^{(d)}(A\overbar T - \overbar A T)\|_{\munderbar 2}/\|\Pi^{(d)}A\overbar T\|_2$ for various projector ranks $d$ and system lengths $N$, in the loop model representation. In this table ``*'' means a highly variable number of order 1. They are likely the result of a 0/0 since the corresponding values in Table \ref{loop_table_AT} are so small. ``0*'' means a number that is less than $10^{-6}$. I estimate the uncertainty in the nonzero numbers to be about $10^{-6}$.}
\label{loop_table_AT_norm}
\end{table}

\begin{table}[H]
\centering
\begin{tabular}{cccc}
\toprule
$d$ & \multicolumn{3}{c}{$N$} \\
\cmidrule(l){2-4}
 & $10$ & 12 & 14 \\
\midrule
1 & 0* & 0* & 0* \\
2 & 0* & 0* & 0* \\
4 & 0* & 0* & 0* \\
5 & 0* & 0* & 0* \\
6 & 0* & 0* & 0* \\
7 & 0* & 0* & 0* \\
9 & 0* & 0* & 0* \\
10 & 0* & 0* & 0* \\
11 & 0* & --- & --- \\
12 & --- & $0.236909$ & $0.134669$ \\
13 & $0.424521$ & $0.236909$ & --- \\
14 & $0.424521$ & --- & --- \\
16 & --- & --- & $1.05087$ \\
17 & --- & $1.99303$ & $1.05087$ \\
18 & $3.79519$ & $1.99303$ & $1.05087$ \\
20 & $3.79519$ & $1.99303$ & $1.05087$ \\
21 & $3.79519$ & $1.99303$ & $1.05087$ \\
23 & $3.80601$ & $1.99303$ & $1.05369$ \\
25 & --- & $1.99874$ & $1.05369$ \\
26 & --- & $1.99874$ & $1.05369$ \\
27 & $3.80601$ & --- & $1.05369$ \\
29 & $3.80601$ & --- & --- \\
30 & --- & $1.99874$ & --- \\
31 & $3.80617$ & --- & $1.05369$ \\
32 & --- & --- & $1.05369$ \\
34 & $3.80617$ & $1.9989$ & $1.05369$ \\
35 & $3.80617$ & --- & --- \\
36 & $3.80617$ & $1.9989$ & --- \\
37 & --- & --- & $1.16385$ \\
38 & --- & $1.9989$ & --- \\
39 & --- & $2.4453$ & --- \\
40 & $3.86286$ & $2.44529$ & $1.16385$ \\
41 & --- & --- & $1.99782$\\
\midrule
no projector & $3.92271$ & $2.17113$ & $1.2513$ \\
\bottomrule
\end{tabular}
\caption{The norm $\|\Pi^{(d)}(A\overbar t - \overbar A t)\|_{\munderbar 2}$ for various projector ranks $d$ and system lengths $N$, in the loop model representation. In this table ``0*'' means a number that is less than $10^{-5}$. I estimate the uncertainty in the nonzero numbers to be about $10^{-4}$.}
\label{loop_table_At}
\end{table}

\begin{table}[H]
\centering
\begin{tabular}{cccc}
\toprule
$d$ & \multicolumn{3}{c}{$N$} \\
\cmidrule(l){2-4}
 & $10$ & 12 & 14 \\
\midrule
1 & 0* & 0* & 0* \\
2 & 0* & 0* & 0* \\
4 & 0* & 0* & 0* \\
5 & 0* & 0* & 0* \\
6 & 0* & 0* & 0* \\
7 & 0* & 0* & 0* \\
9 & 0* & 0* & 0* \\
10 & 0* & 0* & 0* \\
11 & 0* & --- & --- \\
12 & --- & $0.0572678$ & $0.0451944$ \\
13 & $0.0739766$ & $0.0572678$ & --- \\
14 & $0.0739766$ & --- & --- \\
16 & --- & --- & $0.356867$ \\
17 & --- & $0.485697$ & $0.356867$ \\
18 & $0.650097$ & $0.485697$ & $0.356867$ \\
20 & $0.650097$ & $0.485697$ & $0.356867$ \\
21 & $0.650097$ & $0.485697$ & $0.356867$ \\
23 & $0.374967$ & $0.485697$ & $0.213$ \\
25 & --- & $0.278863$ & $0.213$ \\
26 & --- & $0.278863$ & $0.213$ \\
27 & $0.374968$ & --- & $0.213$ \\
29 & $0.374968$ & --- & --- \\
30 & --- & $0.278863$ & --- \\
31 & $0.368193$ & --- & $0.213$ \\
32 & --- & --- & $0.213$ \\
34 & $0.368193$ & $0.278954$ & $0.213$ \\
35 & $0.368194$ & --- & --- \\
36 & $0.368194$ & $0.278954$ & --- \\
37 & --- & --- & $0.240011$ \\
38 & --- & $0.278954$ & --- \\
39 & --- & $0.00225377$ & --- \\
40 & $0.373997$ & $0.00225377$ & $0.240011$ \\
\midrule
no projector & $0.380182$ & $0.302237$ & $0.268341$ \\
\bottomrule
\end{tabular}
\caption{The norm $\|\Pi^{(d)}(A\overbar t - \overbar A t)\|_{\munderbar 2}/\|\Pi^{(d)}A\overbar t\|_2$ for various projector ranks $d$ and system lengths $N$, in the loop model representation. In this table ``0*'' means a number that is less than $10^{-6}$. I estimate the uncertainty in the nonzero numbers to be about $10^{-5}$.}
\label{loop_table_At_norm}
\end{table}

\newpage
\section{Supplemental figures for Chapter \ref{mixing}} \label{mixing_plots}
\begin{figure}[H]
\centering
\includegraphics[width=\textwidth]{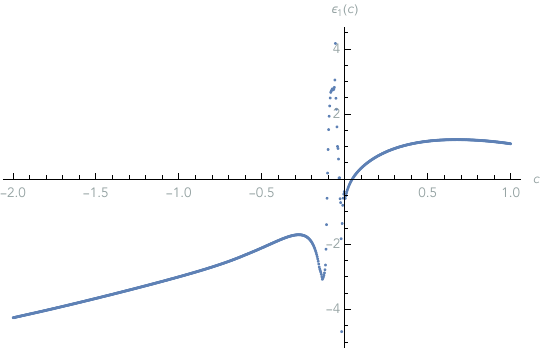}
\caption{Plot of the exponent $\epsilon_1$ as a function of $c$.}
\label{e1}
\end{figure}
\begin{figure}
\centering
\includegraphics[width=\textwidth]{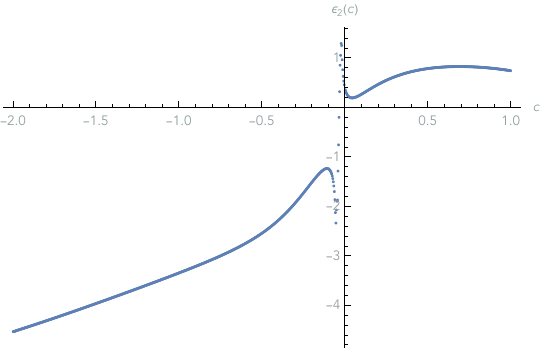}
\caption{Plot of the exponent $\epsilon_2$ as a function of $c$.}
\label{e2}
\end{figure}
\begin{figure}
\centering
\includegraphics[width=\textwidth]{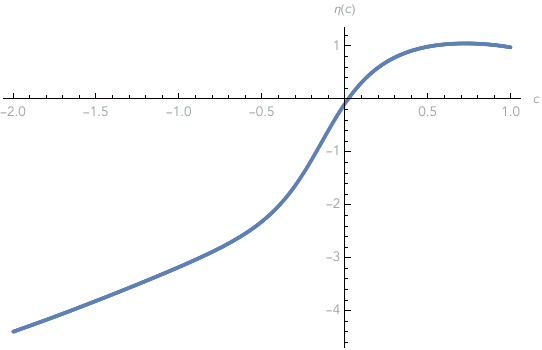}
\caption{Plot of the exponent $\eta$ as a function of $c$.}
\label{eta}
\end{figure}
\begin{figure}
\centering
\includegraphics[width=\textwidth]{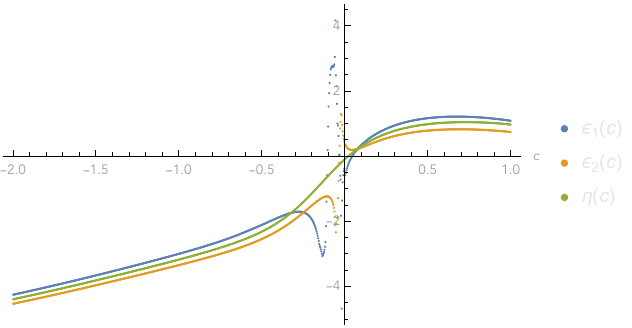}
\caption{Plots of the exponents $\epsilon_1$, $\epsilon_2$, $\eta$ as functions of $c$.}
\label{all}
\end{figure}
\begin{figure}
\centering
\includegraphics[width=\textwidth]{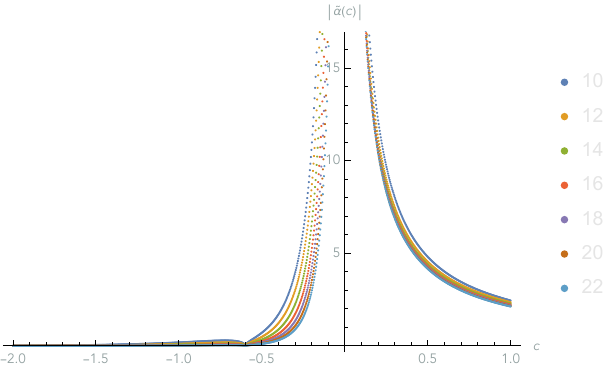}
\caption{}
\label{alpha}
\end{figure}
\begin{figure}
\centering
\includegraphics[width=\textwidth]{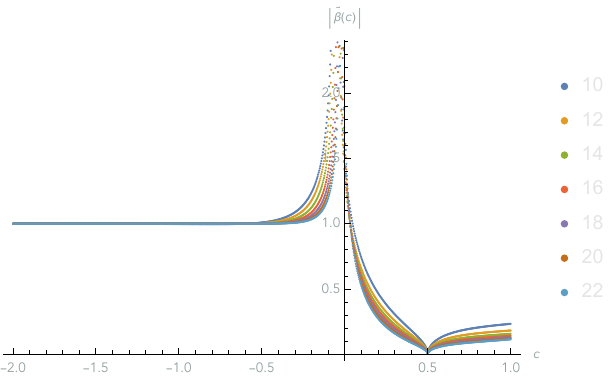}
\caption{}
\label{beta}
\end{figure}
\begin{figure}
\centering
\includegraphics[width=\textwidth]{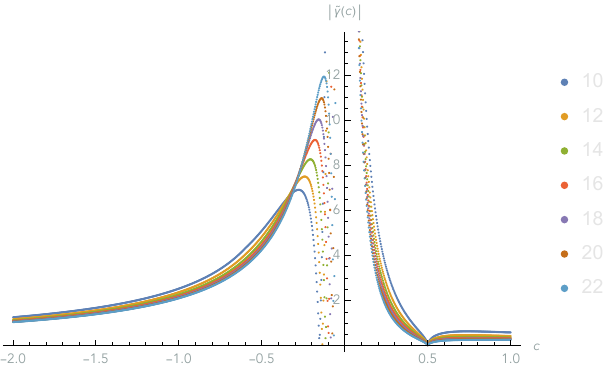}
\caption{}
\label{gamma}
\end{figure}
\begin{figure}
\centering
\includegraphics[width=\textwidth]{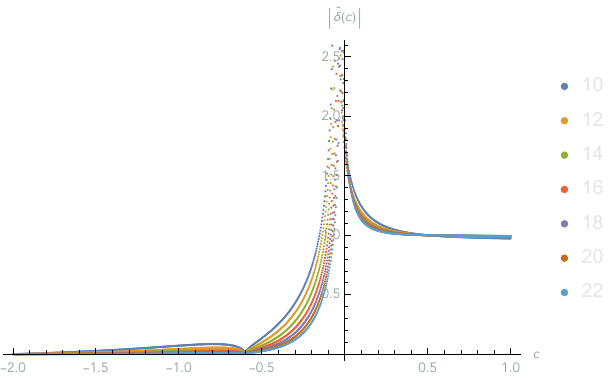}
\caption{}
\label{delta}
\end{figure}
\begin{figure}
\centering
\includegraphics[width=\textwidth]{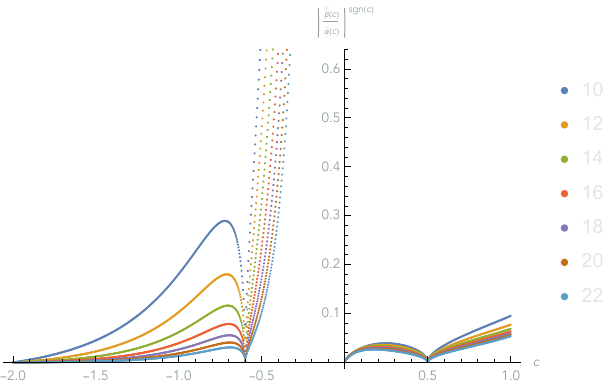}
\caption{}
\label{betaalpha}
\end{figure}
\begin{figure}
\centering
\includegraphics[width=\textwidth]{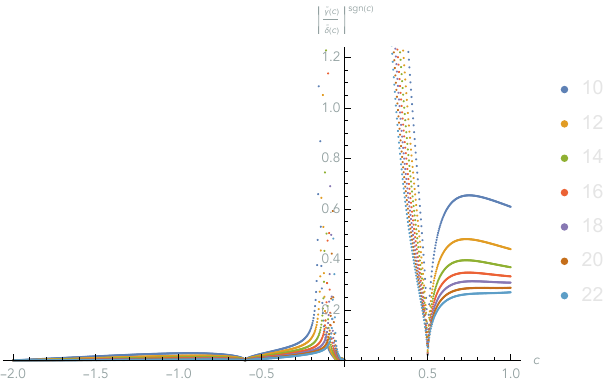}
\caption{}
\label{gammadelta}
\end{figure}
\begin{figure}
\centering
\includegraphics[width=\textwidth]{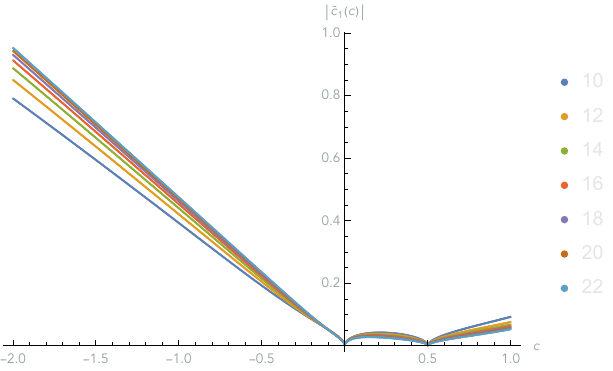}
\caption{}
\label{c1}
\end{figure}
\begin{figure}
\centering
\includegraphics[width=\textwidth]{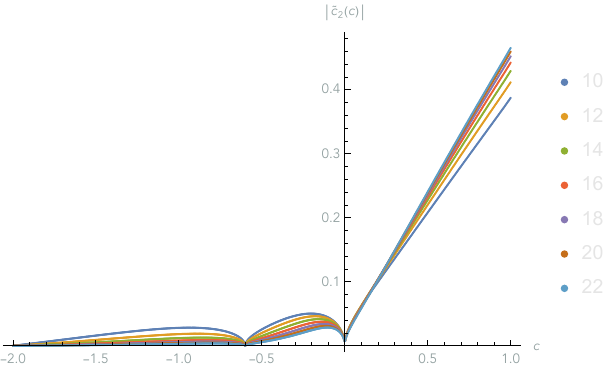}
\caption{}
\label{c2}
\end{figure}
\begin{figure}
\centering
\includegraphics[width=\textwidth]{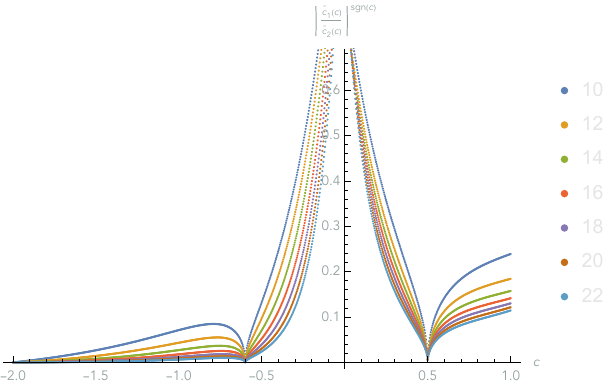}
\caption{}
\label{c1c2}
\end{figure}

\end{document}